\documentclass[a4paper,11pt,oneside]{Thesis}

\usepackage[utf8]{inputenc} 

\usepackage{mathrsfs}
\usepackage{makeidx}

\newcommand \bs{\boldsymbol}
\newcommand \br[1]{\boldsymbol{\mathrm #1}}
\begin{document}
\frontmatter	  % Begin Roman style (i, ii, iii, iv...) page numbering

\title{Selection of a Model of Cerebral Activity for fMRI Group Data
  Analysis}
\authors  {\texorpdfstring
            {\href{merlin.keller@cea.fr}{Merlin Keller}}
            {Merlin Keller}
            }
\addresses  {\groupname\\\deptname\\\univname}  % Do not change this here, instead these must be set in the "Thesis.cls" file, please look through it instead
\date       {\today}
\subject    {}

\maketitle
%% ----------------------------------------------------------------

% \pagestyle{empty}  % No headers or footers for the following pages
% $ $ \eject\pagestyle{fancy}
\setstretch{1.3}  % It is better to have smaller font and larger line spacing than the other way round

% Define the page headers using the FancyHdr package and set up for one-sided printing
\fancyhead{}  % Clears all page headers and footers
\rhead{\thepage}  % Sets the right side header to show the page number
\lhead{}  % Clears the left side page header

\pagestyle{fancy}  % Finally, use the "fancy" page style to implement the FancyHdr headers

%% ----------------------------------------------------------------
% The "Funny Quote Page"
\pagestyle{empty}
$ $ \eject\pagestyle{fancy}
\pagestyle{empty}  % No headers or footers for the following pages

\null\vfill
% Now comes the "Funny Quote", written in italics
% \textit{``Le football est \`a la religion ce que le lexomil est \`a
%   l'h\'ero\"ine pure.''}
% \textit{``Les hémorragies cérébrales sont moins fréquentes chez les joueurs de football. Les cerveaux aussi.''}
% \textit{``Le cerveau, comme le parachute, doit être ouvert pour fonctionner.''}
% \textit{``If the human brain were so simple that we could understand it, we would be so simple that we couldn't.''}
\textit{``As followers of natural science we know nothing of any relation between thoughts and the brain, except as a gross correlation in time and space.''}
\begin{flushright}
% Les Cahiers du Football
% Pierre Desproges
% Pierre Daninos
% Emerson M. Pugh
Sir Charles Scott Sherrington
\end{flushright}

\vfill\vfill\vfill\vfill\vfill\vfill\null
% Funny Quote page ended, start a new page
%% ----------------------------------------------------------------
\eject

% The Abstract Page
\pagestyle{empty}
$ $ \eject\pagestyle{fancy}
\pagestyle{fancy}

\addtotoc{R\'esum\'e}  % Add the "Resume" page entry to the Contents
% \abstract
{
\begin{center}
\huge Résumé
\end{center}
\addtocontents{toc}{\vspace{0em}}  % Add a gap in the Contents, for aesthetics

% The Thesis Abstract is written here (and usually kept to just this
% page). The page is kept centered vertically so can expand into the
% blank space above the title too\ldots

L’imagerie par résonance magnétique fonctionnelle (IRMf) permet d’acquérir des images tridimensionnelles de l’activité cérébrale d’un sujet soumis à une séquence de stimulations sensorielles. L’analyse statistique des ces données permet de détecter les aires cérébrales actives en réponse aux différentes stimulations.

Lorsque plusieurs sujets ont été recrutés pour une expérience, l’analyse de groupe consiste à généraliser les résultats individuels à la population d’intérêt dont sont issus les sujets. La variabilité morphologique du cerveau humain rend cependant la comparaison des images acquises sur les différents sujets problématique

L’approche usuelle pour contrer cette difficulté consiste à recaler les sujets dans un référentiel commun, puis de comparer les cerveau séparément en chaque point de ce référentiel. Cette étape de recalage n’étant jamais parfaite, il en résulte une incertitude sur la localisation spatiale de chaque sujet.

Nous proposons dans un premier temps d’étendre le modèle classique d’analyse de groupe afin de prendre en compte cette incertitude spatiale. Dans un deuxième temps, nous développons à partir de ce modèle une nouvelle approche de détection d’aires cérébrales actives, basée sur des régions d’intérêt prédéfinies plutôt que sur les procédures de seuillage couramment utilisées.

}
\eject
% The Abstract Page
\pagestyle{empty}
$ $ \eject\pagestyle{fancy}
\pagestyle{fancy}

\addtotoc{Abstract}  % Add the "Abstract" page entry to the Contents
\abstract{
\addtocontents{toc}{\vspace{0em}}  % Add a gap in the Contents, for aesthetics

% The Thesis Abstract is written here (and usually kept to just this
% page). The page is kept centered vertically so can expand into the
% blank space above the title too\ldots

This thesis is dedicated to the statistical analysis of multi-subject fMRI data, with the purpose of identifying bain structures involved in certain cognitive or sensori-motor tasks, in a reproducible way across subjects. To overcome certain limitations of standard voxel-based testing methods, as implemented in the Statistical Parametric Mapping (SPM) software, we introduce a Bayesian model selection approach to this problem, meaning that the most probable model of cerebral activity given the data is selected from a pre-defined collection of possible models.

Based on a parcellation of the brain volume into functionally homogeneous regions, each model corresponds to a partition of the regions into those involved in the task under study and those inactive. This allows to incorporate prior information, and avoids the dependence of the SPM-like approach on an arbitrary threshold, called the cluster-forming threshold, to define active regions. By controlling a Bayesian risk, our approach balances false positive and false negative risk control. Furthermore, it is based on a generative model that accounts for the spatial uncertainty on the localization of individual effects, due to spatial normalization errors.

On both simulated and real fMRI datasets, we show that this new paradigm corrects several biases of the SPM-like approach, which either swells or misses the different active regions, depending on the choice of a cluster-forming threshold.

}

% Abstract ended, start a new page
%% ----------------------------------------------------------------

% \setstretch{1.3}  % Reset the line-spacing to 1.3 for body text (if it has changed)
% 
% % The Acknowledgements page, for thanking everyone
\eject
$ $ \eject
\acknowledgements{
\addtocontents{toc}{\vspace{0em}}  % Add a gap in the Contents, for aesthetics
% 

% \input{Acknowledgements}
% Si une th\`ese est l'aboutissement d'un long travail de r\'eflexion personnelle, elle est aussi, et avant toute chose, le fruit de rencontres et de collaborations multiples. Rendre justice \`a toutes les personnes ayant contribu\'e, directement ou indirectement, au pr\'esent travail est un plaisir autant qu'un exercice p\'erilleux; je prie donc par avance ceux que je ne manquerai pas d'oublier ici de bien vouloir m'en excuser.
% 
% Ma gratitude va en premier lieu \`a tous ceux qui m'ont permis, apr\`es six ann\'ees pass\'ees \`a enseigner les math\'ematiques dans les coll\`eges et lyc\'ees de la r\'egion parisienne, de revenir sur les bancs de la fac et de me lancer dans cette aventure. Tout d'abord, je remercie Pascal Massart de m'avoir chaleureusement accueilli au sein du master de Probabilit\'es et Statistiques d'Orsay, et de m'avoir aid\'e par ses nombreux conseils avis\'es \`a n\'egocier cette p\'eriode de transition.
% 
% Listes des gens \`a remercier:
% 
% - Les enseignants du master, et plus particuli\`erement Gilles Celeux, St\'ephane  Robin, Liliane Bel, C\'ecile Durot et Avner Bar-Hen
% - Edouard Duchesnay, qui m'a fait d\'ecouvrir le premier le monde de la neuroimagerie
% - Alexis Roche pour son encadrement. Explorer les directions toujours originales qu'il me proposait dans mes recherches fut un r\'eel plaisir. Je lui suis aussi redevable de la d\'ecouverte du monde merveilleux de Python, du d\'eveloppement collaboratif et du contr\^ole de version (longue vie \`a Nipy!)
% - Marc Lavielle qui a dirig\'e ce travail, 

Mes remerciements vont en premier lieu \`a Alexis Roche pour la disponibilit\'e et la patience dont il a fait preuve dans son encadrement tout au long de ma th\`ese, et pour sa capacit\'e \`a me faire reprendre de la hauteur vis-\`a-vis de mon sujet, les nombreuses fois o\`u je me perdais dans des d\'etails. Merci \'egalement de m'avoir fait d\'ecouvrir le monde merveilleux de Python et du d\'eveloppement collaboratif; longue vie \`a NiPy!

De la m\^eme fa\c con, je remercie Marc Lavielle d'avoir accept\'e de diriger mon travail, avec une rigueur et un soucis de clart\'e qui ont \'et\'e des guides pr\'ecieux dans l'\'elaboration de ma d\'emarche. Merci \'egalement pour les innombrables solutions apportées aux divers probl\`emes méthodologiques que j'ai rencontrés, et qui m'ont permis d'approfondir ma connaissance des algorithmes d'optimisation stochastique.

Merci \`a Elisabeth Gassiat d'avoir accept\'e de pr\'esider mon jury de th\`ese. Un grand merci \`a Mark Woolrich pour sa relecture d\'etaill\'ee du manuscript, ses remarques pertinentes, ainsi que pour avoir affront\'e la  neige et les pannes de train pour traverser la Manche et venir participer \`a ma soutenance!

Mon s\'ejour au laboratoire de neuroimagerie assist\'e par ordinateur (LNAO) fut l'occasion de nouer de nombreuses collaborations, sans lesquelles ce travail n'aurait pas vu le jour. Ma gratitude va en premier lieu \`a Jean-Fran\c cois Mangin pour son accueil au sein de son \'equipe, et son soucis d'assurer les meilleures conditions de travail possibles aux membres de son équipe.

Merci \`a Sébastien, mon prédécesseur, dont le travail a constitué le socle sur lequel je me suis appuyé dans mes recherches. Merci également pour son soutien lors de mon stage de master, et grâce auquel j'ai pu faire mes premières armes dans l'utilisation de SPM et de la fameuse Distance Toolbox!

Merci \`a Bertrand Thirion, pour l'int\'er\^et dont il a fait preuve vis-\`a-vis de mon travail, et dont le point de vue toujours pertinent fut une r\'elle source d'inspiration. Merci aussi pour les barbecues et autres raclettes si conviviaux qui ont ponctu\'es ces trois ann\'ees!

Merci \`a Philippe Ciuciu, grand Bay\'esien et marathonien devant l'Eternel, pour son abondante expertise sur les techniques d'\'echantillonnage stochastique, ainsi que pour la d\'ecouverte en course \`a pied des plus beaux recoins du plateau de Saclay...

Ma gratitude va \'egalement \`a Edouard Duchesnay, par qui toute cette aventure a commenc\'ee, qui m'a initi\'e aux myst\`eres de la neuroimagerie, et donn\'e envie d'en conna\^itre plus. Merci \`a Jean-Baptiste Poline pour son soutien constant et le point de vue toujours pertinent et constructif qu'il a apporté sur mon travail. Merci \'egalement aux h\'eros du service informatique, Dimitri Papadopoulos et Pascal Stokowski, sur qui reposaient la lourde charge d'assurer la continuit\'e d'un r\'eseau informatique capricieux...

Je tiens \'egalement \`a remercier la fine \'equipe des th\'esards, post-docs et assimil\'es dont j'ai eu le plaisir de faire partie, pour l'esprit de stimulation et d'entraide mutuelle, ainsi que pour les nombreuses sorties et soir\'ees
qui permettaient de d\'ecompresser et repartir de plus belle, bravo en particulier aux remarquables organisatrices que furent C\'ecilia et Valdis!

Une d\'edicace sp\'eciale aux `geeks' de la bande, Thomas, Alan, Benjamin, Gaël, Matthieu et les autres, pour les innombrables heures pass\'ees \`a combler patiemment mes lacunes en informatique! Merci aussi \`a Dominique, Denis et Yann d'avoir si souvent r\'epondu \`a mes appels d\'esesp\'er\'es lorsque je n'arrivais pas \`a utiliser les outils maison (suite le plus souvent \`a l'oubli d'une virgule dans une ligne de commande!)

Reprendre des \'etudes apr\`es cinq ann\'ees d'interruption ne s'est pas fait sans difficult\'es, et je suis \'enorm\'ement redevable \`a toute l'\'equipe du master de statistiques d'Orsay, qui m'a permis d'effectuer cette reprise dans les meilleures conditions.

Un grand merci tout d'abord \`a Pascal Massart, qui m'a ouvert les portes du master, puis m'a si bien conseill\'e tout au long de cette transition. Merci aux enseignants pour leur clart\'e et leur p\'edagogie, et plus particuli\`erement \`a Liliane Bel et Gilles Celeux dont la bienveillante attention m'a permis de prendre confiance dans mes capacit\'es \`a faire de la recherche.

Enfin, mes pens\'ees vont \`a ma famille dont le soutien sans faille a \'et\'e indispensable \`a l'accomplissement de ce travail, et vers qui j'ai toujours pu me tourner dans les in\'evitables moments de doutes qui accompagnent tout chercheur, particuli\`erement \`a ses d\'ebuts! Merci tout spécialement \`a Astrig qui fut en premi\`ere ligne durant ces ann\'ees, et qui a notamment accepté que j'emmène mon ordinateur portable en vacances pour rattraper le travail en retard\ldots

Je conclue en dédiant cette thèse à mon neveu Stanley, dont la curiosité débordante, l'imagination foisonnante et l'envie toujours renouvelée de se dépasser ne manquent jamais de me surprendre, et laissent augurer à mon avis d'une brilliante carrière d'inventeur génial!

% The acknowledgements and the people to thank go here, don't forget to include your project advisor\ldots
% 
}
\cleardoublepage  % End of the Acknowledgements
%% ----------------------------------------------------------------

\pagestyle{fancy}  %The page style headers have been "empty" all this time, now use the "fancy" headers as defined before to bring them back

%% ----------------------------------------------------------------
\pagestyle{empty}
% $ $ \eject
\pagestyle{fancy}
\pagestyle{fancy}
\lhead{\emph{Contents}}  % Set the left side page header to "Contents"
\tableofcontents  % Write out the Table of Contents

%% ----------------------------------------------------------------
% \lhead{\emph{List of Figures}}  % Set the left side page header to "List if Figures"
% \listoffigures  % Write out the List of Figures

%% ----------------------------------------------------------------
% \lhead{\emph{List of Tables}}  % Set the left side page header to "List of Tables"
% \listoftables  % Write out the List of Tables

%% ----------------------------------------------------------------
\setstretch{1.5}  % Set the line spacing to 1.5, this makes the following tables easier to read
\eject
\pagestyle{empty}
$ $ \eject\pagestyle{fancy}
\pagestyle{fancy}
\lhead{\emph{Abbreviations}}  % Set the left side page header to "Abbreviations"
\listofsymbols{ll}  % Include a list of Abbreviations (a table of two columns)
{
% \textbf{Acronym} & \textbf{W}hat (it) \textbf{S}tands \textbf{F}or \\
% \textbf{LAH} & \textbf{L}ist \textbf{A}bbreviations \textbf{H}ere \\
\vspace{-1cm}
\textbf{AAL} & \textbf{A}utomated \textbf{A}tlas \textbf{L}abels \\\vspace{-1cm}
\textbf{BOLD} & \textbf{B}lood \textbf{O}xygen \textbf{L}evel \textbf{D}ependent  \\\vspace{-1cm}
\textbf{CSA} & \textbf{C}ortical \textbf{S}ulci \textbf{A}tlas\\\vspace{-1cm}
\textbf{fMRI} & \textbf{f}unctional \textbf{M}agnetic \textbf{R}esonance \textbf{I}maging  \\\vspace{-1cm}
\textbf{FDR} & \textbf{F}alse \textbf{D}iscovery \textbf{R}ate \\\vspace{-1cm}
\textbf{FFX} & \textbf{F}ixed \textbf{FX} {\em (effect)} \\\vspace{-1cm}
\textbf{FPR} & \textbf{F}alse \textbf{P}ositive \textbf{R}ate \\\vspace{-1cm}
\textbf{FSL} & \textbf{F}MRIB \textbf{S}oftware \textbf{L}ibrary \\\vspace{-1cm}
\textbf{FWER} & \textbf{F}amily-\textbf{W}ise \textbf{E}rror \textbf{R}ate \\\vspace{-1cm}
\textbf{GLM} & \textbf{G}eneral \textbf{L}inear \textbf{M}odel\\\vspace{-1cm}
\textbf{HRF} & \textbf{H}aemodynamic \textbf{R}esponse \textbf{F}unction \\\vspace{-1cm}
\textbf{MCMC} & \textbf{M}onte \textbf{C}arlo \textbf{M}arkov \textbf{C}hain\\\vspace{-1cm}
\textbf{MCP} & \textbf{M}ultiple \textbf{C}omparison \textbf{P}rocedure \\\vspace{-1cm}
\textbf{MFX} & \textbf{M}ixed \textbf{FX} {\em (effect)} \\\vspace{-1cm}
\textbf{MH} & \textbf{M}etropolis--\textbf{H}astings\\\vspace{-1cm}
\textbf{NRL} & \textbf{N}eural \textbf{R}esponse \textbf{L}evel\\\vspace{-1cm}
\textbf{ROI} & \textbf{R}egion \textbf{O}f \textbf{I}nterest \\\vspace{-1cm}
\textbf{RFX} & \textbf{R}andom \textbf{FX} {\em (effect)} \\\vspace{-1cm}
\textbf{SAEM} & \textbf{S}tochastic \textbf{A}veraging \textbf{E}xpectation \textbf{M}aximization\\
\textbf{SPM} & \textbf{S}tatistical \textbf{P}arametric \textbf{M}apping \\\vspace{-1cm}
% \textbf{} & \textbf{}\\
% \textbf{} & \textbf{}\\
}
\eject
\pagestyle{empty}
$ $ \eject\pagestyle{fancy}

\pagestyle{fancy}
\notations{
% \addtocontents{toc}  
\addtocontents{toc}{\vspace{0em}}  % Add a gap in the Contents, for aesthetics

% \input{Notations}
% \begin{tabular}{p{0.3\textwidth}p{0.6\textwidth}}
% $\mathcal V = (\br v_1, \ldots, \br v_p) \subset \mathbb R^3$
% &
% Search volume (voxel grid) \\
% $\mathcal V = \mathcal V_1 \cup \ldots \cup \mathcal V_N$
% &
% Priot partition into $N$ regions of interest\\
% $\ell \in \{1, \ldots, N\}^p$
% &
% Region labels ($\forall k,\, \br v_k \in \mathcal V_{\ell_k}$)\\
% $\bs y_i = (y_{i,1}, \ldots, y_{ip})$
% &
% Subject $i$ estimated effects map (observations), \newline
% $i = 1, \ldots, n.$ We also use the notation $\bs y_i(\br v_k) := y_{i,k}$\\
% $\bs s_i^2 = (s_{i,1}^2, \ldots, s_{ip}^2)$
% &
% Subject $i$ estimation variance map (known)\\
% $\bs x_i = (x_{i,1}, \ldots, x_{ip})$
% &
% Subject $i$  effects map (latent variable)\\
% $\br u_i = (\br u_{i,1}, \ldots, \br u_{ip})$
% &
% Subject $i$ registration errors\\
% $\br w_i = (\br w_{i,1}, \ldots, \br w_{i,B})$
% &
% Subject $i$ elementary displacements\\
% $\br v_{k_1}, \ldots, \br v_{k_B}$
% &
% Deformation field control points\\
% $\mathcal K(\, \cdot\, ,\, \cdot\, )$
% &
% Interpolation kernel\\
% $\omega$
% &
% Deformation field regularity\\
% $\sigma_S^2$
% &
% Elementary displacements variance\\
% $\bs \mu = (\mu_1, \ldots, \mu_p)$
% &
% Mean group effect map\\
% $\bs \eta = (\eta_1, \ldots, \eta_N)$
% &
% Regional means\\
% $\bs \nu^2 = (\nu_1^2, \ldots, \nu_N^2)$
% &
% Regional variances\\
% $\bs \sigma^2 = (\sigma_1^2, \ldots, \sigma_N^2)$
% &
% Between-subject variance (one per region)\\
% $\alpha, \beta, \lambda$
% &
% Hyperparemeters of Normal-Inverse Gamma prior\\
% \end{tabular}

% \chapter{Notations}
% \lhead{\emph{Notations}}

\centering
\begin{tabular}{p{0.3\textwidth}p{0.6\textwidth}}
% \hline
$\mathcal V = (\br v_1, \ldots, \br v_d) \subset \mathbb N^3$ \newline
$\mathcal V = \mathcal V_1 \cup \ldots \cup \mathcal V_N$ \newline
$\ell \in \{1, \ldots, N\}^d$ \newline
$\br y_i = (y_{i,1}, \ldots, y_{i,d})$ \newline
 \newline
$\br s_i^2 = (s_{i,1}^2, \ldots, s_{i,d}^2)$ \newline
$\br x_i = (x_{i,1}, \ldots, x_{i,d})$ \newline
$\br u_i = (\br u_{i,1}, \ldots, \br u_{i,d})$ \newline
$\varphi_i$ \newline
$\br w_i = (\br w_{i,1}, \ldots, \br w_{i,B})$ \newline
$\br v_{k_1}, \ldots, \br v_{k_B}$ \newline
$\mathcal K(\, \cdot\, ,\, \cdot\, )$ \newline
$\omega$ \newline
$\sigma_S^2$ \newline
$\bs \mu = (\mu_1, \ldots, \mu_d)$ \newline
$\bs \eta = (\eta_1, \ldots, \eta_N)$ \newline
$\bs \nu^2 = (\nu_1^2, \ldots, \nu_N^2)$ \newline
$\bs \gamma \in \{0, 1\}^N$ \newline
$\br p = (p_1, \ldots, p_N)$ \newline
$\bs \sigma^2 = (\sigma_1^2, \ldots, \sigma_N^2)$ \newline
$\alpha, \beta, \lambda$
&
Search volume (voxel grid) \newline
Partition into $N$ regions of interest\newline
Region labels ($\forall k,\, \br v_k \in \mathcal V_{\ell_k}$)\newline
Subject $i$ estimated effects map (observations), \newline
$1 \leq i \leq n.$ We also use the notation $\bs {\mathrm y}_i(\br v_k) := y_{i,k}$\newline
Subject $i$ estimation variance map (known)\newline
Subject $i$  effects map (latent variable)\newline
Subject $i$ registration errors\newline
Voxel $k$ is displaced to voxel $\varphi_i(k)$ for subject~$i$ \newline
Subject $i$ elementary displacements\newline
Deformation field control points\newline
Interpolation kernel\newline
Deformation field regularity\newline
Elementary displacements variance\newline
Mean group effect map\newline
Regional means\newline
Regional variances\newline
Indicator variable ($\gamma_j = 1$ iff $\eta_j \neq 0$)\newline
Prior probabilities \newline
Between-subject variance (one per region)\newline
Hyperparemeters of the Normal-Inverse Gamma prior\\
% \hline
\end{tabular}
% \caption{List of notations\label{tab:notations}}

}
 % End of the Acknowledgements
%% ----------------------------------------------------------------
% \cleardoublepage  % Start a new page
% \lhead{\emph{Physical Constants}}  % Set the left side page header to "Physical Constants"
% \listofconstants{lrcl}  % Include a list of Physical Constants (a four column table)
% {
% % Constant Name & Symbol & = & Constant Value (with units) \\
% Speed of Light & $c$ & $=$ & $2.997\ 924\ 58\times10^{8}\ \mbox{ms}^{-\mbox{s}}$ (exact)\\
% 
% }

%% ----------------------------------------------------------------
% \cleardoublepage  %Start a new page
% \lhead{\emph{Symbols}}  % Set the left side page header to "Symbols"
% \listofnomenclature{lll}  % Include a list of Symbols (a three column table)
% {
% % symbol & name & unit \\
% $a$ & distance & m \\
% $P$ & power & W (Js$^{-1}$) \\
% & & \\ % Gap to separate the Roman symbols from the Greek
% $\omega$ & angular frequency & rads$^{-1}$ \\
% }
%% ----------------------------------------------------------------
% End of the pre-able, contents and lists of things
% Begin the Dedication page

% \setstretch{1.3}  % Return the line spacing back to 1.3
% 
% \pagestyle{empty}  % Page style needs to be empty for this page
% \dedicatory{For/Dedicated to/To my\ldots}

\addtocontents{toc}{\vspace{2em}}  % Add a gap in the Contents, for aesthetics

\eject
\pagestyle{empty}
$ $ \eject\pagestyle{fancy}
\pagestyle{fancy}
%% ----------------------------------------------------------------
\mainmatter	  % Begin normal, numeric (1,2,3...) page numbering
\pagestyle{fancy}  % Return the page headers back to the "fancy" style

% Include the chapters of the thesis, as separate files
% Just uncomment the lines as you write the chapters

% Introduction

% \input{Chapter1}
\chapter{Introduction}
\lhead{\emph{Introduction}}

\section{Context and objectives}

\begin{figure}[ht!]
\includegraphics[width=\textwidth]{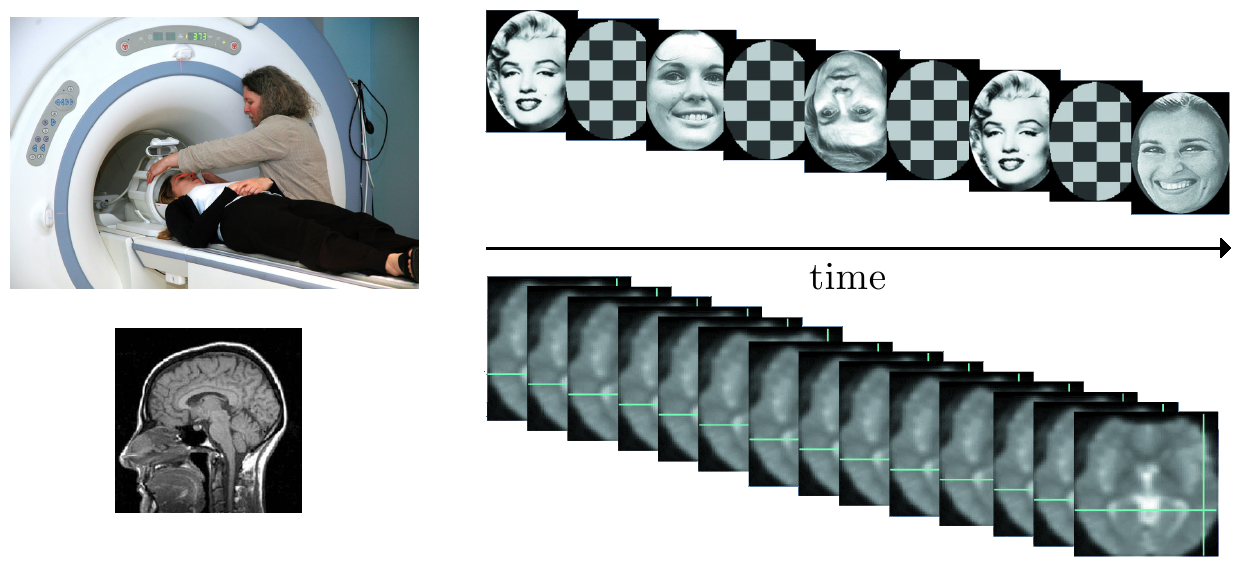}
\caption{\label{fig:fMRI_experiment} A typical fMRI experiment. The subject, lying in the MRI scanner (upper left) is submitted to a series of stimuli (upper right), visual, auditory or other, while a series of 3D images of his/her brain activity is acquired (lower right). A structural image of the subject's brain (lower left) is also acquired, using the same scanner.}
\end{figure}

This thesis is dedicated to the analysis of multi-subject fMRI data. Functional magnetic resonance imaging (fMRI) is a modality of {\em in vivo} brain imaging that allows to measure the variations of cerebral blood oxygen levels induced by the neural activity of a subject lying inside a MRI scanner and submitted to a series of stimuli (see Figure~\ref{fig:fMRI_experiment}). One of the main goals of fMRI group studies, through the analysis of the data acquired on a cohort of subjects, is to identify brain structures involved in certain cognitive or sensori-motor tasks, in a reproducible way across subjects.

Many sources of variability are present in fMRI datasets, making statistical methods necessary to perform such analyses. These include, but are not limited to, the movements of the subject inside the machine, his or her level of attention during the experiment, various physiological signals, such as heartbeats and breathings, etc. Furthermore, the BOLD response itself varies across the different runs of a single experiment, due to habituation and attentional effects. For all these reasons, the measures acquired on each subject are uncertain, and variable should the experiment be reproduced with the same subject. Additionally, each subject reacts differently when exposed to the same stimuli, so a certain variability in the brain activation pattern is also expected across the subjects.

Accounting for these two types of variability, classically referred to as {\em within}-subject and {\em between}-subject, is a well-known problem in the literature on statistical inference for biological data \cite{McCulloch01}. Another challenge, specific to medical imaging, and in particular to neuroimaging, is the important variability of the human brain anatomy, which makes the comparison of brain images from different subjects far from straightforward. 
% This may be interpreted as an additional source of between-subject variability, though historically this has not been the most adopted point of view, as explained below. 
To address this, individual images are most often normalized to a common brain template using nonrigid registration. However, registration is prone to errors (even assuming the existence of point-to-point correspondences between different brains), hence it does not seem reasonable to assume that homologous points are aligned across subjects.

As a consequence of all these fluctuations, the activation pattern inferred from any given fMRI dataset can only be determined with a limited amount of confidence. Ideally, the statistical procedure used to analyse the data should therefore account for the different sources of variability, and assess the uncertainty they induce on the estimated activation pattern.

The purpose of this work is to propose such a statistical analysis framework. This is still an open problem, though the development of appropriate methods for fMRI group inference has been an active field of research for at least two decades. The most widely used approach to address these questions is the co-called mass-univariate, or voxel-based detection, popularized by \cite{Friston97}. It starts with normalizing individual images onto a common brain template, as mentioned above. Next, a $t$-statistic is computed in each voxel to locally assess mean group effects. The candidate regions are then defined as the connected components (clusters) of the resulting statistical map above an arbitrary threshold called the cluster-forming threshold. Only the clusters whose sizes exceed a critical value are reported. This critical size acts as a second threshold, and is generally tuned to control the probability of one or more clusters being detected by chance. For scientific reporting purposes, the detected clusters may finally be related to known anatomical regions based on expert knowledge, or using a digital brain atlas such as the Automated Atlas Label (AAL) \cite{Tzourio-Mazoyer02}.

Though simple and widely applicable, this approach suffers from several shortcomings. One of these is the dependence on an arbitrary cluster-forming threshold. High values of this threshold may result in missing active regions, and low values in merging functionally distinct regions, yielding poor localization power, due to the fact that active voxels cannot be localized within each detected cluster \cite{Hayasaka03}. Furthermore, the absence of activations outside the detected clusters cannot be assessed. Consequently, there is no guarantee that the whole functional network can be recovered. Finally, due to unavoidable intersubject reg\-istration errors, the observed activations are not well-localized, and possibly displaced across distinct functional regions, which may result in blurring the group activation map, and create non-handled false positives.

To date, these limitations of the mass-univariate approach have been tackled separately. This thesis introduces a new approach for fMRI group data analysis that addresses them jointly.

\section{Organization and main contributions}

The remaining of this thesis is organized in six chapters, which we now summarize:

\begin{itemize}

\item In Chapter~\ref{chap:group}, we give a review of current approaches for fMRI group data analysis, with a focus on mass-univariate detection. This was popularized by the Statistical Parametric Mapping (SPM) software package, and we refer to it as SPM-like in the following. Our goal here is not to give a complete list of existing methods, but rather summarize the main directions of research that have been followed up to now. This chapter also contains several contributions we have made to the SPM-like approach. \cite{Keller07,Keller08b} present an implementation of the Gaussian two-level group model using the maximum likelihood ratio test, a nonparametric generalization of which is developed in \cite{Roche07}. We have also participated to another paper \cite{Thirion06a} describing a high-level group analysis approach which models the spatial variability of the activation patterns, and which we have included in this review.

\item In Chapter~\ref{chap:unsupervised}, we propose to overcome the dependence of the SPM-like approach on an arbitrary detection threshold, and its exclusive control over false positives, by using the random threshold approach developed in \cite{Lavielle07}. One of the key ideas of this chapter, which we re-exploit in the following, is to re-interpret the task of detecting activations, traditionally seen from a multiple testing point of view, as a model selection problem. This work has been submitted to the Canadian Journal of Statistics.

\item Chapter~\ref{chap:modeling} proposes a detection method that relaxes the assumption of perfect match. To do this, we extend the classical mass-univariate model by incorporating a set of latent variables representing registration errors, and model them as random deformation fields. 
% In contrast, previous works dealing with spatial uncertainty in fMRI data model where feature-based, meaning that they modeled the spatial variability of high-level features extracted from the individual images. Our work is therefore the first to model spatial uncertainty at the voxel-level. 
Our main results consist in demonstrating the stretching effect of the group estimated activation pattern due to neglecting spatial uncertainty, and its compensation through our approach. A first version of this work was published in \cite{Keller08}.

\item Chapter~\ref{chap:bayesian} introduces a new paradigm for fMRI group data analysis that addresses jointly some key limitations of the SPM-like approach, and was published in \cite{Keller09}. Based on a Bayesian model selection framework, regions involved in the task under study are selected according to the posterior probabilities of a nonzero mean activation, given a pre-defined parcellation of the search volume into functionally homogeneous regions. Thus our approach is threshold-free, while allowing to incorporate prior information, provided that the parcellation is sensible. By controlling a Bayesian risk, our approach balances false positive and false negative risks, with weights than can be tuned depending on the application domain. Importantly, it is based on the same spatial uncertainty model as in Chapter~\ref{chap:modeling}, and thus accounts for the mis-alignment of individual images, due to inevitable registration errors. Hence for each subject, the membership of a voxel to a given region is probabilistic rather than deterministic.
% Based on a pre-defined parcellation of the search volume into functionally homogeneous regions, we select the most probable functional network involved in the task at hand, each functional network corresponding to a partition of the pre-defined regions into involved and inactive, and defining a generative model for the data. This methods adopts a Bayesian framework that controls both false positive and false negative risks. Moreover, registration errors are accounted for, based on the spatial uncertainty model introduced in Chapter~\ref{chap:modeling}. Finally, the use of a pre-defined parcellation allows to incorporate prior information, instead of relying on an arbitrary cluster-defining threshold. 

Results on simulated data show that neglecting spatial uncertainty may result in a bias toward false positives when selecting active regions. This is a consequence of the stretching effect evidenced in Chapter~\ref{chap:modeling}. We also show that this bias can be compensated when modeling spatial uncertainty.

\item In Chapter~\ref{chap:case} we validate our model selection approach on a real fMRI dataset, by successfully recovering the whole network of regions involved in basic number and language processing, in accordance with previous works. The purpose of this chapter is also to illustrate the limitations of the SPM-like approach, which we apply to the same dataset, and the benefits of modeling spatial uncertainty.

\item We conclude in Chapter~\ref{chap:conclusion} by a summary of the main results and discuss perspectives for future work.

Finally, all the methods developed throughout this work where implemented in Python, and are now part of the NIPY open-source neuroimaging analysis package \href{http://neuroimaging.scipy.org/site/index.html}{http://neuroimaging.scipy.org/site/index.html}. Methods concerning the SPM-like approach have also been integrated into the fMRI toolbox of the BrainVisa software, developed at CEA/Neurospin \href{http://www.brainvisa.info/index.html}{http://www.brainvisa.info/index.html}.

\end{itemize}

% Statistical Analysis of fMRI Group Data
\eject
\pagestyle{empty}
$ $ \eject\pagestyle{fancy}
\pagestyle{fancy}

\chapter{fMRI group data analysis: the `SPM-like' approach}\label{chap:group}
\lhead{\emph{fMRI Group Data Analysis: the `SPM-like' Approach}}

\section{Introduction}

The main goal of this chapter is to give a detailed account of the
current approaches to fMRI group data analysis, with a focus on the
approach primarily developed in \cite{Friston97} and popularized by the
Statistical Parametric Mapping (SPM) software, which we will hence refer
to in the following as the `SPM-like' approach. It is the most widely
used method to date, and has served both as the basis for our
research, and the reference to validate our methods.

This approach comprises two main steps. First, each subject's data is processed separately, resulting in a series of summary statistics, such as a statistical map of the subject's brain activity in response to any given contrast of experimental conditions. This step is described in Section~\ref{sec:single}, along with some basic facts on the nature of fMRI data.

Next, the subjects' summary statistics are used as input data for group level analysis. In Section~\ref{sec:mass} we describe how this is accomplished in the `SPM-like' approach, and discuss its advantages and limitations.

Many alternatives have been proposed to overcome the limitations of the `SPM-like' approach. We give an overview of the current literature on this subject in Section~\ref{sec:alternative}, then in Section~\ref{sec:conclusion1} discuss the directions we have decided to follow in our work, and how these relate to the existing approaches.

% \item Provide the necessary background on functional magnetic resonance imaging (fMRI) to understand the
% challenges posed by the analysis of multi-subject fMRI datasets.
% 
% 
% \item Motivate the work presented in the following chapters. Our goal has been to overcome some limitations
% of the `SPM-like' approach, specifically its dependence on an arbitrary threshold to define potentially 
% active brain regions, the exclusive control on false positives ({\em i.e.} the absence of guarantee for 
% not missing an activated region), and the uncertainty on the spatial localization of individual signals, 
% due to registration errors.
% 
% \end{itemize}
% 
% This chapter is organized as follows:
% 
% 

\section{Single-Subject Data Analysis}\label{sec:single}

\begin{figure}
\begin{center}
\includegraphics[width=0.9\textwidth]{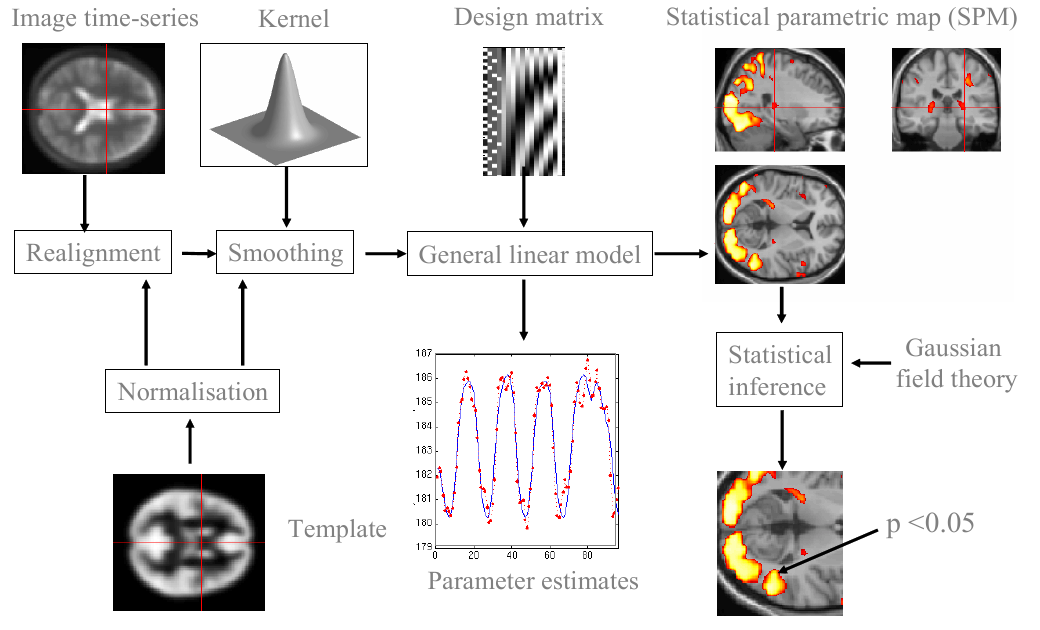}
\end{center}
\caption{\label{fig:individual_data_processing} Pipeline of single-subject data analysis using the SPM-like approach, from \cite{Friston95}. The data is realigned, normalized to a brain template representing a standard space, and smoothed. These pre-processings are described in Sections~\ref{sec:BOLD} and \ref{sec:normalisation}. Next, the time-series are modeled separately in each voxel using the general linear model, whose parameters represent the amplitude of the BOLD response evoked by each type of stimuli (see Section~\ref{sec:glm}). Active brain areas are detected by thresholding a map of test statistics, as explained in Section~\ref{sec:statistical}.}
\end{figure}

\subsection{The blood oxygen level dependent effect}\label{sec:BOLD}

% La fMRI est apparue dans les annees 70!
Since the early nineties, functional magnetic resonance imaging (fMRI) has become one of the principal tools to investigate the functional organization of the human brain. This popularity is due to both the relatively good spatial resolution and low invasiveness of fMRI compared to other functional imaging modalities. For instance, positron-emission tomography (PET) involves exposure to ionizing radiation; electroencephalography (EEG) or magnetoencephalography (MEG) have  significantly less spatial resolution, and involve an inverse-problem for source reconstruction which has no unique solution.

\begin{figure}[!h]
\begin{center}
\includegraphics[width=0.9\textwidth]{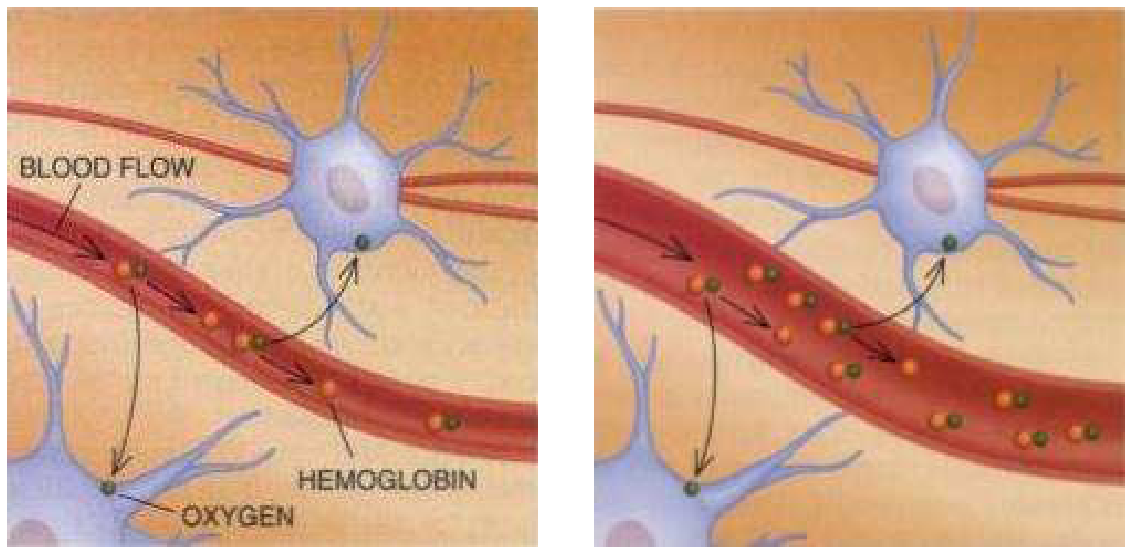}
\end{center}
\caption{\label{fig:bold_effect} Illustration of the BOLD effect, from \cite{Raichle94}. (left) at rest (right) during an activation. Each neural activation induces a local increase in blood flow, that is substantially higher than the increase in oxygen consumption of the nearby neurons. This results in a fall in the concentration in {\em deoxyhemoglobin} (in orange), and a surge in the fMRI signal.}
\end{figure}

Using fMRI, cerebral activity is measured indirectly, through its impact on the vascular network. More precisely, neural activity induces a local increase in blood oxygenation, known as the blood oxygen level dependent (BOLD) effect \cite{Ogawa90} (see Figure~\ref{fig:bold_effect}). Because of the magnetic properties of the oxyhemoglobin molecule (oxyHb), this effect can be measured by the scanner. One may note that this inherently limits the precision of this technique in locating activations, since the surge in blood oxygenation does not necessarily take place at the exact place of the neural activity. The link between these two phenomenons is still an active research area.

\subsection{Data acquisition and preprocessings}\label{sec:preproc}

Most fMRI paradigms are designed to identify brain areas involved in certain cognitive or sensori-motor tasks. They consist in a series of stimuli (visual, auditory, etc.) presented to the subject lying inside the scanner. The primary motive of this thesis lies in the analysis of data arising from such experiments.
% The methods for detecting regions activated in response to each
% given experimental condition are described in Sections~\ref{sec:glm,sec:statistical}.
An alternative which has received much attention lately consists in scanning the subject in absence of external stimuli. The aim of studying such `resting state' data is to identify latent attentional networks, {\em e.g.} using multivariate exploratory techniques \cite{Beckmann04,Perlbarg08}.

In all cases, the data resulting from a fMRI experiment consists in a sequence of three-dimensional (3D) brain images, with a typical spatial resolution of $3 mm^3$ on a standard 3~Tesla (3T) scanner, acquired at the rate of one volume every 2 to 3~seconds. An anatomical image is also acquired using the same scanner, at a higher spatial resolution of about $1 mm^2$ planar resolution (inter-slice resolution may be lower).

%%% Que veut dire 'proper analysis'? Rien. 
Before undergoing %proper 
group analysis, this data 
%%%must be : non (tout le monde ne sera pas d'accord)
is submitted to a number of pre-processing steps. These are briefly summarized hereafter (see \cite{Friston97} for further details). 
%%% En general on commence par le slice-timing... 
Pre-processing usually starts with the temporal realignment, also
termed {\em slice-timing}, of the successive scans to ensure that all
time series inside the volume are sampled at the same time-points,
even though the different slices are acquired at slightly different
moments. Next, the scans are realigned in order to compensate for
subject's motion inside the scanner.
%%% Que veux-tu dire??? 
%%%Finally, movement related effects, such as interpolation artefacts and
%%%magnetic field distortions, are estimated and subtracted from the
%%%data.

After temporal and spatial realignment, the data is
%%% ne pas oublier le recalage fMRI/T1
rigidly registered to the associated anatomical image of the same
subject, and warped using a nonrigid transformation resulting from the
registration of the anatomical image to a brain template, which
represents a standard stereotactic space such as the Talairach space
\cite{Talairach88}. This so-called spatial normalization step has a
key impact on group data analysis, as discussed in
Section~\ref{sec:normalisation}. Note that the different spatial
transformations resulting from motion correction,
functional/anatomical registration, and brain warping, are usually
composed in order to prevent the data from being resampled more than
once after slice timing correction.

Finally, the data is spatially smoothed, in order to enhance the signal-to-noise (SNR) ratio, most often using a Gaussian kernel. The spread of this kernel, measured by its full-width at half maximum (FWHM), usually varies between $5$ and $8~mm$ \cite{Friston97}.

\subsection{GLM analysis of time series}\label{sec:glm}

\begin{figure}
\centering
\includegraphics[width=0.75\textwidth]{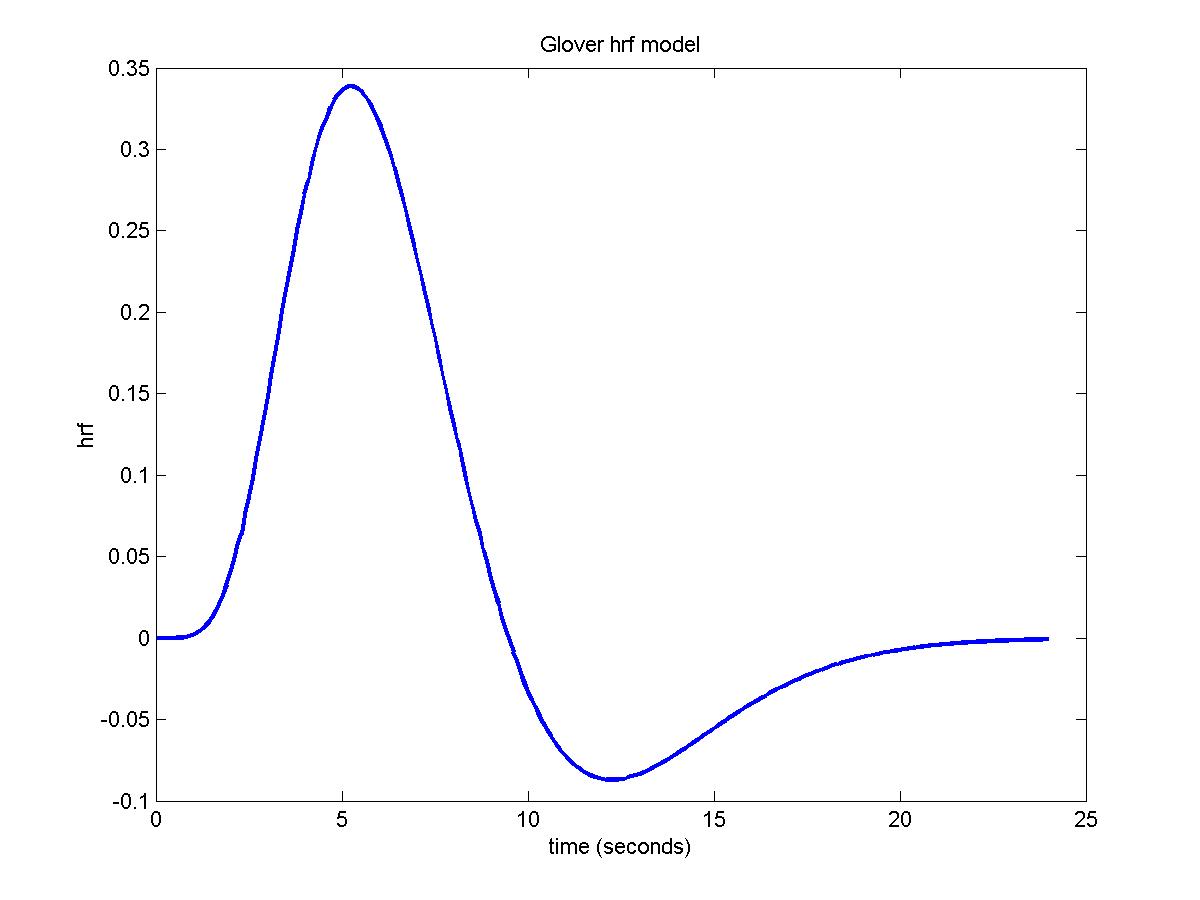}
\caption{\label{fig:hrf} \bf Glover haemodynamic response function (HRF).}
\end{figure}

The pre-processed data of a single subject is most often analysed separately at each 
%%% Pas besoin d'introduire la notion de 'observed point'. Pour info,
%%%voxel signifie 'volume element'
%%%observed point, or {\em voxel},
{\em voxel}
using regression techniques. This approach, often termed {\em massively univariate}
%%% Une petite explication s'impose: 
because voxels are processed independently from one another, 
has been developed primarily in \cite{Friston97, Worsley02}, and is available in many software packages, such as Statistical Parametric Mapping (SPM) or the FMRIB Software Library (FSL).

\subsubsection{BOLD response modeling}

Let us denote $\bs {\rm Y}_k = ({\rm Y}_{k,1}, \ldots, {\rm Y}_{k,T}),$ the time-series observed at voxel~$k,$ for $k=1, \ldots, d,$ where $d$ denotes the number of voxels 
within
%%%inside 
the search volume. 
%%% Tr�s impr�cis:
%%%This volume is usually defined as a mask of the subject's brain,
%%%computed from the anatomical image.
This volume may be defined as the intersection of individual brain
regions extracted from the functional images, or a unique region
delineated in the template space. 
% Defining the search volume boils
% down to defining a mask which has some consequences on group
% analysis, as we will see in Section~\ref{sec:mass}.

% \subsubsection{Voxelwise formulation}
\begin{figure}
\includegraphics[width=\textwidth]{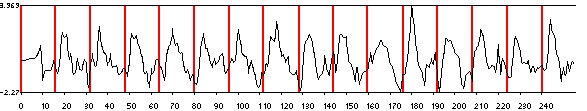}
\caption{\label{fig:time_series} Example of a fMRI time-series observed in a certain voxel. Occurrence times (onsets) of the successive stimuli appear in red (time is measured in seconds).}
\end{figure}

The time-series, illustrated in Figure~\ref{fig:time_series}, reflects the BOLD response at voxel~$k.$ The response to a series of stimulations of the same kind is traditionally modeled as a stationary linear filter with finite impulse response called the HRF. This leads to a general linear model (GLM):

\begin{eqnarray}\label{eq:GLM}
\bs {\rm Y}_k &=& \bs {\rm X} \bs \beta_k + \bs \varepsilon_k.
\end{eqnarray}

Here $\bs {\rm X}$ is the $T \times m$ {\em design matrix} containing the $m$ regressors (one per column), and $\bs \beta_k$ the corresponding parameters of interest. Regressors include the constant vector $\bs 1_T,$ modeling a baseline activation, along with vectors modeling the BOLD responses to the different stimulations, also termed {\em neural response levels} (NRLs). These are defined for each given stimulus type as the (discretized) convolution of the HRF $\bs h$ by a function $\br S$ representing the external stimulation. For event-related fMRI designs, $\br S$ is  a sum of Dirac masses specifying the occurrence times, or {\em onsets} of the successive stimuli, whereas for block-related fMRI designs, $\br S$ is typically a sum 
%%% En pratique, SPM mod�lis un bloc comme une succesion de diracs (un
%%% tous les TR/16). Par lin�arit�, l'amplitude de la r�ponse � un
%%% bloc doit �tre proportionnelle � sa dur�e: �a joue si le paradigme
%%% utilise des blocs de dur�es diff�rentes. 
of boxcar functions, 
specifying the beginning and the end of each stimulation block.

Additional regressors may be added to model 
%%% Il vaut mieux parler de variables de confusion que de param�tres
%%% de nuisance, ici. On n'en est pas encore � d�finir les param�tres
%%% d'int�r�t.
%%%nuisance parameters,
confounds, 
such as a low frequency drift due to 
%%% Les signaux physio ne sont pas des artefacts!!!
physiological signals 
and scanner-related artefacts. This drift can be decomposed on a 
%%% Ce n'est une base que pour les ignards!
%%%basis of periodical (e.g. cosine) functions 
family of functions such as periodical (e.g. sine/cosine) functions or
polynomials \cite{Friston97}.

\subsubsection{Noise modeling}\label{sec:noise_modeling}

In (\ref{eq:GLM}), $\bs \varepsilon_k \in \mathbb
R^d$ models the estimation errors, and is usually defined as a
%%%En anglais on ne dit pas trop 'centered' 
zero-mean Gaussian random variable. If its covariance matrix is spherical, {\em i.e.} if the estimation errors are assumed to
be {\em i.i.d.}, with common variance $\sigma_k^2,$ then the least square estimate and the maximum likelihood estimate of $\bs\beta_k$ coincide. This common estimator is the best linear unbiased estimate (BLUE) of the model parameters, and is found by solving the linear system $\br X' \br Y_k = \br X' \br X \bs\beta_k.$ The variance $\sigma_k^2$ is estimated in this case by the residual mean squares $\frac{1}{T-r}\parallel \br Y_k - \br X \bs\beta_k \parallel^2,$ where $r$ is the rank of the design matrix $\br X.$

However, fMRI time-series are known to be temporally correlated \cite{Boynton96,Zarahn97}.
%%%Si tu dis que l'autocorrelation est 'known to be present' dans les
%%%donnees, il faut mettre des refs la-dessus (regarde dans mon
%%%MICCAI'04 sur ma page web).
Noting $\br V_k$ the covariance matrix of the noise process $\bs\varepsilon_k,$ the BLUE of the model parameters is now obtained by solving the following linear system:
\begin{eqnarray}\label{eq:wightening}
\br X' \br V_k^{-1} \br Y_k &=& \br X' \br V_k^{-1} \br X \bs\beta_k,
\end{eqnarray}
in the case where $\br V_k$ is known. In general though, it is unknown and must also be estimated. As noted in \cite{Donnet06}, this is an ill-posed problem since there are more parameters to be determined than available observations, so constraints must be imposed on $\br V_k.$

Several approaches have been adopted. \cite{Friston97,Worsley02} model $\bs \varepsilon_k$ as first-order auto-regressive (AR(1)) process, so that at each time-point $t=1,\ldots,T$:
\begin{eqnarray}\label{eq:AR}
\varepsilon_{k,t} &=& \rho_k \varepsilon_{k,(t-1)} + \xi_{k,t},
\end{eqnarray}
where $\rho_k$ is a spatially varying autocorrelation parameter, and $\bs \xi_k$ is a Gaussian white noise, with a spatially varying variance $\sigma_k^2.$ In \cite{Friston97}, a simpler version of this model is used, where $\rho_k$ is assumed
uniform across the search volume. In both cases, the variance parameters are estimated by restricted maximum likelihood (ReML), and the estimated covariance matrix $\hat{\br V}_k$ is substituted to the unknown true one in (\ref{eq:wightening}). The precision of the resulting `plug-in' estimate of model parameters depends crucially on the accuracy of the covariance estimate $\hat{\br V}_k.$ The study in \cite{Friston00} suggests that the use of AR(1) with a global auto-correlation parameter may result in biased parameter estimates. Temporal smoothing is then suggested to improve the estimation.

More sophisticated techniques to estimate $\br V_k$ are compared in \cite{Woolrich01}, including a Tukey taper combined with nonlinear spatial smoothing, which performs optimally among the tested strategies. Also, distinct auto-correlation parameters are estimated in each voxel, an option which is not considered in \cite{Friston00}, and may in part explain the observed bias.

An alternative to the above plug-in estimation is to estimate the covariance and regression parameters jointly, by maximizing the full likelihood of the model. This is the approach developed in \cite{Roche04}, where an AR(1) model is fit separately to the time-series acquired in each voxel. This is done iteratively, using an extension of the Kalman filter. This strategy further enables an `online', or real-time, estimation of the model, the parameter estimates being updated as new scans are acquired.

Generalization to higher-level auto-regressive models (AR($p$), with $p \geq 1$ ) is considered in \cite{Penny03}, where a variational Bayes (VB) approximation is used to jointly estimate all model parameters. A very interesting feature of this approach is that it allows to select the order $p$ of the auto-regressive model, separately in each voxel. This is done by maximizing the free energy $\mathcal F(p),$ that is, the lower bound on the marginal likelihood central to the VB approach, easily available from the output of the VB algorithm. On an application to a particular real fMRI dataset, histograms of the optimal values for the AR model order $p$ show that it never exceeds $p = 3.$ In fact, in most voxels it is equal to either 0 or 1. This provides a rough justification for the models described above, which all assume $p = 1.$

% Assuming a fixed HRF, the parameters $\bs \theta_k = (\bs \beta_k, \sigma_k^2, \rho_k)$ can be estimated by the maximum likelihood method:
% \begin{eqnarray}
% \hat {\bs \theta}_k &=& \arg\max_{\bs \theta_k} \mathcal L_{\bs \theta_k}(\bs Y_k),\nonumber
% \end{eqnarray}
% where $\mathcal L_{\bs \theta_k}(\bs Y_k)$ denotes the likelihood of the model defined by~(\ref{eq:GLM},\ref{eq:AR}). However, there is no closed form for $\hat {\bs \theta}_k$ because of the AR covariance noise structure. \cite{Worsley02} propose a weighted least-square method which involves spatially smoothing a map of autocorrelation parameters estimates $\hat \rho_k,$ obtained from the residuals of a first model assuming a spherical variance. The covariance matrix of the observations is thus estimated, and used to pre-whiten the data. 
%%% En fait, Worsley utilise une �tape interm�diaire de clustering des
%%% voxels en fonction du param�tre AR, ce qui lui permet d'utiliser
%%% la m�me covariance pour tous les voxels d'un m�me cluster. 

% \cite{Friston95} use a simpler version where $\rho_k$ is assumed
% uniform across the search volume. To avoid the smoothing or pooling
% of the $\rho_k$'s, and more generally the estimation biases present in
% the above methods, \cite{Roche04} maximize the full likelihood
% associated with the model using an extension of the Kalman filter.

\subsubsection{Limits and alternatives}\label{sec:limits_GLM}

We conclude this review on single-subject data modeling by noting that the formulation in~(\ref{eq:GLM}) relies on the following key assumptions:

\begin{itemize}

\item {\bf Fixed HRF.} Several canonical shapes have been proposed for the HRF, such as the Glover HRF \cite{Glover99}, used in SPM, defined as the difference of two Gamma functions (see Figure~\ref{fig:hrf}). This simplifying assumption may lead to a poor model fit in regions where the true HRF is significantly different from the one used in the estimation. 

\item {\bf Linearity.} The effects of the different stimuli are
  assumed to contribute additively to the overall effect evoked by the
  experimental paradigm, thus justifying the use of a linear
  model. 
%%% D'o� sors-tu ces 20 secondes? Ca me semble une limite vraiment
%%% haute... Qu'appelles-tu 'iner-stimulus gap'? Est-ce le SOA
%%% (Stimulus Onset Asynchrony)? Je te conseille vivement de regarder
%%% la pr�sentation de Patricia Romaigu�re aux JIRFNI 2009 et la page
%%% de Rick Henson dont elle est inspir�e: 
%%%http://imaging.mrc-cbu.cam.ac.uk/imaging/DesignEfficiency
  This assumption may hold for an event-related paradigm if the
  inter-stimulus gap is long enough \cite{Boynton96,Glover99}. Otherwise,
  non-linear effects may occur.

\item {\bf Time-Invariance.} For a given stimulus type, the amplitude of the BOLD response is assumed constant across the successive stimulations, justifying the use of a single regressor for each stimulus type. This neglects so-called habituation effects, such as repetition-suppression, {\em i.e.} a decrease in the BOLD response observed when the same stimulus is presented repeatedly. Modeling such phenomenons using the standard GLM requires to incorporate additional regressors, hence reducing the degrees of freedom. Thus parametric modulation may be preferable, as discussed below.

\end{itemize}

The modeling and analysis of brain haemodynamics is still a field of active research. To cite only a few examples, \cite{Friston00b} suggested modeling non linearities in the BOLD response through Volterra series while remaining in the GLM framework, by adding additional regressors, defined as temporal derivatives of the canonical HRF. To overcome the limitations inherent to the use of a canonical HRF, \cite{Penny03,Woolrich04b} modeled the HRF as a linear combination of a pre-defined set of functions. In \cite{Makni08}, a joint detection-estimation (JDE) framework is developed to estimate at the same time the shape $\bs h$ of the HRF, in a nonparametric way, and the NRLs $\bs \beta,$ in a Bayesian setting. In \cite{Donnet06b} an alternative approach to the statistical modeling of fMRI time series is investigated, based on the physiological model of the blood flow / oxygen consumption coupling proposed in \cite{Buxton97}.

Another limitation, inherent to the mass univariate model, is that it ignores the spatial structure of the data. Several authors, including \cite{Gossl01b,Penny07,Smith07b,Makni08}, have proposed to model the NRLs, according to a spatial mixture model, with three classes, corresponding to null, activated and deactivated (inhibited) voxels. The labels identifying the state of each voxel are further modeled jointly as a hidden Markov random field, which tends to group active voxels in clusters. This has a regularizing effect, since isolated voxels, assimilated to noise, are less likely to be detected.
Such spatial mixture models have also been proposed to model directly the statistical maps resulting from the GLM analysis, as a means to compute a detection threshold (see Section~\ref{sec:thresholding} for a discussion of these methods).

\subsection{Statistical map of brain activity}\label{sec:statistical}

\begin{figure}
\centering
\includegraphics[width=0.6\textwidth]{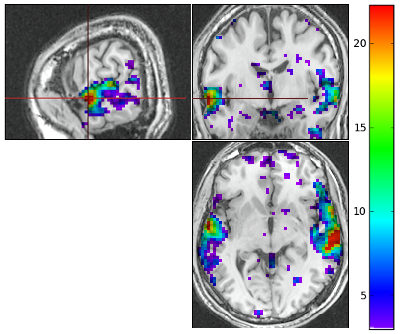}
\caption{\label{fig:activation_map} Individual $t$-score map for the `audio-video' contrast, from the Localizer dataset \cite{Pinel07}, with the the subject's anatomical image in the background. The map is thresholded at $10^{-3},$ uncorrected, meaning that the presence of an activation is tested independently in each voxel at level $10^{-3},$ without accounting for multiple comparisons. Activations are clearly seen in the bilateral temporal regions; many isolated voxels are also detected, which may be false positives.}
\end{figure}

Based on the model~(\ref{eq:GLM}), an effect of interest, such as a difference between stimuli, can be specified by $\br c \bs \beta_k,$ where $\br c$ is a row vector of $m$ contrasts \cite{Worsley02}. Detecting a nonzero effect is then equivalent to testing the null hypothesis that the effect is zero, $\mathcal H_{0,k}\, :\, \br c \bs \beta_k = 0.$ Assuming the data has been pre-whitened as described in the previous section, so that its covariance structure is approximately spherical, it follows from the Neyman-Pearson lemma that the optimal statistic for testing $\mathcal H_{0,k}$ is Student's $t$-statistic:

\begin{eqnarray}\label{eq:t_statistic}
T_k &=& \frac{\br c \hat{\bs \beta}_k}
{ \hat \sigma_k \sqrt{\br c (\tilde{\bs {\rm X}}_k'\, \tilde{\bs {\rm X}}_k)^{-1} \br c'}},
\end{eqnarray}

where $\tilde{\bs {\rm X}}_k$ is the pre-whitened design matrix given by $\bs{\rm V}_k^{-\frac{1}{2}} \bs{\rm X}_k,$ given the covariance matrix $\bs{\rm V}_k\sigma_k^2$ of the data $\bs{\rm Y}_k$. If $\bs{\rm V}_k$ is known, then under $\mathcal H_{0, k},$ $T_k$ follows a Student distribution with $T - (m + 1)$ degrees of freedom (df). Consequently, $\mathcal H_{0, k}$ is rejected at level $\alpha$ if $T_k > t_{\alpha},$ where $t_{\alpha}$ is the $(1 - \alpha)$-th quantile of the Student distribution with $T - (m + 1)$ df. In practice $\bs{\rm V}_k$ is estimated from the data, as explained in Section~\ref{sec:noise_modeling}, so the test level is not exactly $\alpha.$

Computing $T_k$ for all voxels $k = 1, \ldots, d$ results in a statistical map, also termed {\em activation map}, reflecting the subject's activation pattern for the given contrast $\br c.$ Detecting activated regions in this setting is equivalent to testing simultaneously all the null hypotheses $\mathcal H_{0, k}$, the number of which is the number of voxels in the search volume and is typically of the order of $100\, 000$. 
%%% Le pb des comparaisons multiples se pose d�j� � K=2 voxels...

To do so, one must address the multiple comparison problem. For
instance, the naive procedure that consists in rejecting $\mathcal
H_{0, k}$ for all voxels~$k$ such that $T_k > t_{\alpha},$ as
described above, choosing for instance $\alpha=0.001,$ ensures that the
probability of a false detection in each voxel is at most $0.001.$
However, this means that, in the worst-case scenario where all voxels 
are in fact inactive, an average of $\alpha \times 100\, 000 = 100$ false positives
are detected throughout the volume.
This issue is illustrated in Figure~\ref{fig:activation_map}.

The same multiple testing problem arises when detecting group activation pattern using the `SPM-like' approach, as we will see in Section~\ref{sec:test_group_effect}. Thus, we postpone this issue to Section~\ref{sec:multiple}, where we adress it jointly for individual and group data analysis.

\section{Group analysis: The mass univariate approach}\label{sec:mass}

In a typical fMRI study, several subjects are recruited from a population of interest and scanned while submitted to the same series of stimuli. Activation maps associated with a given contrast are obtained for each subject, as described in the previous section, and used as input data for inference at the between-subject level, where the goal is to evidence a general brain activity pattern.

In the following, we describe how this is performed in the `SPM-like' approach, starting with the spatial normalisation step in Section~\ref{sec:normalisation}, which aims to match each individual image to a brain template. The activation maps are then compared on a voxelwise basis, as described in Section~\ref{sec:between}, resulting in the computation of a map of statistics, detailed in Section~\ref{sec:test_group_effect}, to test in each voxel the presence of a positive mean activation across subjects.

\subsection{Spatial normalisation and smoothing}\label{sec:normalisation}

\begin{figure}
\includegraphics[width=\textwidth]{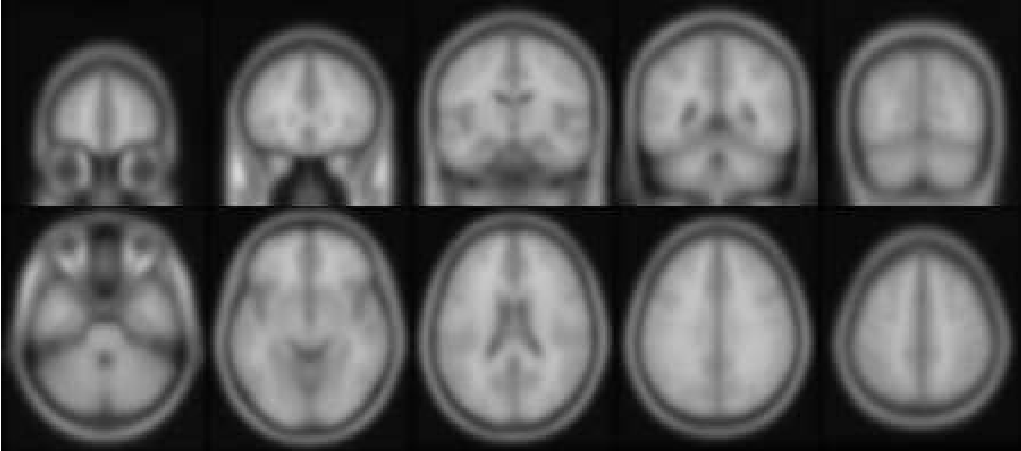}
\caption{\label{fig:template} coronal and axial slices of the MNI template, from \cite{Flandin04}. This is obtained as the average of 152 individual anatomical images, registered to an earlier version of the template (MIN305) using affine transformations.}
\end{figure}

The high morphological variability of the human brain \cite{Brett02}, makes the comparison of cerebral images across subjects problematic.
\begin{figure}
\includegraphics[width=\textwidth]{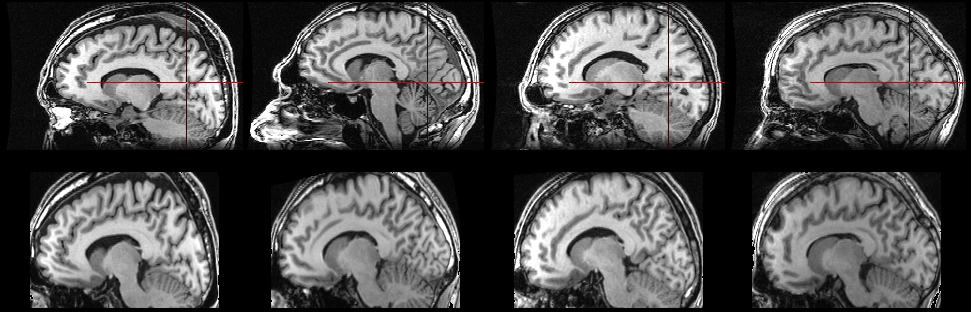}
\caption{\label{fig:normalisation} Illustration of spatial normalisation on a sagittal slice: (top) images, before normalisation; (bottom) after normalisation, using affine transformations. Anatomical differences across subjects are seen to persist, especially in the sulco-gyral geometry.}
\end{figure}

A traditional way to compensate for this variability, as mentioned in Section~\ref{sec:BOLD}, is to register, or normalize, the individual images of all subjects to a common brain template~\cite{Ashburner99}, such as the widely used Montreal Neurological Institute (MNI) template (see Figure~\ref{fig:template}). This is usually done by minimizing a measure of discrepancy between image intensities over a suitable class of spatial transformations (see Chapter~\ref{chap:modeling} for a brief review on registration methods). Comparative studies of several normalization methods can be found in~\cite{Hellier03,Klein09}.

Any location in the brain can then be marked in a standard coordinate system, such as the one developed by \cite{Talairach88}. However, registration is prone to errors (even assuming the existence of point-to-point correspondences between different brains), hence it does not seem reasonable to assume that homologous points are exactly aligned across subjects.

This fact is often used as a motivation to justify a preliminary linear spatial smoothing of the data, with a typical FWHM of $8$ to $12\, mm,$ as a way to increase the overlap of functionally homologous regions over subjects. This smoothing is sometimes applied to the individual activation maps, resulting from the individual data processing, in addition to the first smoothing step used on the raw fMRI time-series to enhance SNR (see section~\ref{sec:BOLD}).

The `SPM-like' approach then compares the individual images on a voxelwise basis, thus making an implicit assumption that each subject is in perfect match with the template. Among the consequences of that assumption, one may anticipate a stretching effect on group activity patterns due to the ``jitter'' induced by inaccurate registration. This effect can only be reinforced by the above-mentioned preliminary smoothing step. Evidence for this stretching is provided in \cite{Keller08}, and will be one of the important results of Chapter~\ref{chap:modeling}.

% \cite{Kriegeskorte06} \cite{Woolrich06}

\subsection{Between-subject modeling}\label{sec:between}

\begin{figure}
\centering
\includegraphics[width=\textwidth]{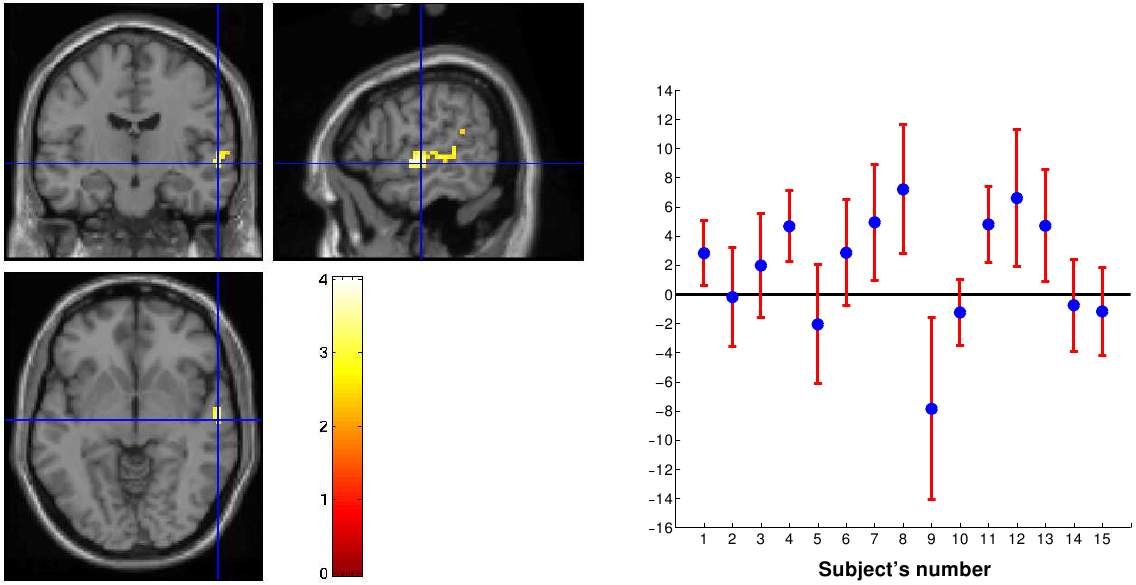}
\caption{\label{fig:MFX} Example of fMRI group data in one voxel,
during a language processing task, from \cite{Meriaux06}.
Left, activations detected using the one-sided mixed-effect statistic, thresholded using a permutation test (see Section~\ref{sec:test_group_effect}), with cross-hair at $(60;15;6)$ Talairach coordinates in mm. Right, plot of the estimated 
effects in the same voxel and associated $70\%$ confidence intervals}
\end{figure}

%%% La notation K serait plus appropri�e que d pour le nombre de
%%% voxels, �tant donn� que k est utilis� pour indicer les voxels.
%>>> J'avais pense a d comme dimension, puisque les donnees vivent dans R^d
Let $\br x_i = (x_{i,1}, \ldots, x_{i,d})$ denote the map of \index{BOLD}BOLD effects in response to a certain contrast
of experimental conditions, for subject~$i = 1\, \ldots, n.$ As seen in Section~\ref{sec:statistical}, a noisy estimate of $\br x_i,$ $\br y_i = (y_{i,1}, \ldots, y_{i,d})$ is available from the analysis of the subject's scans, along with an image of estimation variances $\br S_i^2 = (s_{i,1}^2, \ldots, s_{i,d}^2).$

More precisely, for each subject~$i,$ following the notations introduced in Section~\ref{sec:statistical}, we define in each voxel~$k:$ 
$$
x_{i,k} := \br c \bs \beta_k;
\quad 
y_{i,k} := \br c \hat{\bs \beta}_k;
\quad
s_{i,k}^2 := \hat \sigma_k \sqrt{\br c (\tilde{\bs {\rm X}}_k'\, \tilde{\bs {\rm X}}_k)^{-1} \br c'},
$$
for a certain row contrast vector $\br c.$

Under sufficient degrees of freedom at the within-subject level, it is reasonable to consider $y_{i,k}$ as being normally distributed around $x_{i,k}$ with standard deviate $s_{i,k}$ considered fixed \cite{Worsley02}. To address questions regarding the variability of the effect in a population, the unobserved effects $x_{1,k}, \ldots, x_{n,k}$ are further modeled as independent random variables drawn from an unknown distribution which characterizes the across-subject variability of BOLD responses. When this distribution is assumed Gaussian with unknown mean and variance $(\mu_k, \sigma_k^2),$ we obtain the same hierarchical model as in \cite{Worsley02,Beckmann03a,Meriaux06}:

\begin{itemize}
\item First level (within-subject): 
\begin{eqnarray}
\label{eq:within}
y_{i,k} &=& x_{i,k} + \varepsilon_{i,k}; \quad 
\varepsilon_{i,k} \stackrel{ind.}{\sim}  \mathcal N( 0 , s_{i,k}^2 ),
\end{eqnarray}
\item Second level (between-subject): 
\begin{eqnarray}
\label{eq:between}
 x_{i,k} &=& \mathrm \mu_k + \eta_{i,k}; \quad 
\eta_{i,k} \stackrel{i.i.d.}{\sim} \mathcal N(0, \sigma_k^2),
\end{eqnarray}
\end{itemize}
where the independence sampling assumptions at both levels imply that in each voxel~$k,$ the pairs $(x_{1,k},\, y_{1,k}), \ldots, (x_{n,k},\, y_{n,k})$ are mutually independent conditionally on the population parameters~$(\mu_k, \sigma_k^2).$ By integrating out  the hidden variables $x_{i,k},$ we see that the observed effects are drawn independently but, in general, non-identically from Gaussian distributions:

\begin{eqnarray}\label{eq:heteroscedastic}
y_{i,k} &=& x_{i,k} + \xi_{i,k}; \quad 
\xi_{i,k} \stackrel{ind.}{\sim}  \mathcal N( 0 , s_{i,k}^2 + \sigma_k^2 ).
\end{eqnarray}

That is to say, the observations are generally heteroscedastic unless all first-level deviations $s_{i,k}$ are equal. In this special case, the model boils down to the simple sampling model in \cite{Friston95}, that is computationally attractive but may lack robustness against 
%%%unreliable 
noisy observations
% , as illustrated in Figure~\ref{fig:MFX}
.

We conclude this description of the standard model for fMRI group data with the following remarks:

\begin{itemize}

\item As for individual subject data (see Section~\ref{sec:glm}), the group data is modeled separately in each voxel, in a `massively univariate' fashion. Consequently, possible correlations between neighboring voxels are ignored at this stage. They will be accounted for in the multiple comparison step, discussed in Section~\ref{sec:multiple}.

\item The within-subject model (\ref{eq:within}) is also referred to as a {\em fixed-effect} (FFX) model, since it specifies the unobserved effects $x_{1,k}, \ldots, x_{n,k}$ as the (fixed) parameters of interest. Inference in this model is therefore limited to the cohort of subject scanned during the experience. Likewise, the between-subject model (\ref{eq:between}) is referred to as a {\em random-effect} (RFX) model, since it considers the effects $x_{1,k}, \ldots, x_{n,k}$ as random variables. Finally, the hierarchical model specified by (\ref{eq:within}) and (\ref{eq:between}) is also called a {\em mixed-effect} (MFX), or {\em mixed}, model, since it `mixes' the FFX and RFX models.

\item As noted in \cite{Worsley02}, besides simply inferring on the mean population effect $\mu_k,$ we may also wish to compare two or more populations, and more generally regress the subjects' effects $x_{i,k}$ on a certain set of regressors $\bs z_{i,k}$. To this end, the mean effect $\mu_k$ in (\ref{eq:between}) may be replaced by a linear term $\bs z_{i,k}'\, \bs \mu_k$ in the regressor variables. Though in the following we will focus on the one-sample setting for the sake of simplicity, the methods exposed here are fully adaptable to the general regression setting, except when mentioned otherwise.

\end{itemize}

Several extensions to the standard hierarchical model specified by
(\ref{eq:within}) and (\ref{eq:between}) have been proposed. In
\cite{Woolrich04c} for instance, the uncertainty on first-level 
standard deviates $s_{i,k},$ is accounted for. More specifically, the
general linear model (\ref{eq:GLM}) used to estimate the subject-specific
parameters is directly used at the within-subject level, rather than 
the Gaussian approximation (\ref{eq:within}).

Based on this more realistic model, a Bayesian approach is developed
wherein the posterior distribution of the group mean effect,
$p(\mu_k | y_{1,k},\, \ldots,\,y_{n,k}),$ is estimated, either using 
MCMC techniques, or a faster approximation, assimilating it to a 
noncentral $t$-distribution.

Such an approach may seem computationally intensive at first, since 
the complete data of all subjects must be analyzed together. However, it 
is shown in \cite{Woolrich04c} that the posterior distribution of 
$\mu_k$ depends in fact only on certain summary statistics of the individual 
datasets. These can be computed beforehand, hereby greatly reducing the 
computational complexity of this approach.

%%% Je n'adh�re pas du tout � ce qui suit: tu sugg�res qu'on aurait le
%%% droit de changer de mod�le apr�s avoir d�j� fait une analyse,
%%% c'est mal...
% Moreover, they require individual data to be re-processed each time a
% new contrast is added, or a new subject is scanned.
%%% et en plus c'est faux!

Some authors have further proposed to relax the Gaussian assumption at the between-subject level. In \cite{Roche07}, the distribution of the $x_{i,k}$'s is assumed totally unknown, and a nonparametric maximum likelihood estimation (NMLE) method is developed. 
%%% Je te laisse rajouter la r�f au papier de Woolrich, Neuroimage
%%% (2008), Robust group analysis using outlier inference
In \cite{Woolrich08}, a mixture model consisting of two Gaussian
distributions is used to describe population heterogeneity, one class corresponding to possible outlier subjects.

%%% L'argumentaire est trop faible ici: rejeter ces approches pour des
%%% questions de temps de calcul est d�risoire... 
% Though conceptually attractive, this approach is computationally much
% more intensive compared to the one based on the Gaussian model.
%%% et qui suis-je pour me porter en juge de ces recents developpements?

\subsection{Test of a nonzero group effect}\label{sec:test_group_effect}

Based on the hierarchical model introduced in the previous section, our final goal is to test in each voxel~$k$ the presence of a nonzero mean effect, {\em i.e.}, test $\mathcal H_{0,k} : \mu_k = 0$ versus $\mathcal H_{1,k} : \mu_k \neq 0.$ Besides this {\em two-sided} null hypothesis, one may also wish to test the presence of a positive effect (corresponding to an activation), that is, test the {\em one-sided} null hypothesis $\mathcal H_{0,k} : \mu_k \leq 0$ versus $\mathcal H_{1,k} : \mu_k > 0.$ This test involves the two following steps:

\begin{itemize}
\item Definition of a test statistic $T_k = T_k(y_{1,k}, \ldots, y_{n,k})$
\item Statistical calibration of $T_k,$ {\em i.e.} computing the threshold $u$ such that $P(T_k \geq u | \mathcal H_{0,k}) \leq \alpha,$ for any required level $\alpha.$ The level specifies the probability of falsely rejecting $\mathcal H_{0,k},$ {\em i.e.,} the probability of a type I error, or false positive.
\end{itemize}

\subsubsection{Choice of a test statistic}

In the simple RFX model, {\em i.e.}, when the first-level standard deviates $s_{i,k}$ are implicitly assumed constant, as in \cite{Friston95}, a natural choice is the standard $t$-statistic
$$
T_k = \displaystyle\frac{\bar{\bs {\mathrm y}}_k}{std(\bs {\mathrm y}_k) / \sqrt {n-1}},
$$
where $\bar{\bs {\mathrm y}}_k = \frac{1}{n} \sum_{i=1}^n y_{i,k}$ and $std(\bs {\mathrm y}_k)^2 = \frac{1}{n} \sum_{i=1}^n (y_{i,k} - \bar{\bs {\mathrm y}}_k)^2.$ This yields the uniformly more powerful (UMP) test at any level $\alpha$, both for the one-sided and the two-sided test \cite{Lehmann86}.

In the general MFX model, there is no optimal choice of a test statistic in terms of power. In \cite{Worsley02}, $\sigma_k^2$ is estimated by restricted maximum likelihood (ReML), using an expectation-maximization (EM) algorithm \cite{Dempster77}. This is equivalent to an iterative reweighted least-square procedure to estimate $\mu_k,$ using the weight $(s_{i,k}^2 + \hat \sigma_k^2)^{-1}$ for observation $y_{i,k}.$ In analogy to the RFX case, the following approximate $t$-statistic is then used:
$$
\tilde T_k = \displaystyle\frac{\hat \mu_k}{\hat \sigma_k / \sqrt n}.
$$
A similar approximate $t$-statistic is used in \cite{Woolrich04c}, where $(\mu_k, \sigma_k)$ are estimated in a Bayesian setting by their posterior mean, having marginalized out all other hidden variables using a Monte-Carlo Markov-Chain (MCMC) sampling algorithm.

Another, more systematic, approach, advocated in \cite{Meriaux06,Keller08b}, is to use the maximum likelihood ratio (MLR) for the two-sided test:
$$
R_k = \frac
{\displaystyle\sup_{\mu_k = 0,\ \sigma_k^2\in\mathbb R_+^*} \mathcal L_k(\mu_k,\sigma_k^2)}
{\displaystyle\sup_{\mu_k \neq 0,\ \sigma_k^2\in\mathbb R_+^*} \mathcal L_k(\mu_k,\sigma_k^2)},
$$
where $\mathcal L_k(\mu_k,\sigma_k^2)$ is the likelihood of the model (\ref{eq:heteroscedastic}). For the one-sided test, the following sign modulation is used:
$$
\tilde T_k = sign(\hat \mu_k) \sqrt{R_k}.
$$
The maximum likelihood estimates $(\hat \mu_k, \hat \sigma_k)$ can be computed using an EM algorithm. In \cite{Roche07}, this approach is extended to the nonparametric setting, and it is shown that the nonparametric maximum likelihood estimate of the between-subject distribution is a combination of at most $n$ Dirac masses.

\subsubsection{Statistical calibration}

Voxelwise statistical calibration is straightforward in the simple RFX model in \cite{Friston95}, using the fact that, under $\mathcal H_{0,k},$ $T_k$ follows a Student distribution with $n-1$ degrees of freedom. The two-sided null hypothesis $\mathcal H_{0,k}: \mu_k = 0$ is then rejected if $|T_k| > t_{\alpha/2},$ where $t_{\alpha/2}$ is the $(1 - \alpha/2)$-th quantile of the $t_{n-1}$ distribution, and the one-sided null hypothesis $\mathcal H_{0,k}: \mu_k \leq 0$ is rejected if $T_k > t_{\alpha}.$
% However since the RFX model is misspecified, in that it assumes constant first-level estimation standard errors $s_{i,k},$ it may lead to unexact control over the type I error $\alpha,$ and to biased inference, as shown in \cite{Meriaux04} for instance. This is mainly problematic for small groups, since the Student calibration is always asymptotically exact, owing to the Central Limit Theorem.
The Student distribution is also used as a substitute for the unknown null distribution of the approximate $t$-statistics in \cite{Worsley02} and \cite{Woolrich04c} (see the previous section), using the same degrees of freedom as the standard $t$-statistic. %In this case also there is a risk of biased inference for small groups of observations.
Use of the Student distribution is valid in both cases if the data distribution is Gaussian, or for large sample sizes ($n \to \infty$), owing to the central limit theorem.

An alternative solution to this calibration problem is to use permutation tests \cite{Good05}. They allow exact control on the type I error, under mild assumptions on the sampling distribution, and for any choice of a test statistic. This method of calibration was introduced in the neuroimaging literature by \cite{Holmes96}. It consists in sampling the null distribution of the test statistic by permuting the data, under certain exchangeability hypotheses. Having sampled $N$ values, the threshold of the test is then simply equal to the $[N\alpha]$-th largest sampled value.

The principal limitations of permutation tests are the heavy computations they require, and also their limited applicability. For instance, the universally exact control on false positives does not extend to the test of partial correlations in a multiple regression model. Approximate permutation testing procedures have been proposed in this case \cite{Anderson99b,Cade05}. They have been found empirically to provide a more precise control over false positives than standard parametric tests, in certain cases where the assumptions underlying the latter are not verified. However, no general result exists to support these observations.

% In conclusion, in many fMRI experiments which include only a limited number of subjects, statistical calibration should be done using a permutation test, rather than asymptotically exact parametric distributions. The principal limitations of permutation tests are the heavy computations they require, and also their limited applicability. For instance, they cannot be used to test partial correlations in a multiple regression model.

\section{Multiple comparisons}\label{sec:multiple}

A multiple comparison problem arises when testing several voxels simultaneously. This is the same problem encountered in the analysis of individual subject data (see Section~\ref{sec:statistical}). In its simplest form, it consists in finding a detection threshold~$u$ for a given statistical map $\br T = (T_k)_{1 \leq k \leq d}$ such that the balance between specificity (control over false positives) and sensitivity (control over false negatives) is optimal in a certain sense. Appendix~\ref{sec:generalities} gives a precise definition of these concepts. Furthermore, the chosen multiple comparison procedure (MCP) must also produce results that are useful in terms of neuroscience. Thus the detected activations must be easily linked to known anatomical structures. Moreover, a stringent control over false positives is usually required to avoid erroneous interpretations.

MCP's in neuroimaging can be divided into two main categories. The first one contains voxel-level thresholding procedures, that are presented in Section~\ref{sec:voxel}. However, due to the strict control on false positives, and the large number of tests performed (up to $\approx 100\, 000$ voxels in a whole brain image), these procedures typically lack sensitivity. Another shortcoming is that they detect individual voxels, which are hard to interpret, as they do not constitute relevant biological units.

Another type of MCP's has been developed to overcome these limitations, which we will refer to as cluster-level procedures. First, the statistical map is thresholded as before, but this time at an arbitrary level. Next, connected components, or {\em clusters} of voxels above this threshold are identified. Finally, the presence of activations is tested within each cluster, using a secondary threshold on the cluster's size. Cluster-level procedures are in general more powerful than their voxel-level counterparts, since they entail much less hypotheses to be tested.
% They are also more readily interpretable, since results are reported in terms of regions rather than individual voxels. 
A review of these approaches is presented in Section~\ref{sec:cluster}, and we conclude by discussing the pro's and con's of the different methods.

\subsection{Voxel-level inference}\label{sec:voxel}

We now address the problem of choosing a detection threshold $u$ for a map of voxelwise test statistics $\br T = (T_1, \ldots, T_d),$ controlling a certain error rate. The problem can also be defined as that of finding a threshold $c$ for the map of $p$-values $(p_1, \ldots, p_d).$ Here $\br T$ may stand either for the activation map of an individual subject (see Section~\ref{sec:statistical}), or a group activation map (see Section~\ref{sec:test_group_effect}). In both cases, the null hypotheses considered here, noted $\mathcal H_k = 0$ in Appendix~\ref{sec:generalities}, are the voxelwise null hypotheses $\mathcal H_{0,k}.$

\subsubsection{FWER-controlling procedures}

A lot of attention has been given to the control of the FWER, that is, the probability of one or more false positives, in the neuroimaging literature, as a strict control on false positive is necessary in order for the detected activated brain areas to be reliable. We review here some of the main approaches. We refer to Appendix~\ref{sec:error} for a rigorous definition of this and other multiple test error rates.

\paragraph{The Bonferroni procedure.}

The Bonferroni procedure consists in rejecting every null hypothesis $ \mathcal H_{0,k}$ whose $p$-value is smaller than $\alpha / d,$ where $\alpha$ is the desired upper bound on the FWER, and $d$ the number of tested hypotheses (one per voxel). Its justification relies on no other assumption than subset pivotality, necessary to define the $p$-values (see Appendix~\ref{sec:pivotality}). It is a direct application of Boole's inequality, and the fact that each $p_k$ is uniformly distributed under $\mathcal H_{0,k}:$

\begin{eqnarray}
FWER &=& P[V > 0 | H_{\mathcal M_0}] \nonumber\\
     &=& P [ \cup_{k \in \mathcal M_0} \{p_k < \alpha / d\} | \mathcal H_{\mathcal M_0} ] \nonumber\\
     &\leq& \sum_{k \in \mathcal M_0} P[p_k < \alpha / d | \mathcal H_{0,k}] \nonumber\\
%      &=& \sum_{k \in \mathcal M_0} \alpha / d \\
     &=& d_0 \times \alpha / d \leq \alpha.
\end{eqnarray}

In the above derivation, $d_0 = \sharp \mathcal M_0$ is the number of true null hypotheses. It can be shown that, under independence of the data across voxels, the Bonferroni procedure is near optimal \cite{Ge03}. This means that there is no procedure significantly more powerful having the same level of FWER strong control.

\paragraph{The maxT Procedure.}

In the case of mutually dependent tests, the more general maxT procedure can be applied. It relies on the fact that, under the global null $\mathcal H_{\mathcal M},$ the probability of one or more false detections can be controlled knowing the null distribution of the maximal statistic:

\begin{eqnarray}
P[V > 0 | H_{\mathcal M}] &=& P[\exists k \in \mathcal M,\, T_k > u | \mathcal H_{\mathcal M}] \nonumber\\
                          &=& P\left[\max_{k \in \mathcal M} T_k > u | \mathcal H_{\mathcal M}\right]. \nonumber\\
\end{eqnarray}

Thus, to control $P[V > 0 | \mathcal H_{\mathcal M}]$ at a given level $\alpha,$ the threshold $u$ must be equal to the $(1 - \alpha)$-th quantile of the distribution of the maximal statistic $\max_{k \in \mathcal M} T_k$ under the global null $\mathcal H_{\mathcal M}.$ Note that this method gives weak control on the FWER. Strong control follows under subset pivotality \cite{Westfall93}. The justification, elementary, is given in Appendix~\ref{app:multiple}.

\paragraph{Computing Tail Probabilities.}

\begin{figure}
\includegraphics[width=\textwidth]{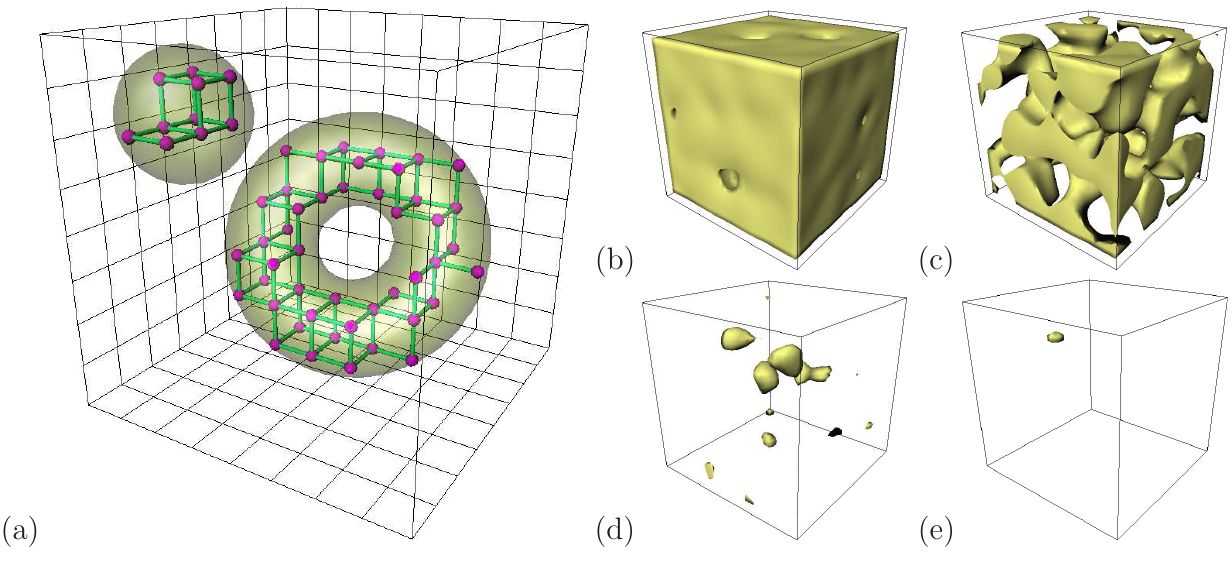}
\caption{\label{fig:Euler_characteristic} Excursion sets of $3D$ random fields, from \cite{Taylor07b}. (a) Ball and torus. The corresponding Euler characteristic is: $\chi = 2 - 1 + 0 = 1.$ (b) Isotropic Gaussian field, with zero mean and variance one, above a threshold $t = -2,$ corresponding to $\chi = 6$ (unseen hollows contribute +1 each) (c) $t = 0,$ $\chi=-6$ (handles dominate) (d) $t = 2,$ $\chi = 14$ (handles disappear) (e) $t = 3,$ $\chi = 1.$}
\end{figure}

The main difficulty in applying the maxT principle is the computation of the tail probabilities $P\left[\max_{k \in \mathcal M} T_k > u | \mathcal H_{\mathcal M}\right].$ These depend strongly on the definition of the test statistics $T_k,$ and on the distribution of the underlying data. Hence, many different approaches have been proposed, that we briefly review here. A detailed study can be found in \cite{Nichols03}.

Parametric approximations were introduced in \cite{Worsley94}. They essentially equate the tail probability $P\left[\max_{k \in \mathcal M} T_k > u | \mathcal H_{\mathcal M}\right]$ with the expectation of the Euler characteristic $\chi$ of the {\em excursion set} $\{k \in \mathcal M, T_k > u\}.$ For a 3D excursion set, $\chi$ is essentially the number of connected components, minus the number of `handles', plus the number of `hollows' (see Figure~\ref{fig:Euler_characteristic}). For a high threshold $u,$ the excursion set is expected to be either empty, or composed of a single cluster with no holes, hence the tail probability is approximately equal to $E[\chi | \mathcal M].$ This last expectation can be estimated using inequalities from random field theory (RFT), that are based on an estimation of the smoothness of the spatial map $\br T.$ In practice, these approximations yield satisfying results for smooth maps, but are overly conservative for non smooth maps, where the Bonferroni correction is optimal \cite{Nichols03}. An approach to bridge this gap is developed in \cite{Worsley05,Taylor07,Taylor07b}, using improved Bonferroni-type inequalities based on the discrete local maxima (DLM) of the statistical map. These lead to bounds on the tail probability evaluated using RFT-type approximations, that are shown to be near optimal at all smoothness levels.

The main drawback of RFT-based approximations is that they rely on heavy parametric assumptions, which are hard to verify in practice, such as the stationarity, Gaussianity and smoothness of the statistical map. They are also restricted to a certain class of test statistics, such as $t$ or $F$-statistics. Finally, their domain of validity is hard to determine, since they assume a `high' threshold $u,$ but it is not clear what this means in practice.

In the context of group data analysis, it is possible to avoid these issues by using a permutation test to tabulate the null distribution of $\max_{k \in \mathcal M} T_k$ \cite{Holmes96}. 
% This approach is detailed in Appendix~\ref{app:permutation}. 
When applicable, this approach has many advantages: It is valid under minimal assumptions concerning the distribution of the data; it can be applied to any choice of the test statistic $T_k;$ finally it is near optimal in terms of statistical power \cite{Nichols03}. Despite of all these virtues, permutation testing has not yet replaced RFT techniques as a standard for fMRI group data analysis, though both are available in SPM and FSL, the most used software packages to date for the analysis of fMRI data. This may in part be explained by the important computation time they require, and also by the limited range of questions they allow to answer. Indeed, as mentioned in Section~\ref{sec:test_group_effect}, their exists no generally exact permutation test of a partial correlation in a multiple regression model.
% fact that RFT is a core technique of SPM and FSL softwares, the most used software packages to date for the analysis of fMRI data.

\subsubsection{FDR-controlling procedures}

\begin{figure}[tb]
\begin{center}
\begin{tabular}{cccc}
\includegraphics[width=0.3\textwidth] {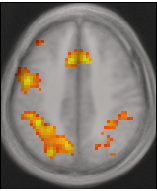} %&
\includegraphics[width=0.3\textwidth] {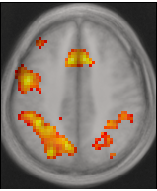} %&
\includegraphics[width=0.3\textwidth] {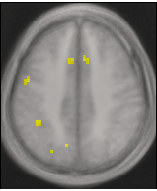} %&
\end{tabular}
\end{center}
\caption{\label{fig:voxel_level} Voxel-level error rate controls compared on the same activation map. {\em Left:} FPR controlled at $0.001$, using voxelwise tests. {\em Middle:} FDR controlled at $0.05$ using the BH procedure. {\em Right:} FWER controlled at $0.05$ using the Bonferroni procedure. The FDR control enables to detect substantially more voxels than the FWER control at the same level. However, $5\%$ of voxels detect by this approach, {\em i.e.} $118$ in this example, are expected to be false positives. In contrast, less than $60$~false positives are detected on the average using a simple voxelwise test (left), for a qualitatively similar result.}
\end{figure}

In spite of the abundant literature devoted to voxelwise FWER-controlling procedures, their application remains limited by their lack of power. This is due both to the strict control they impose on false positives, and to the large number (up to a hundred thousands) of voxels, and therefore of tested hypotheses, present in brain activation maps. More recently, there have been some attempts to control the less stringent FDR criterion instead in order to gain statistical power. This is illustrated in Figure~\ref{fig:voxel_level}, where different error rate controls are compared for a same activation map.

The most famous procedure for controlling the FDR is the Benjamini-Hochberg procedure \cite{Benjamini95}. It consists in sorting the $p$-values in increasing order: $p_{(1)} \leq \ldots \leq p_{(d)}.$ They are then compared to the linear series $\displaystyle \left(\frac{k}{d}\alpha\right)_{1 \leq k \leq d},$ where $\alpha$ is the desired FDR level. Next the following index is computed: $\displaystyle k^{\ast} = \max \{ 1 \leq k \leq d, \, p_{(k)} \leq \frac{k}{d} \alpha \}.$ If for all $k,$ $p_{(k)} > \frac{k}{d} \alpha,$ then no null hypothesis is rejected. Otherwise, all null hypotheses $\mathcal H_{0,k}$ for $k = 1, \ldots, k^{\ast}$ are rejected. \cite{Benjamini95} showed that, under independence of the $p$-values, this procedure had strong control over the FDR, {\em i.e.},
$$
FDR = E \left[ \frac{V}{R}1_{R>0} \right] \leq \frac{d_0}{d}\alpha \leq \alpha.
$$
This proof, along with the simplicity of the algorithm, greatly contributed to popularize this approach. 
% Note that independence of the $p$-values is a special case of subset pivotality.

Several works have been concerned with the use of FDR for neuroimaging data, such as \cite{Pacifico04}, which proposes an extension of of the BH approach to random fields, based on an estimation of the field's smoothness similar to the one developed in \cite{Worsley94}. In \cite{Nichols03}, was expressed the belief that FDR would soon replace FWER as a standard type I error rate in fMRI data analysis. However, relatively few applications have been published to date based on the FDR, though the BH procedure has been included in the reference SPM software. To this we see two possible explanations. The first one is that the BH procedure can be unstable with respect to the $p$-values, which is a problem when analyzing fMRI datasets, known to be very noisy. Hence, while allowing better statistical power, FDR-controlling procedures are also liable to produce more false positives, without being able to discriminate them from true positives, as illustrated in Figure~\ref{fig:voxel_level}. Another one is that, as mentioned earlier, voxels do not constitute relevant units from a cognitive point of view. In a broader sense, this may also be the main reason why voxel-level inference methods are not mainstream in neuroimaging papers concerned with the applications.

\subsection{Cluster-level inference}\label{sec:cluster}

We now turn to cluster-level multiple testing, which is the most widely used approach to detect activated regions given fMRI activation maps. Using the same notations as above, the statistical map $(T_1, \ldots, T_d)$ is first thresholded at a certain height $u.$ However, in this context $u$ does not serve to test the voxelwise hypotheses $\mathcal H_{0,k}.$ Rather, it is used to identify connected components, or {\em clusters}, from the excursion set $\{k \in \mathcal M, T_k > u\}.$ Thus it is referred to as a {\em cluster-defining threshold}, and is a fixed, user-specified quantity. The choice of $u$ is arbitrary; a popular heuristic is to tune it in order to control the voxelwise FPR at a certain level $\alpha,$ that is, allow a proportion $\alpha$ of null voxels to be retained at this stage.

Next, the presence of an activation is tested within all detected clusters $C_1, \ldots, C_M.$ More precisely, for each cluster $C_i,$ the null hypothesis: $\mathcal H_{\mathcal C_i} = \bigcap_{k \in C_i} \mathcal H_{0,k}$ is tested, against the alternative $\bar{\mathcal H}_{\mathcal C_i} = \bigcup_{k \in C_i} \mathcal H_{1,k}.$ The cluster null hypothesis is equivalent to stating that $C_i$ is enclosed in the null set $\mathcal M_0$ defined in Section~\ref{sec:generalities}:  $\mathcal H_{\mathcal C_i} = \{ C_i \subseteq \mathcal M_0 \}.$ The alternative hypothesis~$\bar{\mathcal H}_{\mathcal C_i}$ states that at least one voxel in $C_i$ is activated.

The decision statistic used to test each cluster-level hypothesis is generally the cluster size $\sharp C_i,$ though many other choices are available, as discussed in Section~\ref{sec:thresholding}. Hence, the null hypotheses $\mathcal H_{\mathcal C_i}$ are rejected for all clusters whose sizes exceed a critical value $N,$ tuned to control a certain error rate, that is, for clusters that would be `unusually large' in absence of any activation.

As in voxel-level inference, cluster-level multiple comparison procedures are usually required to have strong control over the FWER, which in this case is the probability of detecting one ore more clusters by mistake:
\begin{eqnarray}
FWER &=& P[V > 0 | \mathcal H_{\mathcal M_0}] \nonumber\\
     &=& P[\exists C_i \subseteq \mathcal M_0,\, \sharp C_i > N | \mathcal H_{\mathcal M_0}]. \nonumber
\end{eqnarray}

Following the maxT principle (see Section~\ref{sec:voxel}), this is usually done by controlling the tail probability of the maximum cluster size, under the global null $\mathcal H_{\mathcal M},$ since the true subset of null hypotheses is unknown. This means tuning $N$ so that
\begin{eqnarray}\label{eq:cluster_FWER}
P \left[ \max_{C_i} \sharp C_i > N | \mathcal H_{\mathcal M} \right] &\leq& \alpha.
\end{eqnarray}
This procedure only provides weak control on the FWER; strong control follows under subset pivotality \cite{Holmes96}. The proof of this result is given in Appendix~\ref{app:multiple}.

Computing the tail probabilities of null distribution of the the maximum cluster size $\max_{C_i} \sharp C_i$ is the principal difficulty of this approach. The principal methods for doing so are reviewed in \cite{Hayasaka03}. As previously, parametric approximations based on RFT theory are available \cite{Worsley02}, as well as exact calibration based on permutation tests, if $\br T$ is a group activation map. These have the same advantages and drawbacks as in the voxel-level setting (see Section~\ref{sec:voxel}). To summarize, though easy and fast to implement, the RFT approach relies on heavy parametric assumptions, and may be overly conservative is the statistical map is not smooth; the permutation testing approach is exact under very mild nonparametric assumption, and is always near optimal~\cite{Hayasaka03}. Its only drawback is that it is computationally intensive.

Cluster-level inference as described above has several key advantages over voxel-level inference. First, the number $M$ of clusters is much smaller than the number $d$ of voxels. Thus, testing cluster-level hypotheses greatly reduces the multiple comparison problem; this explains why cluster detection is in general more powerful than voxel detection. Another advantage is that results are reported in terms of regions rather than voxels, and are therefore easier to interpret from a cognitive point of view. The detected clusters may be related to known anatomical regions based on expert knowledge, or using a digital brain atlas such as the Automated Atlas Label (AAL) \cite{Tzourio-Mazoyer02}.

\begin{figure}[tb]
\begin{center}
\begin{tabular}{cccc}
\includegraphics[width=0.3\textwidth] {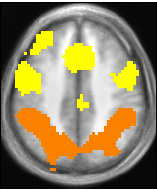} %&
\includegraphics[width=0.3\textwidth] {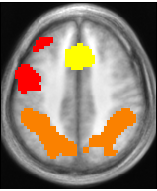} %&
\includegraphics[width=0.3\textwidth] {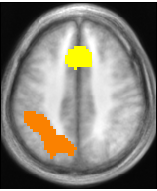} %&
\end{tabular}
\end{center}
\caption{\label{fig:clusters} Clusters detected at different cluster-forming thresholds. Axial slices $z = 37{\rm mm}$ in Talairach, represented with the subjects' mean anatomical image in the background). From left to right, the threshold is tuned to control the false positive rate (FPR) respectively at $10^{-2}$, $10^{-3}$ and $10^{-4}$ uncorrected. Each cluster surviving the FWER controlling-threshold at $5\%$ is represented with a specific color, showing how distinct functional regions are merged. In particular, left and right hemispheres are not segmented.}
\end{figure}

The principal disadvantage of cluster-level inference wish respect to voxel-level inference is its dependence on the cluster-defining threshold, which ultimately defines the detected regions. The fact that a suprathreshold cluster is found significant by the cluster size test only implies that it contains {\em some} active voxels~\cite{Hayasaka03}. Low values of the cluster-forming threshold may result in merging functionally distinct regions, thus yielding poor localization power, while high values may result in missing active regions. This is illustrated in Figure~\ref{fig:clusters}.

\section{Limits of the SPM-like approach}\label{sec:limits}

The `SPM-like' approach has been summarized in Sections~\ref{sec:single}, \ref{sec:mass} and \ref{sec:multiple}. In the following, we will assume that activations are detected at the cluster-level, as described in Section~\ref{sec:cluster}. Though simple, and widely applicable, this approach suffers from several important limitations, which we summarize here:

\paragraph{Arbitrary cluster-forming threshold.} 
As noted in Section~\ref{sec:cluster}, clusters are defined for a certain cluster-defining threshold. The choice of this threshold is arbitrary, even though it is crucial for the resulting inference.

\paragraph{Exclusive control of false positives.}
Error rates are controlled by maximizing the statistical power of the tests while limiting a certain type I error rate at a given level $\alpha$ (see Section~\ref{sec:error}). Consequently, there is no direct control on the amount of type II errors, or false negatives, meaning that the absence of activations outside the detected clusters cannot be assessed, and that there is no guarantee that the whole functional network can been recovered. This partly explains the poor reproducibility of group analyses across datasets \cite{Thirion07}.

\paragraph{Assumption of perfect match between individual brains.}
Due to unavoidable inter-subject reg\-istration errors (see Section~\ref{sec:normalisation}), the observed activations are not well-localized, and possibly displaced across distinct functional regions, which may result in blurring the group activation map and creating unhandled false positives \cite{Keller08}.
\ \\

\section{Alternative approaches}\label{sec:alternative}

Many approaches have been developed to overcome the limitations identified in the previous section, which we review now. Research on this subject has been conducted in different directions, which we have chosen to identify as follows:

To start with, many efforts have been devoted to develop new thresholding techniques to address the multiple comparison problem, both at the voxel and cluster level, and are discussed in Section~\ref{sec:thresholding}. Their goal is to overcome the limitations of the standard MCPs, such as the exclusive control over false positives, or the dependence on a cluster-forming threshold.

Next, to define potentially active regions, as a natural alternative to suprathreshold clusters one may use pre-defined regions of interest (ROIs), related to the question to be answered by the fMRI experiment. For instance, ROIs may be defined as anatomical structures that one expects to be functionally involved in a certain target task. This strategy avoids the problematic choice of a cluster-forming threshold, and provides a way to incorporate prior informations on the regions involved in the task at hand. However, the use of such fixed regions poses several challenges, which we describe in Section~\ref{sec:ROI}.

An increasingly popular approach, termed surface-based analysis, aims at improving the localization of each subject's activity. It consists in projecting for each subject the fMRI data on the cortical surface, segmented from the anatomical images. Spatial normalisation is also performed on these surfaces rather than on the whole brain volume. We describe several procedures for surface-based analysis, and discuss their pro's and con's in Section~\ref{sec:surface}.

Finally, so-called {\em feature-based} approaches represent a very fruitful line of research. They consist in extracting from the individual data certain high-level features, such as local maxima of an activation map, and comparing them across subjects. By performing the group analysis at a higher level than the voxel-level, such approaches naturally reduce the multiple comparison problem, and also provide a way to deal with spatial uncertainty. Several methods are presented in Section~\ref{sec:feature}.

\subsection{Thresholding techniques}\label{sec:thresholding}

We now review a few papers which are representative of the directions which have been explored to ameliorate standard voxel and cluster-level thresholding techniques.

\subsubsection{Joint control over false positive and false negative risks}

\paragraph{Voxel-level Bayesian inference}

Voxel-level inference methods have been proposed that control both false positive and false negative rates. In \cite{Friston02,Friston02b}, the usual voxelwise test is extended using Bayesian inference. Thus, the voxelwise $p$-values $p_k$ are replaced by the posterior probabilities $q_k$ that the mean group effect $\mu_k$ is larger than a certain baseline $b$. The method implemented in \cite{Friston02} has several drawbacks however. First, the meaning of the baseline $b$ is unclear, and so is its choice. Second, the Bayesian inference proposed in \cite{Friston02} is solely aimed at voxel detection. In particular, it provides no equivalent to the cluster-level test (see Section~\ref{sec:cluster}) which is the current standard in fMRI data analysis.

Third, the multiple comparison problem is not addressed. This seems due to a confusion between Bayesian inference and multiple testing. Indeed, it is stated in the abstract that in contrast to `conventional SPMs' ({\em i.e.}, the $p$-value maps $(p_1, \ldots, p_d)$), the posterior probability maps (PPMs) $(q_1, \ldots, q_d)$ `are not confounded by the multiple comparison problem'. However, both SPMs and PPMs address the same problem of selecting active voxels, while jointly limiting the number of false positives and false negatives. The fact that these are measured by frequentist rates (in the case of SPMs) or Bayesian risks (in the case of PPMs) does not change the fact that increasing the number of tested hypotheses mechanically increases the risk of detecting activations by chance. Thus, some form of correction for multiple comparisons must be applied in both cases.

These issues may explain why this approach is scarcely used in practice.

\paragraph{Mixture modeling}

An alternative proposed in \cite{Beckmann03b} is to fit a spatial mixture model to the voxel-wise test statistics, where null, activated and deactivated (inhibited) regions are modeled separately. A Gaussian distribution is used for null, or inactivated, voxels, a Gamma distribution for activated voxels, and a negative Gamma distribution for deactivated voxels. The same model has also been proposed in the context of individual subject fMRI data analysis, to model neural response levels (see Section~\ref{sec:limits_GLM}). A threshold can then be derived, which minimizes a certain classification risk, such as the binary risk, associated to the 0\,-1 loss function, resulting in a `naive Bayes' classifier.

This approach is revisited in \cite{Woolrich05,Woolrich06}, where the status of each voxel is further modeled according to a Markov random field. This allows to account for the spatial structure of fMRI activation patterns, where active voxels tend to be grouped in clusters rather than isolated. The same model has also been used for the modeling of fMRI time-series, as mentioned in Section~\ref{sec:glm}. Active voxels are detected using Bayesian inference, by computing a map of posterior probabilities that each voxel is active, and selecting the most probable state for each voxel, as in \cite{Beckmann03b}.

% It can be noted that other distributions could be chosen to specify this mixture model. The choice of an appropriate distribution remains an open issue, since there has been to our knowledge no assessment of the robustness of the threshold with respect to the choice of a distribution.
% This approach does not require prior tuning of an arbitrary type I error level, but in  Its main drawback is the absence of justification for the choice of the Gamma family to model activated data. Indeed, the choice of another family would lead to a different threshold value.

\subsubsection{Threshold-free cluster enhancement}

In the context of cluster-level inference, dependence on the cluster-forming threshold is addressed in \cite{Smith09}, where an original approach is developed, termed threshold-free cluster enhancement (TFCE). It consists in applying a filter to the statistical map, that has the effect of enhancing the amplitude of extended signals, while reducing less extended ones. This reduces the dependence on the cluster-defining threshold, while at the same time increasing SNR.

It must be noted though that the enhancement filter is defined by two parameters, noted $E$ and $H,$ which need to be tuned by the user, in exchange to the lessened dependence on the cluster-forming threshold. The choice of $E$ and $H$ is however extensively studied in \cite{Smith09}, and the robustness of the approach with respect to this tuning is demonstrated in practice over a wide range of datasets.

When used in association with a multiple testing procedure, another potential shortcoming of TFCE is that the resulting statistical map does not verify subset pivotality (see Appendix~\ref{sec:pivotality} for a precise definition). In other terms, the value of the enhanced statistic in each voxel is not independent from the statistics in other voxels. As a consequence, only weak control is guaranteed for the resulting procedure, but not strong control (see Appendix~\ref{app:multiple} for formal definitions of weak and strong control, and justifications of these assertions). This means that the absence of bias due to TFCE, though supported by simulations, has yet to be demonstrated mathematically.

\subsubsection{Alternative choices of cluster-level summary statistic}

Finally, the choice of a decision statistic for testing cluster-level hypotheses remains a point of debate. The widespread use of the cluster size has often been criticized, as it naturally favors spatially extended signals, and may result in missing small activations, irrespective of their intensity. Thus, several approaches have been developed to combine cluster statistics. In \cite{Poline97b} the cluster size and cluster peak (maximum statistic value) are combined through multivariate rejection regions. In \cite{Hayasaka04b}, a wider panel of possible cluster summary statistics are compared on several datasets, using combining functions. Though promising, these approaches have not been shown to produce significantly different results from the conventional cluster size test. The core problem seems to be that each cluster summary statistic defines a test that is optimally powerful for a certain type of signal (extended, peaked, etc.). Combining different statistics does not produce a more powerful test, but rather changes the type of signal that is detected.

To date, there is to our knowledge no cluster-level inference approach that controls false negative clusters, {\em i.e.}, controls the risk of missing an activated cluster.

\subsection{ROI-Based Analysis}\label{sec:ROI}

\subsubsection{ROIs in fMRI single-subject data analysis}

The use of regions of Interest (ROIs) to segment the brain volume is common in single-subject fMRI data analysis \cite{Poldrack07}. They are often defined as anatomical structures, extracted from the T1-weighted image of the subject under study, or as clusters detected as active from a previous fMRI experiment. fMRI paradigms specifically designed to define ROIs are known as {\em localizers}. They usually consist of simple stimuli, meant to elicit a response from a known functional region, in order to identify its location for the subject under study.

In both cases, ROIs are an attractive alternative to suprathreshold clusters for the definition of candidate activated regions; they avoid the troublesome choice of cluster-defining threshold, and provide a way to include prior information on the regions involved in the investigated task. When defined anatomically, they also provide a statistically sound way to assess the link between brain structure and function, which is one of the fundamental questions fMRI studies aspire to answer.

\subsubsection{ROIs in fMRI group data analysis}

\begin{figure}
\includegraphics[width=\textwidth]{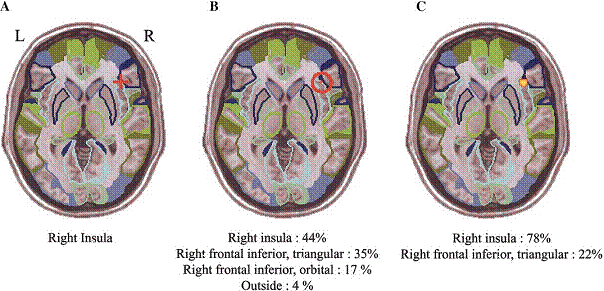}
\caption{\label{fig:AAL} Three procedures to perform the automated anatomical labeling of functional activations, proposed in \cite{Tzourio-Mazoyer02} (the outline of the AAL parcellation is overlaid on the z=5.0 mm axial slice of the single-subject anatomical image used to define it): (A) Local maxima labeling: the red cross indicates the location of the local maximum. (B) Extended local maxima labeling: percentage of overlap between a 10-mm sphere radius centered on the local maximum and the different parcels. (C) Cluster labeling: percentage of overlap between the activation cluster and the different parcels.}
\end{figure}

In spite of these advantages, ROIs are scarcely used as a basis for inference in the context of fMRI group data analysis. Rather, clusters detected as active are empirically compared to anatomical ROIs, usually in the form of an atlas of brain regions, such as the single subject Automated Atlas Labels (AAL) \cite{Tzourio-Mazoyer02}, in a post-treatment step, as illustrated in Figure~\ref{fig:AAL}. This raises several questions, because active voxels cannot be localized within an active cluster (see Section~\ref{sec:cluster}). Thus, when a cluster extends over several atlas regions, there is no guarantee that any of these regions contain any active voxels. Moreover, there is no single way of performing this comparison, so several heuristics are traditionally used, with potentially different answers.

In fact, it seems rather surprising to us that, after all the efforts devoted to develop statistically well-defined activation detection procedures, the crucial step wherein these activations are identified with anatomical structures is performed using post-hoc methods, producing results which cannot be validated or refuted on an objective basis.

A notable exception is found in \cite{Bowman08}, which uses AAL to divide the brain volume in regions assumed to represent distinct functional units. Based on this parcellation, a Bayesian hierarchical approach is developed to identify regions whose functional responses are statistically correlated, a phenomenon known as {\em functional connectivity}. Thus, parameters at the regional level are modeled as a Gaussian vector with arbitrary covariance structure. By estimating the covariance matrix, functional connectivity may be assessed between any couple of ROIs, for the specific task investigated. Moreover, this framework could also be used to test regionwise hypotheses, as in the SPM-like approach, even though this is not done in \cite{Bowman08}.

However, the use of ROIs for fMRI group analysis poses a series of challenges, which may explain why they have not yet been adopted as a standard tool. As noted in \cite{Poldrack07}, a major issue is the important variability of anatomical structures accross subjects, only crudely compensated by the normalisation step (see Section~\ref{sec:normalisation}). Thus, anatomical ROIs are hard to define at the group level, and individual images are never aligned with any given set of ROIs. This may result in false positives, since activations are likely to be displaced accross regions.

\begin{figure}[ht]
\centering
\includegraphics[width=0.5\textwidth]{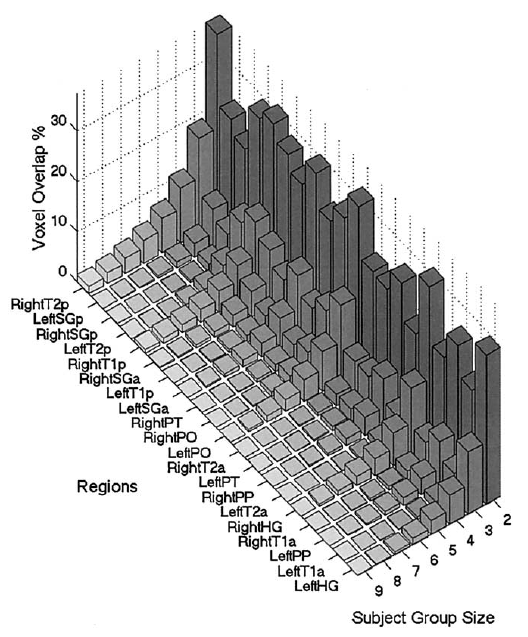}
\caption{\label{fig:ROI_overlap} Mean overlap of individual anatomical ROIs across different subject group sizes in \cite{Nieto03}.}
\end{figure}

This point of view is developed in \cite{Nieto03}. It is shown that the overlap of individual anatomical ROIs across subjects vanishes quickly as the group size increases (see \ref{fig:ROI_overlap}). This observation is used to motivate an alternative analysis of multi-subject neuroimaging data, based on individual ROIs rather than an anatomical atlas. Briefly summarized, the functional data is analyzed separately for each subject-specific ROI. For each ROI, the corresponding summary statistics are compared across subjects. Note that this approach does not require inter-subject registration, since each subject's functional data is analyzed in reference to its own anatomy.

However, as discussed in \cite{Poldrack07}, measuring the functional signal of a single subject within each given ROI is a challenging task. For each subject, there may be only a small proportion of voxels activated inside a given region, so that averaging the signal, as seems to be done in \cite{Nieto03}, may swamp the signal with the noise from the many inactive voxels. In contrast, standard mass-univariate methods gain statistical power by averaging the data across subjects in each voxel. Thus, they may be able to recover activations which would otherwise be lost at the single-subject level.

Another issue in terms of group modeling is that individual anatomical structures may not be present in all subjects. This must be taken into account if one wants to perform a RFX analysis, {\em i.e.} generalize the findings on the cohort of subjects under study to their parent population.

These shortcomings may explain why this methodology has only been demonstrated so far in a fixed-effect (FFX) setting. Though \cite{Nieto03} claim that random-effect (RFX) analyses would yield `similar results', there may be both a sensitivity and a modeling issue in the generalization of this approach.

\subsection{Surface-based analysis}\label{sec:surface}

\subsubsection{Motivation}

The SPM-like approach searches for BOLD activations throughout the whole brain volume. However, neurons are mainly concentrated near the cortical surface (though they are also present in subcortical structures). Moreover, there is evidence supporting the assertion that cortical landmarks, such as sulci and gyri, correspond to boundaries between functionally distinct regions \cite{Brett02}.

Consequently, volume-based detection may be sub-optimally sensitive because statistical power is wasted by searching for activations in the wrong places, such as the white matter. Also, volume-based registration may fail to match cortical structures, which may in part explain the bad overlap between homologous functional regions. Moreover, distinct functional areas which are well separated along the cortical surface may be close in terms of the euclidean distance, due to the highly convoluted topology of the cortical folds. Hence, They may be hard to separate in a fMRI contrast map, whose lower spatial resolution ($3{\rm mm}^2$ against up to $1{\rm mm}^2$) cannot account for such fine details.

Based on these observations, the analysis of functional neuroimaging data based on the cortical surface rather than the entire brain volume has received an increasing attention over the last decade. The goal of such surface-based techniques is to obtain a more precise localization of individual functional areas, and improve their matching across subjects, based on the assumption that functional regions are better identified on the cortical surface by the associated anatomical landmarks than in the 3D Talairach coordinate system.

\subsubsection{Surface-based registration}

\begin{figure}[ht!]
\includegraphics[width=\textwidth]{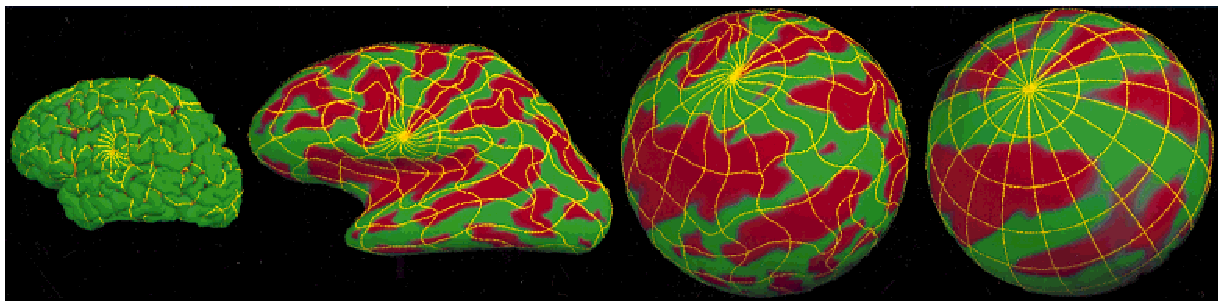}
\caption{\label{fig:fischl} Unfolding of the cortical surface onto a sphere (in \cite{Fischl99b}).}
\end{figure}

A method for surface-based registration is presented in \cite{Fischl99b}. For each individual, the procedure starts by reconstructing the cortical surface from a structural MRI image. This surface is then inflated, and transformed into a sphere using a registration algorithm that minimizes metric distortions, as illustrated in Figure~\ref{fig:fischl}. In this fashion, the folding pattern of the subject's cortex is mapped using a polar coordinate system, so that to each point corresponds a measure $C$ of curvature, with negative and positive values indicating gyral and sulcal regions respectively. This folding pattern is then aligned with a canonical, average folding pattern (Figure~\ref{fig:fischl}, to the right).

The advantages of this approach with respect to volume-based registration are demonstrated in \cite{Fischl99b}: The overlap of cortical structures, such as the central sulcus, are greatly improved, and sensitivity in the detection of fMRI activation is observed. However, no clue is provided regarding how the fMRI data is projected on the cortical surface, though this step is bound to be a challenge in this type of analysis.

\begin{figure}
\includegraphics[width=\textwidth]{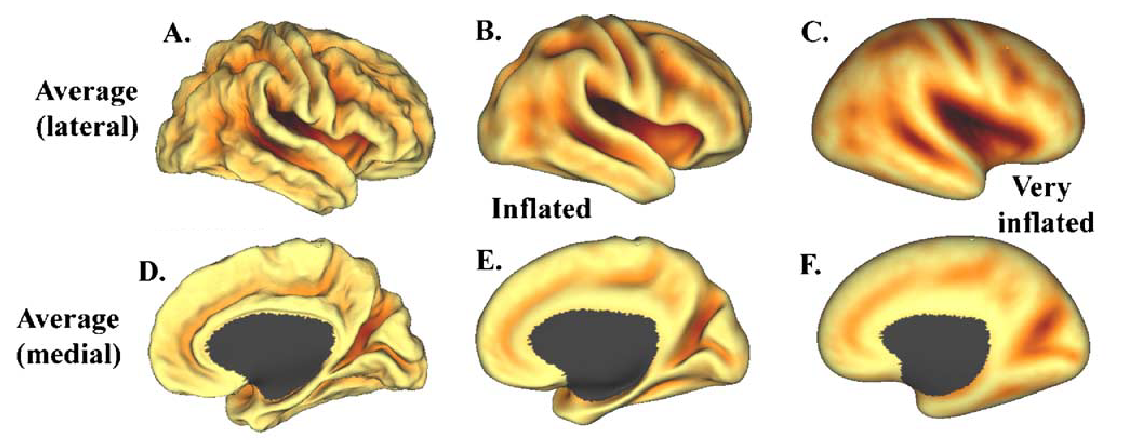}
\caption{\label{fig:VanEssen} Average cortical surfaces of 12 subjects, registered to the atlas in \cite{VanEssen05}. A – C. Lateral views of the average surface, both with and without inflation. D – F. Similar to A – C, but showing the medial views. The average surface preserves the main features of the sulcal anatomy, showing a clear improvement over volume-based registration using affine transformations (see Section~\ref{sec:normalisation}).}
\end{figure}

A similar approach is developed in \cite{VanEssen05}, involving a flattened representation of the cortical surface, which is mapped onto a sphere and matched to an atlas of the cortical folding pattern. However, a more sophisticated atlas is constructed here, obtained by matching anatomical landmarks extracted from each individual's cortex. As in \cite{Fischl99b}, this approach significantly improves the alignment of cortical structues (see Figure~\ref{fig:VanEssen}). The issue of mapping fMRI data to the cortical surface is addressed by simply assigning to each node of the mesh representing the cortex the voxel containing it, though \cite{VanEssen05} mentions $7$ (at least) other algorithms that could be used instead.

\subsubsection{Projection of fMRI data onto the cortical surface}

Mapping the fMRI data onto the cortical surface thus seems to be one of the main difficulties of surface-based analysis. Indeed, because of the low resolution of fMRI data, it is impossible to assign each voxel to a particular point of the cortical surface. In practice, no reference projection method exists, and the choice of a particular strategy may influence the results. A recent development on this subject can be found in \cite{Operto08}, wherein fMRI data is projected onto the cortical surface, using interpolation kernels informed by the subject's anatomy. This approach is shown to be more robust to registration errors than other, less sophisticated approaches. However, the authors agree that matching of the fMRI data volume with the cortical ribbon remains largely an open issue, due mainly to the low resolution of current functional images, and the ensuing partial volume effects.

In conclusion, surface-based analysis of fMRI data seems a promising alternative to volume-based approaches, by providing a better match between the cortical structures of different subjects, and consequently increasing the overlap of homologous functional regions. However, this increased anatomical precision seems to come at the cost of a degraded match between functional and anatomical data. Thus, in both cases the uncertainty on the localization of individual activations persists, and should be taken into account when conducting inference at the group level.

\subsection{Feature-based approaches}\label{sec:feature}

As mentioned in Section~\ref{sec:surface}, the SPM-like approach detects activations throughout the whole brain volume, by comparing the individual images in a voxelwise fashion. The motivation behind surface-based analysis is that voxels are not relevant entities from a physiological point of view.

This criticism of voxel-based comparisons leads to a broader class of methods, which we refer to as {\em feature-based}. The principle they rely on consists in extracting high-level features from the individual data, and matching them across subjects. By choosing relevant features that summarize the information contained in the subject's data, these approaches may be expected to provide results that are easier to interpret, and gain statistical power by reducing the number of multiple comparison. 
% Furthermore, these features may include spatial coordinates that can be used to model the spatial variability of the inferred activation pattern, thus providing a way to address registration uncertainties.
For instance, the ROI-based analysis of group fMRI data developed in \cite{Nieto03} is feature-based, the features being the summary statistics measuring the BOLD response for each subject-specific ROI.

The following examples have been chosen to illustrate the diversity of feature-based approaches. These methods are best classified according to the choice of a feature set.

\subsubsection{Functionally homogeneous parcels}

\begin{figure}
\includegraphics[width=\textwidth]{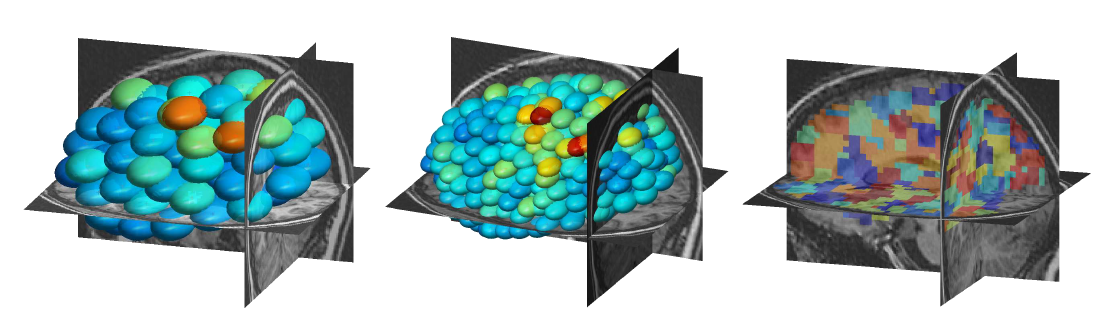}
\caption{\label{fig:anatomo_functional_parcels} Examples of anatomo-functional parcellations for a grasping task in \cite{Flandin04}. From left to right: mixture model with $100$ Gaussian components, $500$ components, corresponding maximum {\em a posteriori} parcellation.}
\end{figure}

The idea of parcelling the brain volume into parcels that are homogeneous both anatomically and functionally is investigated in \cite{Flandin04}. A first approach is developed, based on the anatomical data of a single subject, to compute a Voronoi tesselation of the search volume into $K$ connected components. These tesselations are then used to reduce the dimensionality of the subject's fMRI data, by averaging the time courses (see Section~\ref{sec:glm}) within each parcel. The $K$ average time-courses are then used as input for the SPM-like approach, instead of the $d$ original time-courses.

In particular, the presence of activation is tested at the parcel rather than the voxel level, using the methodology presented in Section~\ref{sec:voxel}. As a consequence, the multiple comparison problem is reduced, and the greater sensitivity of this approach is demonstrated compared to the standard voxelwise detection procedure, even though the data is spatially smoothed in the last case, another way of reducing the data's resolution.

Going one step further, \cite{Flandin04} then generalizes this approach by introducing a parcellation into clusters containing voxels with similar BOLD responses, based on both the anatomical and functional MRI data of the subject under study. Thus, in addition to reducing the data's dimension, the proposed procedure also defines units that are relevant in terms of neuroscience, since they may be interpreted as functionally homogeneous areas.

Finally, this approach is extended to the context of fMRI group data analysis in the following way. The individual images are first normalized to a given template, to ensure proximity between homologous functional areas across subjects (perfect alignement of the images is not assumed however). Then, the fMRI time-courses of the different subjects are concatenated and analyzed as if they belonged to a single subject, using the anatomo-functional parcellation described above. This creates group-level clusters regrouping voxels with similar fMRI time-courses. A nice feature of this approach is that it implicitly defines clusters at the subject level that are automatically matched, clusters from two distinct subjects being homologous if they belong to the same group-level clusters.

This approach provides an elegant way to deal with inter-subject anatomical variability, but has several limitations. Indeed, because it includes no inter-subject model, this approach can only characterize the group of subjects under study, {\em i.e.} perform a fixed-effect analysis. Additionally, single-subject clusters are not necessarily spatially connected, making them more difficult to interpret as functional units. Moreover, there is a risk that clusters defined according to some measure of fMRI time-course similarity may be influenced by the main effect studied in the experiment, which in general will dominate the fMRI signal. Hence, such a clustering may not work well when analyzing more subtle effects (such as interactions between different experimental factors).

These shortcomings are addressed in \cite{Thirion06f}, which revisits the approach in \cite{Flandin04} by adding a group level to the model defining the subject parcels. According to this model, subject parcels are no longer forced to be subsets of group-level parcels, or {\em cliques}. The relaxation of this constraint allows the definition of spatially connected parcels, such that all subjects are represented in each clique. Furthermore, the data is clustered according to the vector of estimated effects rather than the raw fMRI time-courses, which means that it is no more dominated by the main effect.

As in \cite{Flandin04}, presence of activity is tested for each clique~$C,$ by averaging for each subject~$i$ the map of estimated effects over the subject-cluster homologous to $C$ and performing a $t$-test on the resulting values. Such an approach is shown to outperform conventional voxelwise tests. There is some concern however about the control this approach offers on false positives. Indeed, because subject clusters and cliques are estimated from the whole dataset, they cannot be considered spatially independent. Thus, there is no guarantee that the vector of clique-level $t$-statistics verifies the subset pivotality, hence that this procedure has strong control over the false positive risk considered.

In a recent development based on similar ideas \cite{Tucholka08b}, the same type of multi-subject anatomo-functional parcellations is derived, combined to a surface-based approach (see Section~\ref{sec:surface}). More precisely, A first segmentation of the cortical surface into gyri defines distinct anatomical regions. Each region is then segmented into a certain number of cliques, represented for each subject by a parcellation into an equal number of clusters. The total number of cliques is optimized using a cross-validation criterion, providing an insight into the number of functionally distinct regions involved in the fMRI paradigm under study.

\subsubsection{Spatial mixture modeling}

Following the same idea of grouping voxels with similar responses that is at the core of clustering approaches, several mixture models have been proposed for the statistical analysis of fMRI data, with the specific goal of detecting activated regions. This idea was first pursued in \cite{Vaever-Hartvig00}, which considered representing the BOLD activation pattern as a superposition of Gaussian shaped `bumps', in addition to a uniform background level. The number of bumps was determined automatically, using a stochastic geometry model. More recently, \cite{Penny03b} defined a mixture model to represent the map of BOLD effects, each component corresponding to an activated cluster, modeled using a similar Gaussian shaped response surface, plus an additive noise. This idea was further extended in \cite{kim06b}, which used an infinite Bayesian mixture model, based on a Dirichlet process prior, to deal with the problem of selecting the number of components.

The above works are all concerned with single-subject fMRI data analysis, and use a parametric surface response to model activation patterns. In a recent work, \cite{Xu09} uses a similar Bayesian spatial mixture modeling approach for fMRI group data analysis, using different levels of hierarchy to model individual activation centers and population-level activation centers. The spatial variability of individual centers around the population centers is explicitly modeled, allowing to account for mis-registrations.
%  Moreover, the parametric assumptions concerning the shape of the response surface are relaxed, and voxels belonging to a common cluster are modeled as i.i.d. Gaussian variables. 
An original feature of this work is that it adopts a Bayesian model averaging point of view, integrating out the number of mixture components, rather then estimating it and characterizing each cluster. Model averaging is implemented through a reversible-jump MCMC algorithm.

\subsubsection{Critical points of the individual activation maps}

\begin{figure}
\centering
\includegraphics[width=0.75\textwidth]{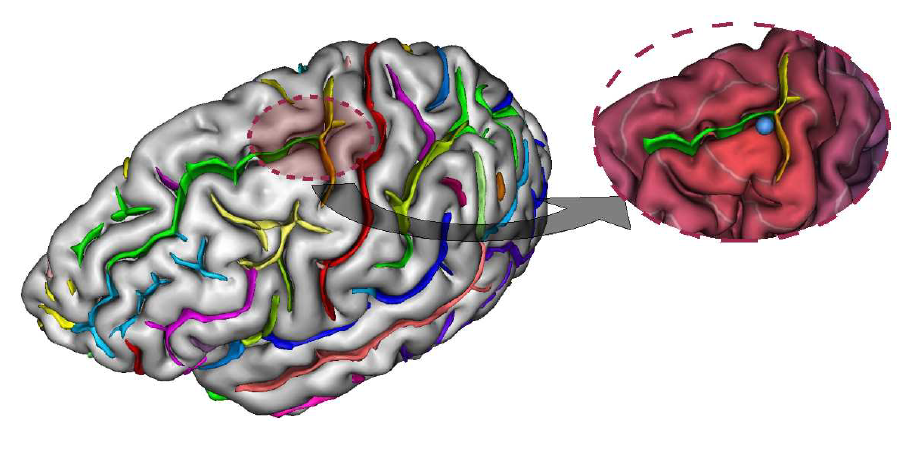}
\caption{\label{fig:triangulation} Within labelled sulci framework, the activation landmark of computation (blue ball on the picture) in the prefrontal lobe is localized near the intersection of the three sulci: superior precentral, marginal precentral and superior frontal (from \cite{Tucholka08}).}

\end{figure}

While in the previous approaches the individual features are defined, and possibly estimated, as latent variables of a certain model describing the data, it is also possible to directly extract them as {\em e.g.} salient features of the activation maps of the individual subjects, and build a group model from these. This is exactly what is done in \cite{Thirion07c}, where features are defined as the critical points of the individual activation maps, such as local maxima and minima, and the associated level sets. These are used to represent individual activation centers, and are modeled using a Bayesian spatial mixture model similar to those described above. In particular, a Dirichlet process prior is used to define an infinite mixture model to deal with the estimation of the number of classes, as in \cite{kim06b}. Each class defines a separate group-level activity cluster, and the set of such clusters provides a description of the activation pattern at the population level. These group-level clusters are shown to be more reproducible across datasets than standard clusters obtained with the standard SPM-like analysis.

Along similar lines, \cite{Tucholka08} propose an original approach to localize the local maxima extracted from the individual activation maps of a group of subjects with known anatomical landmarks, using triangulation techniques. The local maxima are first matched across subjects, and viewed as functional landmarks. Then, for each subject, the distance of each landmark to three user-chosen neighbouring sulci (see Figure~\ref{fig:triangulation}) is computed, then averaged over subjects. Given a new subject, the position of the homologous functional landmark is then predicted by triangulation, based on these average distances. This is compared with the position predicted by the average coordinates in the standard space, and shown to be much more accurate. This nice result shows that brain anatomy and function are linked, and suggests that the standard coordinate system may not be optimal to localize functional areas.

\subsubsection{Scale-space blobs}

\begin{figure}
\includegraphics[width=\textwidth]{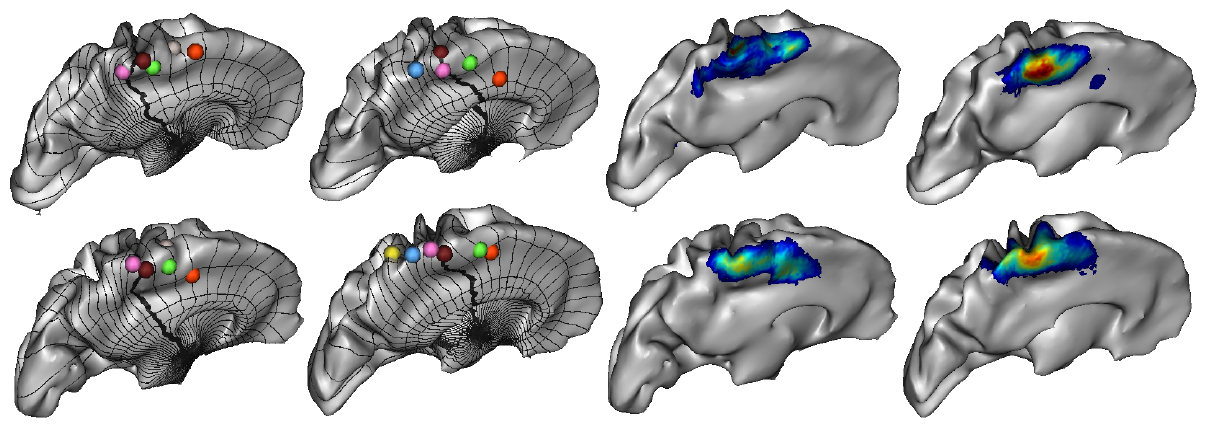}
\caption{\label{fig:scale_space_blobs} Reproducible activations from a right-motor contrast for 4 subjects, in \cite{Gregory08b}. (left) labeled after maximization (right) thresholded individual statistical $t$-maps ($p < 0.05$ corrected).}
\end{figure}

We conclude this review of feature-based approaches with the fMRI group data analysis method proposed initially in \cite{Coulon01} for the analysis of 3D activation maps and revisited in \cite{Gregory08b} in the setting of surface-based analysis. In the latter, the individual fMRI data are first projected on the cortical surface, as explained and for the reasons exposed in Section~\ref{sec:surface}. The critical points are then extracted from the individual $t$-maps as in \cite{Thirion07c,Tucholka08}. The key point of this approach however is the scale-space representation of these features it then constructs. This adds to the spatial structure linking the critical points another level of hierarchy, obtained by considering different levels of spatial smoothing. This description allows to account for patterns that appear at different levels of resolution, based on the idea that objects of interest happen at many different scales.

This complex structure is coded for each subject by a graph, whose nodes are referred to as {\em scale space blobs}, and represent the features of interest. Group analysis is performed by embedding these graphs in an encompassing graph, and matching the blobs across subjects. The optimal matching is found by optimizing an objective function which accounts for the proximity of each pair of matched blobs, and the likelihood that these blobs correspond to an active brain area. This is performed in a Bayesian framework, the field of labels coding matches between blobs being modeled as hidden Markov field, that is estimated by {\em a posteriori} maximization.

The final representation of the functional reponse of the group of subjects consists in a list of labels representing activations that are reproducible across subjects, and for each subject a set of scale-space blobs representing instances of these activations, as illustrated in Figure~\ref{fig:scale_space_blobs}.

As in \cite{Flandin04} and \cite{Thirion06}, this creates a `structural' description of the group of subjects, with informations that may be exploited both at the group and at the subject level. Results are however not generalizable to the general population, so that this approach can be categorized as performing a fixed effect analysis.

\section{Conclusion}\label{sec:conclusion1}

The review we have just presented, though far from exhaustive, suggests the wide variety of approaches that have been developed for the statistical analysis of fMRI data during the last decades, to address the limitations of the `SPM-like' approach identified in Section~\ref{sec:limits}: dependence on an arbitrary cluster-defining threshold, exclusive control of false positives, and the assumption of a perfect match between the different brains. However, to date these issues have been addresses separately. 

The goal of the present work is to propose a new procedure for fMRI group data analysis, addressing them jointly. To position ourselves with respect to the existing methods, we start by discussing their similarities and divergences. We have found three key points that summarize this comparison:

\begin{itemize}

\item First, all the described approaches agree on one point, namely that group inference should be performed at a higher level than the single voxel. To quote \cite{Gregory08b}: ``The voxels are only the acquisition space, and have never had any anatomical meaning other than the simple localization provided by spatial normalization.''

Thus, the group activation pattern is always described in terms of higher-level features, such as suprathreshold clusters, ROIs, group activation centers, parcels or multiscale blobs.

\item Secondly, most of the methods presented here rely on preliminary spatial normalization to align subjects, using either volume or surface-based registration. A notable exception is \cite{Nieto03}, which relies solely on the anatomical structures extracted in each subject to compare them. It is generally accepted that perfect alignment of the anatomies is never attained in practice (see Section~\ref{sec:normalisation}). In spite of this, to date feature-based approaches have been the only ones to account for the resulting spatial variability of the activation patterns, through explicit modeling.

\item Finally, apart from the ROI analysis framework presented in Section~\ref{sec:ROI}, none of the proposed approaches provide an assessment of the link between detected activations and anatomical brain structures, though this question is at the core of most fMRI studies. An exception is \cite{Tucholka08}, which shows that activation foci are better localized with respect to neighbouring sulci than than in the standard coordinate system (see Section~\ref{sec:feature}).

\end{itemize}

Based on our review and these final remarks, it seems to us that the ROI framework has a certain number of key advantages, in relation to the issues we want to address. It is a simple and flexible tool that allows to define regions of potential activity at the group level based on expert knowledge rather than an arbitrary threshold. Furthermore, joint control on false positive and false negative risks is not only possible, but conceptually simple in this setting, by adopting a Bayesian framework. Moreover, as shown in \cite{Bowman08}, it also provides a means of investigating the functional connectivity of distant brain regions.

Feature-based approaches also constitute an appealing starting point for our work, as they have demonstrated their ability to account for imperfect matches between subjects. However, we feel that using a high-level description of the single subjects estimated effects maps would not necessarily help us attain our goal. The main reason for that is that we are not interested in describing each subject, but rather the functional network involved in the task at hand, in a reproducible way across subjects. Furthermore, we have no idea how to define these features, since there is an infinity of possible choices, that may all be at the expense of discarding valuable information. Thus it appears to us that adopting a feature-based approach in our case would only add un-necessary complexity.

% To start with, how do we choose the relevant features that summarize the information of each subject? There are infinite possiblities, and each choice may be at the expense of discarding useful information.

% Moreover, performing group analysis based on high-level features is at least as difficult as at the voxel level, since we face the exact same issues: individual features are not necessarily aligned across subjects, and specifying a model to perform random effect analysis on these objects can be quite difficult, as can be judged from the examples presented in Section~\ref{sec:feature}.

% More fundamentaly, we are not interested in a high-level description of each subject, but rather of the functional network involved in the task at hand, and that is reproducible across subjects. With this in mind, it appears that, though we want to model some form of spatial variability, adopting a feature-based approach without a precise idea of the choice of a feature would only add un-necessary complexity.

Thus, we propose to extend the current state-of-the-art ROI analysis methodology, by relaxing the assumption that the individual images are in perfect match on a voxelwise basis. This can be done by explicitly modeling spatial variability, as is common in feature-based procedures, but this time at the voxel-level. 
\cleardoublepage
$ $ \eject\pagestyle{fancy}
% \input{Chapter3_revised}
% \chapter{Advances on Thresholding Techniques}\label{sec:advances}
% 
% \section{Permutation Tests for Multiple Comparisons}
% 
% \section{Generalizing Cluster Definition}
% \section{Multivariate Rejection Regions}
% \section{Illustration on Real Data}

\chapter{Random thresholds for linear model selection, revisited. Application to fMRI data analysis.}\label{chap:unsupervised}
\lhead{\emph{Random thresholds for linear model selection, revisited.}}

% Work submitted to {\em The Canadian Journal of Statistics}

\section*{Abstract}

In \cite{Lavielle07}, a random thresholding method is introduced to select the significant, or non null, mean terms among a collection of independent random variables, and applied to the problem of recovering the significant coefficients in non ordered model selection. We introduce a simple modification which removes the dependency of the proposed estimator on a window parameter while maintaining its asymptotic properties. A simulation study suggests that the modified estimator performs better at low signal to noise ratios, where the original one is unstable with respect to the window parameter. An application of the method to the problem of activation detection on functional magnetic resonance imaging (fMRI) data is discussed.

\section{Introduction}

A popular approach to activation detection in fMRI data analysis consists in thresholding a statistical map of brain activity \cite{Friston97}.

The choice of a detection threshold is usually addressed by controlling at a user-fixed level a certain type I error rate, such as the family wise error rate (FWER) \cite{Nichols03} or the false discovery rate (FDR) \cite{Pacifico04, Benjamini95}. It can be argued however that the choice of a level, which ultimately defines the detected regions, is arbitrary, as their is no safe guideline to what an `optimal' level of false detections should be.

% An alternative is to control both type I and type II error rates, for instance by fitting a spatial mixture model to the data, where null, activated and deactivated (inhibited) regions are modeled separately. The Gamma-Gaussian mixture model (GGM) in \cite{Beckmann03b} is most often used in this context. It uses a Gaussian distribution for null, or inactivated, data, a Gamma distribution for activated data, and a negative Gamma distribution for deactivated data. A threshold can then be derived, which minimizes a certain classification risk, such as the binary risk, associated to the 0\,-1 loss function, resulting in a `naive Bayes' classifier.

% Although this last approach does not require prior tuning of an arbitrary error level, there is no justification for the choice of the Gamma family to model activated data, and the choice of another family would lead to a different threshold value. So the quest for a default, or objective, threshold, relying on as minimal assumptions as possible, is still on.

We consider here as an alternative the random threshold method recently developed in \cite{Lavielle07}, which consists in estimating the number of non null mean terms among a collection of independent random variables. It is based on a random centering of the partial sums of the ordered observations.
% , and requires parametric assumptions on the null data only, contrary to the GGM approach above. 
Consistency of the proposed estimator can be shown, using L-statistics techniques.

Though requiring no prior tuning of a type I error rate, this method still depends on a user-chosen window width, and has so far been demonstrated on high signal to noise (SNR) simulated data only, whereas fMRI data is known to be very noisy. This paper describes a simple modification of the procedure which makes it totally unsupervised; the modification simply consists in replacing the fixed window width by a varying one.

The article is organized as follows: In Section~\ref{sec:method}, we review the random threshold method in \cite{Lavielle07}, and introduce the varying window extension. In Section~\ref{sec:simu_rand_thresh}, a simulation study compares the random threshold procedure to the standard type I error rate control methods described above. Application to fMRI data analysis is discussed in Section~\ref{sec:fmri}.

\section{Method}\label{sec:method}

Assume we observe $y_i = \mu_i + \epsilon_i$, where variables $\epsilon_i$ are centered, independent and identically distributed with common cumulative distribution function (cdf) $F_{\epsilon}$.

\subsection{Original random threshold procedure}

% \newline

\subsubsection{Testing the presence of significant coefficients}\label{sec:test_coefficients}

We start by recalling the procedure for testing if all the $\mu_i$'s are null or not, given the observations ${y_i;\ 1 \leq i \leq n}$,  that is, testing the null hypothesis $\mathbf{H_0} : \mu_i \equiv 0\ \mathrm{for}\ i = 1, \ldots, n$ against the alternative $\mathbf{H_1} : \exists I \subseteq \{1, \ldots, n\}\ |\ \forall i \in I, \mu_i > 0$. The procedure is defined as follows :

\begin{itemize}

\item[i)] Order the observations $|y_{(1)}| \geq |y_{(2)}| \geq \ldots \geq |y_{(n)}|$.

\item[ii)] For $i = 1, \ldots, n$, let $X_{(i)} = - \log(1 - F_{|\epsilon|}(|y_{(i)}|))$.

\item[iii)] Let $T_j = \sum_{i= 1}^n X_{(i)}$ and $Q_j = \mathbb{E}_{\mathbf{H_0}} (T_j | T_n)$.

\item[iv)] Define the test statistic $D_n = \max_j |T_j - Q_j| / \sqrt{n}$. The null hypothesis is rejected if $D_n > d_{\alpha}$, where $d_{\alpha}$ ensures that the level of the test is at most $\alpha$.

\end{itemize}

The conditional expectation $\mathbb{E}_{\mathbf{H_0}} (T_j | T_n)$ is easily computed due to the fact that under the null hypothesis, $(X_{(i)})_{1 \leq i \leq n}$ is an ordered sequence of exponential random variables, and to the following result, whose proof is 
% in \cite{Lavielle07}:
omitted here:

\begin{Proposition}\label{prop:cond_mean}

Assume $X_{(1)}, \ldots, X_{(n)}$ is an ordered sequence of Exp(1) random variables, with $X_{(1)} \geq \ldots \geq X_{(n)}.$ For any $1 \leq j \leq n,$ let $T_j = \sum_{i= 1}^n X_{(i)}.$ Then, for any $1 \leq j \leq K \leq n$:

\begin{eqnarray}
\mathbb{E}_{\mathbf{H_0}} (X_{(i)}) & = & \sum_{\ell=i}^n \frac{1}{\ell} \nonumber\\
\mathbb{E}_{\mathbf{H_0}} (T_j) & = & j + j \sum_{\ell=j}^n \frac{1}{\ell} \nonumber\\
\mathbb{E}_{\mathbf{H_0}} (T_j | T_n) & = & \frac{\mathbb{E}_{\mathbf{H_0}} (T_j)}{\mathbb{E}_{\mathbf{H_0}} (T_n)} T_n\nonumber.
\end{eqnarray}

\end{Proposition}

\subsubsection{Selecting the significant coefficients}\label{sec:procedure}

Upon rejection of the null hypothesis, the following task consists in selecting the significant coefficients. The procedure 
% in \cite{Lavielle07}
 for doing so can be interpreted in a data-dependent `multiple hypothesis testing' setting, as described hereafter. Consider the null hypothesis $\mathbf{H_0}$ as defined in Section~\ref{sec:test_coefficients}, and the set of alternative hypotheses:

\begin{itemize}
\item[] $\mathbf{H_1(k)} :$ for any $i \leq k,\ \mu_{(i)} > 0,$ and $\mu_{(k+1)} = \ldots = \mu_{(n)} = 0.$
\end{itemize}

Denote $\mathbb{E}_k$ the expectation under $\mathbf{H_1(k)}$ (instead of $\mathbb{E}_{\mathbf{H_1(k)}}$). The procedure first computes the $X_{(i)}$'s using the same steps i) and ii) as in Section~\ref{sec:test_coefficients}, then adds the following steps:

\begin{itemize}

\item[iii)] Let $K_n$ be some positive integer. For $1 \leq k \leq n - K_n$ and $1 \leq j \leq K_n,$ compute:
\begin{eqnarray}
T_{k,j} & = & \sum_{i = k + 1}^{k + j} X_{(i)} \nonumber\\
Q_{k,j} & = & \mathbb{E}_k (T_{k,j} | T_{k,K_n}) \nonumber\\
\eta_k & = & \max_{1 \leq j \leq K_n} |T_{k,j} - Q_{k,j}| / \nonumber\sqrt{n}.
\end{eqnarray}

\item[iv)] Let $\hat{k}_n = \mathrm{argmin}_{1 \leq k \leq K_n} \eta_k.$

\end{itemize}

As in the preceding section, $Q_{k,j}$ can easily be computed using Proposition~\ref{prop:cond_mean}. $\eta_k$ can also be defined from the centered partial sums $(T_{k,j} - Q_{k,j})$ using the $\ell_p$ norm for $1 \leq p < \infty$ rather than the $\ell_{\infty}$ norm, by setting $\eta_k = n^{-p/2 - 1} \sum_{j = 1}^{K_n} |T_{k,j} - Q_{k,j}|^{p}.$

\subsubsection{Unknown distribution extension}\label{sec:unknown}

The above procedure can be extended to the case where the distribution $F_{\epsilon}$ of the $\epsilon_i$'s is a parametric distribution $F_{\epsilon}(\cdot\, ;\, \theta^{\star}),$ but where $\theta^{\star}$ is unknown. %\cite{Lavielle07}.

For $0 \leq k \leq n-1,$ let $\hat{\theta}_k = \hat{\theta}(y_{k+1}, \ldots, y_n)$ be an estimator of $\theta.$ Let $F_{|\epsilon|}(\cdot\, ;\, \theta^{\star})$ be the distribution of the $|\epsilon_i|$'s. For any $\theta \in \Theta,$ let $X_i(\theta) = -\log \left( 1 - F_{|\epsilon|}(|y_{(i)}\, ;\, \theta|) \right)$ and $T_{k,j}(\theta) = \sum_{i = k+1}^{k + j} X_{(i)}(\theta).$ Then the following procedure is defined :
%  in \cite{Lavielle07}:

\begin{itemize}
\item[i)] Let $K_n \leq [(1 - b)n]$ be some positive integer. For $[a\, n] \leq k \leq n - K_n:$
    \begin{enumerate}
    \item let $\hat{\theta}_k = \hat{\theta}(y_{k+1}, \ldots, y_n),$
    \item for $i = 1, \ldots, n,$ let $X_{(i)}(\hat{\theta}_k) = -\log \left( 1 - F_{|\epsilon|}(|y_{(i)}|\, ;\, \hat{\theta}_k) \right),$
    \item for $1 \leq j \leq K_n,$ compute
    \begin{eqnarray}
    T_{k,j}(\hat{\theta}_k) & = & \sum_{i = k + 1}^{k + j} X_{(i)}(\hat{\theta}_k) \nonumber\\
    Q_{k,j}(\hat{\theta}_k) & = & \mathbb{E}_k (T_{k,j}(\hat{\theta}_k) | T_{k,K_n}(\hat{\theta}_k)) \nonumber\\
    \eta_k(\hat{\theta}_k) & = & \max_{1 \leq j \leq K_n} |T_{k,j}(\hat{\theta}_k) - Q_{k,j}(\hat{\theta}_k)| / \sqrt{n}.\nonumber
    \end{eqnarray}
    \end{enumerate}
\item[ii)] Let $\hat{k}_n = \mathrm{argmin}_{a\, n \leq k \leq b\, n} \eta_k(\hat{\theta}_k).$
\end{itemize}

In the applications presented in Section~\ref{sec:simu_rand_thresh} and Section~\ref{sec:fmri}, we consider Gaussian noise: $F_{\epsilon}(\cdot\, ;\, \sigma^2) = \mathcal{N}(\cdot\, ;\, 0, \sigma^2)$ and estimate $\theta = \sigma^2$ by the usual mean squares estimator: $\hat{\theta}_k = \frac{1}{n - k} \sum_{i = k + 1}^n y_{(i)}^2.$

\subsection{Varying window extension}\label{sec:varying}

The procedures defined in the preceding sections depend on a parameter $K_n$ which can be interpreted as a window width, since $\eta_k$ is a function of $X_{(k + 1)}, \ldots, X_{(k + K_n)}.$ Tuning this parameter may be difficult, as seen in Section~\ref{sec:simu_rand_thresh}. This issue can be avoided by re-defining $\eta_k$ as a function of $X_{(k + 1)}, \ldots, X_{(n)},$ thus replacing the fixed width $K_n$ by a varying width $n - k,$ which requires no prior tuning. We define the following procedure:

\begin{itemize}

\item[iii)] Let $\kappa_n$ be a lower bound on the number of null coefficients. For $1 \leq k \leq n - \kappa_n$ and $1 \leq j \leq n - k,$ compute:
\begin{equation}\label{eq:procedure}
\begin{array}{ccc}
T_{k,j} & = & \sum_{i = k + 1}^{k + j} X_{(i)} \\
Q_{k,j} & = & \mathbb{E}_k (T_{k,j} | T_{k,n - k}) \\
\eta_k & = & \max_{1 \leq j \leq n - k} |T_{k,j} - Q_{k,j}| / \sqrt{n - k}.
\end{array}
\end{equation}

\item[iv)] Let $\hat{k}_n = \mathrm{argmin}_{1 \leq k \leq n - \kappa_n} \eta_k.$

\end{itemize}

In other terms, $\eta_k$ would be strictly equal to the test statistic $D_n$ defined in Section~\ref{sec:test_coefficients}, if the sequence $(X_{(i)})_{1 \leq i \leq n}$ where replaced by the subsequence $(X_{(i)})_{k + 1 \leq i \leq n},$ \emph{i.e.}, the null set of variables under $\mathbf{H_1(k)}.$ As in the previous section, $\ell_p$ norms can be used instead of the $\ell_{\infty}$ norm by setting $\eta_k = (n - k)^{-p/2 - 1} \sum_{j = 1}^{n - k} |T_{k,j} - Q_{k,j}|^{p}.$

Notice that $\hat{k}_n$ is independent of $\kappa_n,$ as long as $\eta$ reaches its global minimum on $\{1, \ldots, n - \kappa_n\}.$ %$K_n$ may thus be interpreted in this new setting as the minimal number of $X_{(i)}$'s required to compute $\eta_k.$
It is also immediate that the varying window extension, which we present here for the procedure in Section~\ref{sec:procedure}, can be applied to the unknown distribution extension in Section~\ref{sec:unknown}.

\subsubsection{Asymptotic properties}\label{sec:consistency}

The estimator of the number of significant coefficients presented in Section~\ref{sec:procedure} is consistent. This is the main result in \cite{Lavielle07}, and it can be extended to the varying window setting. We start by recalling the following asymptotic framework:
% in~\cite{Lavielle07}:

\begin{itemize}
\item[{\bf AF1}] There exists $t^{\star} \in (0, 1)$ and a subset $I_{k_n^{\star}}$ of $\{1, \ldots, n\},$ with $k_n^{\star} = [t^{\star} n]$ and $|I_{k_n^{\star}}| = k_n^{\star},$ such that $\mu_i \neq 0$ if $i \in I_{k_n^{\star}}.$ For all other index, $\mu_i = 0.$
\item[{\bf AF2}] For any $i \in I_{k_n^{\star}},$ $|\mu_i| \geq \alpha_n,$ for a certain sequence $(\alpha_n)$ with $\alpha_n \to \infty$ (see \cite{Lavielle07} for further details).
\item[{\bf AF3}] $\kappa_n / n \to c$ such that $0 < c < 1 - t^{\star}.$
\end{itemize}

We then have the following result:

\begin{Theorem}\label{theo:consistency_rand_thresh}

Let $\hat{k}_n$ stand for the estimator defined in Section~\ref{sec:varying}. Under assumptions {\bf AF1}, {\bf AF2}, {\bf AF3}, and appropriate Von-Mises type conditions on the cdf $F_{\epsilon}$ of the errors (see \cite{Lavielle07} for further details), $\hat{k}_n$ is consistent in the sense that:

\begin{equation}
\label{eq:consistency}
P \bigg( \bigg| \frac{\hat{k}_n}{n} - t^{\star} \bigg| > u_n \bigg) \to 0,
\end{equation}

for any positive decreasing sequence $(u_n)$ such that $\sqrt{n} u_n \to \infty.$

\end{Theorem}

This result can be refined by deriving an upper bound, which we do not detail here, on the convergence rate of the probability in Equation~(\ref{eq:consistency}), for a particular choice of sequence $(u_n).$ The proof of this Theorem is given in Appendix~\ref{app:proof}.

Consistency also holds in the unknown distribution case (see Section~\ref{sec:unknown}). Consider the following assumptions on the cdf $F_{|\epsilon|}$ of the $|\epsilon_i|$'s:

\begin{itemize}
\item[{\bf F1.}] $F_{|\epsilon|}$ is two times differentiable as a function of $\theta$ with \textit{a.e.} strictly positive derivative at $\theta = \theta^{\star}.$
\item[{\bf F2.}] $\theta^{\star}$ belongs to some compact set $\Theta$ and there exists under $H_{k_n^{\star}}$ a consistent estimator $\hat{\theta_{k_n^{\star}}} = \hat{\theta}(y_{k_n^{\star}+1}, \ldots, y_n)$ of $\theta^{\star}.$
\item[{\bf F3.}] There exists $(a,b)$ such that $0 < a < t^{\star} < b < 1$ and a Lipschitz continuous function $\tilde{\theta}$ defined on $[a, b]$ such that, under $H_{k_n^{\star}},$ $(\hat{\theta}_{[tn]})$ converges in  probability uniformly on $[a,b]$ to $\tilde{\theta}(t).$
\end{itemize}

Then we have the following result:
% , whose proof can be found in \cite{Lavielle07}:

\begin{Theorem}\label{theo:consistency_rand_thresh2}
Let $\hat{k}_n$ stand for the estimator defined in Section~\ref{sec:varying}. Assume {\bf F1}, {\bf F2} and {\bf F3}. Let $(u_n)$ be any positive decreasing sequence  such that $\sqrt{n} u_n \to \infty.$ Under the asymptotic framework defined by {\bf AF1}, {\bf AF2}, {\bf AF3},
\begin{equation}
% \label{eq:consistency2}
P_{H_1(k_n^{\star})} \bigg( \bigg| \frac{\hat{k}_n}{n} - t^{\star} \bigg| > u_n \bigg) \to 0.\nonumber
\end{equation}
\end{Theorem}

The proof of this result in the varying window setting follows the same lines as that of Theorem~\ref{theo:consistency_rand_thresh}, and is not detailed here.

\begin{figure}[hb!]
\centering
\includegraphics[width=0.75\textwidth]{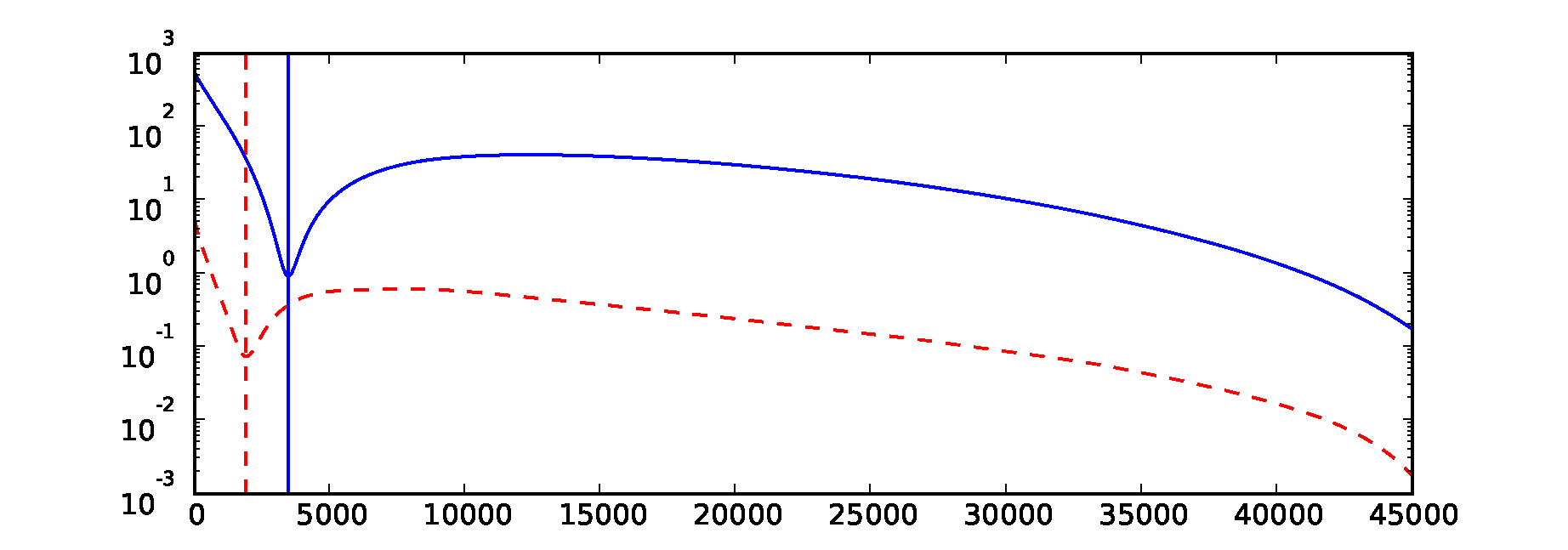}
\caption{\label{fig:simu_cuves} \bf The sequence $\eta_k$ for $0 \leq \mu_i \leq 4$} 
{\em Dashed line:} fixed window width $K_n = 5\, 000;$ {\em Solid line:} varying window width.
\end{figure}

\section{Simulations}\label{sec:simu_rand_thresh}

\subsection{Experiment summary}

\begin{table}[ht!]
\begin{center}
\caption{\label{tab:risk} \bf 
Number of misclassified voxels (false positives and false negatives) for different thresholding methods.} 
% \begin{tabular}{|c||r|r||r||r|}
\begin{tabular}{|c|rrrr|}
\hline
 & \multicolumn{1}{c}{$K_n = 15\, 000$}
 & \multicolumn{1}{c}{$K_n = 35\, 000$}
 & \multicolumn{1}{c}{var. window}
 & \multicolumn{1}{c|}{GGM} \\
\hline
\vspace{-5mm}
% \hline
$0 \leq \mu_i \leq 4$ &     7477 $\pm$ 193 &  7148 $\pm$ 228 & 7120 $\pm$ 237 & 11016 $\pm$  \ \,666 \\
\vspace{-5mm}
% \hline
$1 \leq \mu_i \leq 5$ &    5079 $\pm$ 243 & 4736 $\pm$ 250 & 4706 $\pm$ 255 & 8306 $\pm$ 1862 \\
\vspace{-5mm}
% \hline
$2 \leq \mu_i \leq 6$ &    2561 $\pm$ 185 & 2385 $\pm$ 182 & 2381 $\pm$ 184 & 1901 $\pm$  \ \,125 \\
\vspace{-5mm}
% \hline
$3 \leq \mu_i \leq 7$ &    930 $\pm$ 118 &  892 $\pm$ 120 &  891 $\pm$ 120 &  766 $\pm$  \ \,102 \\
% \vspace{-5mm}
% \hline
$4 \leq \mu_i \leq 8$ &    297 $\pm$ \ \,55 &   298 $\pm$ \ \,68 &   297 $\pm$ \ \,67 &   262 $\pm$ \ \,\ \,52 \\
\hline
\end{tabular}\\
\end{center}
Results are obtained over 100 replications, and given in the form: mean $\pm$ std. deviate.
\end{table}

In this experiment, we have repeatedly simulated $n = 50\, 000$ Gaussian random variables with $\mu_i$ distributed uniformly on $[a, b]$ for $1 \leq i \leq 10\, 000,$ and $\mu_i = 0$ for $10\, 001 \leq i \leq 50\, 000.$ $(\epsilon_i)$ is a sample from the $\mathcal{N}(0, 1)$ distribution.

$(a,b)$ was chosen successively equal to $(0, 4),\ (1, 5),\ (2, 6),\ (3, 7),\ (4, 8).$ $100$ datasets where simulated for each of these values. On each simulated dataset, we then used the procedure described in Section~\ref{sec:unknown} (the unknown distribution extension), assuming a Gaussian cdf for $F_{|\epsilon|}$ with unknown variance, and compared the results for different window widths $K_n = 15\,000,\ 35\,000.$ We also used the varying window extension, as described in Section~\ref{sec:varying}, applied to the unknown variance setting, and the Gamma-Gaussian mixture (GGM) modeling procedure in \cite{Beckmann03b}.

Our choice of simulation parameters was guided by the fact that $50\, 000$ is roughly the number of voxels in a whole-brain fMRI activation map. $8$ can be seen as an upper bound on the maximal signal intensity, since in the real data we analyzed (see Section~\ref{sec:fmri}), the maximum $z$-score was equal to $6.82$ for the individual subject activation map, and $5.08$ for the between-subject activation map.

\begin{figure}[ht!]
\begin{center}
\includegraphics[width=0.75\textwidth]{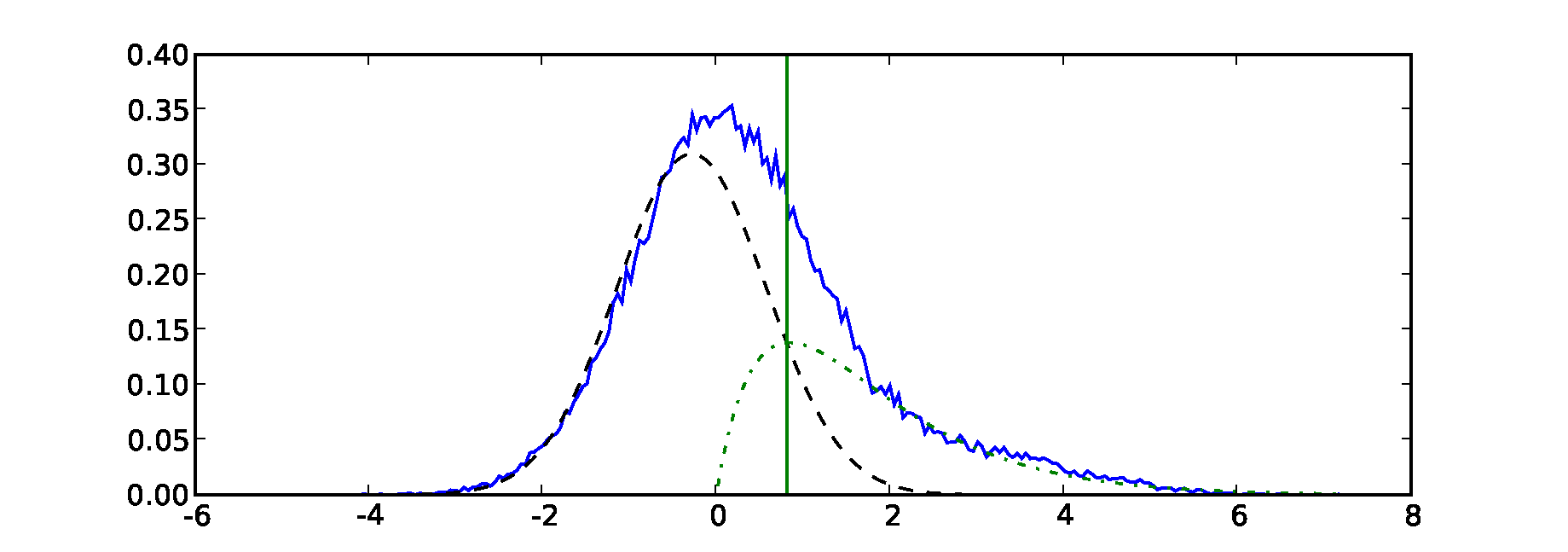} \\
\includegraphics[width=0.75\textwidth]{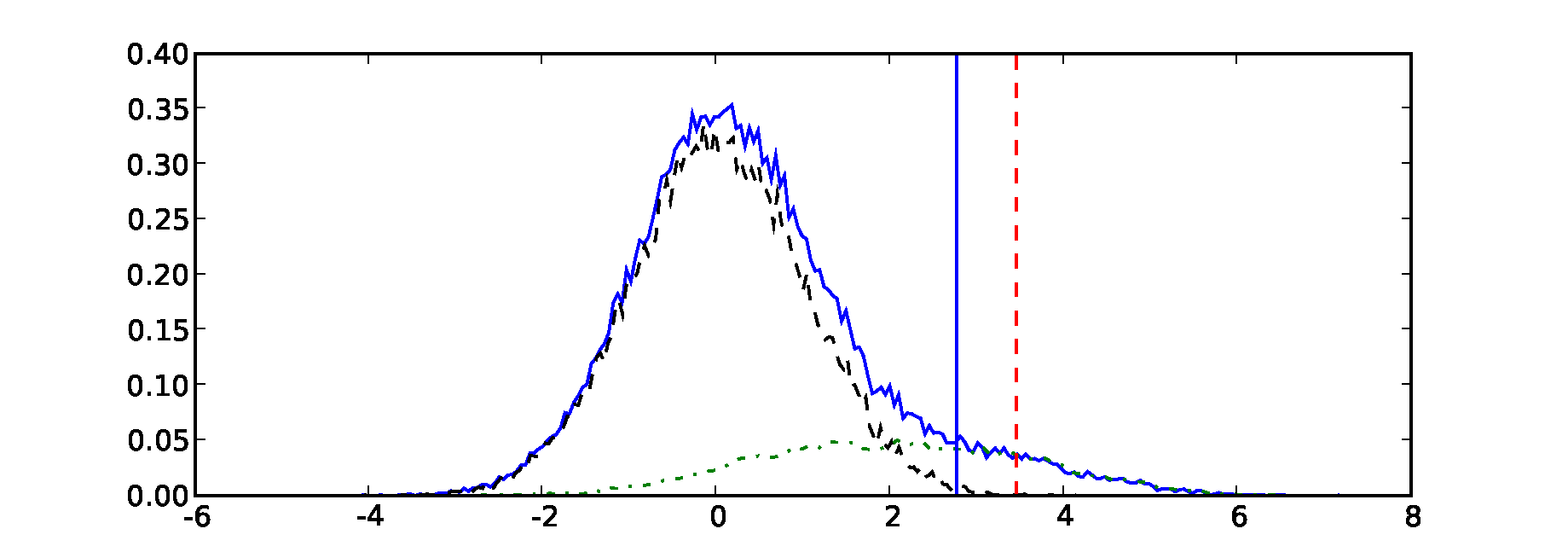}
\end{center}
\caption{\label{fig:simu_pdf} \bf Different thresholding strategies for $0 \leq \mu_i \leq 4.$}
\emph{Top}: Gamma-Gaussian mixture fit (solid curve: data, dashes: Gaussian, dash-dots: Gamma). \emph{Bottom}: Random thresholds with $K_n = 5\, 000$ (dashed line) and varying $K_n$ (solid line). Data corresponds to the solid curve, null data to the dashes and non null data to dash-dots.
\end{figure}

\subsection{Results and discussion}

\begin{table}[ht!]
\centering
\caption{\label{tab:fpr} 
\bf False positive rates for different thresholding methods. }
\begin{center}
\begin{tabular}{|c|rrrr|}
\hline
 & \multicolumn{1}{c}{$K_n = 15\, 000$}
 & \multicolumn{1}{c}{$K_n = 35\, 000$}
 & \multicolumn{1}{c}{var. window}
 & \multicolumn{1}{c|}{GGM} \\
\hline
\vspace{-5mm}
% \hline
$0 \leq \mu_i \leq 4$ & 0.001 $\pm$ 0.001 & 0.002 $\pm$ 0.001 & 0.002 $\pm$ 0.001 & 0.219 $\pm$ 0.019 \\
\vspace{-5mm}
% \hline
$1 \leq \mu_i \leq 5$ & 0.001 $\pm$ 0.001 & 0.002 $\pm$ 0.001 & 0.002 $\pm$ 0.001 & 0.182 $\pm$ 0.062 \\
\vspace{-5mm}
% \hline
$2 \leq \mu_i \leq 6$ & 0.002 $\pm$ 0.001 & 0.003 $\pm$ 0.001 & 0.003 $\pm$ 0.001 & 0.009 $\pm$ 0.003 \\
\vspace{-5mm}
% \hline
$3 \leq \mu_i \leq 7$ & 0.002 $\pm$ 0.001 & 0.002 $\pm$ 0.001 & 0.002 $\pm$ 0.001 & 0.004 $\pm$ 0.001 \\
% \hline
$4 \leq \mu_i \leq 8$ & 0.001 $\pm$ 0.001 & 0.001 $\pm$ 0.001 & 0.001 $\pm$ 0.001 & 0.002 $\pm$ 0.001 \\
\hline
\end{tabular}\\
\end{center}
Results are obtained over 100 replications, and given in the form: mean $\pm$ std. deviate.
\end{table}

We compared the different approaches through the error rates they achieved on the different datasets. We considered the binary risk, \textit{i.e.}, the number of misclassified voxels, as a measure of how well the null and non null sets where separated (see Table~\ref{tab:risk}). We also computed the false positive rate (FPR) (see Table~\ref{tab:fpr}).

As could be expected, differences between methods were most observed when the data was simulated with a low signal-to-noise ratio (SNR), (first and second line of each table).
In this case, the Gamma-Gaussian mixture (GGM) model is clearly misspecified since it cannot account for negative observations in the non-null set, as illustrated in Figure~\ref{fig:simu_pdf}. This leads to risk and false positive rate (FPR) values much higher than those obtained by the other thresholding methods.

For higher SNRs, the different approaches gave similar results, even though the GGM model yielded slightly lower risks than the different random thresholds, at the cost of higher FPR values.

On the whole, the different random thresholding procedures gave very similar results in all cases, and yielded mean FPR values that were systematically below $0.003,$ and also much lower than those found by GGM fit.

In conclusion, this experiment suggests that the random thresholding approach provides a very good control on false positives, and yields reasonable risk values at all SNRs, with a slight advantage for the varying window extension. In contrast, the GGM fit appears to yield more false positives, and performs poorly at low SNRs.

% 
% Interestingly, the different versions of the random thresholding procedure are seen to systematically under-estimate the number of significant coefficients. Also, the fixed-window procedure is seen to be unstable with respect to the window width, as the downward bias increases for small values of $K_n,$ especially at low SNR. Finally, the use of a varying window, which avoids the instability issues, also seems to compensate for the bias, since it systematically recovered more significant coefficients than the fixed-window versions.
% 
% In conclusion, the varying-window random threshold appears in this study to be more stable at low SNR than both the fixed-window random threshold, which is dependent on $K_n,$ and the threshold derived by GGM modeling, which behaves erratically in presence of too many negative observations in the non-null subset.

\section{fMRI data}\label{sec:fmri}

We used an event-related fMRI protocol involving a
% relatively large
cohort of 37 right-handed subjects.
% group of 15 right-handed subjects.
The participants were presented
with a series of stimuli or were engaged in tasks such as passive
viewing of horizontal or vertical checkerboards, left or right click
after audio or video instruction, computation (subtraction) after
video or audio instruction, sentence listening and reading. Events
occurred randomly in time (mean inter stimulus interval: 3s), with ten
occurrences per event type, and ten event types in total.

The subjects gave informed consent and the protocol was approved by
the local ethics committee. Functional images were acquired on a
General Electric Signa 1.5T scanner using an Echo Planar Imaging
sequence (time of repetition = $2400$~ms, time to echo = $60$~ms,
matrix size = $64 \times 64$, field of view = $24$~cm$^2$). Each
volume consisted of $34$ $64 \times 64$ $3$~mm-thick axial contiguous slices. A
session comprised $130$ scans. Anatomical T1 weighted images were
acquired on the same scanner, with a spatial resolution of
$1\times1\times1.2$~mm$^3$. Finally, the cognitive performance of the
subjects was checked using a battery of syntactic and computation
tasks.

First-level analyses were conducted using SPM5 
\href{http://www.fil.ion.ucl.ac.uk}{http://www.fil.ion.ucl.ac.uk}. 
Data were submitted successively to motion correction, slice timing 
and normalization to the~MNI template. For each subject, BOLD contrast 
images were obtained from a fixed-effect analysis on all sessions. 
Group analyses were restricted to the intersection of all subjects' 
whole-brain masks, comprising $43\, 367$~voxels.

We considered the $t$-score maps computed for different contrasts of experimental conditions.
These were first converted to $z$-score maps, to obtain approximatively
Gaussian statistics in inactivated voxels. Using these maps as input data, we then compared, as in the simulation study, the detection thresholds obtained by Gamma-Gaussian mixture modeling (GGM), fixed-window random thresholding and the varying-window extension using in the last two cases the unknown variance extension of the method (see Section~\ref{sec:unknown}). For simplicity, we only present here the results obtained for a fixed window equal to $K_n = 15\,000.$

\begin{figure}
\centering
\begin{minipage}{1.0\textwidth}
\includegraphics[width=0.23\textwidth]{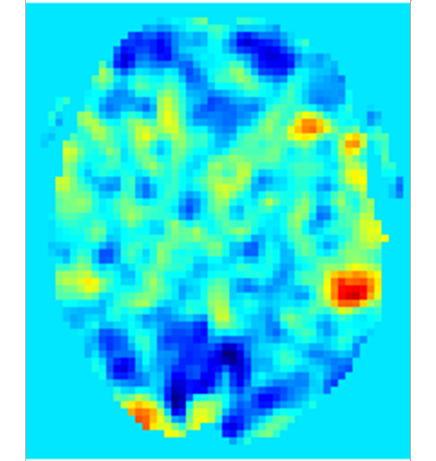}\hfill
\includegraphics[width=0.23\textwidth]{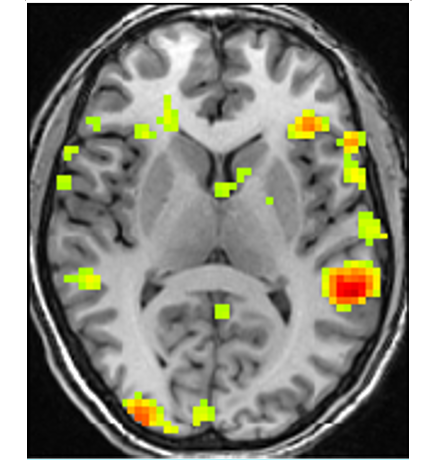}\hfill
\includegraphics[width=0.23\textwidth]{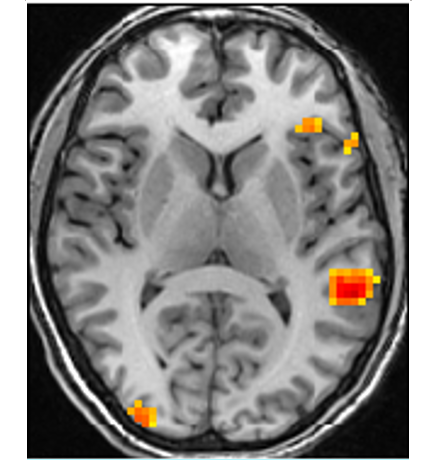}\hfill
\includegraphics[width=0.23\textwidth]{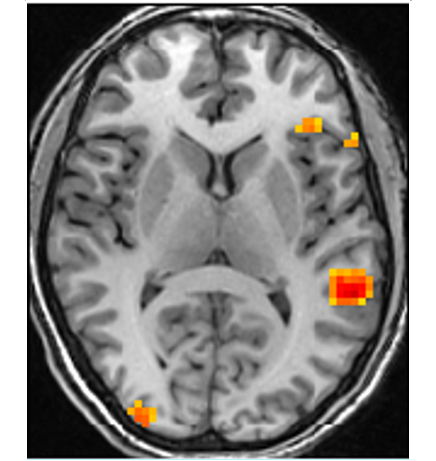}
\end{minipage}
\caption{\label{fig:intra_maps} \bf Axial slice from a $z$-score map for the "sentence--checkerboard" contrast.} From left to right: Unthresholded, thresholded by GGM model fit, varying-window and fixed-window ($K_n = 15\, 000$) random thresholding. Detected activations are superimposed on the subject's anatomical image.
\end{figure}

\subsection{Individual subject activation map}

Our first illustration concerns the activation map of a single subject, for the ``sentence--checkerboard'' contrast. This contrast subtracts the effect of viewing horizontal and vertical checkerboards from that of reading video instructions, thus allowing to detect brain regions specifically implicated in the reading task.

% \begin{figure}
% % \begin{tabular}{c}
% \centering
% \includegraphics[width=0.75\textwidth, height=0.33\textwidth]{sujet10_con_31_curves.png}
% % \end{tabular}
% \caption{\label{fig:intra_curves} The sequence $\eta_k$ using fixed window width $K_n = 15\, 000$ (dashed line) and a varying window width (solid line), on an individual activation map.}
% \end{figure}

Figure~\ref{fig:intra_maps}, left, shows an axial slice from the $z$-score map before thresholding. Activations are clearly seen in Wernicke's and Broca's areas (right and upper right), which are known to be involved in language processing (see \cite{Price00}, for instance). The detection threshold found by GGM fit for the $z$-score map ($2.03$) is much lower than those found by the random threshold procedure, both with a varying window ($3.19$) and a fixed window ($3.33$). We note that these thresholds follow the same order found in the simulation study.

The random thresholds with fixed and variable windows yield very similar activation maps in this case, which seem to capture the activated regions seen in the raw map. In contrast, the much lower threshold found by mixture modeling detects several smaller clusters, some of which may be false positives.

\subsection{Group activation map}

In this second example, we consider a group activation map, specifically a map of $t$-statistics computed from the
individual contrast maps of 15 subjects, thus enabling to infer regions of positive mean effects in the parent population.
Our choice of limiting the number of subjects, rather than using the whole cohort, was driven by the fact that many fMRI
studies are conducted on groups of less then 20 subjects. The remaining subjects were used
to assess the variability of the thresholds found by the different approaches with respect to the choice of the subgroup,
as described in Section~\ref{sec:reproducibility}.

We report results from the ``calculation--sentences'' contrast, which
subtracts activations due to reading or hearing instructions from the
overall activations detected during the mental calculation tasks. This
contrast may thus reveal regions that are specifically involved in the
processing of numbers.

% \begin{figure}
% % \begin{tabular}{c}
% \centering
% \includegraphics[width=0.75\textwidth, height=0.33\textwidth]{con29_curves.png}
% % \end{tabular}
% \caption{\label{fig:inter_curves} The sequence $\eta_k$ using fixed window width $K_n = 15\, 000$ (dashed line) and a varying window width (solid line), on inter-subject fMRI data.}
% \end{figure}

Figure~\ref{fig:intra_maps}, left, shows an axial slice from the activation map before thresholding, with clear activations in the bilateral anterior cingulate (upper middle), bilateral parietal (lower left and right) and right precentral (upper right) regions, all known to be involved in number processing, as explained in \cite{Pinel07}.

Though sorted in the same order as previously, the varying window random threshold ($2.49$) is now roughly at equal distances from the threshold found by GGM modeling ($1.79$) and the fixed window random threshold ($3.06$).

The three methods detected activations in the regions described above, though the fixed window random threshold seemed to miss some
activations, and the GGM approach further detected smaller clusters, some of which may be false positives.

\begin{figure}
\centering
\begin{minipage}{1.0\textwidth}
\includegraphics[width=0.23\textwidth]{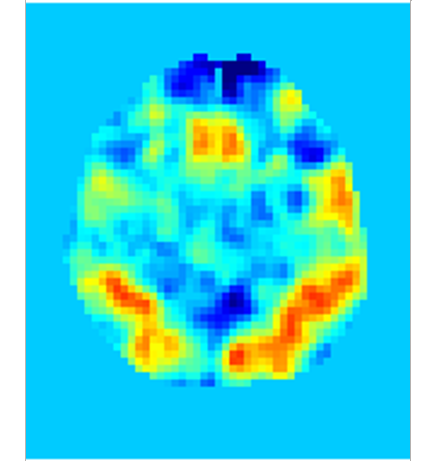}\hfill
\includegraphics[width=0.23\textwidth]{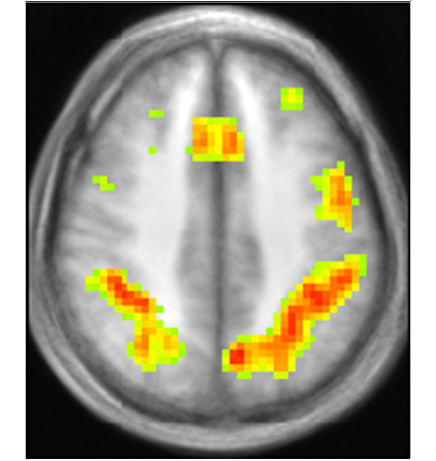}\hfill
\includegraphics[width=0.23\textwidth]{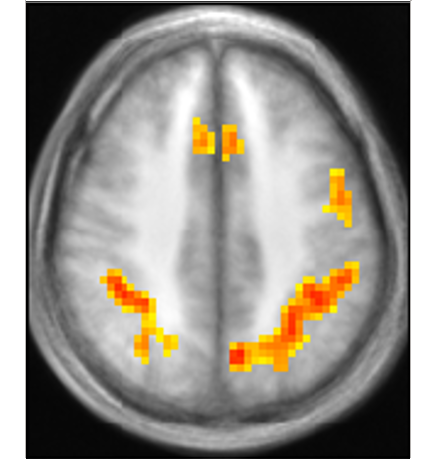}\hfill
\includegraphics[width=0.23\textwidth]{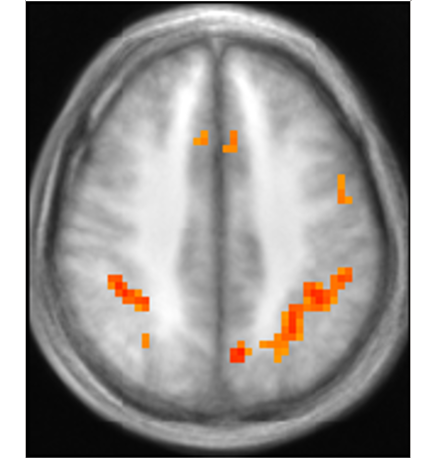}
\end{minipage}
\caption{\label{fig:inter_maps} \bf Axial slice from the group activation $z$-score map for the "calculation--sentence" contrast.} From left to right: Unthresholded, thresholded by GGM model fit, varying-window and fixed-window ($K_n = 15\, 000$) random thresholding. Detected activations are superimposed on the mean anatomical image of all subjects.
\end{figure}

\subsection{Reproducibility study}
\label{sec:reproducibility}

In both experiments described above, it seemed that the varying window threshold found a good compromise between the GGM fit, which selected possibly inactive voxels, and the fixed window random threshold, which seemed to be overly conservative. However, we cannot conclude from these observations alone that the varying window threshold is `better' than the other thresholds.

To compare them on a more objective basis, we studied the variability of the different thresholds with respect to the choice of a subgroup. More precisely, we repeatedly and randomly selected a group of 15 distinct subjects from the cohort of 38, computed the corresponding group activation map for the computation task, and thresholded it by all three methods. We then computed the empirical mean and variance of each sample of threshold values. These provided unbiased estimates for each threshold's mean and variance, with respect to the data's sampling distribution:

% we selected 15 distinct subjects randomly from the cohort of 38 involved in the experiment, and thresholded the corresponding group activation maps by all three methods. Repeating this procedure 100 times, we obtained sample values for the three random thresholds. We then computed the empirical variance of each sample, which provided an unbiased estimate of the threshold's variance with respect to the sampling distribution of the subjects' data:

\begin{Proposition}\label{prop:unbiased}

Let $Y_i = (y_{i1}, \ldots, y_{in})$ be the activation map for subject $i \in \{1, \ldots, N\},$ let $J \subseteq \{1, \ldots, N\}$ be any subgroup of subjects, and $T = T(Y,J) = T(\{Y_i\}_{i \in J})$ any statistic computed from the activation maps indexed by $J$. Define:
\begin{eqnarray}
\hat {\mathbb E}(T) &=& \frac{1}{N!} \sum_{\pi \in \Pi_n} T(\{Y_{\pi(i)}\}_{i \in J}), \nonumber\\
\hat {\mathbb V}(T) &=& \frac{1}{N!} \sum_{\pi \in \Pi_n} (T(\{Y_{\pi(i)}\}_{i \in J}) - \hat {\mathbb E}(T))^2\nonumber,
\end{eqnarray}
where $\Pi_N$ is the set of all permutations on $\{1, \ldots, N\}.$
If the $Y_i$'s, for $i \in \{1, \ldots, N\},$ are independent random variables following a common distribution $\mathcal P$ on $\mathbb R^n,$ then $\hat {\mathbb E}(T)$ and $\hat {\mathbb V}(T)$ are unbiased estimates of the expectation and variance of $T,$ respectively.

\end{Proposition}

\paragraph{Proof.} This is an immediate consequence of the first two lemmas from Section~6 in \cite{Strasser99}.
\newline

Note that we used Monte-Carlo estimates of the quantities defined in Proposition~\ref{prop:unbiased} by using $100$ random permutations, since performing exhaustive permutations was intractable.

As can be seen in Table~\ref{tab:variance}, The GGM yields the most variable threshold, and the varying window is slightly less variable than the fixed window random threshold. Thus the varying window variant yields the most stable threshold among those tested here.

\begin{table}
\centering
\caption{\label{tab:variance} \bf Mean and variance estimates of the thresholds found by different methods for the computation task.}
\begin{tabular}{|c|ccc|}
\hline
Method
%  & \multicolumn{1}{}{GGM}
%  & \multicolumn{1}{}{fix. window}
%  & \multicolumn{1}{}{var. window} \\
 & GGM & fix. window & var. window \\
\hline
\vspace{-5mm}
Mean
 & 1.92 & 2.96 & 2.54 \\
% \hline
Variance
 & 0.76 & 0.12 & 0.08 \\
\hline
\end{tabular}
\end{table}

% GGM_th = 0.87048077785609124
% fix_th = 0.35255778311247926
% var_th = 0.28765047658637632

% However, the GGM approach detected isolated voxels, which may be false positives, and failed to segment properly the bilateral cingulate regions. These were merged into a single region extending over the inter-hemispheric space

\section{Discussion}

By introducing a simple modification to the random procedure proposed in \cite{Lavielle07}, we have obtained an entirely unsupervised procedure for recovering non null mean terms from a collection of independent random observations, based solely on a parametric model of the null terms. Importantly, our modification, which requires no prior tuning, conserves the consistency properties of the original procedure.

Simulation results suggest that the random threshold approach has very good properties in terms of type I error rate control. While the fixed window seems overly conservative, and unstable with respect to the window parameter at low signal to noise ratios, the varying window extension, which avoids instability, also seems to find a better compromise between specificity and sensitivity.

The first results on real fMRI data are encouraging and seem to confirm the good stability achieved by the varying window random threshold, with respect to the other tested  strategies.

A more thorough investigation is needed to confirm these preliminary results. Reproducibility of the regions found by random thresholding, rather than just the threshold values, can be assessed by resampling techniques similar to those used here, as described in \cite{Thirion07}, and compared to mixture modeling, as well as to false positive control strategies. Finally, studying the asymptotic properties of the random threshold estimator when null mean and non-null mean terms are not well separated is also very important in view of the applications.

% The `incidental' model, where the $\mu_i$ are unknown constants, could be replaced for such a study by a `structural' model, where the $\mu_i$ are independent random variables with unknown common point density function (pdf) $f_{\mu},$ The observations $y_i$ would then following a mixture of the $f_{\mu}$ by the error pdf $f_{\epsilon}.$ If $f_{\epsilon}$ is symmetric and $\mu_i > 0$ \textit{a.e.}, then the non null mean terms $y_i$'s are stochastically greater then the null mean terms.
% Modeling Spatial Uncertainty
\eject
\pagestyle{empty}
$ $ \eject\pagestyle{fancy}
\pagestyle{fancy}
\chapter{Modeling spatial uncertainty}\label{chap:modeling}
\lhead{\emph{Modeling Spatial Uncertainty}}

\section*{Abstract}

This chapter presents an extension of the mass univariate detection approach for group fMRI data, which relaxes the assumption of perfect match between the effect maps of the different subjects. A set of hidden variables is introduced in the classical mass univariate model, which represent registration errors, and are modeled as random deformation fields. The group mean effect map is estimated in a Bayesian setting by its posterior expectation. 
This is evaluated numerically, using a Metropolis-within Gibbs algorithm to sample from the posterior density of all hidden variables. We also show the consistency of the posterior density of the model parameters.

Using simulations, we evidence a stretching effect of the estimated activation pattern when the registration errors are unaccounted for, causing neighboring activations to be merged. This stretching effect is substantially reduced when registration errors are modeled. When applied to real fMRI data, our method yields group effect maps under spatial uncertainty that are both smoother and more contrasted than under no spatial uncertainty, an effect that cannot be reproduced by linear isotropic smoothing. These results are obtained in spite of the slow mixing of our posterior sampling algorithm, suggesting some space for improvement.

\section{Introduction}

In this chapter, we relax the assumption of perfect match between individual brains, which is one of the limitations of the SPM-like approach identified in Section~\ref{sec:alternative}. To this end, we incorporate in the mass univariate model specified by (\ref{eq:within}) and (\ref{eq:between}) a set of hidden variables, representing spatial normalisation errors, which are modeled as multivariate random fields. A first implementation of this idea has been published in~\cite{Keller08}.

In the following, the individual effect maps are seen as warped and noisy versions of the group activation pattern to be estimated. Our goal here is not to solve the registration problem but rather to develop an inference strategy that accounts for registration errors. When performing registration, the spatial transformations are the parameters of interest. In our case, these transformations are viewed as nuisance parameters to be integrated out during the analysis. Also, they are modeled as residual deformation fields, which persist after the actual registration has taken place. This precludes the use of global linear transformations, such as translations or rotations, to model the deformations. Rather, these are seen as local, zero-mean displacements, with no pre-determined form.

There exists an extensive literature on image registration. General reviews on this subject can be found in \cite{Brown92,Zitova03}. More specific reviews include \cite{Van-den-Elsen93,Maintz98}, on the registration of medical images, and \cite{Toga99}, on registration in brain imaging.
% There has a been a large diversity of registration methods proposed in the literature, depending on the nature of the data and the application envisioned. These 
% Registration methods are classically divided into two broad categories, known as {\em feature-based} and {\em area-based} or {\em intensity-based} \cite{Zitova03}. In the first case, the images are aligned by matching high-level features, such as points, lines, surfaces, etc., which summarize their essential characteristics. In the second case, A global measure of similarity between the two images is optimized, the most popular being the correlation coefficient \cite{Svedlow78}. There is an obvious parallel here with the division of fMRI group analysis methods into feature-based and voxel-based, as discussed in Section~\ref{sec:alternative}. Indeed it seems that many ideas originating in the domain of registration have percolated to other branches of image analysis, including neuroimaging.
% Of particular interest in our case are intensity-based methods, as discussed in Section~\ref{sec:conclusion1}. These may be further classified according to the definition of similarity measure, the type of deformations considered, and the optimization method used. As discussed above, we are interested in nonlinear local deformations, such as the class of {\em radial basis functions}, most often represented by thin-plate splines (TPS) \cite{Bookstein91}.
Because we wish to model registration errors, we are interested in a statistical formulation of the registration problem. 
% We are also interested in a statistical formulation of the registration problem, that we could use to model registration errors. 
In \cite{McGillem76}, a first step in this direction is taken, where the two images to be aligned are considered as identical up to a deformation and an additive Gaussian noise, justifying the use of the variance as a measure of overlay quality. More generally, in addition to such a dissimilarity measure, the criterion to be optimized by the registration procedure may contain a so-called regularization or penalty term, which imposes constraints on the transformation \cite{Hajnal01}. 
% The penalty term is particularly important in nonlinear registration to prevent nonsensical transformations to be estimated by overfitting the data. 
This penalty term may take many different forms. In the framework of pattern theory \cite{Grenander93}, deformations are modeled in a Bayesian setting, in which case the penalty is interpreted as a prior density.

Furthermore, the problem we address here is closely related to template estimation, also referred to as {\em atlas construction} in the medical image analysis literature \cite{Subsol95,Joshi04}. 
% This is usually performed iteratively, each iteration consisting in registering individual images to a current version of the template, then computing the average of the registered images to update the template. 
A simple atlas construction method consists in iterating a 'registration step' and an 'averaging step': on each iteration, individual images are registered to a current version of the template, and then averaged in order to update the template.
For instance, the widely used MNI anatomical template was constructed in this fashion (see Section~\ref{sec:normalisation}). However, there is no guaranty that this scheme yields a statistically consistent estimate of the mean image. More recently, \cite{Allassonniere07,Allassonniere08,Lepore08,Sabuncu08} have proposed to construct a Bayesian estimate of the template, given prior distributions on both the deformation fields and the template parameters. Consistency of the template's MAP estimator is shown in \cite{Allassonniere07}.
% a statistical approach to the problem of estimating an unknown template based on deformed and noisy estimation, in a setting very similar to ours. As above, a Bayesian model is proposed, with priors on both the deformation field and the data model, conditional on the deformation. A consistency proof for the template's MAP estimator is provided in this setting.
% Noting the problem inherent in (\ref{eq:nongenerative}), they use the more standard formulation where the deformation $\br u$ is applied to the template, rather than the data. This is also the form we use here.

This chapter is organized as follows: in Section~\ref{sec:observation}, we introduce spatial uncertainty in the mass univariate model specified by (\ref{eq:within}) and (\ref{eq:between}), under the form of unknown spatial deformation fields. The deformation fields are modeled in Section~\ref{sec:deformation}. From this model, we derive in Section~\ref{sec:estimation} a Bayesian estimate of the template $\bs \mu,$ given a certain prior distribution on the model parameters, in an approach similar to \cite{Allassonniere07}. Section~\ref{sec:simulations} illustrates on several simulation studies how techniques that ignore spatial variability may result in blurring and stretching activation patterns, and how this effect can be reduced by our approach. Finally, results on real fMRI data are presented in Section~\ref{sec:fmri_data}.

\section{Observation model}\label{sec:observation}

Following the notations introduced in Section~\ref{sec:between}, $\br x_i = (x_{i,1}, \ldots, x_{i,d})$ is the map of \index{BOLD}BOLD effects of subject~$i = 1, \ldots, n,$ in response to a certain contrast of experimental conditions; $\bs {\mathrm y}_i = (y_{i,1}, \ldots, y_{i,d})$ is the noisy estimate of $\br x_i,$ available from the analysis of the subject's scans (see Section \ref{sec:single}), and $\br s_i^2 = (s_{i,1}^2, \ldots, s_{i,d}^2)$ an image of estimation variances. Recall that, under a sufficient number of acquired scans, the estimation error $\bs \varepsilon_i$ can be assumed Gaussian,  so that:

\begin{eqnarray}\label{eq:within2}
\bs {\mathrm y}_i &=& \br x_i + \bs \varepsilon_i; 
\quad \bs \varepsilon_i \sim \mathcal N\left(\bs 0, \mathrm{diag}(\br s_i^2)\right),
\end{eqnarray}

where $\mathrm{diag}(\br s_i^2)$ denotes the diagonal matrix whose diagonal is given by $(s_{i,1}^2, \ldots, s_{i,d}^2).$ As in \cite{Keller08}, we extend the mass-univariate Gaussian between-subject model described in Section~\ref{sec:between} by incorporating spatial uncertainty so that at voxel~$k:$

\begin{eqnarray}\label{eq:between2}
\br x_i(\br v_k) &=& \bs \mu(\br v_k + \br u_{i,k}) + \xi_{i,k}; 
\quad \bs \xi_i \sim \mathcal N\left(\bs 0, \sigma^2\mathbf I_d\right).
\end{eqnarray}

Here, we note $\br x_i(\br v_k) = x_{i,k}$ to emphasize that it is a spatial map; $\bs \mu \in \mathbb R^d$ is the map of mean population effects; the vector $\br u_{i,k} \in \mathbb R^3$ is a hidden variable that models the subject-to-atlas registration error for subject~$i$ at voxel~$k$; and the $\xi_{i,k}, 1\leq i\leq n,$ model the between-subject variability of the effect at voxel~$k.$ Finally, we define $\bs \mu(\br v_k + \br u_{i,k})$ by discrete interpolation, as being equal to the mean population effect in the nearest voxel $\br v_{k'} = \arg \min_{\br v_l} || \br v_l - (\br v_k + \br u_{i,k})||.$ In practice, this is done by rounding each coordinate of $\br u_{i,k}$ toward the nearest integer. We will also note in the following $\varphi_i$ the function that maps each voxel index~$k$ to the corresponding displaced voxel index~$k'$ for subject~$i,$ so that $\bs \mu(\br v_k + \br u_{i,k}) = \mu_{\varphi_i(k)}.$

(\ref{eq:between2}) generalizes the classical mass univariate model (\ref{eq:between}), which corresponds to the special case where the registration errors $\br u_{i,k}$ are neglected. Note however that the between-subject variance~$\sigma^2$ is uniform over the search volume in our formulation, whereas it is voxel dependent in the classical setting. Indeed, we found from practical experience that a voxel dependent variance raised overfitting issues when modeling registration errors, and resulted in degenerate estimates in certain voxels. We used a uniform variance as the simplest form of constraint to avoid this overfitting, at the price of a possible over-simplification. In Section~\ref{sec:regionalized}, we introduce a regionalized version of Equation (\ref{eq:between}) where $\sigma^2$ is allowed to vary across functionally distinct regions. More sophisticated methods could be considered, for instance by modeling the variance map $\sigma_ k^2$ as a smooth random field, and constitute a possible direction for future work.

\section{Deformation field model}\label{sec:deformation}

As a standard way of representing nonlinear local deformations \cite{Zitova03}, we use Gaussian splines to model the displacements $\br u_{i,k}.$ Specifically, elementary displacements~$\br w_{i,b}$ are defined in a limited number of fixed control points $\{\br v_{k_b}, b = 1, \ldots, B\}$. These are then interpolated to the whole brain image using a radial basis function~$\mathcal K$,

\begin{eqnarray}\label{eq:deformation}
\br u_{i,k} &=& \sum_{b = 1}^B \mathcal K(\br v_k, \br v_{k_b}) \br w_{i,b},
\end{eqnarray}

where $\mathcal K(\br v_k, \br v_{k_b}) = \exp-\{\| \br v_k - \br v_{k_b} \|^2 / 2 \omega^2\},$ and $\omega$ is a user-chosen parameter controlling the displacement field's smoothness.
% This is illustrated in Figure~\ref{fig:elementary}.

The $\br w_{ib}$'s are modeled as a {\em i.i.d.} Gaussian variables, with spherical covariance matrix $\sigma^2_S \mathbf I_3,$ where $\sigma_S$ models the standard registration error:

\begin{eqnarray}\label{eq:elementary}
\pi(\br w_i | \sigma_S^2) &\propto& 
% \left\{ 
\prod_{b = 1}^B \mathcal N(\br w_{i,b}; \bs 0, \sigma^2_S \mathbf I_3).
% \right\}.
\end{eqnarray}

\section{Posterior mean estimate}\label{sec:estimation}

We estimate the population effect map $\bs \mu$ by its posterior mean:
$$
\mathbb E[\bs\mu | \br y] = \int \bs\mu \pi(\bs \mu | \br y) d\bs\mu,
$$
where $\pi(\bs \mu | \br y) \propto \pi(\bs \mu) f(\br y| \bs \mu)$ is the posterior density of $\bs\mu$ relative to a given prior density $\pi(\bs \mu).$ To define this prior, $\bs\mu$ is modeled according to:
$$
\mu_k = \eta + \chi_k; \quad
\chi_k \sim \mathcal N(0, \nu^2),
$$
where $\eta$ represents the mean activation across voxels, and $\nu^2$ the variance of the activation pattern. This hierarchical prior can be seen as an instance of the regional model developed in Chapter~\ref{chap:bayesian}, in the special case of a single functionally homogeneous region. We then define a prior density on these hyperparameters, as well as on the spatial uncertainty parameter $\sigma_S^2,$ as explained in Section~\ref{sec:prior}.

The integral in the above display cannot be computed analytically, so we resort to MCMC strategies instead to produce a sample $\bs \mu_1, \ldots, \bs \mu_G$ from the posterior density $\pi(\bs \mu | \br y),$ yielding the following Monte-Carlo estimate:
$$
\hat {\bs\mu} = G^{-1} \sum_{g=1}^G \bs\mu_g.
$$
Sampling the posterior density in the model with spatial uncertainty raises some technical issues. The simplest strategy would be to use a Gibbs sampler \cite{Geman84}, to generate a sequence of samples from the joint posterior density of all hidden variables $\br x,$ $ \br w,$ $ \bs \mu,$ $ (\eta, \nu),$ $ \sigma^2,$ and $\sigma_S^2.$ This is done by partitioning the variables into blocks, then sampling successively each block conditionally on all others. However, it turns out that the conditional distribution of each elementary displacement $\br w_{ib}$ has no closed-form, and cannot be directly sampled. Therefore, we use the more general Metropolis-Hastings (MH) algorithm \cite{Hastings70}. This involves the choice of a proposal density to generate candidate values $\br w_{ib},$ which are then accepted with a certain rate. Thus, we use a {\em Metropolis-within-Gibbs} algorithm \cite{Tierney94} that generalizes the standard Gibbs sampler by the incorporation of MH iterations. Technical details on this algorithm can be found in appendix~\ref{app:MCMC}, for the more general model introduced in Chapter~\ref{chap:bayesian}.

As is often the case with Bayesian estimation, the posterior distribution $\pi(\bs \mu | \br y)$ has very good frequentist properties. In particular, it is {\em consistent}, {\em i.e.,} it concentrates on the true value $\bs\mu_0$ of the mean population effect map, assuming the data is indeed distributed under our generative model $f(\br y|\bs\mu).$ Thus $\hat {\bs\mu},$ as well as any other Bayes estimator based on this posterior distribution, are reasonable choices to estimate $\bs \mu_0.$

\begin{Theorem}[{\textbf Consistency of the Posterior Distribution}]
\label{theo:consistency}
Let $\bs\theta = (\bs\mu, \sigma^2, \sigma_S^2)$ denote the parameter vector of the model defined by (\ref{eq:within2}), (\ref{eq:between2}), (\ref{eq:deformation}), and (\ref{eq:elementary}), and $\br y^{(n)} = \br y = (\br y_i)_{1\leq i\leq n}$ the data vector, where $n$ is the number of observed activation maps. Then for all $\varepsilon > 0$:
$$
\mathbb P
[ 
\parallel \bs\theta - \bs\theta_0 \parallel > \varepsilon | \br y^{(n)}
] 
\stackrel{n \to \infty}{\longrightarrow} 0,
$$
for any value of the true parameter $\bs\theta_0,$ except possibly on a set of $\pi$-measure $0.$

\end{Theorem}

\paragraph{Proof.} 

This derives from Doob's theorem (see \cite{Vandervaart00}, Chapter~10, for instance). Doob's theorem states that in an identifiable parametric model and for any prior distribution $\pi$, the sequence of posterior densities is consistent for $\pi$-almost every parameter value. The identifiability of our model is demonstrated in Appendix~\ref{app:identifiability}.

However, Doob's theorem applies only to independent, identically distributed (iid) variables. In contrast, the observations $\br y_i$ in (\ref{eq:within2}) are defined conditional on the first-level variances $\br s_i^2,$ which are different for each subject, and so are not identically distributed. This can be arranged through the device introduced in \cite{Meriaux06}, where the $\br s_i^2$ are considered as part of the data vector, and modeled as a sample from an unspecified density $f(\br s_i^2),$ independent from all other variables. This has no effect on the posterior density, and under this assumption the observations $(\br y_i, \br s_i^2)_{1\leq i\leq n}$ are iid, so that the result holds $\square$

\section{Simulations}\label{sec:simulations}

\subsection{1D simulations}\label{sec:example}

We now illustrate our model of spatial uncertainty on a simplistic one-dimensional (1D) example. Our purpose here is to give some insight on how deformations are modeled, and how they can impact the estimation of the activation pattern $\br\mu,$ which in this case is simply a real-valued function.

\subsubsection{Description of the dataset}\label{sec:dataset_1D}

\begin{figure}[ht!]
\centering
\includegraphics[width=\textwidth]{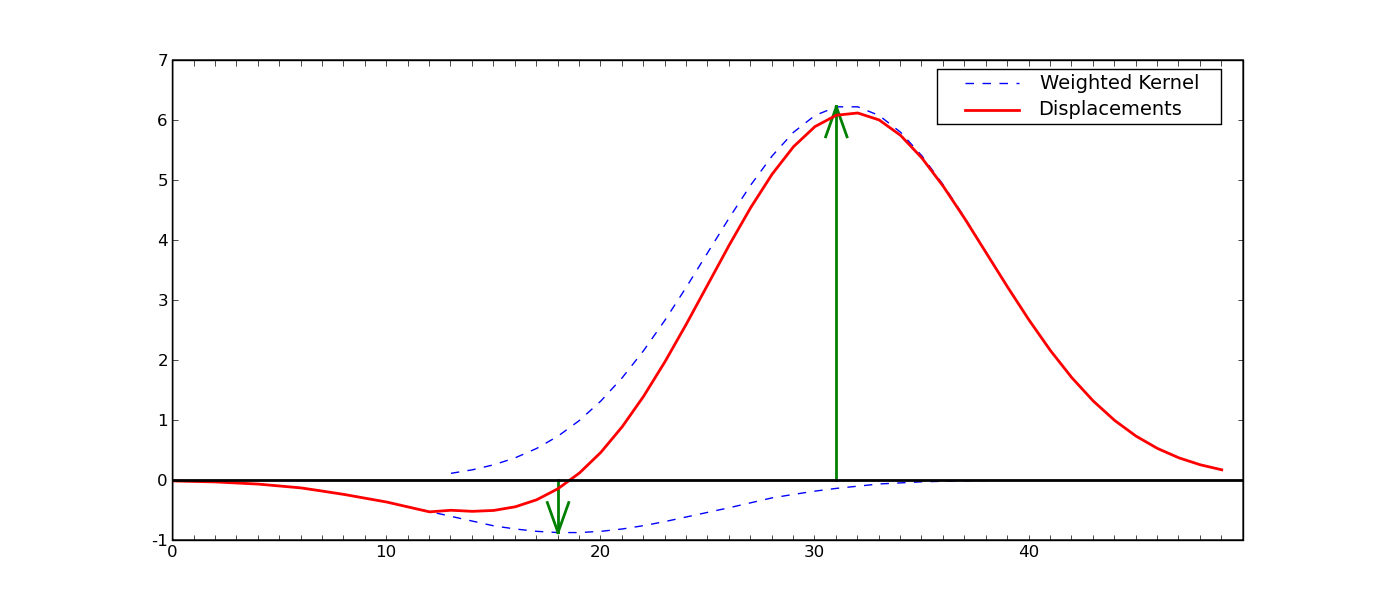}
\caption{\label{fig:elementary_displacements} Illustration of the deformation field model. Elementary displacements (arrows) in pre-specified control points are interpolated to all other points using a radial kernel (dashed lines). The displacement field (solid line) is the sum of the resulting weighted kernels. }
\end{figure}

The voxels, aligned and equally spaced in this example, are represented by the integers $k = 1, \ldots, 50.$ We simulated a deformation field according to model (\ref{eq:deformation}), with a standard displacement of $\sigma_S = 3$ voxels, and a kernel width of $\omega=6.5$ voxels. We also empirically set the distance between adjacent control points to $2 \times \omega = 13.0$ voxels, with no control points at at less then $2.5 \times \omega = 16.25$ voxels from the voxel set boundaries. In the present case, this resulted in two control points only, as illustrated in Figure~\ref{fig:elementary_displacements}.

\begin{figure}[ht!]
\centering
\caption{\label{fig:1D_data} Illustration of the model with spatial uncertainty on 1D data.}
\includegraphics[width=\textwidth]{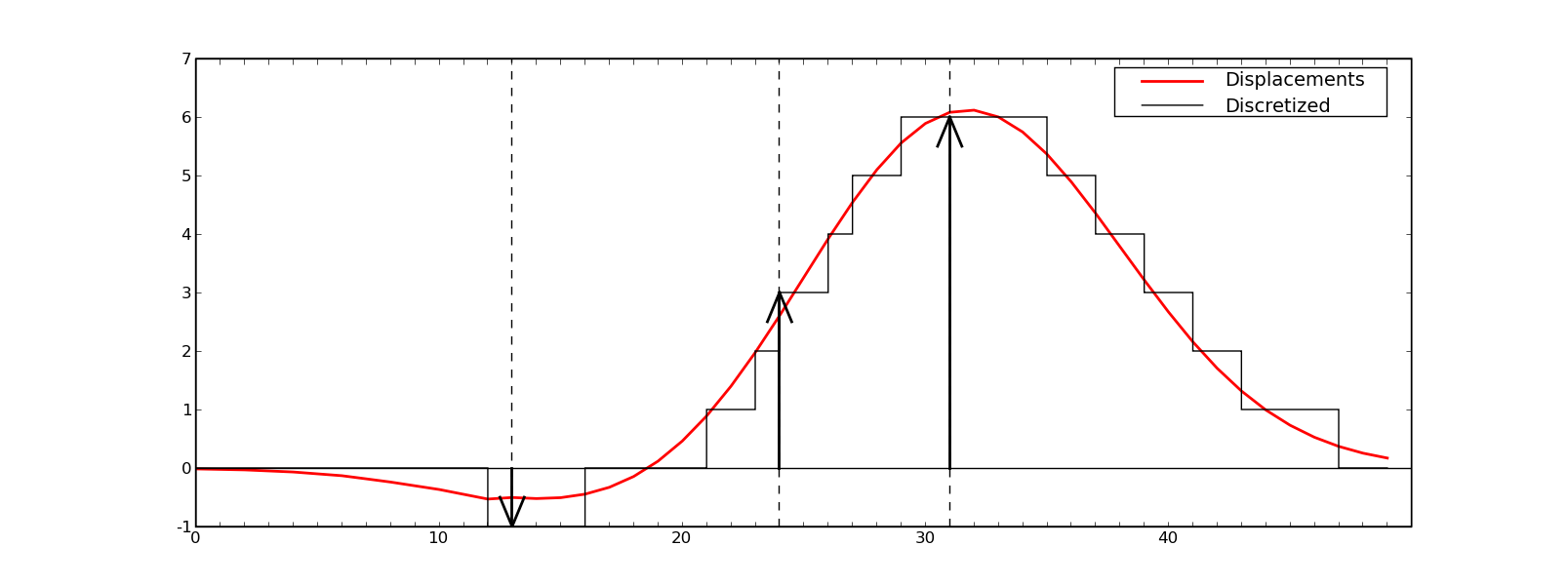}
\small Deformation field (corresponding to $\br u_{i,k}$ in (\ref{eq:between})). For instance, $[u_{i,13}] = -1,$ $[u_{i,24}] = +3,$and $[u_{i,31}] = +6.$
\includegraphics[width=\textwidth]{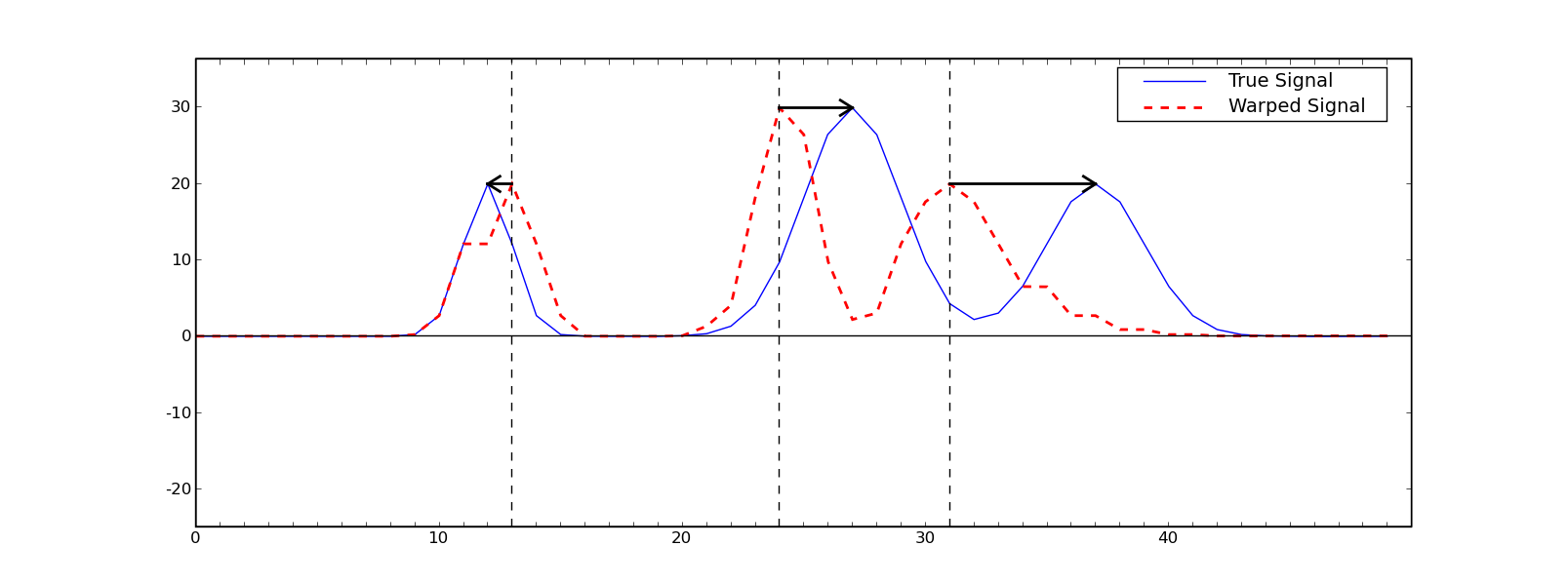}
\small Effect of the deformation field on the signal $\br \mu$ (solid line). For instance, the warped signal at point $k = 24$ is equal to the true signal at point $24 + [u_{i,24}] = 24 + 3 = 27.$
\includegraphics[width=\textwidth]{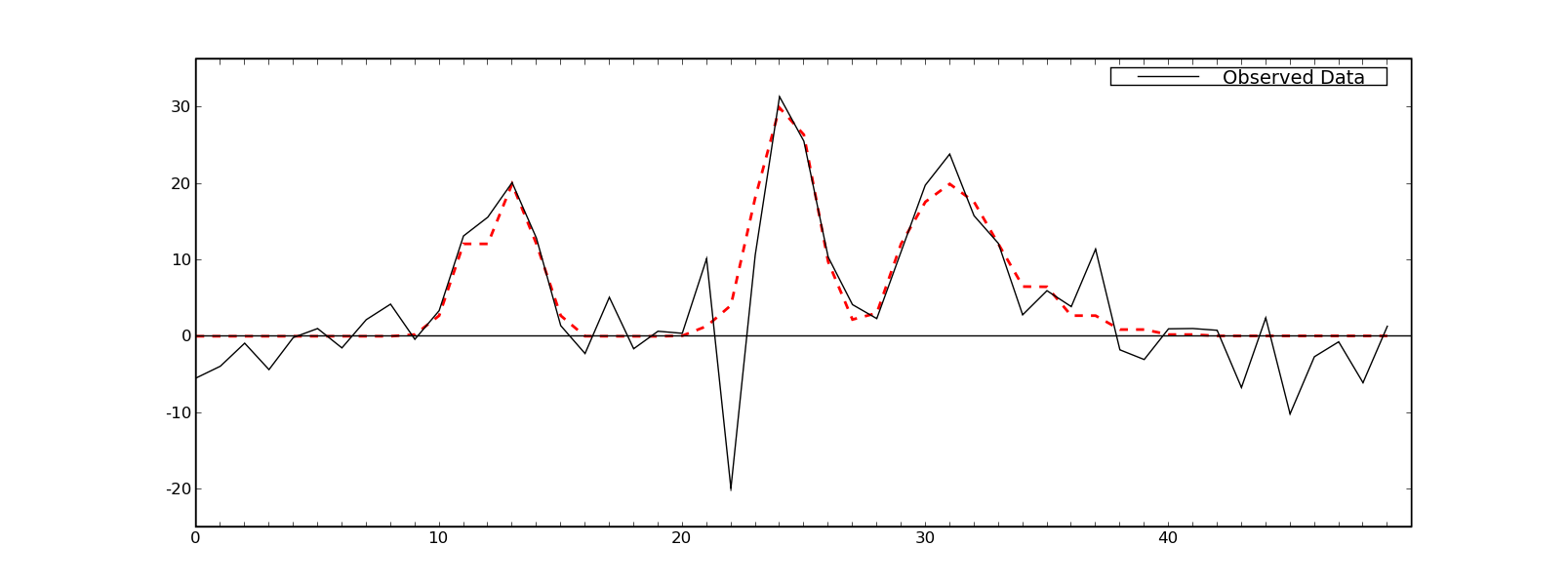}
\small Data $\br y_i$ (solid line), obtained by adding Gaussian noise to the warped signal.
\end{figure}

This displacement field was used to warp a signal defined as the sum of three Gaussian `bumps', representing the mean effect map $\bs \mu,$ following (\ref{eq:between}), (see Figure~\ref{fig:1D_data}, middle). Heteroscedastic noise with variance $\sigma^2 + s_{i,k}^2$ was then added to this warped signal to produce a synthetic observation $\br y_i$ (see Figure~\ref{fig:1D_data}), according to (\ref{eq:within2}), (\ref{eq:between2}). We chose $\sigma = 1.0,$ and the individual standard deviates $s_{i,k}$ where generated as independent standard normal variables, multiplied by a noise level of $\varepsilon = 4.0.$

\subsubsection{Methods compared}

We generated $n = 40$ observations as described above, and tried to recover the original signal $\bs \mu$ from these warped, noisy versions. To do this, we computed the posterior mean $\mathbb E[\bs \mu|\br y],$ as explained in Section~\ref{sec:estimation}. We compared the estimate obtained in the full model with that obtained in the model without spatial uncertainty, setting $\sigma_S^2=0.$ In both cases, the posterior mean was averaged over $100$ Gibbs iterations, following $100$ `burn-in' iterations which were discarded.

\begin{figure}[ht!]
\caption{\label{fig:std_plot_1D} Posterior sampling of different parameters in the 1D model with and without spatial uncertainty}
\begin{minipage}{0.5\textwidth}\centering
\includegraphics[width=\textwidth]{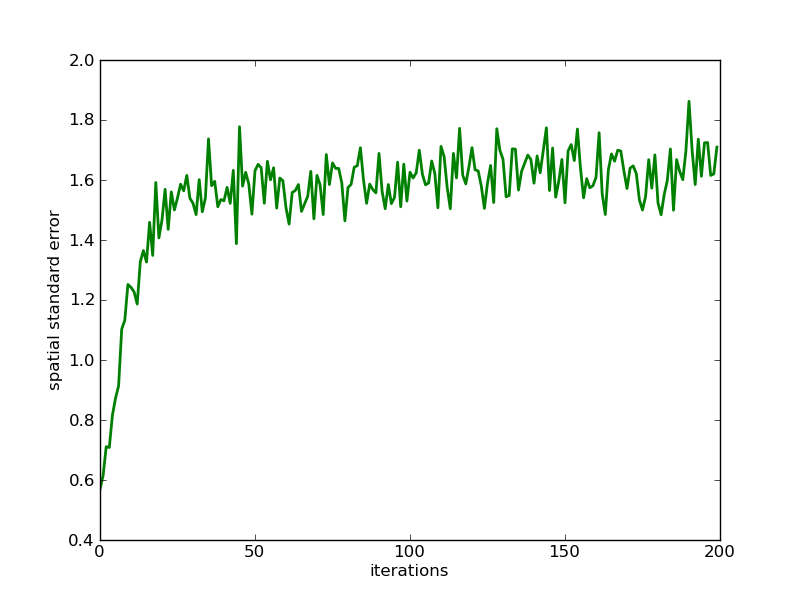}\\
$\sigma_S$
\end{minipage}\hfill
\begin{minipage}{0.5\textwidth}\centering
\includegraphics[width=\textwidth]{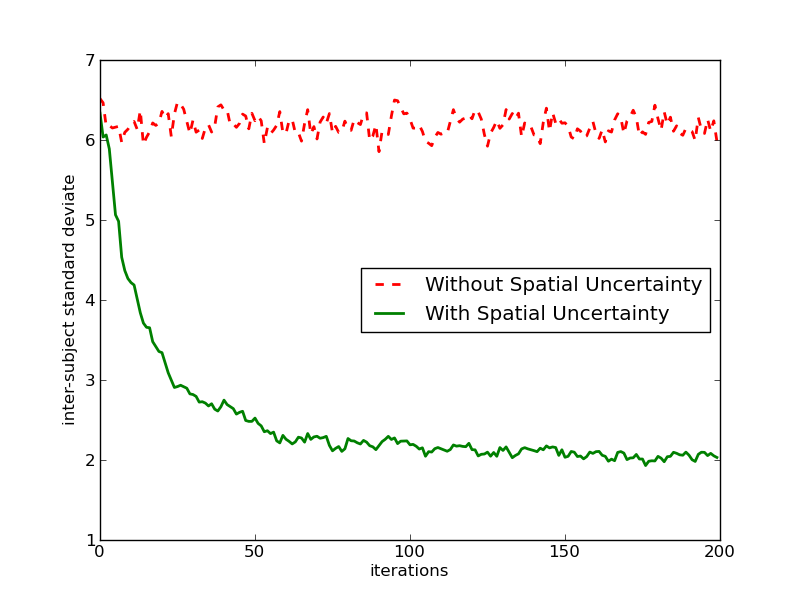}\\
$\sigma$\\
\end{minipage}\\
\begin{minipage}{0.5\textwidth}\centering
\includegraphics[width=\textwidth]{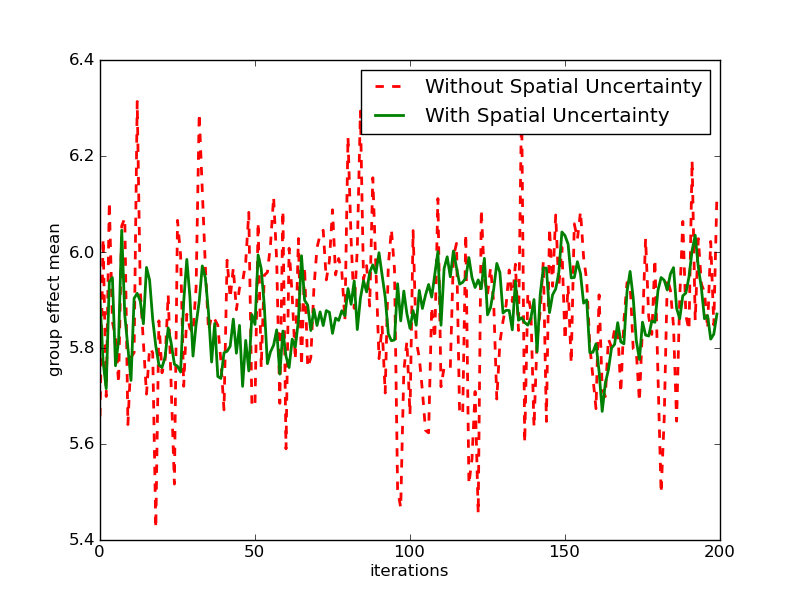}\\
$\eta$
\end{minipage}\hfill
\begin{minipage}{0.5\textwidth}\centering
\includegraphics[width=\textwidth]{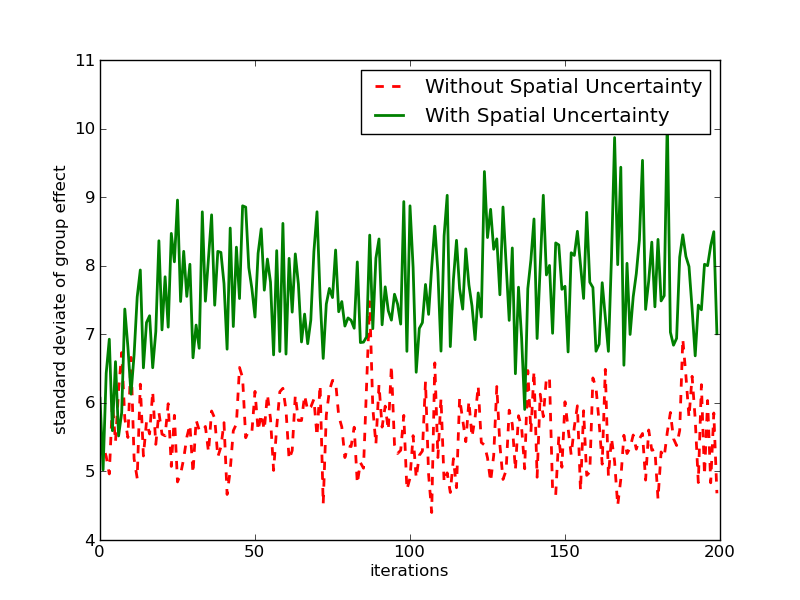}\\
$\nu$
\end{minipage}
\end{figure}

It can be seen in Figure~\ref{fig:std_plot_1D} that, when sampling the posterior distribution of the full model, the Markov chain quickly reached a stable state, after approximately 100 iterations. However, the resulting posterior mean estimate of the spatial incertitude parameter, $\hat {\sigma}_S = 1.6,$ was lower than the true value, $\sigma_S = 3.0.$ Conversely, the posterior mean of the between-subject standard deviate, $\hat \sigma = 2.0$ was higher than the actual value, $\sigma = 1.0.$

A possible explanation is that the chain was trapped next to a mode of the posterior distribution, away from the global maximum. Escaping from this mode would theoretically happen after a sufficient number of iterations, but there is no upper bound on the CPU time this would require. This illustrates the difficulty of sampling from the joint distribution of the deformation fields, even in this simplistic 1D example with two control points per field.

Alternatively, this bias in the variance estimates may also be a consequence of the so-called {\em shrinkage} (or {\em regularizing}) effect, {\em i.e.} a compromise between estimator bias and variance induced by the prior distribution. Specifically, the model on spatial displacements $\br w$ (\ref{eq:elementary}), together with the prior on the spatial standard error $\sigma_S$ (\ref{eq:prior_var}), favor small displacements, thus preventing unreasonable distortions of the individual images which would artificially increase the likelihood by overfitting the data.

Supporting this explanation, the convergence plot $\sigma$ (Figure~\ref{fig:std_plot_1D}, upper right, solid line), shows that its sampled values rapidly decrease at first, while the spatial incertitude parameters augments (Figure~\ref{fig:std_plot_1D}, upper left). This suggests that the Markov Chain is able to find a significantly better match between the individual images by spatially deforming them, $\sigma$ being a measure of their global dissimilarity (in contrast, $\sigma$ values sampled in the model without spatial uncertainty (Figure~\ref{fig:std_plot_1D}, upper right, dashed line) do not vary significantly from their initial value). Then, both $\sigma$ and $\sigma_S$ settle in a stable state, presumably because an equilibrium has been met between a good match across the images (low $\sigma$) and small, regular deformations (low $\sigma_S$).

Additional indicators of the enhanced match between individual images, achieved when modeling spatial uncertainty, is provided by the convergence plot of the group effect map's mean $\eta$ (Figure~\ref{fig:std_plot_1D}, bottom left, solid line) which is less variable than in the model without spatial uncertainty (dashed line). Also, the resulting map $\hat{\bs\mu}$ is more contrasted, as can be seen from the higher sampled values of its variance $\nu$ (Figure~\ref{fig:std_plot_1D}, bottom right).

\begin{figure}
\includegraphics[width=\textwidth]{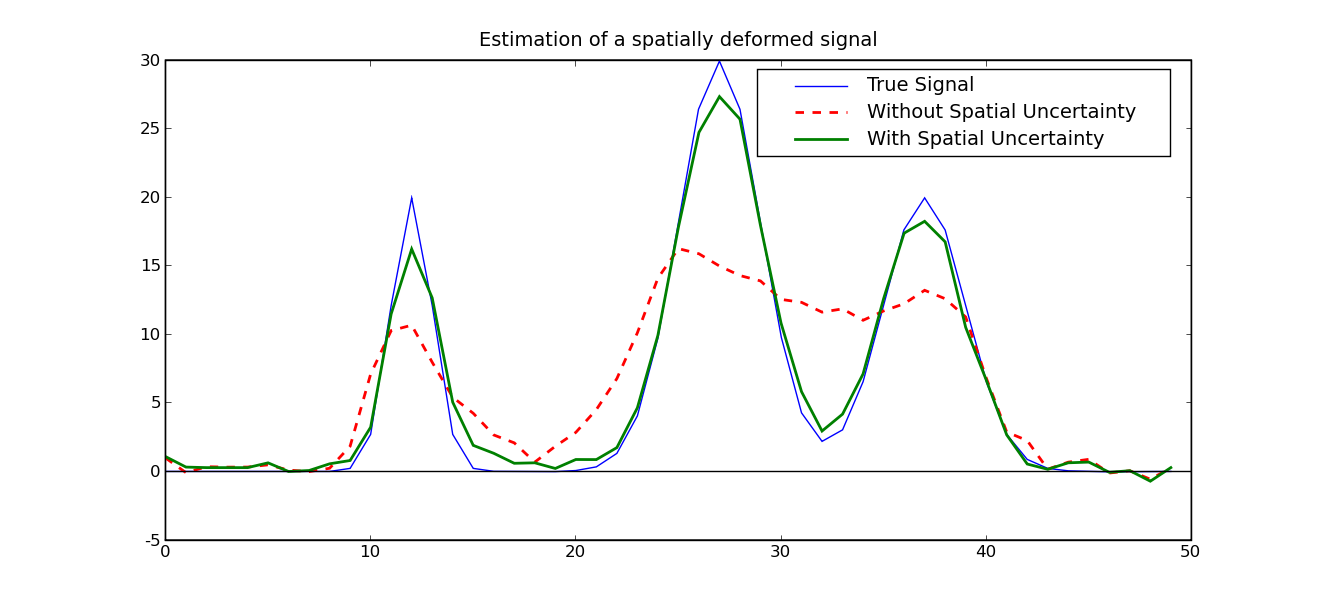}
\caption{\label{fig:signal_estimates}  Posterior estimates of $\bs\mu$ in the 1D-model, with and without modeling spatial uncertainty.}
\end{figure}

Finally, we can see from Figure~\ref{fig:signal_estimates} that the posterior mean estimate $\hat{\bs\mu}$ under the full model successfully recovered the three different modes present in the original signal. In contrast, as could be expected beforehand, the estimate which neglected the signal's spatial variability behaved poorly. As a result, regions of positive signal were swelled, and the two nearby peaks merged in a single mass.

\subsubsection{Sensitivity study}

Next, we investigated the reproducibility of the above results, and their sensitivity to various features of the simulated data. We measured the quality of each estimate $\hat{\bs\mu}$ by its mean-square error $MSE(\hat{\bs\mu}) = \frac{1}{d} \sum_k (\hat{\mu}_k - \mu_k)^2.$

\paragraph{Influence of noise and deformation field smoothness.}

In a first experiment, we simulated the data for different noise levels $(\varepsilon = 4.0, 8.0, 12.0, 16.0)$ and different values of the deformation field smoothness parameter $(\omega = 5.0, 6.5, 8.0).$ For each of the $4 \times 3$ possible combinations of these quantities, we generated $100$ datasets, which we used to estimate $\bs\mu,$ using the two methods described above. In particular, the deformation field model used for this posterior estimation assumed a field smoothness of $\omega_{est} = 6.5,$ so that it alternatively under or over estimated the true field smoothness. This allowed to investigate the behavior of our sampling scheme when using a mis-specified deformation field model, which is inevitable when analysing real fMRI data.

\begin{figure}
\includegraphics[width=\textwidth]{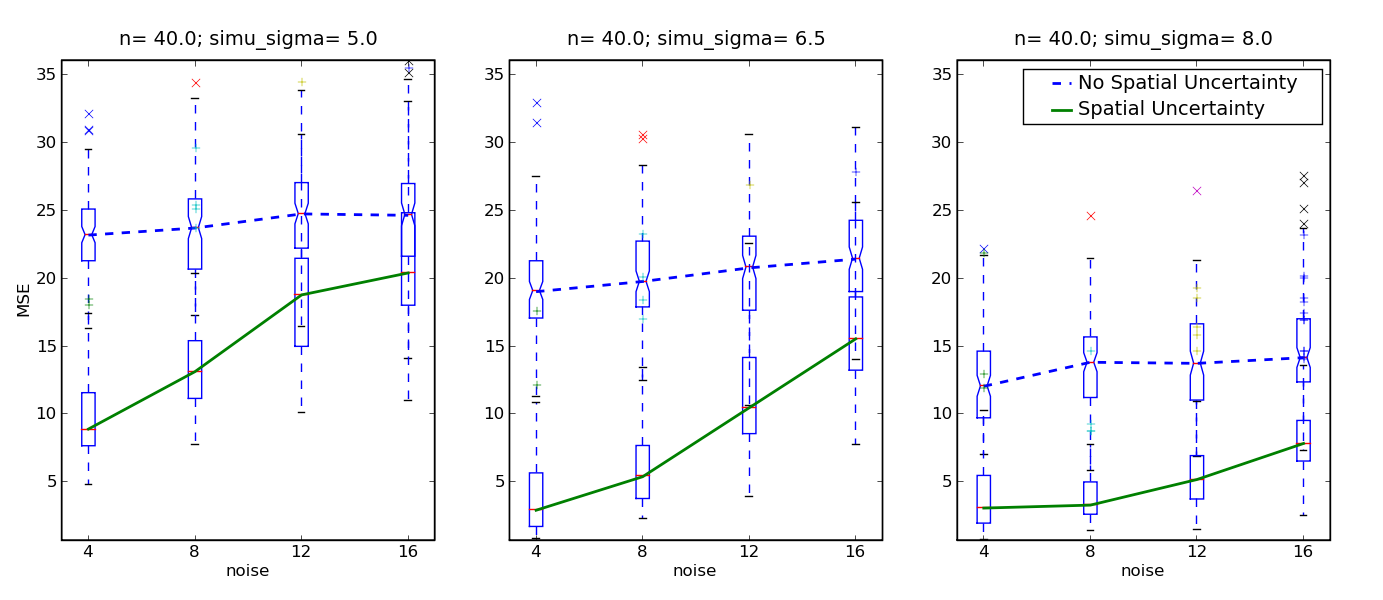}
\caption{\label{fig:noise_simu_sigma_1D} Sensitivity of the posterior mean estimate to noise and deformation field smoothness on 1D data.}
\end{figure}

Results of this experiment are presented in Figure~\ref{fig:noise_simu_sigma_1D}. It can be seen that both posterior estimates are very sensitive to noise, since the MSE increases with the noise level $\varepsilon$ in all cases. Also, the difference in MSE values between the estimates with and without spatial uncertainty is clearly seen at the lowest noise level considered $\varepsilon = 4.0.$ However, it tends to disappear at the highest noise level $\varepsilon = 16.0.$ Deformation field smoothness, measured here by the parameter $\omega,$ also seems to be a critical parameter, with estimations that are better for smoother fields. % Indeed, when it is low, so that the estimation model overestimates it (Figure~\ref{fig:noise_simu_sigma_1D}, left), then we see that the MSE obtained by modeling spatial uncertainty is higher than, when $\omega_{sim}$ is high, so that the estimation model underestimates it (Figure~\ref{fig:noise_simu_sigma_1D}, right).

\paragraph{Influence of Sample Size.}

\begin{figure}
\centering
\includegraphics[width=0.5\textwidth]{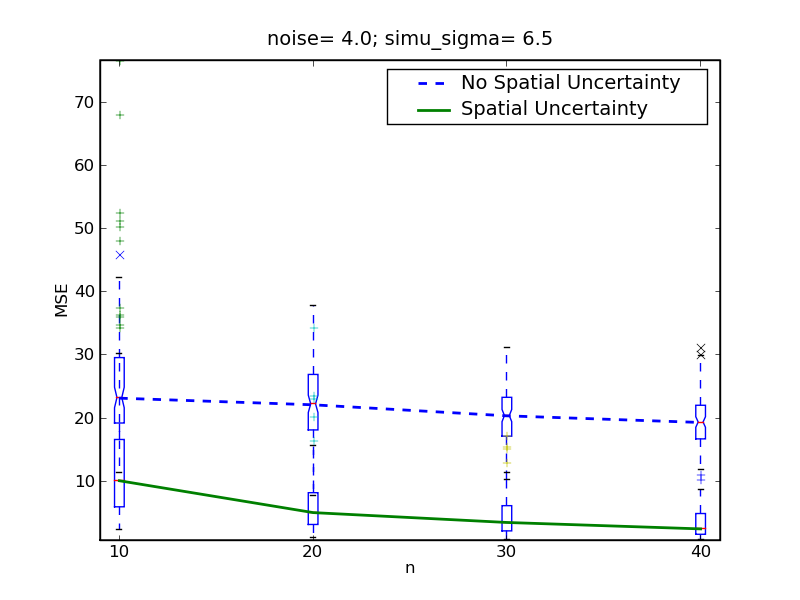}
\caption{\label{fig:MSE_1D} Sensitivity of the posterior mean to sample size on 1D data.} %\small Mean-Square Error (MSE) for the estimation of signal ($\bs \mu$), in the model with spatial uncertainty (solid line) and without (dashed line). The lines represent the median MSE across 10 replications, ploted against the number of observed images $n.$ Dispersion of MSE values is represented by the boxplots.
\end{figure}

In a second experiment, we studied how the performance of the posterior mean estimate varied when applied to datasets, simulated as described in Section~\ref{sec:dataset_1D}, of increasing sizes $n = 20.0,\, 30.0,\, 40.0,\, 50.0.$ As the sample size increases, the posterior distribution of the full model concentrates around the true value of the parameters, according to Theorem~\ref{theo:consistency}, unlike the posterior distribution of the model without spatial uncertainty, whose asymptotic behavior is unknown. Hence, we expect the difference in MSE between the two corresponding estimates of $\bs\mu$ to increase with sample size. This is precisely what we observe in Figure~\ref{fig:MSE_1D}. 
% In particular, these results suggest that spatial uncertainty is worth modeling especially for datasets of at least $20$ images.

\subsection{3D simulations}\label{sec:3D_simu}

In this second numerical experiment, we apply our method to 3D simulated data, and show that the observations on simulated 1D data are confirmed, specifically, that standard voxelwise techniques (not accounting for spatial uncertainty) may lead to overestimating the size of positive effected regions. We also show that this undesirable ``stretching effect'' may be reduced by modeling spatial uncertainty.

\subsubsection{Data simulation}

A synthetic dataset was generated as follows. We defined a volume of $24 \times 32 \times 32$ voxels, containing two spherical activated regions, with uniform intensity value~$5$ (the background was set to $0$) and a fixed diameter of $7$~voxels.

\begin{figure}[ht!]
\includegraphics[width=0.33\textwidth]{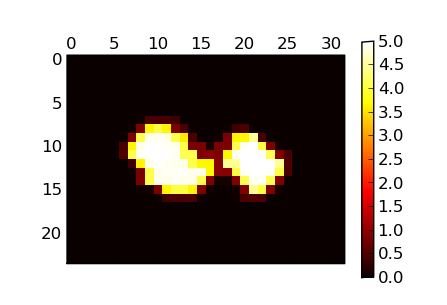}\hfill
\includegraphics[width=0.33\textwidth]{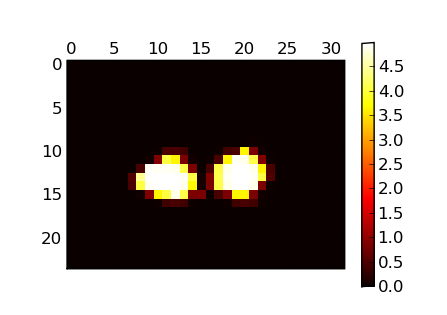}\hfill
\includegraphics[width=0.33\textwidth]{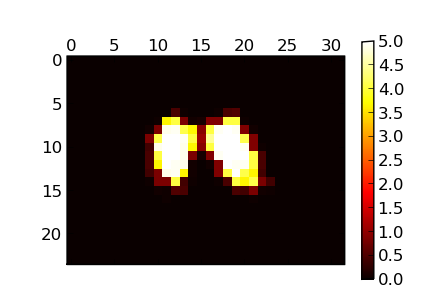}\\
\includegraphics[width=0.33\textwidth]{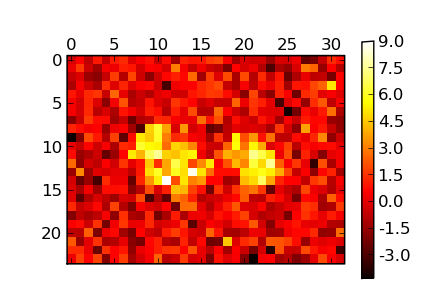}\hfill
\includegraphics[width=0.33\textwidth]{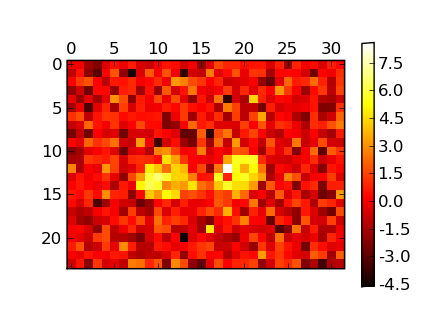}\hfill
\includegraphics[width=0.33\textwidth]{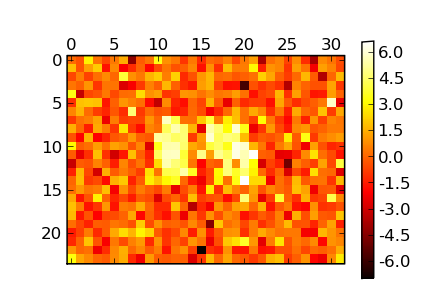}\\
\caption{\label{fig:toy_dataset} Examples of simulated data.  (top) Deformed signals. (bottom) Data (deformed, noisy signals).}
\end{figure}

This idealized activation, interpreted as the mean population effect map $\bs \mu,$ was slightly smoothed (using a Gaussian kernel with standard deviation $0.5$~voxels) to emulate the partial volume effect present in real fMRI data.  It was then deformed according to a displacement field $\br u,$ simulated under the model described in Section~\ref{sec:deformation}, with one control point in each voxel. The standard displacement was taken equal to $\sigma_S = 2.0$ voxels and the field smoothness parameter was set to $\omega=4.0$ voxels.

As for 1D simulations, independent heteroscedastic Gaussian noise was then added to each voxel $\br v$, the variance of which was taken equal to $1 + \br s^2(\br v)$, with $(\br s/\varepsilon)^2(\br v)\sim\chi^2(1)$, $\varepsilon$ being the noise level, set to $1.0$ in this example.  A total of $n=40$ pairs $(\br y_i, \br s^2_i)$ of effect and variance maps were sampled in this fashion, as illustrated in Figure~\ref{fig:toy_dataset}, and constitute a sample from the hierarchical model in Section~\ref{sec:observation} .

\subsubsection{Methods compared}

Based on the synthetic observations $\br y_1, \ldots, \br y_n,$ and the hierarchical model in Section~\ref{sec:observation}, we estimated the original signal $\bs \mu$ by its posterior mean $\mathbb E [ \bs \mu | \br y],$ as described in Section~\ref{sec:estimation}. Importantly, the estimation model used to compute $\hat{\bs \mu}$ differed from the one used to simulate the data, in that the deformation fields were defined by two control points, instead of one control point in each voxel. This modification was introduced to verify that the signal could be correctly estimated, even though the model was mis-specified, as is unavoidable in real-world datasets. Finally, $\bs \mu$ was estimated in the model without spatial uncertainty, setting $\sigma^2_S = 0, \br u \equiv 0,$ allowing to verify the presence of a stretching effect due to unaccounted spatial uncertainty, as was observed on the 1D example (Section~\ref{sec:example}). In order to compare objectively the different methods, we again measured the performance of each estimate $\hat {\bs \mu}$ by its mean-square error: $MSE = \frac {1} {d} \sum_{k} (\hat \mu_k - \mu_k)^2.$

\begin{figure}[ht!]
\caption{\label{fig:std_plot_3D} Posterior sampling of different parameters in the 3D model with and without spatial uncertainty}
\begin{minipage}{0.5\textwidth}\centering
\includegraphics[width=\textwidth]{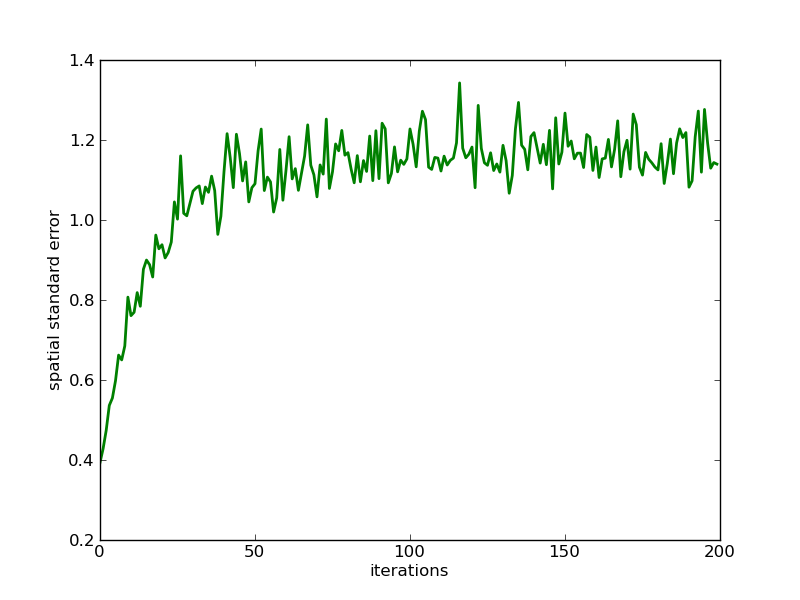}\\
$\sigma_S$
\end{minipage}\hfill
\begin{minipage}{0.5\textwidth}\centering
\includegraphics[width=\textwidth]{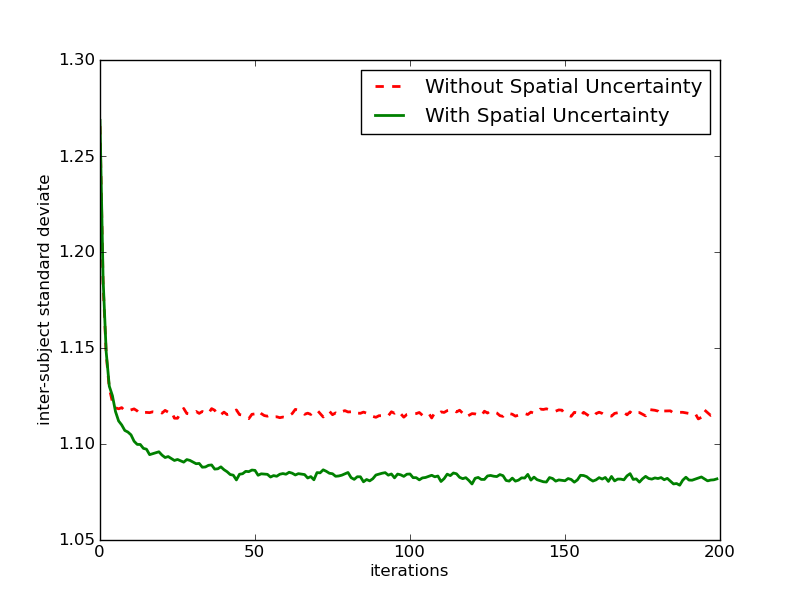}\\
$\sigma$\\
\end{minipage}\\
\begin{minipage}{0.5\textwidth}\centering
\includegraphics[width=\textwidth]{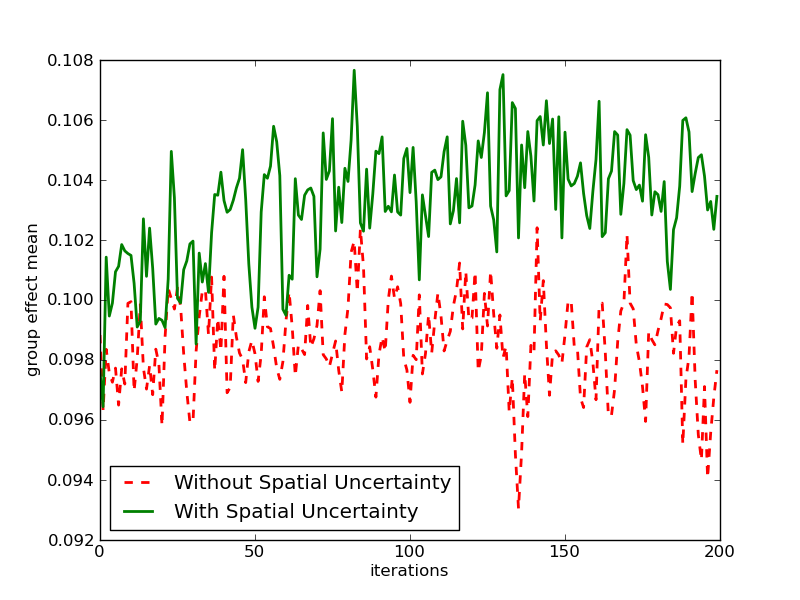}\\
$\eta$
\end{minipage}\hfill
\begin{minipage}{0.5\textwidth}\centering
\includegraphics[width=\textwidth]{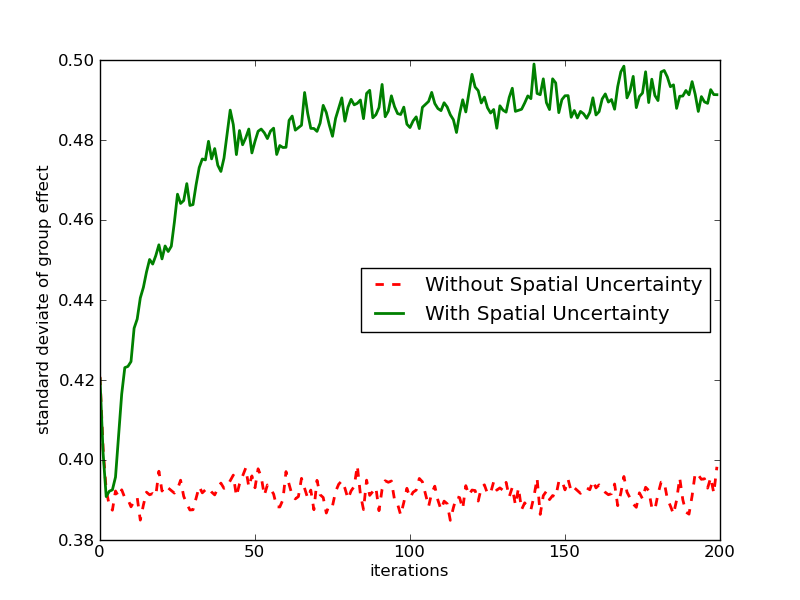}\\
$\nu$
\end{minipage}
\end{figure}

As in the 1D case, we can see in Figure~\ref{fig:std_plot_3D} that the Markov chain quickly settles in a stable state. A similar srinkage effet is suggested from the estimated spatial standard error $\hat{\sigma}_S = 1.2$ lower than the true value, $\sigma_S = 2.0.$ The estimated between subject standard error $\hat\sigma = 1.1,$ is however closer now to its true value $\sigma = 1.0.$ This may be because the 3D data used in this simulations include many more voxels ($18\, 432$) than the length of the previous $1D$ data ($50$ points), and thus allow a much better estimation of certain model parameters (though $\sigma_S$ remains conservatively biased).

\begin{figure}[ht!]
\includegraphics[width=0.33\textwidth]{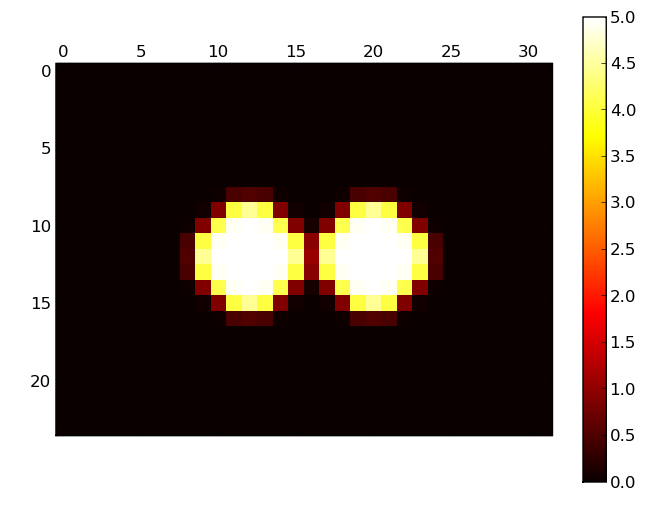}\hfill
\includegraphics[width=0.33\textwidth]{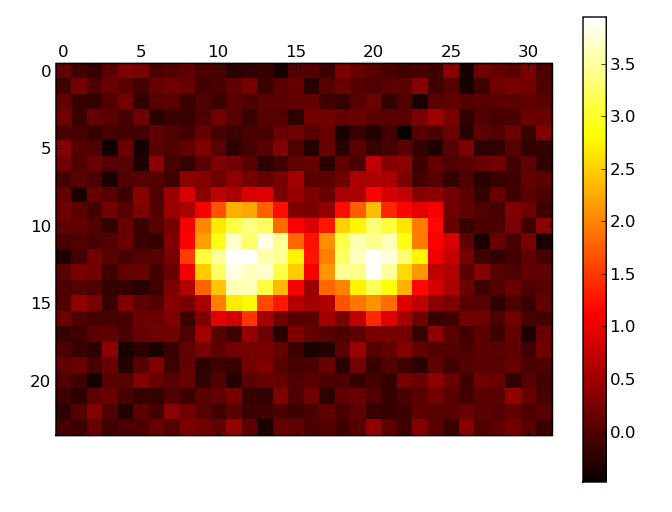}\hfill
\includegraphics[width=0.33\textwidth]{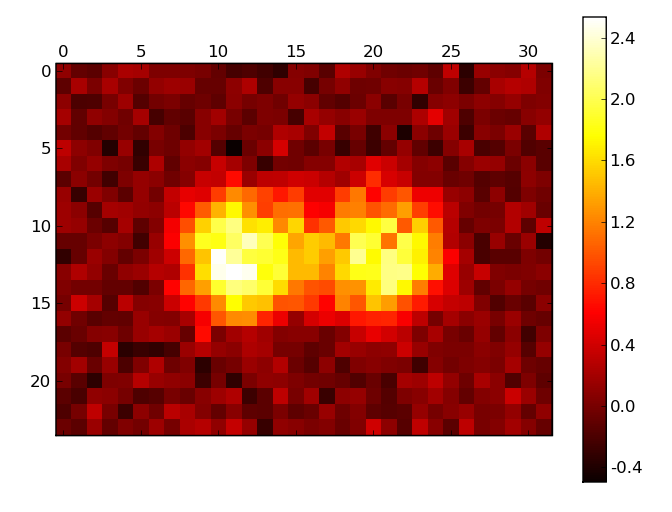}
\caption{\label{fig:posterior_estimates} Posterior estimates of $\bs \mu$ in the 3D-model, with and without spatial uncertainty.}
\end{figure}

As previously, the posterior estimate of the group effect is significantly more accurate using the full model than its version without spatial uncertainty, as can be seen in Figure~\ref{fig:posterior_estimates}. As expected, the latter (right) is visibly more spread out spatially than the original signal (left), and the two activated regions are merged into a single blurry region. In contrast, the signal estimated in the model with spatial uncertainty (center), correctly recovers the two separate activation spheres, which also appear less blurred. These impressions are confirmed by the fact that its MSE ($0.11$) is lower than the one obtained without spatial uncertainty ($0.19$).
$\hat\sigma = 2.1$

\subsubsection{Sensitivity study}

Next, we investigated the reproducibility of the above results, and the sensitivity of the posterior estimates with respect to various simulation parameters.

\paragraph{Influence of Noise and Deformation Field Regularity.}

We studied the influence of both the regularity of the deformations and the amount of observation noise. Thus, the field regularity parameter used to simulate the data, $\omega_{sim},$ was chosen in $[3.0, 4.0, 5.0].$ Also, we generated the observation variance $s_{i,k}$ for subject~$i$ at voxel~$k$ according to $\displaystyle \frac {s_{i,k}} {\epsilon}^2 \sim \chi^2(1),$ with $\epsilon,$ controlling the noise level, varying in $[1.0, 2.0, 3.0, 4.0].$

For all $3 \times 4$ combinations of these parameters, $100$ synthetic datasets where generated and used to estimate $\bs \mu,$ both with and without spatial uncertainty, as described above. In all cases, the estimation model used two control points to define each deformation field, and a field regularity parameter $\omega_{est}$ fixed to $4.0.$ It was therefore mis-specified with respect to the model used to simulate the data, for which the deformation fields where defined with one control point in each voxel, and varying values of the regularity parameter $\omega_{sim}.$

\begin{figure}
\includegraphics[width=\textwidth]{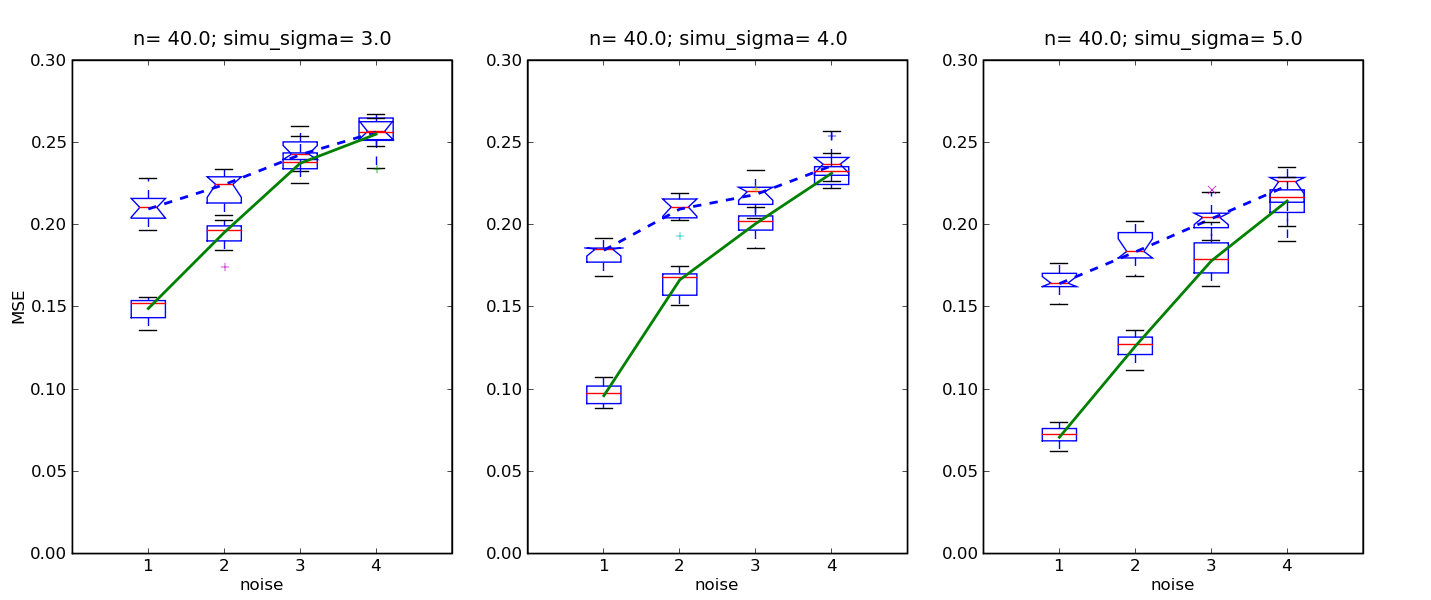}
\caption{\label{fig:noise_simu_sigma_3D} Sensitivity of posterior estimation to noise and deformation field smoothness. Mean-Square Error (MSE) for the estimation of signal ($\bs \mu$), in the model with spatial uncertainty (solid line) and without (dashed line). The lines represent the median MSE across 10 replications, ploted against the noise level $\epsilon.$ Dispersion of MSE values is represented by the boxplots. Each sub-figure corresponds to a different value of the deformation field smoothness parameter $\omega_{sim},$ which increases from left to right.}
\end{figure}

The results of the sensitivity analysis are presented in Figure~\ref{fig:noise_simu_sigma_3D}. As in the $1D$ case, both approaches are sensitive to observation noise, and the MSE 
% obtained from the estimation under spatial uncertainty 
is seen to decrease when the deformation field smoothness $\omega$ increases. 
% In contrast, the estimation under no spatial uncertainty appears fairly insensitive to $\omega.$

\paragraph{Influence of sample size.}

\begin{figure}
\begin{center}
\includegraphics[width=0.5\textwidth]{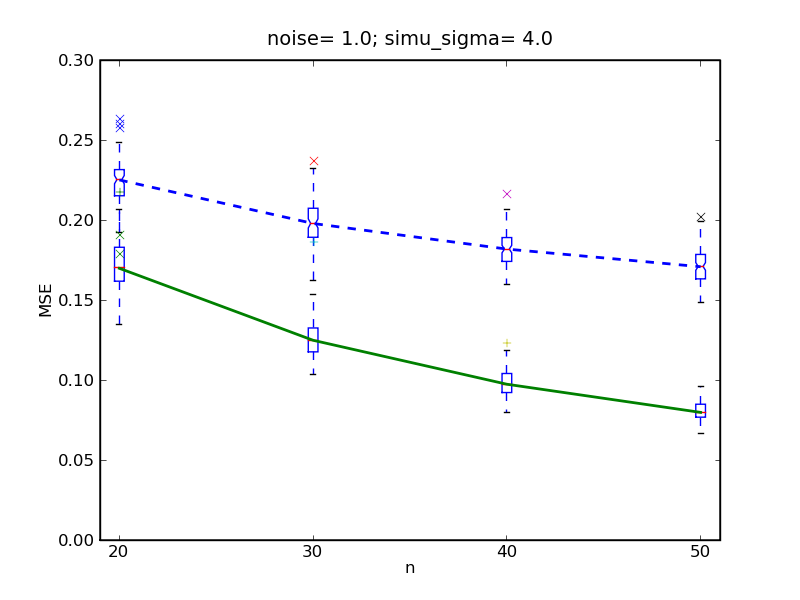}
\caption{\label{fig:MSE_3D} Sensitivity of posterior estimation to sample size.} 
\end{center}
\small Mean-Square Error (MSE) for the estimation of signal ($\bs \mu$), in the model with spatial uncertainty (solid line) and without (dashed line). The lines represent the median MSE across 100 replications, ploted against the number of observed images $n.$ Dispersion of MSE values is represented by the boxplots.
\end{figure}

Finally, we studied the behavior of our approach when the sample size varied, choosing $n=20, 30, 40, 50. $ (see Figure~\ref{fig:MSE_3D}). As in the $1D$ case, the drop in MSE due to spatial uncertainty modeling is seen to increase with the data size.

\subsection{Conclusion}

The simulations presented in the above sections~\ref{sec:example} and \ref{sec:3D_simu} show how the modeling of unknown deformations can lead to significant improvements in the estimation of a spatially warped signal.

The sensitivity to noise of our estimation method is also highlighted in these illustrations. This sensitivity may be over-estimated here due to the slow mixing of our sampling algorithm, which in particular under-estimated the spatial uncertainty parameter. Nevertheless, this suggests that modeling spatial uncertainty is most useful in cases where the data is not too noisy, and motivates the use of a moderate preliminary smoothing step when processing the individual fMRI datasets (see Section~\ref{sec:BOLD}).
% Such smoothing may be acceptable when estimating the population mean effect map. 
% When testing presence of activations within interest regions however, it may however create false positives by spreading activations across regions. 
We would however recommend nonlinear smoothing strategies, such as cortical surface-constrained filtering \cite{Andrade01b} or anisotropic diffusion \cite{Kim05}, in order to limit the blurring effect inherent to Gaussian and other linear filters. 

% Additionally, we note that our sampling algorithm behave sub-optimally, since it under-estimated the standard spatial error~$\sigma_S^2.$ Likewise, the sensitivity to noise we have observed may in part be inflated due to the limits of our algorithm
% We stress that the smoothing we advocate here has the sole purpose of increasing the signal-to-noise ratio. We do not adhere to the additional motivations for smoothing evoked in \cite{Friston95}, in particular the fact that it compensates for the mis-alignment of homologous functional regions across subjects, since we already deal with this issue in a much more satisfying way.

% In Chapter~\ref{chap:bayesian}, we introduce a regionalized model, which allows to gain statistical power from the functional homogeneity within each parcel, thus reducing the need for smoothing.

% Deformation field smoothness, measured here by the parameter $\omega_{sim},$ also seems to be a critical parameter. Indeed, when it is low, so that the estimation model overestimates it too much (Figure~\ref{fig:noise_simu_sigma_3D}, far left), then we see that modeling spatial uncertainty has little if any benefits. Conversely, when $\omega_{sim}$ is high, so that the estimation model underestimates it (Figure~\ref{fig:noise_simu_sigma_3D}, far right), then a significant drop in the MSE can be observed due to spatial modeling, especially at high SNRs.

Finally, our simulations suggest that the field regularity parameter $\omega_{est}$ of the estimation model should ideally be not larger than the actual deformation field smoothness to optimally account for spatial uncertainty. It should also not be too small to avoid over-fitting and computational issues (since more control points would be necessary to define less regular fields). However, tuning $\omega_{est}$ remains an issue in real-life applications, since we see no way of estimating the actual regularity of the deformations, and their distribution may be very far from that specified in Section~\ref{sec:deformation}.

\section{fMRI data}\label{sec:fmri_data}

We conclude this chapter on spatial incertitude modeling by an illustration on real fMRI data. We used the Localizer dataset \cite{Pinel07}, which is described in Section~\ref{sec:fmri}, restricted to a group of 38 right-handed subjects. Moderate preliminary smoothing by a $5mm$~Gaussian kernel was first used, in order to increase the SNR, as justified in the previous section. Then, each subject's data was pre-processed and analyzed using SPM5, as explained in Section~\ref{sec:single}.

For each contrast of interest $\br c$ considered, the subject's BOLD effect maps $\br y_i$ and estimation variance maps $\br s_i^2$ where calculated, following the notations in Section~\ref{sec:between}. These were used as input data to estimate the mean effect map $\bs \mu,$ as described in Section~\ref{sec:estimation}. As in the simulation studies, we estimated $\bs \mu$ both in the model with spatial uncertainty, and in the model without spatial uncertainty, setting $\sigma_S = 0.$ The deformation field smoothness was set to $\omega=4.5$ voxels, and each deformation field was defined by $60$ equally spaced control points.

$100$ iterations of the Gibbs sampler were used to average the posterior mean, following $100$ steps which were discarded, corresponding to a burn-in period. This took approximately $1$~hour and $40$~minutes, on a PC with $2.8$~GHz clock rate.

\subsection{Number processing task}

\begin{figure}[ht!]
\centering
\includegraphics[width=0.5\textwidth]{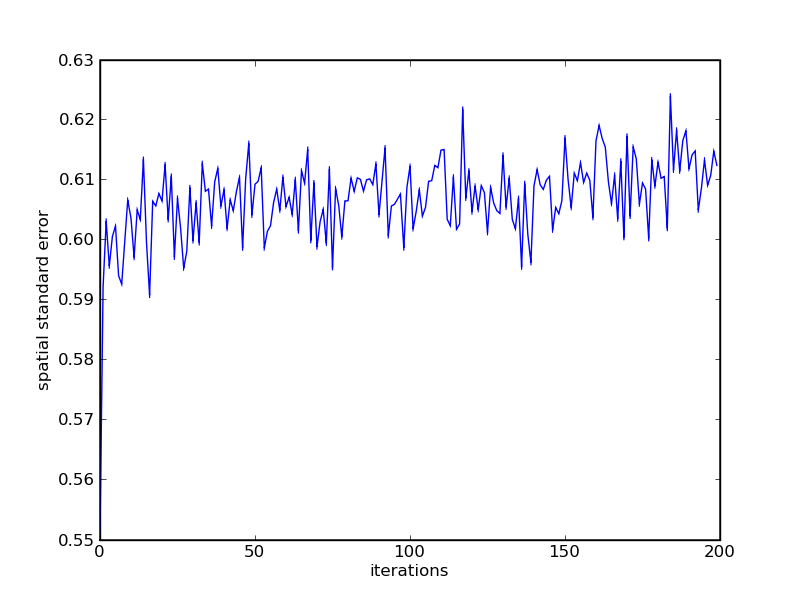}
\caption{\label{fig:std_plot_con0029_smoothed} Posterior sampling of $\sigma_S$ on real fMRI data.}
\end{figure}

Behavior of the Markov chain is illustrated in Figure~\ref{fig:std_plot_con0029_smoothed}, by the sampled values of the standard spatial displacement $\sigma_S.$ At first glance, it seems to settle almost instantly in a stable state, presumably next to a local mode of the posterior distribution.

However, the average value of $\sigma_S$ increases slightly across the iterations, suggesting that the chain is in fact progressing extremely slowly across the parameter space, and is still far from its stationary distribution. This implies that the Monte-Carlo approximation to the posterior mean, $\hat \sigma_S = 0.61$~voxels, corresponding to $4.3{\rm mm},$ is probably lower than the true posterior mean.

\begin{figure}[ht!]
\centering
\caption{\label{fig:posterior_mean_con0029_smoothed} Posterior estimates of $\bs \mu$ for a number processing task.}
\includegraphics[width=0.32\textwidth]{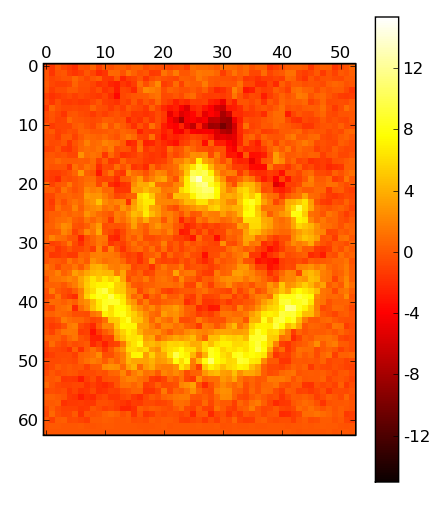}
\includegraphics[width=0.32\textwidth]{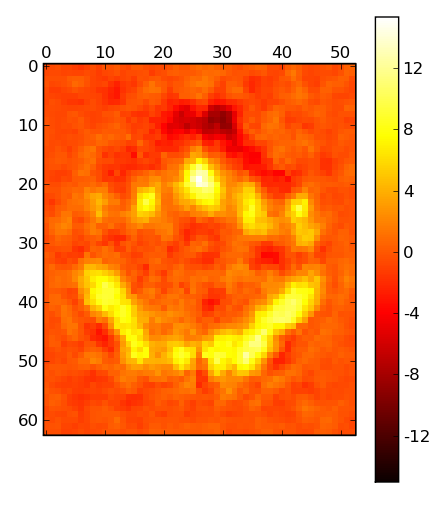}
\includegraphics[width=0.32\textwidth]{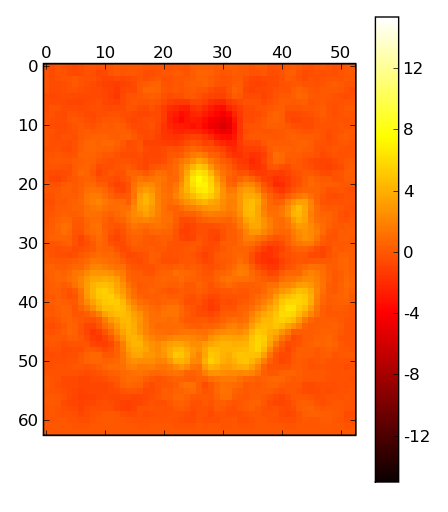}
\begin{tabular}{p{0.32\textwidth}p{0.32\textwidth}p{0.32\textwidth}}
No Spatial Uncertainty &
Spatial Uncertainty &
No Spatial Uncertainty \newline + Smoothing
\end{tabular}
\end{figure}

Even so, the effect of spatial modeling on the posterior mean estimation is clearly visible in Figure~\ref{fig:posterior_mean_con0029_smoothed}. The posterior mean effect map under spatial uncertainty (middle) is clearly smoother than the map obtained under no spatial uncertainty (left). It is also slightly more contrasted, with values ranging from $-14.9$ to $15.7$ against $-13.8$ to $12.7$ when not accounting for spatial uncertainty.

This regularization effect can be explained by the fact that
% The first one is our choice of a hierarchical prior for $\bs \mu$ (see Section~\ref{sec:estimation}) which introduces correlations between all voxelwise mean effects $\mu_k.$ However, the same prior is used fort the estimate without spatial uncertainty, with no visible effect. 
the mean effect map is averaged across a large number of displacements, sampled according to their posterior likelihood, resulting in a smoothing effect. Thus, it is natural to wonder whether spatial modeling does not act merely as a spatial filter, comparable to the linear and isotropic smoothing customarily applied to fMRI data.

We investigated this question by smoothing the posterior mean effect map obtained without spatial uncertainty, using a Gaussian filter with FWHM equal to $4,3mm,$ according to the standard spatial displacement estimated in the model with spatial uncertainty. As can be seen in Figure~\ref{fig:posterior_mean_con0029_smoothed}, right, the result is very different from the spatially uncertain posterior mean effect map, in that the image is smoother, but with less contrast, the values ranging from $-12.3$ to $11.6.$

Thus, the regularization induced by our spatial modeling approach is seen to be highly anisotropic, as it enhances the salient features of the effect map, while reducing the background noise. This effect is reminiscent of other Bayesian hierarchical modeling approaches to fMRI data analysis discussed in Sections~\ref{sec:glm} and \ref{sec:thresholding}, such as \cite{Gossl01b,Woolrich05,Woolrich06,Smith07b,Penny07,Makni08}. In all of these, a similar anisotropic smoothing effect is obtained by explicitly modeling correlations between neighboring voxels through a hidden discrete-valued Markov field. This Markov field describes the {\em state} of each voxel, {\em i.e.}, whether it is active, inactive, or de-activated. Yet another approach can be found in \cite{Harrison08}, which uses a non-stationary spatial regularisation prior on the effect map, which promotes regularization while preserving activation contours.

However, it is worth noting that all the above methods are concerned with single-subject analysis only, and promote regularization by imposing constraints on the subject's activation map. This is very different from our spatial uncertainty model (\ref{eq:between2}), which is defined at the between-subject level. Indeed, the regularization of the group effet map~$\bs\mu$ is essentially a by-product of integrating out unknown spatial deformations~$\br u$ applied to the individual activation maps. The posterior distributions of these deformations are more sharply peaked in regions were the individual images are easily matched (such as at the border of an activation region), than in regions where the matching is unclear (such as the background). This explains how the anisotropic smoothing effect is obtained. Their is little chance that the exact same effect could be achieved by only modeling the group effect map~$\bs\mu,$ according to one of the approaches cited above, which would ignore correlations between individual images.

\subsection{Language processing task}

\begin{figure}[ht!]
\centering
\includegraphics[width=0.5\textwidth]{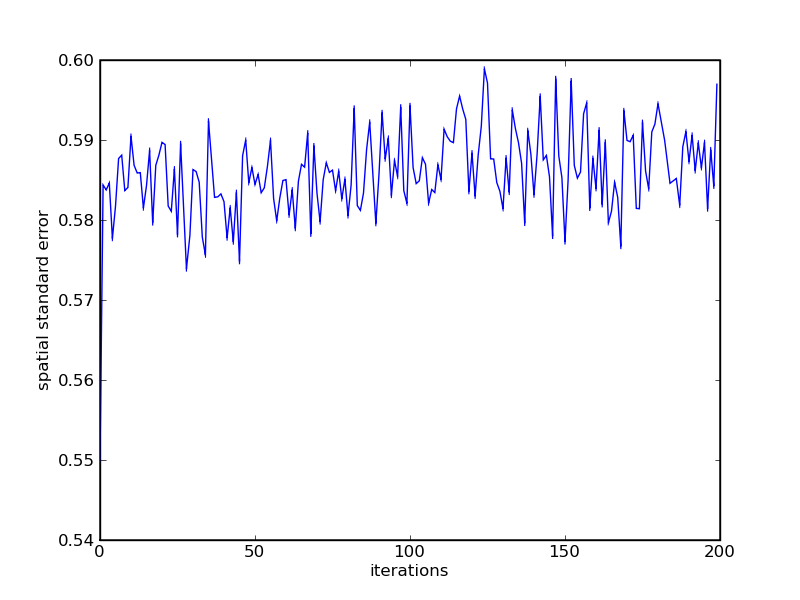}
\caption{\label{fig:std_plot_con0031_smoothed} Posterior sampling of $\sigma_S$ on real fMRI data.}
\end{figure}

We obtained similar results when analyzing data from a language processing task. The estimate of the standard spatial displacement $\hat \sigma_S = 0.59,$ corresponding to a FWHM of $4mm,$ was only slightly smaller, and the mixing of the Markov chain is also seen to be very slow from Figure~\ref{fig:std_plot_con0031_smoothed}.

\begin{figure}[ht!]
\centering
\caption{\label{fig:posterior_mean_con0031_smoothed} Posterior estimates of $\bs \mu$ for a language processing task.}
\includegraphics[width=0.32\textwidth]{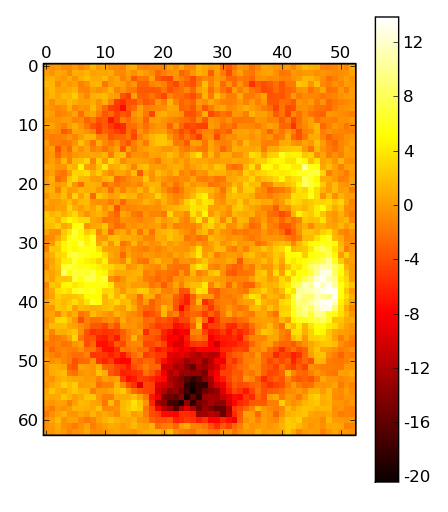}
\includegraphics[width=0.32\textwidth]{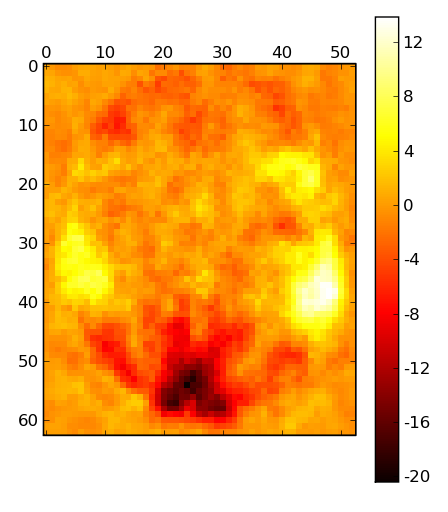}
\includegraphics[width=0.32\textwidth]{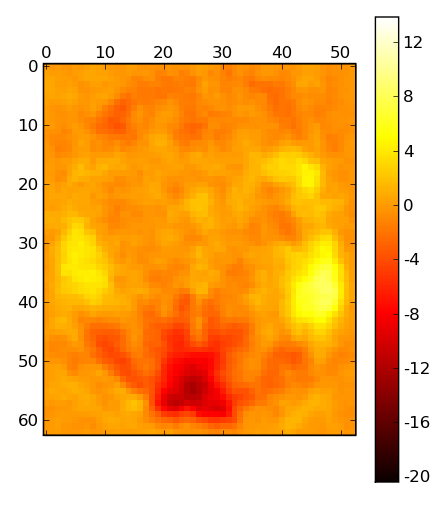}
\begin{tabular}{p{0.32\textwidth}p{0.32\textwidth}p{0.32\textwidth}}
No Spatial Uncertainty &
Spatial Uncertainty &
No Spatial Uncertainty \newline + Smoothing
\end{tabular}
\end{figure}

The gain in contrast and regularizing effect due to spatial modeling is also apparent from Figure~\ref{fig:posterior_mean_con0031_smoothed}, and we noticed the same marginal increase in contrast, with extremal values of $-33.4$ and $20.0$ against $-31.6$ and $19.0$ This combined effect could not be reproduced by a simple isotropic smoothing with the same FWHM value of $4mm.$

\section{Conclusion}\label{sec:conclusion2}

We have devised an approach for Bayesian inference on a spatially distorted signal, which can be applied to fMRI data. On simulated data, this approach estimated the original signal from noisy, warped observations with significantly more accuracy than when ignoring spatial uncertainty. The latter resulted in a blurred estimate, and the merging of neighboring signal peaks. The benefits of spatial modeling were also assessed on a large number of simulated datasets, with different values for the noise level, sample size and deformation field smoothness

Applied to real fMRI data, our method yielded smoother and more contrasted posterior mean effect maps when modeling spatial uncertainty. This effect could not be reproduced by naive linear isotropic smoothing.
% This is also to our knowledge the first time that a deformation template model is demonstrated on 3D data, since all the examples we have found in the literature either deal with 1D or 2D datasets, or with high-level features extracted from 3D data. This may be due to limitations in CPU time, however on complex datasets ($n=38$~images, $d\approx150\, 000$~voxels, $60\times 38$~control points), our Gibbs sampler ran in less than $2$~hours (though we were limited to $200$~iterations).
These results are very encouraging, and illustrate the benefits of accounting for the spatial uncertainty present in neuroimaging data through proper statistical modeling rather than isotropic smoothing, as is the common heuristic. The main difficulty we have encountered is the slow mixing rate of the Markov chain under spatial uncertainty, resulting in a conservative estimate of spatial uncertainty. This leaves space for further improvement, through the exploitation of more sophisticated sampling techniques. Thus, the proposal density for the Metropolis Hastings algorithm used to sample the elementary displacements is clearly sub-optimal; testing alternative proposals appears a promising line of work. Another possible strategy would be to design a {\em tempering} scheme to `flatten' the landscape of the posterior distribution \cite{Marin07}, which would allow the Markov chain to move more freely around the parameter space.

% Bayesian Model Selection for Activation Detection 
\eject
\pagestyle{empty}
$ $ \eject\pagestyle{fancy}
\pagestyle{fancy}
% \input{Chapter5}
% \chapter{Bayesian Detection of Functional Networks}\label{chap:bayesian}
\chapter{A Bayesian model selection approach to the detection of functional networks}\label{chap:bayesian}
\lhead{\emph{A Bayesian model selection approach to the detection of functional networks}}

\section*{Abstract}

In this chapter, we introduce a new paradigm for ROI-based fMRI group data analysis that overcomes certain limitations of the SPM-like approach. Using a Bayesian model selection framework, the functional network associated with a certain cognitive task is selected according to the posterior probabilities of mean regional activations, given a pre-defined parcellation of the brain. Thus our approach is threshold-free, while allowing to incorporate prior information, provided that the parcellation is sensible. Furthermore, by controlling a Bayesian risk, our approach balances false positive and false negative risks. Finally, it is based on the same spatial uncertainty model as in Chapter~\ref{chap:modeling}, and thus accounts for the mis-alignment of individual images, due to inevitable registration errors. As a consequence, the assignment of each voxel to a region is random rather than deterministic, and differs from one subject to another.

Results on simulated data show that badly localized effect can cause inactive regions to be detected by mistake. This bias toward false positives is reduced when modeling spatial uncertainties. However, the posterior probabilities estimates in the model with spatial uncertainty are numerically unstable, presumably because of the slow mixing of the Metropolis-Hastings algorithm used to simulate the displacement fields. We therefore propose a more stable approximate procedure, which consists in fixing the displacement fields to their most probable value {\em a posteriori}. This does not entirely reduce the bias toward false positives, which is further compensated through an additional penalty on model fit. The final procedure is validated on a simulated dataset.

\section{Introduction}

In the previous chapter, we have dealt with the assumption of perfect match between individual brains. This assumption was relaxed, by incorporating into the mass univariate model specified by (\ref{eq:within}) and (\ref{eq:between}) (see Section~\ref{sec:between}) a set of hidden variables, representing spatial normalisation errors. These are modeled as multivariate random fields, as described in Section~\ref{sec:deformation}.

\begin{figure}[ht!]
\begin{center}
\includegraphics[width=0.5\textwidth] {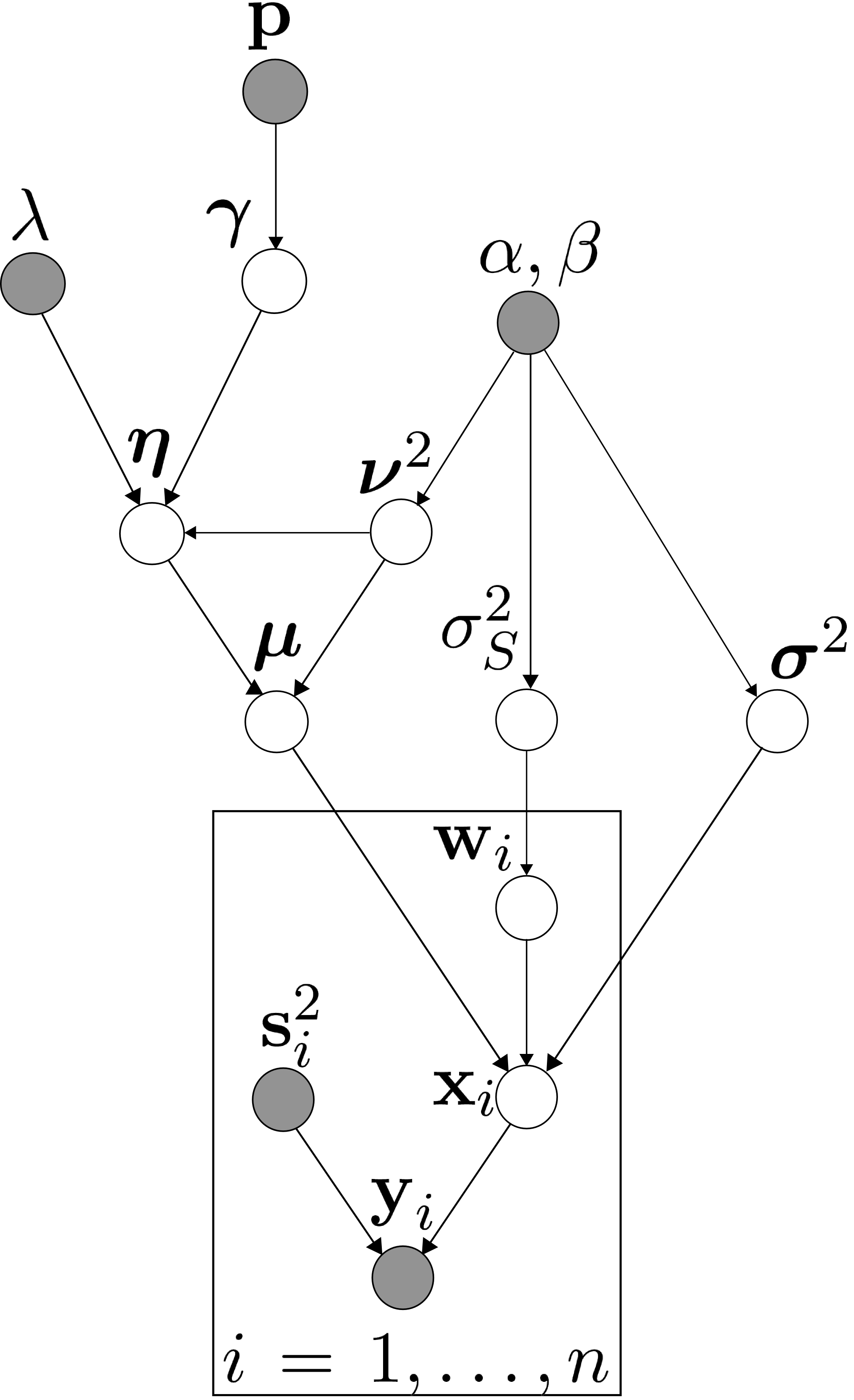}
\end{center}
\caption{\label{fig:graph_model} Directed acyclic graph (DAG) of the full hierarchical model. Gray circles correspond to fixed quantities (hyperparameters and observations), white circles to random quantities (latent variables). The conditional distribution of the subject-specific hidden effects $\br x_i$ and their estimation $\br y_i,$ are described in Section \ref{sec:spatial_model}, while the elementary displacements $\br w_i$ are modeled in Section~\ref{sec:deformation}; the distribution of the population mean effect $\bs\mu,$ conditional on the regional parameters $(\bs\eta, \bs\nu^2),$ is defined in Section~\ref{sec:regionalized}. Finally, the prior distribution on the model parameters $(\bs\eta, \bs\nu^2, \bs\sigma^2, \sigma_S^2)$ and the indicator variable $\bs\gamma,$ is specified in Section~\ref{sec:prior}.}
\end{figure}

We now define a more general hierarchical model for the data, based on this modeling of spatial uncertainty, and on a pre-defined parcellation of the brain volume into regions that are assumed functionally homogeneous, as described in Section~\ref{sec:regionalized}. This model is conveniently represented by its directed acyclic graph (DAG) \cite{Jordan03}, as illustrated in Figure~\ref{fig:graph_model}, showing its conditional dependence structure.

Based on this model, we define each region as {\em involved} in the task under study if it contains a nonzero mean effect, and {\em inactive} otherwise. Thus the unknown functional network we wish to recover is defined as a partition of regions into involved and inactive. Each partition defines a different generative model for the data. Therefore, selecting the right functional network can be seen as a model selection problem. In Section~\ref{sec:network}, we propose to do so in a Bayesian framework, rating each candidate network in terms of its posterior probability, given a prior distribution over all model parameters, defined in Section~\ref{sec:prior}.

Because regions are defined beforehand, this method is threshold-free, while taking advantage of the prior knowledge about functionally homogeneous regions, provided that the parcellation is sensible. By controlling a Bayesian risk, our approach balances false positives and false negatives, with relative weights that may be tuned depending on the application.

The idea of using pre-defined regions of interest (ROIs) for fMRI group data analysis has already been exploited in \cite{Bowman08}, but under the implicit assumption that individual images are perfectly registered. As shown in Section~\ref{sec:simulations}, activations estimated by mass univariate detection methods that neglect the spatial uncertainty caused by inevitable registration errors are swelled. This `stretching' effect may in turn cause inactive regions to be detected by error, a fact illustrated in Section~\ref{sec:simu}.

One of the main contributions of our work is to relax this assumption of perfect registration, by combining the spatial uncertainty model presented in Chapter~\ref{chap:modeling} with the regional response model in Section~\ref{sec:regionalized}. Thus for each subject, the membership of a voxel to a given region is probabilistic rather than deterministic. Because the membership probability accounts for the subject's own functional data, this effectively allows to de-weight the contribution of individual activations to regions they are likely not to belong to, but rather have been projected onto due to spatial normalization errors. In Section~\ref{sec:simu}, we show that accounting for this uncertainty indeed makes it possible to reduce the risk of falsely selecting a region as active. In contrast, using a probabilistic atlas, such as found in FSL, would not provide such subject-specific information, but rather inform on the uncertainty in the definition of the region itself.

The main technical difficulty of our approach is the computation of the posterior probability of each functional network in the model under spatial uncertainty, because it involves evaluating complex integrals on high dimensional spaces. Section~\ref{sec:marginal} describes how these integrals can be evaluated numerically, using Monte-Carlo Markov Chain techniques. However, these techniques turn out to be too computer intensive and numerically unstable to be used in practice. Instead, in Section~\ref{sec:independence} we introduce an approximation based on posterior modes, which requires much less computer time, and is more stable numerically. However, this approximation is less efficient in compensating the stretching effect due to displaced activations, so that the bias toward false positives, though reduced, is still present. In Section~\ref{sec:additional_penalty}, we compensate for this residual bias by incorporating an additional penalty on model fit, calibrated on simulated data. The final procedure is validated on a simulated dataset in Section~\ref{sec:phantom}, and we conclude by a discussion in Section~\ref{sec:discussion}.

\section{Regional response model}\label{sec:regionalized}

Our approach to functional network selection is based on an \textit{a priori} partition of the search volume $\mathcal V$ into $N$ disjoint regions of interest $\mathcal V = \mathcal V_1\, \cup\, \ldots\, \cup\, \mathcal V_N,$ assumed to be homogeneous functional areas. More precisely, we re-define the population mean effects $\mu_k$ as spatially independent Gaussian random variables, identically distributed within each region.

Thus, for all region~$j=1, \ldots, N,$ and for all voxel~$k$ such that $\br v_k \in \mathcal V_j:$
\begin{eqnarray}\label{eq:regional}
\mu_k &=& \eta_j + \chi_k; \quad\chi_k \stackrel{i.i.d.}{\sim} \mathcal N(0, \nu_j^2),
\end{eqnarray}
In other terms, $\mu_k,$ previously defined as a voxelwise fixed effect (see Section~\ref{sec:between}), is now expressed as the sum of a regional fixed effect $\eta_j,$ representing the average BOLD response in region~$j,$ and a voxelwise random effect $\chi_k$ representing the variability of the response across voxels.

The same idea of modeling fMRI data based on fixed parcels containing voxels with similar BOLD responses can be found in \cite{Bowman08} (see Section~\ref{sec:ROI}). Furthermore, in this work the regional means are modeled jointly as a Gaussian vector, with arbitrary covariance matrix, allowing to measure functional correlations between distant regions, as well as between voxels from a same region. Though modeling these correlations is not our primary goal, and has not been considered here for simplicity, it does constitute a promising direction in which our approach could be extended.

\subsection{Generative model without spatial uncertainty}

We start by assuming that the estimated effect maps $(\br y_i)_{1\leq i\leq n}$ are perfectly aligned. Under the assumptions of the regional model (\ref{eq:regional}), the within-subject (\ref{eq:within}) and between-subject (\ref{eq:between}) models defined in Section~\ref{sec:between}, we specify a generative model for the the data, with the further assumption that the between-subject variance~$\sigma_k^2$ in (\ref{eq:between}) is now uniform within each region~$j.$ As discussed in Section~\ref{sec:observation}, this modification will be necessary when modeling spatial uncertainty in the next section, to avoid overfitting the data.

Thus, for all subject~$i=1,\ldots,n,$ all region~$j=1,\ldots,N$ and all voxel~$k$ such that $\br v_k \in \mathcal V_j:$
% \begin{eqnarray}\label{eq:regional_obs}
% y_{i,k} &=& \eta_j + \chi_k + \xi_{i,k} + \varepsilon_{i,k}; \\
% \xi_{i,k} &\stackrel{i.i.d.}{\sim}& \mathcal N(0, \sigma_j^2) ;\nonumber\\
% \varepsilon_{i,k} &\stackrel{ind.}{\sim}& \mathcal N(0, s_{i,k}^2),\nonumber
% \end{eqnarray}
\begin{eqnarray}\label{eq:regional_obs}
y_{i,k} &=& x_{i,k} + \varepsilon_{i,k}; \quad \varepsilon_{i,k} \stackrel{ind.}{\sim} \mathcal N(0, s_{i,k}^2) \\
x_{i,k} &=& \mu_k + \xi_{i,k}; \quad \xi_{i,k} \stackrel{i.i.d.}{\sim} \mathcal N(0, \sigma_j^2), \nonumber\\
\end{eqnarray}
% where $\xi_{i,k}$ is the between-subject, voxelwise, random effect, and $\varepsilon_{i,k}$ the within-subject, voxelwise, random effect, assuming the mutual independence of both noise processes.
where we assume the mutual independence of the noise processes $\xi_{i,k}$ and $\varepsilon_{i,k}.$ 

Using the expression of $\mu_k$ in (\ref{eq:regional}), we can re-write the above model as:
\begin{eqnarray}
y_{i,k} &=& \eta_j + \chi_k + \xi_{i,k} + \varepsilon_{i,k}. \nonumber \\
\end{eqnarray}
This formulation corresponds to a mixed-effect analysis of variance (MFX-ANOVA) model with a single factor, whose $N$~levels correspond to the different regions. In fact, since the noise processes are assumed independent across voxels, it is simply the aggregation of $N$~independent models, one for each region~$j.$

This generalizes the mass univariate model, defined in Section~\ref{sec:between}, which corresponds to the limiting case $N=d,$ that is, when regions reduce to single voxels, under the identifiability constraint: $\nu_j^2 \equiv 0.$ In this case, the regional mean $\eta_j$ coincides with the population mean effect $\mu_k,$ and the regional between-subject variance $\sigma_j^2$ becomes voxel dependent. 
% Besides, the unobserved effects $x_{i,k},$ defined in Section~\ref{sec:between}, can be recovered from the above display as: $x_{i,k} = \eta_j + \chi_k + \xi_{i,k}.$

\subsection{Generative model with spatial uncertainty}\label{sec:spatial_model}

We now relax the assumption that the individual images are perfectly normalized, so that the between-subject model~(\ref{eq:between}) is replaced by its generalization to spatial uncertainty~(\ref{eq:between2}), introduced in Chapter~\ref{chap:modeling}. Spatial normalization errors are defined for each subject~$i$ as a spatial deformation field~$\br u_i$, controlled by a set of hidden variables~$\br w_i,$ as explained in Section~\ref{sec:deformation}.

As in the previous section, we combine this observation model with the regional model (\ref{eq:regional}), and assume the between-subject variance~$\sigma^2$ to be region-dependent. It turns out that, conditionally on the hidden displacements variables~$\br w_i,$ the data is still distributed according to a MFX-ANOVA model, independently across regions. However, (\ref{eq:regional_obs}) must be adapted to account for observations being displaced across regions.

Consequently, for all subject~$i=1,\ldots,n,$ all region~$j=1,\ldots,N,$ we have for all voxels~$k$ such that $\br v_k + \br u_{i,k} \in \mathcal V_j:$
% \begin{eqnarray}\label{eq:cond_regional_obs}
% y_{i,k} | \br w_i&=& \eta_j + \bs\chi(\br v_k + \br u_{i,k}) + \xi_{i,k} + \varepsilon_{i,k}; \\
% \xi_{i,k} | \br w_i &\stackrel{i.i.d.}{\sim}& \mathcal N(0, \sigma_j^2) ;\nonumber\\
% \varepsilon_{i,k} | \br w_i &\stackrel{ind.}{\sim}& \mathcal N(0, s_{i,k}^2).\nonumber
% \end{eqnarray}
\begin{eqnarray}\label{eq:cond_regional_obs}
y_{i,k} &=& x_{i,k} + \varepsilon_{i,k}; \quad \varepsilon_{i,k} \stackrel{ind.}{\sim} \mathcal N(0, s_{i,k}^2) \\
x_{i,k} &=& \bs\mu(\br v_k + \br u_{i,k}) + \xi_{i,k}; \quad \xi_{i,k} | \br w_i \stackrel{i.i.d.}{\sim} \mathcal N(0, \sigma_j^2). \nonumber
\end{eqnarray}

Again, using (\ref{eq:regional}), this can be re-written as:
\begin{eqnarray}
y_{i,k} | \br w_i&=& \eta_j + \bs\chi(\br v_k + \br u_{i,k}) + \xi_{i,k} + \varepsilon_{i,k}. \nonumber\\
% \xi_{i,k} | \br w_i &\stackrel{i.i.d.}{\sim}& \mathcal N(0, \sigma_j^2) ;\nonumber\\
% \varepsilon_{i,k} | \br w_i &\stackrel{ind.}{\sim}& \mathcal N(0, s_{i,k}^2).\nonumber
\end{eqnarray}

% We note that, according to (\ref{eq:between2}), the unobserved effects $x_{i,k}$ can be recovered from the above display as: 
% $$
% x_{i,k} = \br x_i(\br v_k) = \eta_j + \bs\chi(\br v_k + \br u_{i,k}) + \xi_{i,k}.
% $$
Thus, the main difference with the case of no spatial uncertainty is that the mean of any given observation~$y_{i,k}$ now depends on the region~$j$ toward which the voxel~$k$ is displaced, conditional on $\br w_i,$ rather than the region it belongs to. The same observation holds for the conditional variance $\sigma_j^2$ of the hidden effects~$x_{i,k}.$ Finally, because the set of voxels displaced to region~$j$ fluctuates with $\br w,$ the data is no longer independent across regions, after integrating out $\br w.$

% In the following, we will distinguish the vector of hidden variables: 
% $$
% \br z = (\br x, \br w, \bs\mu) \in \mathbb R^{n(d + B) + d}
% $$
% from the parameter vector:
% $$
% \bs\theta = (\bs \eta, \bs\nu^2, \bs\sigma^2, \sigma_S^2) \in \mathbb R^{3N + 1}.
% $$
% Strictly speaking, this distinction is purely arbitrary in the Bayesian context, since all these quantities are unobserved, random variables. However, it will be useful 

This model generalizes all the ones introduced in the previous sections and chapters of this work. Indeed, the regional model without spatial uncertainty is the special case $\sigma_S = 0,$ while the mass univariate model in Section~\ref{sec:between}, is obtained for $\sigma_S = 0$ and $N=d.$ Finally, the model introduced in Chapter~\ref{chap:modeling} corresponds to the case of a single region ($N=1$).

% The same idea of modeling fMRI data hierarchically, based on fixed parcels containing voxels with similar BOLD responses can be found in \cite{Bowman08} (see Section~\ref{sec:ROI}). However, our approach is more general in that it accounts for registration errors, and consequently the fact that individual fMRI data may not be aligned with the pre-defined ROIs. As such the model in \cite{Bowman08}, corresponds to the special of no spatial uncertainty ($\sigma_S = 0$). Conversely, in \cite{Bowman08}, the regional means are modeled jointly as a Gaussian vector, with arbitrary covariance matrix, allowing to measure functional correlations between ROIs. In this sense, it generalizes our model, which corresponds to the special case where this covariance matrix is assumed spherical.

\section{Bayesian model selection framework}\label{sec:network}

Based on the regional response model, we now propose an answer to the basic question underlying most fMRI paradigms, namely, that of selecting the subset of regions that are involved in the task at hand. More precisely, we define region~$j$ as:

\begin{itemize}

\item {\em active} if $\eta_j > 0;$

\item {\em negatively active} if $\eta_j < 0;$

\item {\em inactive}  if $\eta_j = 0,$

\end{itemize}

and define a region~$j$ as {\em involved} if it is either active or negatively active. From a hypothesis testing perspective, we wish to test, for each~$j=1, \ldots, N,$ $\mathcal H_{0,j}:$ `$\eta_j = 0$' versus $\mathcal H_{1,j}:$ `$\eta_j \neq 0$'.

Alternatively, recovering the subset of regions that are involved in the considered task can be seen as a model selection problem, by introducing the indicator variable $\bs\gamma = (\gamma_1, \ldots, \gamma_N) \in \{0,1\}^N,$ such that for all region~$j$: $\gamma_j = 0$ means that $\eta_j = 0,$ and $\gamma_j = 1$ means that $\eta_j \neq 0.$ Each of the $2^N$ possible values of the indicator variable $\bs\gamma$ then defines a different generative model $\mathfrak M_{\bs\gamma}$ describing the data $\br y,$ corresponding to a different subset of explanatory variables $\bs \eta_{\bs\gamma} = (\eta_j)_{j,\, \gamma_j=1}.$ In the following, we will refer to $\bs\gamma$ as the {\em functional network} variable, since it defines the network of regions functionally engaged in the task under study.

According to these notations, our task consists in selecting the `best' (in a certain sense) model $\mathfrak M_{\bs\gamma}$ to describe the data $\br y,$ {\em i.e.}, the best segmentation of the regions into involved and inactive. Equivalently, it can be formulated as an estimation problem, with $\bs\gamma$ as the interest parameter. The Bayesian approach to this problem consists in considering the unknown quantities $\bs\gamma$ and $\bs\theta = (\bs \eta, \bs\nu^2, \bs\sigma^2, \sigma_S^2)$ as random variables, and defining a prior density $\pi(\bs\theta | \bs\gamma) \pi(\bs\gamma),$ representing the prior knowledge on their possible values.

Note that, from a frequentist viewpoint, limiting $\bs\theta$ to the variables cited above makes sense, since these are the unknown, but fixed, quantities, we wish to infer, as opposed to $\br z = (\br x, \br w, \bs\mu),$ which are unknown random variables considered as nuisance factors, and marginalized out during the inference. In the Bayesian setting we have adopted on the other hand, this opposition has no real justification, since both $\bs\theta$ and $\br z$ are vectors of hidden random variables, whose posterior distribution we wish to determine. However, we will see that these notations are convenient in the context of the model selection algorithm proposed in Section~\ref{sec:Chib}. In particular, the expression of the complete density $f(\br y, \br z| \bs\theta)$ is that of a curved exponential model, allowing the evaluation of the MAP estimates of $\bs\theta$ using an efficient MCMC-SAEM algorithm, as described in Appendix~\ref{app:SAEM}.

Based on the prior $\pi(\bs\theta | \bs\gamma) \pi(\bs\gamma)$ and on the likelihood function~$f(\br y | \bs\theta)$ of the full hierarchical model specified by (\ref{eq:cond_regional_obs}), (\ref{eq:deformation}) and (\ref{eq:elementary}), the posterior distribution of $\bs\theta$ under each model $\mathfrak M_{\bs\gamma}$ can be computed according to Bayes' theorem:

\begin{eqnarray}\label{eq:post}
\pi(\bs\theta |\br y, \bs\gamma)
&=&
\frac
{f(\br y | \bs\theta) \pi(\bs\theta | \bs\gamma)}
{\int f(\br y | \bs\theta) \pi(\bs\theta | \bs\gamma) d\bs\theta},
\end{eqnarray}
where 
\begin{eqnarray}\label{eq:evidence}
m(\br y | \bs\gamma) &=& 
\int f(\br y | \bs\theta) \pi(\bs\theta | \bs\gamma) d\bs\theta
\end{eqnarray}
is the {\em marginal likelihood}, or {\em evidence}, of model $\mathfrak M_{\bs\gamma}.$ This quantity is central to Bayesian model selection, since the posterior distribution of the indicator variable can be expressed as:

\begin{eqnarray}
\label{eq:post_prob}
\pi (\bs \gamma | \br y) 
&=& 
\frac
{m (\br y | \bs \gamma) \pi (\bs \gamma)}
{\sum_{\bs\gamma} m (\br y | \bs \gamma) \pi (\bs \gamma)}.
\end{eqnarray}

Assuming that $m (\br y | \bs \gamma)$ can be computed for all possible values of $\bs\gamma,$ the most probable functional network given the data can be selected, according to:
\begin{eqnarray}
\hat{\bs \gamma} &=& \arg\max_{\bs \gamma} \pi (\bs \gamma | \br y),\nonumber\\
 &=& \arg\max_{\bs \gamma} m (\br y | \bs \gamma) \pi (\bs \gamma).\nonumber
\end{eqnarray}

Besides being intuitive, this procedure has a number of attractive features. For instance, because $\bs\gamma$ takes only discrete values, its MAP estimate $\hat{\bs \gamma}$ is also the Bayesian estimator associated to the $0-1$ loss function $\ell(\bs \gamma, \tilde{\bs \gamma}) = \sum_j \bs 1_{\bs \gamma_j \neq \tilde{\bs \gamma}_j},$ {\em i.e.}, it minimizes over all possible networks $\tilde{\bs\gamma}$ the Bayesian risk:

\begin{eqnarray}
E[ \ell(\bs \gamma, \tilde{\bs \gamma}) | \br y]
&=&
\sum_{\bs \gamma \in \{0, 1\}^N} \ell(\bs \gamma, \tilde{\bs \gamma}) \pi (\bs \gamma | \br y).
\end{eqnarray}

From a multiple testing perspective, this means that $\hat{\bs \gamma}$ minimizes the {\em a posteriori} expected sum of type I (false positive) and type II errors (false negative) errors. Other loss functions could be used, to give different weights to false positive and false negative risks. Thus our Bayesian decision-theoretic framework answers one of the limitations of the SPM-like approach, which only controls type I risks.

% One may also give different weights to false positive and false negative risks through the weighted loss
% $$
% \ell_{\rho}(\bs \gamma, \hat{\bs \gamma})
% =
% \sum_j
% \rho \mathbf 1 \{ \bs \gamma_j < \hat{\bs \gamma}_j \}
% + (1 - \rho) \mathbf 1 \{ \bs \gamma_j > \hat{\bs \gamma}_j \},
% $$
% where $\rho \in (0,1)$ controls the relative weight given to false positives; values next to $1$ lead to an almost exclusive control on false positives, and values next to $0$ lead to control mainly false negatives, while $\rho = 0.5$ is equivalent to the $0-1$ loss. This answers one of the limitations of the SPM-like approach which only controls false positive risks.

\section{Prior specification}
\label{sec:prior}

% The posterior probability of a given network $\bs \gamma$ is given by:
% where $\pi (\bs \gamma)$ is the prior probability assigned to $\bs \gamma,$ and $m (\br y | \bs \gamma)$ the {\em marginal likelihood}, or {\em evidence} of the model specified by $\bs \gamma:$
% \begin{eqnarray}
% \label{eq:evidence}
% m (\br y | \bs \gamma) &=& \int f (\br y | \bs \theta, \bs \gamma) \pi (\bs \theta | \bs \gamma) d\bs \theta,
% \end{eqnarray}
% for a certain prior distribution $\pi (\bs \theta | \bs \gamma)$ on the model parameters, $\bs \theta = (\bs \eta, \bs \nu^2, \bs \sigma^2, \sigma_S^2).$ 

We now address the choice of a prior distribution for the functional network $\bs\gamma,$ and the model parameters $\bs\theta.$ Under limited information about their possible values, it is natural to use a weak prior. This choice is problematic in generalized linear mixed models (GLMMs), because Jeffreys prior may be intractable, and the standard scale invariant (improper) prior may lead to an improper posterior \cite{Hobert96}. Instead, we have adopted normal priors on the effects and inverse-Gamma priors for variance components. These are practical choices, their conditional conjugate form making them amenable to posterior sampling. Our prior is specified as follows:

\subsection{Regional means}

For each region~$j,$ we define:

\begin{eqnarray}\label{eq:prior_mean}
\eta_j | \nu_j^2, \gamma_j = 0 &\sim& \delta(0) \\
\eta_j | \nu_j^2, \gamma_j = 1 &\sim& \mathcal N(m, \nu_j^2 / \lambda),\nonumber
\end{eqnarray}

where $\delta(0)$ is the Dirac mass in 0, meaning that in inactive regions, the regional mean vanishes almost surely; in the other regions, the prior mean is set to $m = 0$ so as not to bias the inference towards positive or negative effects and, the scale parameter is set to $\lambda = 10^{-3}$, which may be interpreted as the weight given to the prior mean $m$ with respect to one observation.

\subsection{Variance components}

All variance components share the same prior:
\begin{eqnarray}\label{eq:prior_var}
\pi(\sigma_S^2, \bs \nu^2, \bs \sigma^2) 
&=& 
\mathcal{IG} (\sigma_S^2 ; \alpha, \beta)
\prod_{j=1}^N \big\{
\mathcal{IG} (\nu_j^2 ; \alpha, \beta)
\mathcal{IG} (\sigma_j^2 ; \alpha, \beta)
\big\}
\end{eqnarray}
where  $\mathcal{IG} (\alpha, \beta)$ is the Inverse-Gamma distribution with parameters $(\alpha,\beta)$, and density function
$$
\mathcal{IG} (\, z\, ;\alpha,\beta) = \frac{\beta^{\alpha}}{\Gamma(\alpha)}z^{-\alpha-1}\exp\left(\frac{-\beta}{z}\right).
$$ 

Tuning $\alpha$, and $\beta$ is a difficult task in absence of prior information. A popular choice is to use a ``just'' proper prior in absence of prior knowledge, given for instance by $\alpha = \beta = 10^{-3}$ \cite{Spiegelhalter96}. This common practice has raised some concerns \cite{Natarajan98}, because they can result in poorly behaved posterior sampling schemes, due to the near impropriety of the resulting posterior.

Thus we chose values which reflected the limited amount of knowledge available on the interest parameters, while defining truly proper priors, setting $\alpha=3, \beta=20.$  This means for instance that in each region~$j$, the population variance $\sigma_j^2$ has a prior mean of $10$, and a prior variance of $10^2,$ which seems reasonable from practical experience. The same considerations hold for the regional variances $\nu_j^2,$ and the spatial variance parameter $\sigma_ S^2.$

Several alternative priors are proposed in \cite{Natarajan00}, including a uniform shrinkage prior, and an approximate Jeffreys prior. However, we found the results obtained by the standard priors satisfying enough not to consider these more complex strategies.

\subsection{Indicator variables}

We define independent Bernoulli priors for the indicators variables:

\begin{eqnarray}\label{eq:prior_network}
\pi(\bs\gamma) &=& \prod_{j=1}^N \mathcal B(\gamma_j; 1, p_j),
\end{eqnarray}

meaning that region~$j$ has a prior probability of $p_j$ of being activated. In the following, we use the default choice $p_j \equiv 0.5,$ but prior information on the state of regions can easily be included at this stage, such as results of previous analyses.

\section{Evaluating the marginal likelihood}\label{sec:marginal}

We now address the computation of the marginal likelihood $m(\br y | \bs\gamma),$ defined by (\ref{eq:evidence}).
% , of the hierarchical model (\ref{eq:within2}), (\ref{eq:between2}), (\ref{eq:deformation}), (\ref{eq:elementary}), (\ref{eq:regional}), (\ref{eq:prior_mean}), (\ref{eq:prior_var}), and (\ref{eq:prior_network}). 
As is often the case, this is a difficult task, because it requires integrating the probability density function (pdf) $f(\br y| \bs\theta),$ defined on a high dimensional space, with respect to the prior density $\pi(\bs\theta | \bs\gamma),$ which cannot be done analytically in our case. 
% Using a Laplace approximation would require computing the Hessian matrix of the log un-normalized posterior density $\log f(\br y | \bs\theta)\pi(\bs\theta | \bs\gamma).$ However, this function cannot be computed in the model with spatial uncertainty.

Thus, we turn to numerical approximation strategies. The main challenge in evaluating $m(\br y |\bs\gamma)$ is that the integration must be done with respect to the prior density $\pi(\bs\theta | \bs\gamma),$ but the values significantly contributing to the integral are concentrated around the high density points of the posterior $\pi(\bs\theta | \br y, \bs\gamma).$ 
% Thus, the naive Monte-Carlo approach consisting in sampling the pdf under the prior yields an estimate which, though unbiased, has such a high variance that it is useless. 
We start by briefly reviewing some of the main strategies that have been developed to deal with this classical problem, and discuss their relevance to the present case.

\subsubsection{Importance sampling}

A naive way of estimating $m(\br y|\bs\gamma),$ justified by (\ref{eq:evidence}), would be to generate a sample $(\bs\theta_g)_{1\leq g\leq G}$ from the prior distribution $\pi(\bs\theta|\bs\gamma),$ and deduce the following Monte-Carlo estimate:
$$
\hat m(\br y|\bs\gamma)
= G^{-1} \sum_{g=1}^G f(\br y | \bs\theta_g).
$$

As mentioned earlier, the significant values of the likelihood function $f(\br y | \bs\theta)$ are concentrated on a small region of the support of $\pi(\bs\theta|\bs\gamma),$ so most of the sampled values $f(\br y | \bs\theta_g)$ are likely to be close to zero. Hence, this estimate, though unbiased, would have such a high variance that it would be useless.

A natural alternative to sampling under the prior, known as {\em importance sampling}, is to use a proposal distribution $q(\bs\theta)$ instead, noting that:
$$
m(\br y | \bs\gamma) 
= \int \frac{f(\br y | \bs\theta) \pi(\bs\theta | \bs\gamma)}{q(\bs\theta)} q(\bs\theta) \bs d\theta.
$$
Thus, if $(\bs\theta_g)_{1\leq g\leq G}$ is a sample from $q(\bs\theta),$ an unbiased estimate of $m(\br y |\bs\gamma)$ is given by:
$$
\hat m_q(\br y | \bs\gamma) = 
G^{-1} \sum_{g=1}^G \frac{f(\br y | \bs\theta_g)\pi(\bs\theta_g| \bs\gamma)}{q(\bs\theta_g)}.
$$
The likelihood function $f(\br y | \bs\theta)$ is unknown in the model with spatial uncertainty, but this method could still be applied, replacing $\bs\theta$ by $(\bs\theta, \br w),$ since the density conditional on the displacement parameters $\br w$ can be computed explicitly (see Appendix~\ref{app:likelihood}).

Thus, $\hat m_q(\br y|\bs\gamma)$ is still an unbiased estimator of $m(\br y|\bs\gamma).$ Its variance, which depends on the proposal, is given by:
$$
V \hat m_q(\br y | \bs\gamma)
= \frac{1}{G} \int \left(\frac{f(\br y | \bs\theta)\pi(\bs\theta | \bs\gamma)}{q(\bs\theta)} - m(\br y|\bs\gamma) \right)^2 q(\bs\theta) d\bs\theta.
$$
In particular, the variance tends to zero as $q(\bs\theta)$ gets closer to the posterior density $\pi(\bs\theta | \br y, \bs\gamma).$ Thus, the main difficulty of this approach is to choose $q(\bs\theta)$ `close to' $\pi(\bs\theta | \br y, \bs\gamma)$ which can indeed be hard when little is known on the shape of the posterior density, as is the case here.

A generalization of importance sampling is bridge sampling \cite{Meng96}, based on the following identity:
$$
m(\br y | \bs\gamma) =
\frac
{\int f(\br y | \bs\theta) \pi(\bs\theta | \bs\gamma) h(\bs\theta) q(\bs\theta) \bs d\theta}
{\int q(\bs\theta) h(\bs\theta) \pi(\bs\theta | \br y, \bs\gamma) d\bs\theta},
$$
valid for any choice of functions $h$ and $q.$ Given a sample $(\bs\theta_g)_{1\leq g\leq G}$ from $q(\bs\theta)$ as before, and a sample $(\bs\theta_j)_{1\leq j\leq J}$ from the posterior density $\pi(\bs\theta | \br y, \bs\gamma)$ (in our case, replacing $\bs\theta$ by $(\bs\theta, \br w),$ this can be done by by the Metropolis-with Gibbs algorithm in Appendix~\ref{app:MCMC}), an unbiased estimate estimate of $m(\br y |\bs\gamma)$ is:
$$
\hat m_{BS}(\br y | \bs\gamma) = 
\frac
{G^{-1} \sum_{g=1}^G f(\br y | \bs\theta_g) \pi(\bs\theta_g| \bs\gamma) h(\bs\theta_g)}
{J^{-1} \sum_{j=1}^J q(\bs\theta_j) h(\bs\theta_j)}.
$$
This method requires the choice of an additional function $h(\bs\theta),$ and the algorithm reduces to importance sampling when $h(\bs\theta)\equiv 1.$ \cite{Meng96} show that certain choices of $h$ can reduce the variance of the classical importance sampling estimate, and indicate iterative strategies to choose $h.$ However, the choice of a proposal~$q$ remains an issue in our case, and it is not clear how this method performs for `bad choices' (what happens for instance if $q$ is chosen equal to the prior?). Thus, it is not clear how importance Monte-Carlo sampling strategies could be applied in a simple way to the present problem.

\subsubsection{Harmonic mean estimator}

The harmonic mean estimator (HME) in \cite{Raftery07} constitutes a very simple strategy to estimate the marginal likelihood from the output of any posterior sampling scheme, based on the {\em harmonic mean identity}:
$$
\frac{1}{m(\br y | \bs\gamma)}
=
\int\frac
{1}
{f(\br y | \bs\theta)}
\pi(\bs\theta | \br y, \bs\gamma)
\bs d\theta.
$$
Thus, given a sample $(\bs\theta_j)_{1\leq j\leq J}$ from the posterior density, an unbiased estimate of the marginal likelihood is given by:
$$
\hat m_{HM}(\br y | \bs\gamma) = 
\left(
J^{-1} \sum_{j=1}^J 
\frac
{1}
{f(\br y | \bs\theta_j)}
\right)^{-1}.
$$
As with importance sampling, this method could easily be applied in our case, replacing $\bs\theta$ by $(\bs\theta, \br w),$ and sampling their joint posterior density using the sampling scheme described in Appendix~\ref{app:MCMC}.

In spite of its appealing simplicity, the HME is known for its lack of numerical stability, as measured by the variance: ${\rm Var} [f(\br y | \bs\theta)^{-1} | \br y],$ making its applicability hazardous. Indeed, this variance is determined by the second moment:
\begin{eqnarray}
{\rm E}[f(\br y | \bs\theta)^{-2} | \br y] &=& 
\int\frac
{1}
{f(\br y | \bs\theta)^2}
\pi(\bs\theta | \br y, \bs\gamma)
\bs d\theta \nonumber\\
&=& 
\frac{1}{m(\br y | \bs\gamma)}
\int\frac
{\pi(\bs\theta | \bs\gamma)}
{f(\br y | \bs\theta)}
\bs d\theta.
\end{eqnarray}
For the above integral to be finite, the prior distribution~$\pi(\bs\theta | \bs\gamma)$ must have lighter tails than the density $f(\br y | \bs\theta)$ (viewing the latter as a function of $\bs\theta$). Intuitively, this means that the HME is numerically stable when the prior is more sharply peaked, {\em i.e.}, provides more information on the parameter of interest, than the data! \cite{Raftery07} show for instance that this is not the case in the simple Gaussian model, with the usual conjugate inverse Gamma-Gaussian prior distribution on the Gaussian mean and variance (as defined in Section~\ref{sec:prior}).

\cite{Raftery07} indicate possible ways of stabilizing $\hat m_{HM}(\br y | \bs\gamma),$ which consist in replacing $f(\br y | \bs\theta_j)$ by a marginalized version $f(\br y | h(\bs\theta_j)),$ for an arbitrary function~$h.$ It is shown that in some cases the modified estimator has finite variance. However, the choice of $h$ is strongly model-dependent, and we have found in our case no such convenient function ensuring finite variance for the HME.

\subsubsection{Variational Bayes}

The variational Bayes (VB) framework \cite{Beal03} can be used to compute a lower bound on the marginal likelihood. This bound originates in an identity similar to that upon which the importance sampling approach is based,
$$
m(\br y | \bs\gamma) 
= \int 
q(\br z, \bs\theta) \frac{f(\br y, \br z, \bs\theta | \bs\gamma)}{q(\br z, \bs\theta)} 
d\br z d\bs\theta,
$$
valid for any proposal distribution $q(\br z, \bs\theta),$ with $\br z = (\br x, \bs\mu, \br w).$ Taking the logarithm on both sides, and applying Jensen's inequality yields (thanks to the concavity of the logarithm)
$$
\log m(\br y| \bs\gamma)
\geq\int
q(\br z, \bs\theta) 
\log
\frac{f(\br y, \br z, \bs\theta | \bs\gamma)}{q(\br z, \bs\theta)}
d\br z d\bs\theta.
$$
Thus we obtain a lower bound on the marginal likelihood, noted $\mathcal F_{\bs\gamma}(q)$ in the following. This inequality becomes an equality for $q(\br z, \bs\theta) = \pi(\br z, \bs\theta | \br y, \bs\gamma),$ but of course this proposal cannot be used directly, because its normalizing constant is unknown, since it is precisely the marginal likelihood we wish to compute.

The goal of VB approaches is to maximize $\mathcal F_{\bs\gamma}(q)$ with respect to $q,$ restricting the latter to a class $\mathcal C$ of functions such that the maximization may be performed efficiently. Typically, $\mathcal C$ is defined as a set of functions which are factored across several blocks of latent variables, for instance $\mathcal C = \{q |\, q(\br z, \bs\theta) = q_{\br z}(\br z)\, q_{\bs\theta}(\bs\theta)\}.$ In this case, $\mathcal F_{\bs\gamma}(q)$ can be maximized using the VBEM algorithm, which alternates maximizations over $q_{\br z}(\br z)$ and $q_{\bs\theta}(\bs \theta),$ until convergence to a local maximum. It can be shown that iteration~$t$ of the algorithm is given by
\begin{eqnarray}
q^{(t+1)}_{\br z}(\br z)
&\propto&
\exp \left[
\int 
q^{(t)}_{\bs\theta}(\bs \theta)\,
\log
f(\br y, \br z, \bs\theta | \bs\gamma)\,
d\bs\theta
\right]\label{eq:z_proposal}\\
q^{(t+1)}_{\bs\theta}(\bs \theta)
&\propto&
\exp \left[
\int 
q^{(t+1)}_{\br z}(\br z)\,
\log
f(\br y, \br z, \bs\theta | \bs\gamma)\,
d\br z
\right]\label{eq:theta_proposal}.
\end{eqnarray}

Several problems arise at this point. Even if the integral in (\ref{eq:z_proposal}) can be evaluated explicitly, their is little chance that the resulting expression of $q^{(t+1)}_{\br z}(\br z)$ corresponds to a known distribution in the model with spatial uncertainty. Further factorizing $q^{(t+1)}_{\br z}(\br z)$ does not resolve this issue, because it is due to the complicated relation between spatial displacements and the other variables, as explained in Appendix~\ref{app:MCMC}. Because of that, the integral in (\ref{eq:theta_proposal}) has no closed form, making the determination of $q^{(t+1)}_{\bs\theta}(\bs \theta)$ problematic.

Thus, using a variational Bayes bound to approximate the marginal likelihood in the model with spatial uncertainty would most probably either require further approximations, or hybrid strategies combining MCMC and VB approaches \cite{Forbes07}. Though we do not dismiss the use of variational Bayes, it does not seem clear for the moment what advantage it would have over the other approaches, or how it could be implemented efficiently in our case.

\subsubsection{Reversible jump MCMC}

Another alternative would be to use the reversible jump MCMC (RJ-MCMC) algorithm \cite{Green95}, which consists in sampling $\bs\gamma$ along with the other parameters. This approach allows to `jump' from one model specified by $\bs\gamma$ to another, effectively changing the size of the parameter vector from one iteration to another. Thus only the models most relevant given the data are visited, rather than all the models, and a single run of the RJ-MCMC algorithm is necessary for the whole model selection procedure. The marginal are not computed directly in this approach; instead, the posterior probabilities of each model $\mathfrak M_{\bs\gamma}$ is estimated directly by the proportion of total iterations spent in $\mathfrak M_{\bs\gamma}$ by the sampling algorithm.

Though attractive, this technology has several drawbacks. \cite{Chib01} indicate that the algorithm can be quite complicated to tune in order to promote mixing across spaces of varying dimension. Also, each new model introduced in the sampling scheme must include a subset of the existing models, which artificially increases the parameter space. Finally, as noted in \cite{Marin07}, the parameters of interest within each model (and the posterior probabilities $\pi(\bs\gamma | \br y) $) may be poorly estimated if the algorithm keeps jumping from one model to another. Hence, it is often necessary to use hybrid strategies, which alternate reversible jump and classical MCMC iterations. Finally, for each model to be well estimated, it seems that the RJ-MCMC would unavoidably require at least as many iterations as the sum of iterations needed to fit each model using more conventional MCMC techniques. Thus, it seems to us that the RJ-MCMC algorithm is less appropriate for model selection than for model averaging, and in particular to obtain estimates of the parameter $\bs\gamma$ which take into account the uncertainty on the choice of a model. An example of such an application in the context of fMRI group data analysis is given in \cite{Xu09}, which is presented in Section~\ref{sec:feature}. However, this is not our primary goal here.

\subsection{Chib's approach}\label{sec:Chib}

We have finally opted for Chib's method \cite{Chib95,Chib01}, which allows to compute the marginal likelihood from the output of virtually any posterior sampling scheme. It is efficient, simple and applicable in most cases. This approach is based on the basic marginal identity (BMI):
\begin{eqnarray}\label{eq:BMI}
m(\br y | \bs\gamma) 
&=& 
\frac 
{f(\br y | \bs \theta^\ast) \pi(\bs\theta^\ast | \bs\gamma)} 
{\pi(\bs\theta^\ast | \br y, \bs\gamma)},
\end{eqnarray}
which simply expresses the fact that the marginal likelihood is the normalizing constant of the posterior distribution. It is valid for any particular value $\bs\theta^\ast$ of the parameter, but it is advised to choose a high density point, such as the posterior mean or posterior maximum (MAP), where the different densities are more likely to be well estimated than in tails. In our model the MAP cannot be computed analytically, but can be numerically evaluated using the MCMC-SAEM algorithm detailed in Appendix~\ref{app:SAEM}, while the posterior mean can be obtained from the Metropolis within Gibbs (MH-Gibbs) algorithm described in Appendix~\ref{app:MCMC}.

% An interesting feature of (\ref{eq:BMI}) worth noting is that it clearly shows the link between Bayesian model selection and penalized likelihood approaches \cite{Akaike74,Scwhartz78,Massart03}. Indeed, it is the product of the likelihood $f(\br y| \bs\theta^\ast )$ in a high-density point $\bs \theta^\ast,$ reflecting the goodness of fit of the model to the data, by the prior/posterior ratio 
% ${\pi(\bs\theta^\ast | \bs\gamma)} 
% {\pi(\bs\theta^\ast | \br y, \bs\gamma)}^{-1},$
% which may be interpreted as a penalty on the model complexity. This can be compared to the BIC penalty \cite{Scwhartz78}, which in this case is equal to: $BIC(\bs\gamma) = -\frac{1}{2}\log(nd)(N_1(\bs\gamma) + 2N + 1),$ where $N_1(\bs\gamma) = \sum_j \gamma_j$ is the number of active regions. Indeed, $BIC(\bs\gamma$ is constructed so as to be asymptotically equivalent to $\log {\pi(\bs\theta^\ast | \bs\gamma)} {\pi(\bs\theta^\ast | \br y, \bs\gamma)}^{-1}$ under certain conditions.

Computationally, the main advantage of (\ref{eq:BMI}) is that it expresses the marginal likelihood $m(\br y | \bs\gamma)$ in terms of the posterior density $\pi(\bs\theta^\ast | \br y, \bs\gamma),$ rather than the prior density, providing a way to use posterior sampling strategies, such as the Gibbs sampler, to estimate $m(\br y | \bs\gamma).$ Indeed, the posterior density can be written as
\begin{eqnarray}\label{eq:marginal}
\pi ( \bs\theta^\ast | \br y, \bs\gamma )
&=& \int \pi ( \bs\theta^\ast | \br z, \br y, \bs\gamma) \pi(\br z | \br y, \bs\gamma) d\br z,
\end{eqnarray}
where $\br z = (\br x, \br w, \bs \mu).$ Thus, given a sample $(\bs\theta_1, \br z_1, \ldots, \bs\theta_J, \br z_J)$ of the joint posterior density $\pi( \bs\theta, \br z | \br y, \bs\gamma)$, obtained {\em e.g.} by a Gibbs sampler, the posterior density can be computed by {\em Rao-Blackwellisation} \cite{Gelfand90}:
\begin{eqnarray}\label{eq:Rao-Blackwell}
\hat \pi(\bs\theta^\ast | \br y, \bs\gamma)
&=& \frac{1}{J} \sum_{j=1}^J \pi(\bs\theta^\ast | \br z_j , \br y, \bs\gamma),
\end{eqnarray}
since the $\br z_j$'s are asymptotically drawn from the marginal $\pi (\br z | \br y, \bs\gamma).$ The obtained estimate is simulation consistent, {\em i.e.} it converges to the true value when $J \to \infty.$

This approach works nicely in the special case of no spatial uncertainty ($\sigma_S^2 = 0, \br w = \bs 0$), because then the likelihood function $f(\br y| \bs\theta^\ast)$ is available in closed form (simply apply the formulas in Appendix~\ref{app:likelihood} with $\br w = \bs 0$). Furthermore, since in this case the regions are independent, as noted previously in Section~\ref{sec:regionalized}, only $2N$ models need to be estimated instead of $2^N.$

\subsection{Likelihood under spatial uncertainty}\label{sec:likelihood_spatial}

Unfortunately, this appealingly simple method cannot be directly applied to the model with spatial uncertainty, because the conditional density $f(\br y | \br w, \bs \theta^\ast)$ only is available in closed form, but not the likelihood function:

\begin{eqnarray}\label{eq:likelihood}
f(\br y | \bs \theta^\ast) &=& \int f(\br y | \br w, \bs \theta^\ast) \pi(\br w | \bs \theta^\ast) d\br w.
\end{eqnarray}

% Furthermore, regions are not independent anymore, due to spatial uncertainty, as explained in Section~\ref{sec:regionalized}. 
% This last issue is dealt with in Section~\ref{sec:independence}, where the status of each region is inferred separately, based on an independence approximation.

The difficulty in calculating (\ref{eq:likelihood}) is essentially the same as in calculating the marginal likelihood: it is an integral on a high dimension space, with respect to the prior distribution $\pi(\br w | \bs \theta^\ast),$ whereas significantly contributing values of the integrand are concentrated around the modes of $\pi(\br w | \bs \theta^\ast, \br y).$ Therefore, we may apply Chib's method again, by writing the following alternative identity:

\begin{eqnarray}\label{eq:BMI2}
f(\br y | \bs \theta^\ast) 
&=& 
\frac
{f(\br y | \br w^\ast, \bs \theta^\ast) \pi(\br w^\ast | \bs \theta^\ast)}
{\pi(\br w^\ast | \bs \theta^\ast, \br y)},
\end{eqnarray}

valid for any value $\br w^\ast$ of the elementary displacements. A high density value of the posterior ordinate $\pi(\br w | \bs \theta^\ast, \br y)$ is however advisable, for accurate numerical evaluation of (\ref{eq:BMI2}). Thus a reasonable choice is the conditional maximum {a posterior} $\br w_{MAP}^\ast = \arg\max_{\br w} \pi( \br w | \bs \theta^\ast, \br y ),$ which can be evaluated using the simulated annealing (SA) algorithm, as described in Appendix~\ref{app:SA}. A simpler alternative, requiring less computations, is the posterior mean $\hat{\br w} = \mathbb E[ \br w | \br y ],$ directly available from the output of the MH-Gibbs sampler in Appendix~\ref{app:MCMC} as the average of sampled  $\br w$ values. However, it may be distant from the principal mode of $\pi(\br w^\ast | \bs \theta^\ast, \br y),$ and result in a less stable estimator of (\ref{eq:BMI2}).

Having chosen $\br w^\ast,$ both $f(\br y | \br w^\ast, \bs \theta^\ast)$ and $\pi(\br w^\ast | \bs \theta^\ast)$ are available in closed form, but not the posterior ordinate $\pi(\br w^\ast | \bs \theta^\ast, \br y).$ This difficulty cannot be solved using the approach in \cite{Chib95}, which assumes that the hidden variables can be decomposed into blocks with conditional densities available in closed form. In the present case, $\br w^\ast$ can be decomposed into blocks consisting of single elementary displacements $\br w_{ib},$ but the conditional density of each block has no analytical expression, and must be sampled using a Metropolis-Hastings step, as described in Appendix~\ref{app:SA}. A generalization of the method in \cite{Chib95} to this setting is developed in \cite{Chib01}. It consists in factorizing the posterior ordinate across blocks, according to:
\begin{eqnarray}
\pi(\br w^\ast | \bs \theta^\ast, \br y)
&=&
\prod_{i=1}^n
\prod_{b=1}^B
\pi(\br w_{ib}^\ast | \br w_{-ib}^\ast, \bs \theta^\ast, \br y),\nonumber
\end{eqnarray}
where we write $\br w_{-ib}$ to denote the blocks preceding $ib$ in the lexical order, that is, the collection of blocks $\br w_{i'b'}$ where $i' \leq i,$
$b' \leq b,$ and $(i',b')\neq(i,b).$

Each reduced posterior ordinate $\pi(\br w_{ib}^\ast | \br w_{-ib}^\ast, \bs \theta^\ast, \br y)$ can then be evaluated from the output of a reduced run of the multiple block MH algorithm, conditional on $\br w_{-ib},$ and an additional run where $\br w_{ib}$ is added to the conditioning set. Justification and further details on the calculation of the reduced ordinates and the likelihood function are given in Appendix~\ref{app:Chib}.

\section{Comparing different parcellations}\label{sec:parcellations}

The choice of a particular parcellation is an important issue, as our whole decision framework rests upon it. Yet, the definition of functionally homogeneous regions in the human brain remains an open issue. In practice, mis-specified parcellations may cause activated areas to cross several parcels, resulting in reduced sensitivity, and difficulty in interpreting the detected pattern.

Though resolving this issue is beyond the scope of this work, we remark that the same Bayesian model selection formalism used to select the functional network $\bs \gamma$ can be used to compare different candidate parcellations. This is done by considering the parcellation as a random variable to be estimated, rather than a fixed quantity. Thus, two given parcellations $\mathscr P = \{\mathcal V_1, \ldots, \mathcal V_N\}$ and $\mathscr P' = \{\mathcal V'_1, \ldots, \mathcal V'_N\}$ may be compared through their posterior odds:
\begin{eqnarray}\label{eq:parcellation_posterior}
\frac {\pi(\mathscr P | \br y)} {\pi(\mathscr P' | \br y)}
&=&
\frac {m(\br y | \mathscr P)} {m(\br y | \mathscr P')} \times \frac {\pi(\mathscr P)} {\pi(\mathscr P')},
\end{eqnarray}
where $\pi(\mathscr P)$ is the prior probability assigned to parcellation $\mathscr P,$ and $m(\br y | \mathscr P)$ the evidence for $\mathscr P,$ given by:
\begin{eqnarray}
m(\br y | \mathscr P)
&=&
\sum_{\bs \gamma \in \{0,1\}^N} m(\br y | \bs \gamma, \mathscr P) \pi(\bs \gamma | \mathscr P).
\end{eqnarray}
Here $\bs \gamma$ is the vector indicating possible functional networks, based on parcellation $\mathscr P,$ $m(\br y | \bs \gamma, \mathscr P)$ is the marginal likelihood defined by (\ref{eq:evidence}) and $\pi (\bs \gamma | \mathscr P)$ the prior probability of network $\bs \gamma,$ defined by (\ref{eq:prior_network}).

Thus, selecting the most adequate parcellation among a set of candidates, to infer the functional pattern associated with a certain functional task, is a straightforward application of the variable selection framework introduced in the previous section. Furthermore, applied to anatomically defined regions, it provides a tool to investigate the link between brain anatomy and function.

% More generally, the mechanism described here to compare parcellations can be used to select among any classes of competing models. For instance, one may ask if the spatial uncertainty our method aims to compensate is really present in the data, that is, if the model with spatial uncertainty is more likely given the data than the model conditional on $\sigma_S = 0.$ This question can be answered by comparing $m(\br y | \mathscr P)$ to the evidence $m(\br y | \mathscr P, \sigma_S = 0)$ in the `spatially certain' submodel.

\section{2D toy example}\label{sec:simu}

We now illustrate our model selection approach on a simulated dataset. 
% We limited ourselves to 2D~data, because estimating the posterior ordinate of the elementary displacements, as described in the previous section, takes more than an hour on 3D~datasets, even when the search volume is only $20$~voxels wide. The sensitivity analysis we wanted to perform would have taken several days at least, instead of a few hours using 2D~data.
In Section~\ref{sec:3D_simu}, our goal was to evidence a stretching effect of the activations due to spatial variability, and its compensation through appropriate modeling. Going one step further in our analysis, we now show that this stretching effect may result in a bias toward false positives when testing the presence of activations within pre-defined regions. Furthermore, we show that this bias can be corrected when spatial uncertainty is accounted for.

We defined a synthetic activation pattern, within a 2D search volume of $24\times24$~voxels, consisting of a central activated disc, with uniform intensity value~$5$ (the background was set to $0$) and a diameter of $7$~voxels.

\begin{figure}[ht!]
\includegraphics[width=0.3\textwidth]{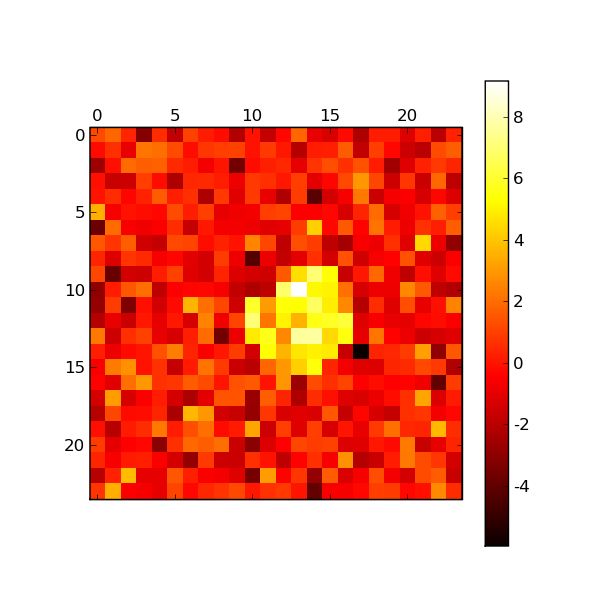}
\hfill
\includegraphics[width=0.3\textwidth]{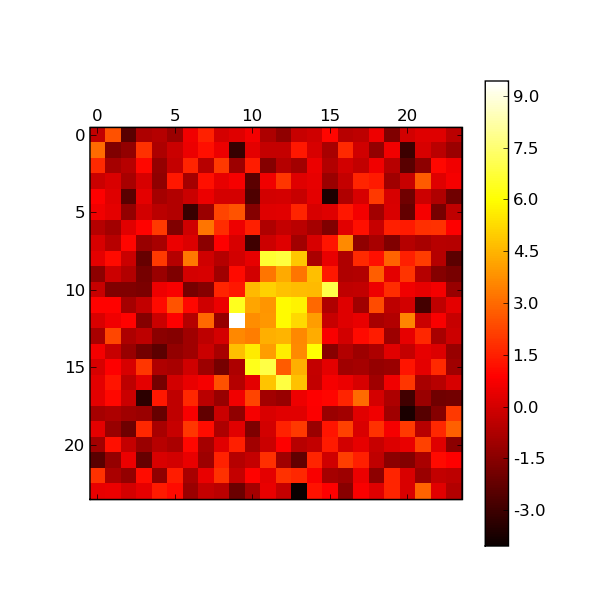}
\hfill
\includegraphics[width=0.3\textwidth]{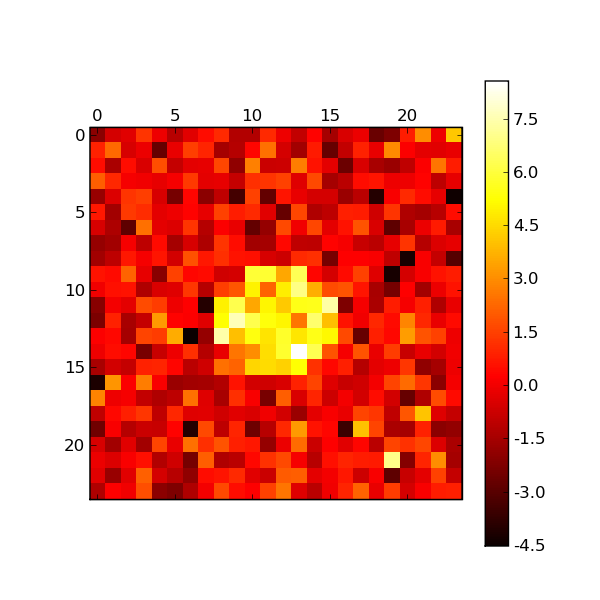}
\caption{\label{fig:toy_dataset_2D} Three different simulated effect effect maps $\br y_i$}
\end{figure}

To simulate the data, this activation was deformed according to a displacement field $\br u,$ simulated under the model described in Section~\ref{sec:deformation}, with one control point in each voxel. The standard displacement was taken equal to $\sigma_S = 1.0$ voxels and the field smoothness parameter was set to $\omega=4.0$ voxels. Independent heteroscedastic Gaussian noise was then added to each voxel $\br v$, with variance equal to $1 + \br s^2(\br v)$, where $\br s^2(\br v)/\varepsilon\sim\chi^2(1)$, $\varepsilon$ being the noise level, set to $1.0$ in this example.  A total of $n=30$ pairs $(\br y_i, \br s^2_i)$ of effect and variance maps were sampled in this fashion, as illustrated in Figure~\ref{fig:toy_dataset_2D}, and constitute a sample from the hierarchical model in Section~\ref{sec:observation} .

\subsubsection{Data analysis}

The search volume was divided in two regions, corresponding to the the background ($\mathcal V_1$) and the active disc ($\mathcal V_2$). In this elementary situation, only $4$~activation models are in competition, each of them corresponding to a value of the indicator variable~$\bs\gamma \in \{0,1\}^2.$ We applied the Bayesian model selection approach described in \ref{sec:network} to recover the parcels with a nonzero mean population effect, both with and without modeling spatial uncertainty. In the model with spatial uncertainty, deformation fields were specified using a single control point in the center of the search volume, and the same regularity parameter $\omega = 4$ used to simulate the data.

More precisely, the marginal likelihood in each model was estimated as explained in Section~\ref{sec:marginal}. Thus, the posterior density of the model parameters was maximized over $500$~iterations of the MCMC-SAEM algorithm (see Appendix~\ref{app:SAEM}), following $500$ `burn-in' iterations. Next, the posterior density of the resulting MAP estimate was computed using $1000$~Gibbs iterations, following a burn-in period of $100$~iterations. This was sufficient to obtain the marginal likelihood in the model without spatial uncertainty, applying (\ref{eq:BMI}).

In the model with spatial uncertainty, the density of the deformation fields, conditional on the MAP estimate of model parameters, was maximized using $500$~iterations of the SA algorithm (see~Appendix~\ref{app:SA}). Finally, this density was computed as in section~\ref{sec:likelihood_spatial}. As explained in Appendix~\ref{app:Chib}, each reduced run was sampled using a number of iterations inversely proportional to the number of sampled blocks, and used $3000$~iterations for the single-block run. Computing the marginal likelihood of each model under spatial uncertainty took approximately $9mn$ on a PC with a clock rate of $1.33GHz,$ against $20s$ seconds without spatial uncertainty. To quantify the variance of the Monte-Carlo estimates of the marginal likelihood values, we repeated all calculations $10$~times on the same dataset.

\subsubsection{Results}

\begin{figure}[ht!]
\includegraphics[width=0.33\textwidth]{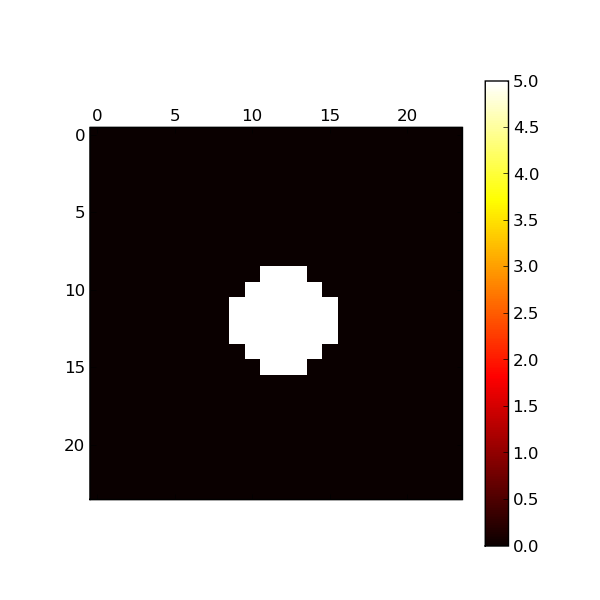}\hfill
\includegraphics[width=0.33\textwidth]{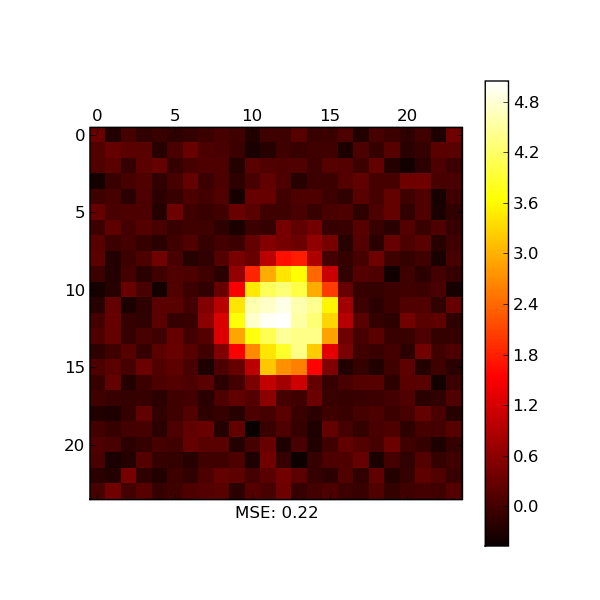}\hfill
\includegraphics[width=0.33\textwidth]{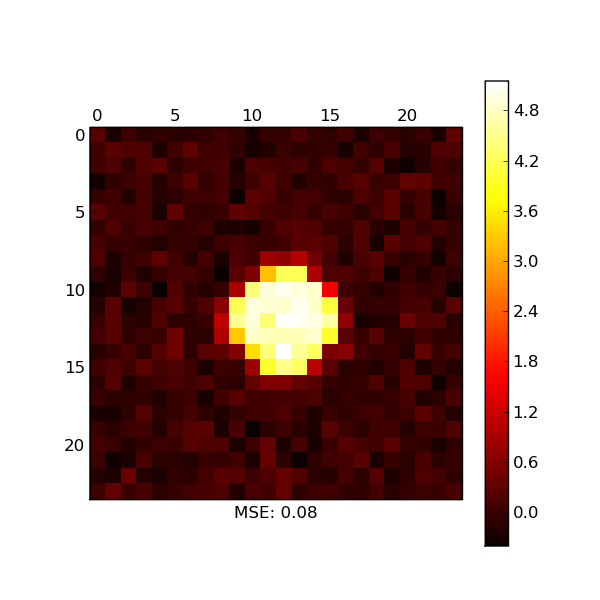}
\caption{\label{fig:posterior_estimates_2D} Posterior estimates of $\bs \mu$ (left), with (right) and without (center) modeling spatial uncertainty.}
\end{figure}

Presence of the stretching effect can be checked in Figure~\ref{fig:posterior_estimates_2D}, where posterior estimates of the signal from one trial are shown. These were computed by averaging over the four models the posterior mean conditional on the MAP parameter estimates $\hat{\bs\theta}_{\bs\gamma}$, according to: $\hat{\bs\mu} = \sum_{\bs\gamma} \mathbb E[\bs\mu | \br y, \hat{\bs\theta}_{\bs\gamma}] \pi(\bs\gamma|\br y)$ (using the posterior mean $\mathbb E[\bs\mu | \br y, \bs\gamma]$ would have made more sense, but was not directly available from the output of our algorithm).

The mean and standard deviate over $10$~trials of the marginal likelihood values of each model, are given in Table~\ref{tab:2D_simu}. As expected, when spatial uncertainty is unaccounted for, the maximum marginal likelihood is attained for $\bs\gamma = (1,1),$ that is, activations are detected in the background, due to the stretching of the estimated activate disc. These values are numerically stable, as can be checked from their standard deviates.

\begin{table}[ht!]
\centering
\begin{tabular}{c|rr}
$\bs\gamma$
& $m(\br y|\bs\gamma, \sigma_S = 0)$
& $m(\br y|\bs\gamma)$\\
\hline
$(0,0)$ & $-87882.68 \pm 0.12$ & $-33655.6 \pm 21.6$\\
$(1,0)$ & $-87868.39 \pm 0.17$ & $-33663.9 \pm 12.7$\\
$(0,1)$ & $-87748.58 \pm 0.10$ & $\bs{-33644.0} \pm 13.2$\\
$(1,1)$ & $\bs{-87733.29} \pm 0.18$ & $-33651.3 \pm 16.1$
\end{tabular}
\caption{\label{tab:2D_simu} Log marginal likelihood values computed on 2D simulated data, over $10$~trials. Results are given in the form: mean $\pm$ std. deviate.}
\end{table}

The stretching effect is compensated in the model without spatial uncertainty, where the maximum marginal likelihood is achieved on the average for the correct model $\bs\gamma = (0,1).$ Moreover, the marginal likelihood values are much higher, indicating the better fit obtained by modeling spatial uncertainty. However, the differences between the marginal likelihoods of the different models are now swamped in the variability of the Monte-Carlo estimates, so that the procedure does not systematically select the correct model on each trial.

This variability comes from the difficulties encountered in sampling the conditional density of the elementary displacements, as described in~\ref{app:MCMC}. As a result, the Markov Chain is very sluggish in its exploration of the space of all possible deformation fields, and tends to get trapped in local maxima, an issue already noted in Sections~\ref{sec:simulations} and \ref{sec:fmri_data}. Because on each trial, the Markov chain gets trapped around a different mode, the resulting estimates of the marginal likelihood are highly variable. Increasing the number of iterations did not solve this issue, though in theory it would allow after sufficient time the chain to escape from each local mode and explore exhaustively the sampling space. 
% This strongly suggests that the random-walk MH step used to sample the elementary displacements is suboptimal, and points to the necessity of developing alternative proposal densities which more closely resemble the target conditional density.

In conclusion, this numerical experiment demonstrates the potential benefits of modeling spatial uncertainty when testing regional hypotheses on fMRI data, using the model selection approach developed in Section~\ref{sec:network}. However, this approach turns out to be numerically unstable when modeling spatial uncertainty, even on the overly simplified toy dataset used here, with 2D data simulated under a high signal to noise ratio, and using smooth deformation fields with known smoothness.

\section{Approximate inference using posterior modes.}\label{sec:independence}

% So far, we have noted the difficulties implied by the computation of the marginal likelihood under spatial uncertainty. An additional issue is the fact that it must be computed for all $2^N$ models to be compared. We intend to use for real-world applications an anatomical brain atlas, such as the AAL \cite{Tzourio-Mazoyer02}, which contains over $100$ regions. The time required to process all the resulting models would be astronomically large, given that treating one model takes at least several minutes, as seen from the example in the previous section.
% 
% \subsubsection{Posterior mode approximation}

To address the above mentioned numerical instability issue, we consider approximating the marginal likelihood $m(\br y | \bs\gamma)$ by the following likelihood, conditional on a particular value $\hat{\br w},$ which is the same for all networks $\bs\gamma:$
\begin{eqnarray}\label{eq:approximation}
m(\br y | \hat{\br w}, \bs\gamma) &=&
\int f(\br y | \hat{\br w}, \bs\theta) \pi(\bs\theta|\bs\gamma).
\end{eqnarray}
By doing this, we avoid the difficult integration with respect to the displacements.
% , and reduce the number of model to be estimated to $2N,$ as explained below.
This may be interpreted in terms of the Laplace approximation (see \cite{Robert07} for instance), which consists in approximating the unnormalized posterior density $m(\br y | \br w, \bs\gamma) \pi(\br w | \bs\gamma)$ of $\br w$ by an unnormalized Gaussian, obtained from a second order Taylor expansion of $\log m(\br y | \br w, \bs\gamma) \pi(\br w | \bs\gamma)$ around its mode $\hat{\br w}.$

Using this approximation would be overly expensive, because the posterior mode $\hat{\br w}$ would need to be computed for each $2^N$ possible values of~$\bs\gamma,$ an unfeasible task for real datasets, since each model would need at least several minutes to be processed. Moreover, the validity of a Taylor expansion of $\log m(\br y | \br w, \bs\gamma) \pi(\br w | \bs\gamma)$ is questionable in our case, since it is only piecewise continuous as a function of $\br w.$

Hence, we simplify the Laplace approximation, replacing $m(\br y | \br w, \bs\gamma) \pi(\br w | \bs\gamma)$ by a Dirac mass in the posterior mode $\hat{\br w}$ instead of a Gaussian, and further assuming $\hat{\br w}$ to be the same for all networks $\bs\gamma.$

% In practice, $\hat{\br w}$ may be chosen as the mode of the posterior density $\pi(\br w | \br y, \bs\gamma^\ast)$ for a particular network $\bs\gamma^\ast.$ However, this would unavoidably bias the selection procedure towards values of $\bs\gamma$ close to $\bs\gamma^\ast.$ Instead, we use the following conditional posterior mode in the model without parcellation, defined in Chapter~\ref{chap:modeling}:
We define $\hat{\br w}$ as the following conditional posterior mode in the model without parcellation, defined in Chapter~\ref{chap:modeling}:
\begin{eqnarray}
\hat{\bs\theta} &=& \arg\max_{\bs\theta} \pi(\bs\theta| \br y); \\
\hat{\br w} &=& \arg\max_{\br w} \pi(\br w | \br y, \hat{\bs\theta}).
\end{eqnarray}
We use the mode conditional on the MAP parameters~$\hat{\bs\theta},$ rather than the unconditional mode, for simplicity, since it can be estimated using the SAEM and SA algorithm previously introduced (see Appendices~\ref{app:SAEM} and \ref{app:MCMC}).

The conditional likelihood (\ref{eq:approximation}) can be evaluated exactly as the marginal likelihood under no spatial uncertainty. Indeed, because voxels are independent conditional on $\hat{\br w},$ this expression can be factorized across regions:
$$
m(\br y | \hat{\br w}, \bs \gamma) =
\prod_{j=1}^N m(\br y^j | \hat{\br w}, \gamma_j).
$$
where $\br y^j = \{ y_{i,k}; \br v_k + \br u_{i,k} \in \mathcal V_j\}$ is the subset of observations corresponding to voxels displaced into region~$j.$ Note that this set is random, since it is a function of $\br u_{i,k},$ itself a function of the elementary displacements~$\br w,$ as defined by (\ref{eq:deformation}). The conditional likelihood of region~$j$ is equal to:
$$
m(\br y^j | \hat{\br w}, \gamma_j) =
\int f(\br y^j | \hat{\br w}, \bs\theta_j) \pi(\bs\theta_j|\gamma_j) d\bs\theta_j,
$$
where $\bs\theta_j = (\eta_j, \nu_j^2, \sigma_j^2)$ is the parameter vector for region~$j.$

These conditional likelihoods may be evaluated separately, or equivalently, (\ref{eq:approximation}) can be computed for $\bs\gamma = \bs 0_N$ and $\bs\gamma = \bs 1_N.$ This can be done by Chib's method (see Section~\ref{sec:Chib}), writing:
\begin{eqnarray}\label{eq:BMI3}
m(\br y | \hat{\br w}, \bs \gamma) &=&
\frac {f( \br y | \hat{\br w}, \hat{\bs \theta}_{\bs\gamma}) \pi (\hat{\bs \theta}_{\bs\gamma} | \bs\gamma)}
{\pi (\hat{\bs \theta}_{\bs\gamma} | \br y, \hat{\br w}, \bs\gamma)},
\end{eqnarray}
where we choose $\hat{\bs \theta}_{\bs\gamma}$ as the conditional MAP estimate: $\hat{\bs \theta}_{\bs\gamma} = \arg\max_{\bs\theta} \pi(\bs\theta | \br y, \hat{\br w}, \bs\gamma).$ The posterior density $\pi (\hat{\bs \theta}_{\bs\gamma} | \br y, \hat{\br w}, \bs\gamma)$ is obtained by the Rao-Blackwell method, as explained in Section~\ref{sec:Chib}, while the exact expression of the density $f( \br y^j | \hat{\br w}, \bs \theta)$ is given in Appendix~\ref{app:likelihood}.

Finally, the posterior density $\pi(\bs\gamma | \br y)$ of the functional network is approximated by $\pi(\bs\gamma | \br y, \hat{\br w}),$ which can be factorized across regions into:
\begin{eqnarray}
\pi(\bs\gamma | \hat{\br w}, \br y)
&=&
\prod_{j=1}^N
\pi(\gamma_j | \br y^j, \hat{\br w}).
\end{eqnarray}
This posterior density is entirely determined by the posterior probabilities that each region is involved,
$$
P_j = \pi(\bs\gamma_j = 1 | \br y^j, \hat{\br w}) = 1 - \pi(\bs\gamma_j = 0 | \br y^j, \hat{\br w}),
$$
obtained from the conditional likelihoods as
\begin{eqnarray}\label{eq:post_prob_approx}
P_j
&=&
\left(
1 + \frac
{m(\br y^j | \hat{\br w}, \gamma_j = 0)}
{m(\br y^j | \hat{\br w}, \gamma_j = 1)}
\times\frac
{\pi(\gamma_j = 0)}
{\pi(\gamma_j = 1)}
\right)^{-1}.
\end{eqnarray}

Intuitively, conditioning on $\hat{\br w}$ may be seen as a pre-processing step wherein the functional images are registered to a template $\bs\mu,$ which is estimated at the same time. Thus, it provides a way to compensate for spatial normalization errors based on the functional data.
However, 
% this approach has several limitations. First, 
because the posterior variability of the elementary displacements around their mode is neglected, we anticipate good results only when this variability is reduced, that is when the data provides enough information for the elementary displacements to be well estimated. 
% Second, the simulated annealing algorithm used to estimate the conditional MAP displacement field $\hat{\br w}$ uses the same proposal to sample candidate displacements as the MH-Gibbs algorithm in Appendix~\ref{app:MCMC}, so it is likely to inherit the same numerical instability, and be prone to local maxima.

\subsection{2D toy example}\label{sec:simu_2D_approx}

We applied the above posterior mode approximation to the same dataset used in Section~\ref{sec:simu}
% to validate the exact model selection approach
. Our goal was to validate the approximation, and evaluate its numerical stability.

\subsubsection{Data analysis}

We used our posterior mode approximation to compute the posterior probabilities~$P_j$ of positive mean activations within each region~$j=1,2.$ More precisely, the MAP estimate $\hat{\bs\theta}$ of model parameters was first obtained without parcelling the search volume, using $500$~iterations of the SAEM algorithm, following $500$ burn-in iterations. Then, the conditional MAP estimate $\hat{\br w}$ of the displacement fields was computed using $500$ iterations of the SA algorithm. The conditional likelihoods $m(\br y | \hat{\br w}, \bs \gamma)$ were then computed both under the null model, defined by $\gamma_j \equiv 0,$ and under the full model, defined by $\gamma_j \equiv 1.$ In each case, the conditional MAP estimate $\hat{\bs \theta}_{\bs\gamma}$ of model parameters was obtained from $1\,000$~iterations of the SAEM algorithm (including $500$ burn-in iterations), and the conditional posterior $\pi (\hat{\bs \theta}_{\bs\gamma} | \br y, \hat{\br w}, \bs\gamma)$ was computed from the output of $1\,000$ iterations of a MH-Gibbs algorithm, following a burn-in of $100.$ We then computed the log Bayes factors:
$$
B_j = \displaystyle\log\frac
{m(\br y^j | \hat{\br w}, \gamma_j = 1)}
{m(\br y^j | \hat{\br w}, \gamma_j = 0)},
$$
from which the $P_j$ were directly deduced as
$$
P_j
= 
\left(
1 + e^{-B_j}
\right)^{-1},
$$
following (\ref{eq:post_prob_approx}), adopting a uniform prior $\pi(\gamma_j = 0) = \pi(\gamma_j = 1)$ for the state of each region. We compared these results with those obtained under no spatial uncertainty, setting $\br w = \bs 0$ and $\sigma_S^2 = 0,$ noting that, in this case, the procedure is exact and equivalent to the one tested in Section~\ref{sec:simu}. All computations were re-run $10$ times to assess numerical stability.

\subsubsection{Results}

\begin{figure}[ht!]
\includegraphics[width=0.33\textwidth]{Chapter5_signal.png}\hfill
\includegraphics[width=0.33\textwidth]{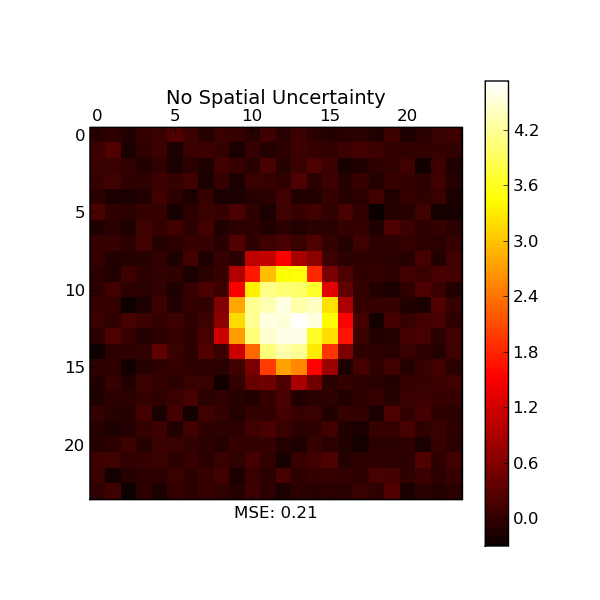}\hfill
\includegraphics[width=0.33\textwidth]{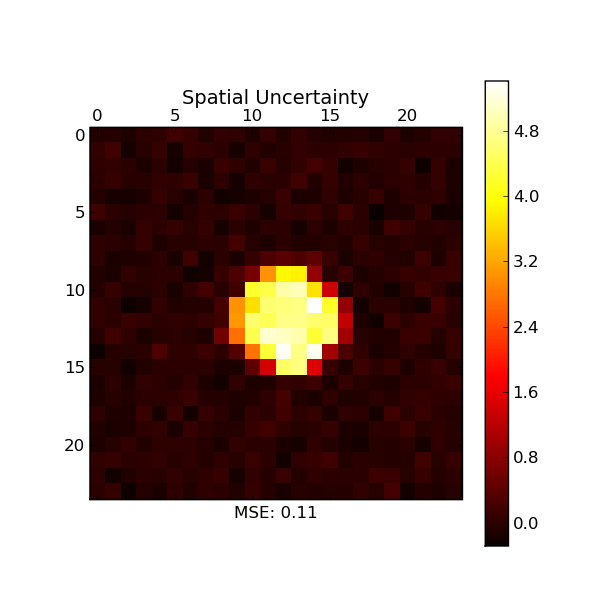}
\caption{\label{fig:posterior_estimates_2D_approx} Posterior estimates of $\bs \mu$ (left), with (right) and without (center) modeling spatial uncertainty, using the posterior mode approximation}
\end{figure}

Presence of the stretching effect can be checked in Figure~\ref{fig:posterior_estimates_2D_approx}, where posterior estimates of the signal from one trial are shown. As previously, these were computed by averaging over the four models the posterior mean conditional on the MAP parameter estimates and the MAP elementary displacements, according to: $\hat{\bs\mu} = \sum_{\bs\gamma} \mathbb E[\bs\mu | \br y, \hat{\bs\theta}_{\bs\gamma}, \hat{\br w}] \pi(\bs\gamma|\br y, \hat{\br w}).$ Also, it can be seen that the posterior mode~$\hat{\br w}$ provides in the present case a good estimate of the unknown displacements. Indeed, the posterior estimate of $\bs\mu$ conditional on $\hat{\br w}$ has a mean square error (MSE) of $0.11,$ almost as low as that of the unconditional estimate tested in Section~\ref{sec:simu}, which was equal to $0.08.$

\begin{table}[ht!]
\begin{center}
\begin{tabular}{c|rr}
& Spatial uncertainty & No spatial uncertainty\\
\hline
Region $j$ & $B_j$  & $B_j$ \\
\hline
$1$ (background) & 
$-7.16 \pm 0.27$ & $29.23 \pm 0.48$ \\
$2$ (disc) & 
$3.21 \pm 0.75$ & $51.89 \pm 0.01$\\
\end{tabular}
\end{center}
\caption{\label{tab:2D_simu_approx} Results of the approximate model selection procedure on a single 2D simulated dataset, over $10$~trials.} Results are given in the form: mean $\pm$ std. deviate.
\end{table}

The mean and standard deviates of the Bayes factor for each region is given in Table~\ref{tab:2D_simu_approx}. Results for the model without spatial uncertainty are identical to those found in the previous experiment, since the procedures are in this case equivalent. They are also quantitatively similar in the model with spatial uncertainty: the background is correctly classified as inactive, with a negative log Bayes factor $B_1,$ and the disc as active, with $B_2 > 0.$ Furthermore, the variance of the log Bayes factors is much reduced with respect to that of the log marginal likelihoods (see Table~\ref{tab:2D_simu}), showing that our approximation provides better numerical stability.

In conclusion, this experiment shows that our posterior mode approximation works well in a favorable setting, were it leads to the same conclusions as the exact procedure, with a much reduced numerical variability. These good results can be attributed to the fact that the data is simulated with a high signal to noise ratio and warped using smooth deformation fields with known regularity, so that the posterior density of $\br w$ is likely to be sharply peaked around its mode.

\subsection{3D toy example}\label{sec:simu_3D_approx}

\begin{figure}[ht!]
\includegraphics[width=0.33\textwidth]{Chapter5_signal.png}\hfill
\includegraphics[width=0.33\textwidth]{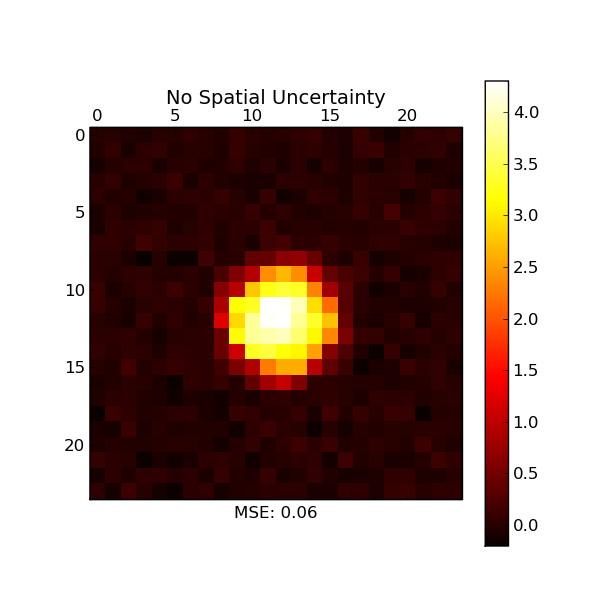}\hfill
\includegraphics[width=0.33\textwidth]{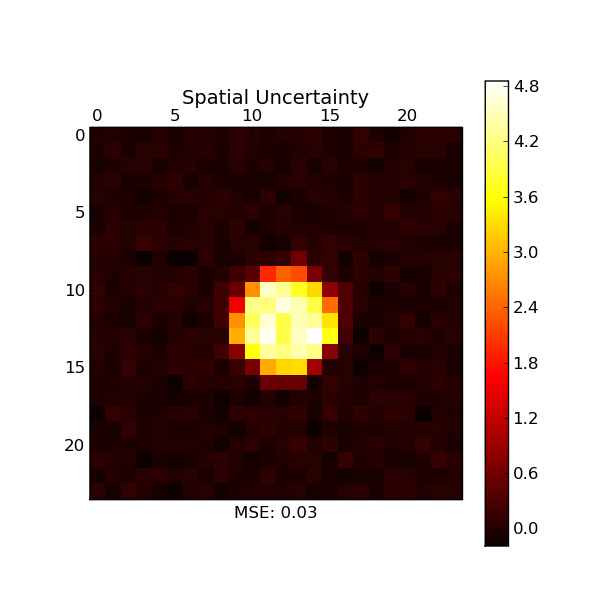}
\caption{\label{fig:posterior_estimates_3D_approx} Posterior estimates of $\bs \mu$ (left), with (right) and without (center) modeling spatial uncertainty, using the posterior mode approximation}
\end{figure}

\begin{table}[ht!]
\begin{center}
% \begin{tabular}{c|rrrrr}
% & \multicolumn{4}{c}{Spatial uncertainty}\\
% \hline
% Region $j$ & $B_j^1$ & $B_j^2$ & $B_j^3$ & $B_j$ & $\tilde B_j$ \\
% \hline
% $1$ (background) & 
% $74.58 \pm 2.94$ & $-8.38 \pm -7.67$ & $4.99 \pm 0.07$ & $61.22 \pm 2.18$ & $-9.63 \pm 2.08$ \\
% $2$ (sphere) & 
% $333.83 \pm 36.08$ & $2.17 \pm 2.06$ & $3.07 \pm 0.33$ & $323.1 \pm 33.91$ & $5.96 \pm 1.01$ \\
% \hline\hline
% & \multicolumn{4}{c}{No spatial uncertainty}\\
% \hline
% Region $j$ & $B_j^1$ & $B_j^2$ & $B_j^3$ & $B_j$ & $\tilde B_j$ \\
% \hline
% $1$ (background) & 
% $190.54 \pm 3.43$ & $-7.37 \pm 1.38$ & $4.91 \pm 0.03$ & $178.27 \pm 2.05$ & $-0.84 \pm 1.17$ \\
% $2$ (sphere) & 
% $570.67 \pm 0.02$ & $-0.05 \pm 0.01$ & $3.5 \pm 0.01$ & $567.13 \pm 0.01$ & $30.69 \pm 0.01$ \\
% \end{tabular}
\begin{tabular}{c|rr}
& Spatial uncertainty & No spatial uncertainty\\
\hline
Region $j$ & $B_j$ & $B_j$ \\
\hline
$1$ (background) & 
$39.55 \pm 1.2$ & $80.47 \pm 0.05$ \\
$2$ (sphere) & 
$264.55 \pm 6.82$ & $237.06 \pm 0.01$ \\
\end{tabular}
\end{center}
\caption{\label{tab:3D_simu_approx} Results of the approximate model selection procedure on a 3D simulated dataset, over $10$~trials.} 
% The data is simulated with a deformation field smoothness of $\omega=4$~voxels, and a noise level of $\varepsilon=1.$ Results are given in the form: mean $\pm$ std. deviate.
\end{table}

% We now test the robustness of the posterior mode approximation with respect to noise and deformation field regularity, on a 3D dataset. We expect its performances to decrease at low SNRs, and for less regular deformation fields, since in these cases the posterior density is less concentrated around its mode, making our approximation less reliable.
The previous results are encouraging, but does our posterior mode approximation work as well on 3D data? To answer this, we used a dataset simulated exactly as in Section~\ref{sec:simu}, except that the 2D~$24\times24$ search volume was replaced by a 3D~$24\times24\times24$ search volume, and the central disc was replaced by a sphere with same diameter ($7$ voxels).

We used our posterior mode approximation to compute the Bayes factors for both regions (background and sphere), using the algorithm detailed in Section~\ref{sec:simu_2D_approx}. We compared these results with those obtained under no spatial uncertainty, setting $\br w = \bs 0$ and $\sigma_S^2 = 0,$ and repeated all calculations over $10$~trials to assess numerical stability. Each trial took approximately one hour on a PC laptop with a clock rate of $1.33GHz.$
Results for the 3D dataset are given in Table~\ref{tab:3D_simu_approx}. As in Section~\ref{sec:simu_2D_approx}, we report the log Bayes factor values $B_j$ for each region~$j.$ 

In terms of estimation, the posterior mode approximation again gives satisfying results, as seen in Figure~\ref{fig:posterior_estimates_3D_approx}, with a mean-square error ($0.03$) lower than that of the estimate in the model with no spatial uncertainty ($0.06$). In terms of model selection, results are less satisfying, since the background is selected as active, with a positive Bayes factor, both with and without spatial uncertainty.

One possible explanation is that estimating 3D displacement fields requires more information than is provided by the data (which seemed sufficient in the 2D case), so that our posterior mode approximation, though it reduces the spreading effect due to the mis-localization of individual activations, does not reduce it enough in order for the background to be found inactive. Consequently, the Bayes factor conditional on the most probable displacements is lower than when displacements are fixed to zero, but still positive.

\subsection{Additional penalty on model fit}\label{sec:additional_penalty}

The above illustrations suggest that approximating the marginal likelihood by fixing the displacements to their most probable value works well on 2D data, but on 3D data fails to entirely compensate the warping of individual images and the ensuing swelling of the estimated activations. Consequently, the systematic bias toward false positives, found when registration errors are not modeled, is reduced but still present when using the posterior mode approximation.

This bias can be seen as a form of data overfit, the number of parameters needed to model the data being overestimated, due to likelihood values inflated by displaced activations. Note that this bias would be automatically compensated when using the true log marginal likelihood $\log m(\br y | \bs\gamma)$ rather than the conditional approximation $\log m(\br y | \hat{\br w}, \bs\gamma),$ because it would contain an additional term penalizing the overfit due to modeling spatial displacements:
$$
\log m(\br y | \bs\gamma) = \log m(\br y | \hat{\br w}, \bs\gamma) + \log \frac{\pi(\hat{\br w})}{\pi(\hat{\br w}| \br y, \bs\gamma)}.
$$
However, the previous example has illustrated the fact that computing this penalizing term requires heavy MCMC calculations (specifically, the posterior ordinate $\pi(\hat{\br w}| \br y, \bs\gamma)$), and in our case resulted in a far too important numerical variability to be of ay use. As a surrogate to this currently unatainable exact quantity, we consider compensating the bias by modifying $B_j,$ adding a penalty to model fit, measured by the log likelihood ratio
\begin{eqnarray}\label{eq:LR_j}
\displaystyle LR_j
&=& \log\frac
{f(\br y^j | \hat{\br w}, \hat{\bs\theta}_{1j})}
{f(\br y^j | \hat{\br w}, \hat{\bs\theta}_{0j})},
\end{eqnarray}
where $\hat{\bs\theta}_{kj} = \arg\max_{\bs\theta_j} \pi(\bs\theta_j|\gamma_j=k)f(\br y^j| \hat{\br w}, \bs\theta_j)$ for $k=0,1.$ This quantity is available as an output of the algorithm which computes the log Bayes factor $B_j,$ since, following (\ref{eq:BMI3}) it can be written as:
\begin{equation}\label{eq:Bayes_factor}
\begin{array}{ccccccc}
B_j
&=&
\displaystyle \log\frac
{m(\br y^j | \hat{\br w}, \gamma_j = 1)}
{m(\br y^j | \hat{\br w}, \gamma_j = 0)}&&&&\\
&=&
\displaystyle \log\frac
{f(\br y^j | \hat{\br w}, \hat{\bs\theta}_{1j})}
{f(\br y^j | \hat{\br w}, \hat{\bs\theta}_{0j})}
&+&
\displaystyle \log\frac
{\pi(\hat{\bs\theta}_{1j} | \gamma_j = 1)}
{\pi(\hat{\bs\theta}_{0j} | \gamma_j = 0)}
&+&
\displaystyle \log \frac
{\pi(\hat{\bs\theta}_{0j} | \br y^j, \hat{\br w}, \gamma_j = 0)}
{\pi(\hat{\bs\theta}_{1j} | \br y^j, \hat{\br w}, \gamma_j = 1)}\\
&=&
\displaystyle LR_j
&+&
\displaystyle D_j.
&&
\end{array}
\end{equation}

We expect $LR_j$ to be always positive. Indeed, the MAP estimates $\hat{\bs\theta}_{kj},$ used to define it in (\ref{eq:LR_j}) are likely to be very close to the maximum likelihood (ML) estimates, defined by $\hat{\bs\theta}_{kj}^{ML} = \arg\max_{\bs\theta_j} f(\br y^j | \br w, \bs\theta_j),$ given that we have chosen a prior $\pi(\bs\theta|\gamma_j)$ that is as minimally informative as possible (see Section~\ref{sec:prior}). Hence, it is reasonable to expect $LR_j$ to be close to the log maximum likelihood ratio, a quantity that is always positive for nested models. Corroborating this, we have always observed in practice positive values for $LR_j.$ Hence we have not felt the need to use the ML estimates instead of the MAP estimates of the parameter values, though this would be a possible alternative.

Using the positivity of $LR_j,$ we define a lower bound on the Bayes factor $B_j$ by:

\begin{eqnarray}\label{eq:stabilization}
\tilde B_j =  c\times LR_j + D_j,
\end{eqnarray}
for a certain scale factor~$c \in (0,1),$ which determines the added penalty. Based on this quantity, we define the following lower bound on the posterior probability~$P_j$ of a nonzero mean activation within region~$j:$

\begin{eqnarray}\label{eq:post_prob_approx2}
\tilde P_j
&=&
\left(
1 - \tilde B_j
\times\frac
{\pi(\gamma_j = 0)}
{\pi(\gamma_j = 1)}
\right)^{-1}.
\end{eqnarray}

$\tilde P_j$ underestimates the probability of each region being involved in the task at hand. Thus, the bias it introduces is always conservative.
% , and can be interpreted as a way of increasing the penalty on model complexity, represented here by $B_j^2 + B_j^3.$

\subsubsection{Calibration of the additional penalty on simulated data}\label{sec:calibration}

\begin{figure}[ht!]
\includegraphics[width=0.5\textwidth]{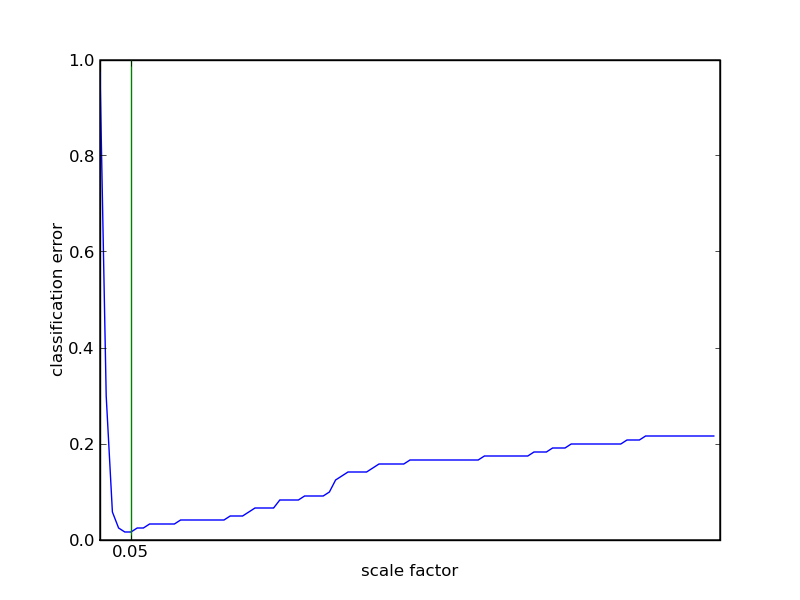}\hfill
\includegraphics[width=0.5\textwidth]{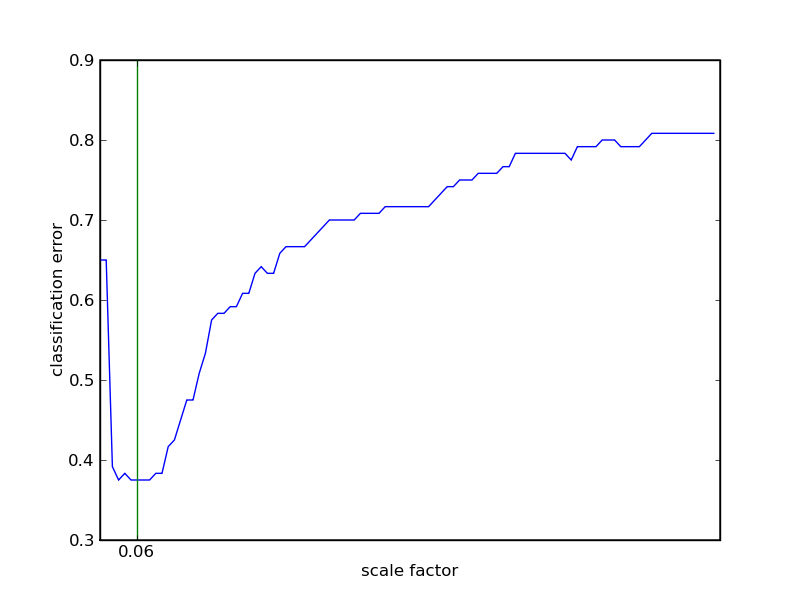}\hfill
\caption{\label{fig:scale_factor} Optimization of the additional penalty on simulated datasets, for the posterior mode approximation using the model with spatial uncertainty (left), and using the model with no spatial uncertainty (right).}
\end{figure}

Ideally, we would like to choose the factor~$c$ small enough so that it will compensate for data overfit, but not too small, for active regions not to be missed. We perform this calibration using intensive calculations, by estimating the optimal value from a collection of datasets simulated under different parameter values.

As in Section~\ref{sec:simu_3D_approx}, we defined a synthetic activation pattern, within a search volume of $24\times24\times24$~voxels, consisting of a central activated sphere, with uniform intensity value~$5$ (the background was set to $0$) and a diameter of $7$~voxels.

The data was then simulated using a field smoothness parameter $\omega=3,4$ or $5$ and a noise level $\epsilon=1,2,3$ of $4.$ For each of the $3\times4$ possible combinations of these parameters, we simulated $10$ different datasets comprising $n=30$ images, to which we applied the procedure in \ref{sec:independence} to compute Bayes factor values $B_j = LR_j + D_j$ for regions~$j=1,2,$ using the same algorithm as in Sections~\ref{sec:simu_2D_approx} and \ref{sec:simu_3D_approx}.

% More precisely, the MAP estimate $\hat{\bs\theta}$ of model parameters was first obtained without parcelling the search volume, using $500$~iterations of the SAEM algorithm, following $500$ burn-in iterations. Then, the conditional MAP estimate $\hat{\br w}$ of the displacement fields was computed using $500$ iterations of the SA algorithm. The conditional likelihoods $m(\br y | \hat{\br w}, \bs \gamma)$ were then computed both under the null model, defined by $\gamma_j \equiv 0,$ and under the full model, defined by $\gamma_j \equiv 1.$ In each case, the conditional MAP estimate $\hat{\bs \theta}_{\bs\gamma}$ of model parameters was obtained from $1\,000$~iterations of the SAEM algorithm (including $500$ burn-in iterations), and the conditional posterior $\pi (\hat{\bs \theta}_{\bs\gamma} | \br y, \hat{\br w}, \bs\gamma)$ was computed from the output of $1\,000$ iterations of a MH-Gibbs algorithm, following a burn-in of $100.$

% We used our posterior mode approximation to compute the Bayes factors for both regions (background and sphere), using exactly the same number of iterations as in Section~\ref{sec:simu_2D_approx}, and assuming the same value for the deformation field smoothness, $\omega_{est} = 4.$ The procedure was re-run $10$ times to assess numerical stability. We compared these results with those obtained under no spatial uncertainty, setting $\br w = \bs 0$ and $\sigma_S^2 = 0.$ The processing of each dataset took approximately one hour on a PC with a clock rate of $1.33GHz.$

Next, for all $c = 0,\, 0.01,\, 0.02,\, \ldots,\, 1,$ we computed the modified criterion~$\tilde B_j(g,c) = c \times LR_j(g) + D_j(g)$ for all $120$ datasets, indexed by $g=1, \ldots, 120$ and measured the proportion of corresponding mis-classified regions, given by:
$$
R(c) = \sum_{g=1}^{120} \left\{ \bs 1_{\tilde B_1(c,g) > 0} + \bs 1_{\tilde B_2(c,g) < 0} \right\}.
$$
Finally, we chose the value of $c$ which minimized $R(c).$

As illustrated in Figure~\ref{fig:scale_factor}, left, a minimum of $2$ misclassified regions (out of $420$) was obtained for $c=0.05.$ The surprisingly low number of classification errors obtained after optimizing $c$ may suggest that it is the introduction of this factor, rather than the posterior mode approximation, that is effective in correcting data overfit.

To verify this assertion, we also computed for each dataset the Bayes factors~$B_j$  using the model without spatial uncertainty, computed the modified criterions~$\tilde B_j$ for different values of $c$ and chose the value optimizing the classification rate. As illustrated in Figure~\ref{fig:scale_factor}, the minimum number of mis-classified regions, obtained in this case for $c=0.06,$ was much higher, and equal to $45.$ This indicates that the good classification score achieved using the model with spatial uncertainty is due to the combination of the posterior mode approximation, as well as the additional penalty.

\subsection{Phantom activations}\label{sec:phantom}

\begin{figure}[hb!]
  \begin{center}
    \includegraphics[width=0.75\textwidth] {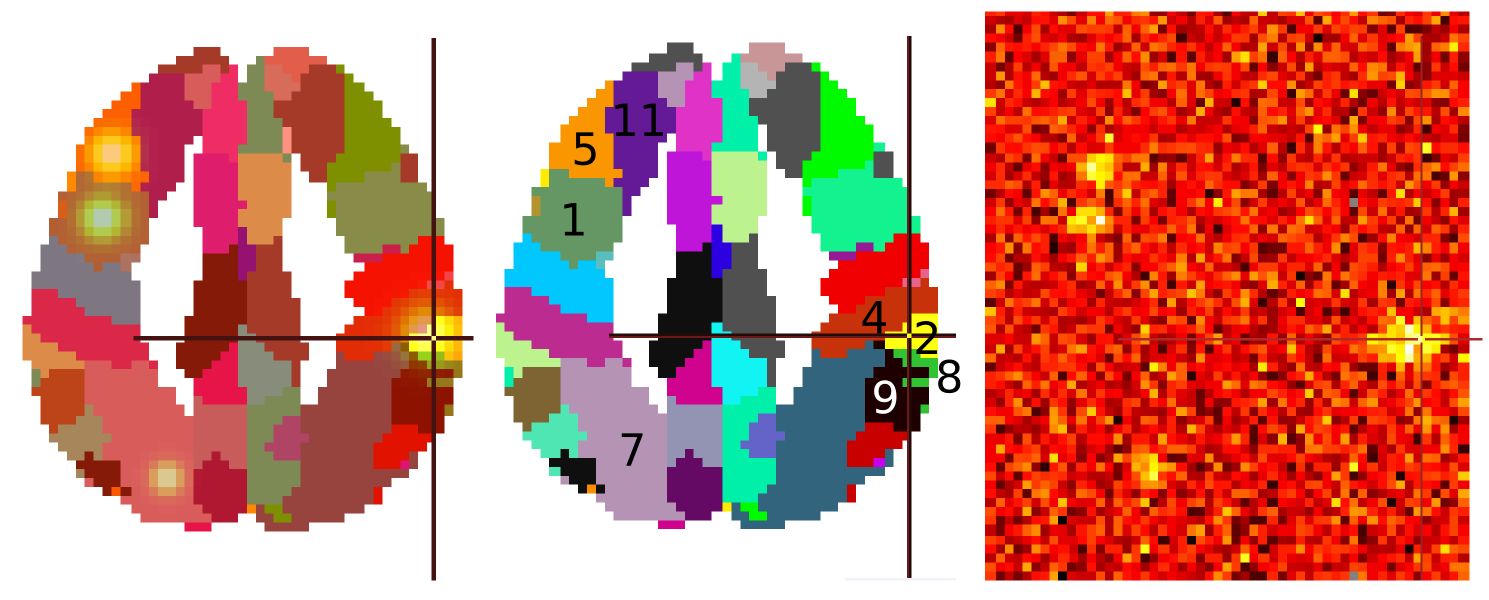}
  \end{center}
    \caption{\label{fig:simulation_phantom} Data simulated using phantom activations.
        Slice $z=37{\rm mm}$ in Talairach space.
        From left to right:
        Synthetic activation pattern (with CSA atlas in the background);
        CSA atlas (numbers correspond to region index in Table~\ref{tab:regions_phantom});
        simulated data example.}
%         $t$-score map, 
%         and posterior mean estimate of the mean population effect map.
\end{figure}

We conclude this chapter by applying our final procedure to a dataset which presents some similarities to real-life situations, and in particular, which was not simulated under the full hierarchical model (illustrated in Figure~\ref{fig:graph_model}) used to analyze the data. 

\subsubsection{Data simulation and analysis}

To start with, we defined a `life-size' search volume of $45\times 62\times 52$ voxels, corresponding to the actual dimensions of true fMRI images. Then, we designed an artificial activation pattern, based on the cortical sulci atlas (CSA), developed in~\cite{Perrot08}, and which we used to analyze real fMRI data (see Chapter~\ref{chap:case}). Each activation was defined in terms of a peak location, from  which the signal decreased radially according to a Gaussian kernel. Each signal peak was taken equal to $5.$ We placed two activations in neighboring regions, one at the intersection of several regions, and a smaller one inside the largest atlas region (see Figure~\ref{fig:simulation_phantom}, region~$7$).

$n = 40$ images were generated by warping this map according to the deformation model defined by (\ref{eq:deformation}), with one control
point in each voxel, choosing $\omega = 4$~voxels and $\sigma_S = 2.0$~voxels. Homoscedastic noise was then added to each image
according to (\ref{eq:cond_regional_obs}), with $\sigma_j^2 \equiv 1$, and the $s_{i,k}^2$'s generated as independent chisquare variables.

We then applied our Bayesian model selection algorithm based on the posterior mode approximation described in Section~\ref{sec:independence} to this dataset, to compute for all regions $j = 1, \ldots, N,$ the Bayes factor testing the presence of a nonzero regional mean
% $\eta_j = \frac{1}{\sharp V_j} \sum_{v_k \in V_j} \mu_k$, 
$\eta_j,$ still using the deformation model defined in (\ref{eq:deformation}), except this time control points were restricted to a grid with regular spacing equal to~$\gamma$ along each axis. We used the algorithm detailed in Section~\ref{sec:simu_2D_approx} (except that $j=1,\ldots,N$ in our case, with $N=124$).

Finally, lower bounds~$\tilde P_j$ on the posterior probabilities of nonzero regional means where computed, as explained in Section~\ref{sec:additional_penalty}, using an additional penalty controlled by a factor~$c,$ which we tuned to the optimal value derived in~\ref{sec:calibration}. This penalty was used for the Bayes factor computed both in the model with and without spatial uncertainty.

%  We also used the SPM-like~approach, described in Chapter~\ref{chap:group}.
%, according to the implementation in \cite{Meriaux06}.
% The cluster-forming threshold was tuned to control the voxel-level false positive rate (FPR) at $10^{-3}$ uncorrected. Clusters with corrected $p$-values less than $5\%$ in the cluster-size test were reported, and labeled according to their maximum statistic.

\subsubsection{Results}

\begin{figure}[!ht]
\begin{center}
\includegraphics[width=0.75\textwidth]{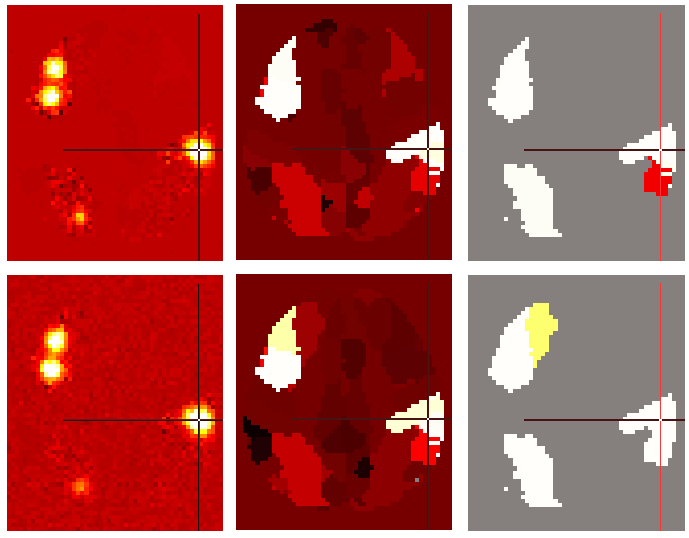}
\end{center}
\caption{\label{fig:results_phantom} Statistical maps obtained by the model selection approach on 3D simulated data.
Axial slice $z=37{\rm mm}$ in Talairach space. First row contains results for the model with spatial uncertainty, bottom row for the model without. From left tor right: posterior estimate of the mean effect map~$\bs\mu,$ the regional mean effect~$\bs\eta,$ and lower bounds~$\tilde P_j$ on the probabilities of a nonzero mean activation, restricted to detected regions ($\tilde P_j>0.$).}
\end{figure}

\begin{table}[!ht]
\begin{center}
\begin{tabular}{lrr|rr|r}
& $\tilde P_j$ & $\hat{\eta}_j$ 
& $\tilde P_j$ & $\hat{\eta}_j$ & $\bar{\bs\mu}_j$\\
Region & \multicolumn{2}{c}{Spatial uncertainty}
& \multicolumn{2}{|c|}{No spatial uncertainty}&\\
\hline
1:
& 1.00        & 0.53
& 1.00        & 0.41       & 0.56 \\
2:
& 1.00        & 0.26
& 1.00        & 0.19       & 0.27 \\
3:
& 1.00        & 0.24
& 1.00        & 0.19       & 0.27 \\
4:
& 1.00        & 0.25
& 1.00        & 0.20        & 0.26 \\
5:
& 1.00        & 0.07
& 1.00        & 0.04       & 0.06 \\
6:
& 1.00        & 0.26
& 1.00        & 0.21       & 0.26 \\
7:
& 1.00        & 0.04
& 1.00        & 0.03       & 0.05 \\
8:
& 1.00        & 0.33
& 0.99        & 0.23       & 0.34 \\
9:
& 0.69        & 0.07
& 1.00        & 0.04       & 0.06 \\
10:
& \textit{ 0.46}        & \textit{ 0.04}
& 1.00        & 0.06       & 0.05 \\
11:
& \textit{ 0.00}       & \textit{ 0.00}
& 0.93       & 0.02       & 0.01 \\
\end{tabular}
\end{center}
\caption{\label{tab:regions_phantom} Results of the model selection approach on 3D simulated data.
Reported regions have an approximate posterior probability of being involved 
in the task greater than $0.5.$ $\hat{\eta}_j$ is the posterior 
estimate of the regional mean effect. $\bar{\bs\mu}_j$ is the average
over region~$j$ of the group mean effect map $\bs\mu.$}
% Asterisks (*) mark regions 
% that are found significant at $5\%$ by the SPM-like approach (corrected
% cluster-level inference, cluster-forming threshold set to ${\rm
% FPR}=10^{-3}$), and correspond to the middle activation map in
% Figure~\ref{fig:clusters_con31}.}
\end{table}

% No spatial: 0.08, spatial: 0.01

Results from this simulation study are illustrated in Figure~\ref{fig:simulation_phantom}. We estimated the mean effect map $\bs\mu$ by its conditional posterior mean, averaged over models:
$$
\hat{\bs\mu} = 
\sum_{\bs\gamma}
\mathbb E(\bs\mu | \br y, \hat{\br w}, \hat{\bs\theta}_{\bs\gamma})\tilde\pi(\bs\gamma|\br y, \hat{\br w}).
$$
where $\tilde\pi(\bs\gamma|\br y, \hat{\br w})$ is noted using a tilde to remind that it is computed using the lower bound~$\tilde P_j$ instead of the posterior probability. It can be seen from Figure~\ref{fig:simulation_phantom}, left, that the map obtained using the spatial uncertainty is less noisy, especially in the background, indicating the better fit obtained conditional on the most probable displacements {\em a posteriori}.

Figure~\ref{fig:results_phantom} shows that both algorithms detected similar regions, but that the model with no spatial uncertainty was less conservative. In particular, the latter detected activations in region~$11,$ simply because it was located next to region~$5$ (using the labels from Figure~\ref{fig:simulation_phantom}) which contained one of the simulated activation peak. Conversely, this region was considered inactive (with $\tilde Pj = 0.00$) using the model with spatial uncertainty.

This tendency is confirmed in Table~\ref{tab:regions_phantom}, where we can see that two more regions were detected in the model ignoring spatial displacements. Both methods detected activations in region~$9,$ which was contaminated by activations from neighboring regions~$8$ and $2,$ though the approximate probability is lower in the model with spatial uncertainty.

Finally, we computed the following estimates of the regional means:
$$
\hat{\eta}_j = 
\sum_{\bs\gamma}
\hat{\eta}_{\bs\gamma}\tilde\pi(\bs\gamma|\br y, \hat{\br w}),
$$
where $\eta_{\bs\gamma}$ is the MAP estimate of $\eta,$ conditional on $\br w$ and $\bs\gamma.$ Still from table~\ref{tab:regions_phantom}, it can be seen that, as an estimate of the average mean effect within region~$j,$ $\bar{\bs\mu}_j = \sum_{k\in \mathcal V_j} \mu_k,$ $\hat{\eta}_j$ is more accurate using spatial uncertainty than not. In fact, the average relative error 
$\epsilon =
\frac{1}{N} \sum_j|\hat{\eta}_j - \bar{\bs\mu}_j|/\bar{\bs\mu}_j
$
was equal to $1\%$ for the spatial uncertainty model against $8\%$ for the other. This confirms the fact that our posterior mode approximation did provide a better fit to the data than obtained without accounting for displacements.

\section{Discussion}\label{sec:discussion}

Throughout this chapter, we have investigated the possibility of testing the presence of a nonzero mean effect within each region of a pre-defined parcellation of the search volume, while accounting for individual images being spatially deformed, according to unknown displacement fields. We proposed a Bayesian model selection approach to this problem, entailing the computation of the marginal likelihood for each possible observation model, specified by a partition of all regions into `involved' (containing a nonzero mean effect) and `inactive'.

We have shown how Monte-Carlo estimates of these marginal likelihoods can be obtained using MCMC techniques. However, this theoretically sound solution fails in practice, because the Monte-Carlo estimates of the marginal likelihoods are unstable numerically.
% This is because our posterior sampling scheme tends to get stuck in local modes of the posterior density of the deformation fields, which must be integrated out to compute the marginal likelihood of each given model. This behavior was observed on a very simple 2D simulated dataset.
% Additionally, the number of possible models increases exponentially with the number of regions, making calculations rapidly intractable. 
% This is due to the fact that the regions are not independent, when modeling spatial uncertainty. In contrast, the number of distinct models in the model without spatial uncertainty is linear in the number of regions.
We then considered an approximate procedure, which consists in assimilating the unknown spatial displacements to their most probable values {\em a posteriori}.
% , so that the marginal likelihoods can be computed as in the model without spatial uncertainty. 
% We investigate this approximation using numerical experiments. 
This approximation 
% makes calculations tractable, and 
substantially reduces the numerical instability, however
% when computing the Bayes factors, {\em i.e.} the marginal likelihood ratios, corresponding to each region. Furthermore, the resulting model selection procedure 
% and gave satisfying results when applied to 2D simulated data.
%  since it selects the correct model, presumably because the displacement fields can be reliably estimated in this case.
% On the contrary, when applied to 3D simulated data, 
we found that it
% this approximation 
re-introduces a certain amount of bias toward false positives, which is systematically present when 
% computing the Bayes factor 
using the model which ignores spatial displacements. We compensated this residual bias by introducing an additional penalty, calibrated on a large number of simulated datasets. This last approximation gave satisfying results when validated on a final synthetic 3D dataset.

There is clearly some space for improvement here, in particular concerning the Monte-Carlo estimation of the exact marginal likelihood. We have seen that its numerical instability stems from the random-walk Metropolis-Hastings step used to sample the elementary displacements, which has a high rejection rate, and results in the Markov Chain getting stuck in local modes of the posterior density. Thus, a promising direction for future work would be to test alternative proposal densities, in view of ameliorating our posterior sampling scheme.

Another alternative we have not yet explored would be to integrate $\br x$ and $\bs\mu$ analytically and sample directly from the joint posterior distribution $\pi(\bs\theta, \br w | \br y, \bs\gamma),$ using two alterning Metropolis-Hastings steps. This would potentially speed-up the convergence of the Markov chain, and moreover allow to compute the unconditional MAP estimate of $\br w,$ using simulated annealing. This is required in Section~\ref{sec:independence}, and we have so far used a conditional MAP estimate as a surrogate.

Such ameliorations would allow to investigate the behavior of the exact approach in more complex situations than provided by the simplistic 2D datasets studied in this chapter. This would also provide an additional way of validating our approximate approach, by comparing the results obtained by both methods.

However, we point that the exact approach would nevertheless be associated with a high computational complexity, since the number of possible models increases exponentially with the number of regions, due to the fact that regions are dependent when modeling spatial uncertainty. In contrast, the number of distinct models in the model without spatial uncertainty, or under the posterior mode approximation, is linear in the number of regions. Thus, the use of approximate techniques, such as the one we developed here, seems unavoidable in view of practical applications.

This raises additional questions concerning the generability of the additional penalty $c$ tuning, which we have done here using a colection of simulated datasets. These datasets where generated in order to reflect some features of the real datasets we intended to analyze (number of observations, inter-subject variance). The resulting penalty was found to work well when applied to real data. However, it may very well be that the optimal value for $c$ varies with respect to those parameters we chose to fix during our simulation study, and may need to be adapted to different situations, such as a much smaller or a much bigger sample size. These questions need to be investigated thoroughly in order for the method proposed here to be widely applicable.

% Case Studies
\cleardoublepage
\chapter{Application to real fMRI data}\label{chap:case}
\lhead{\emph{Application to real fMRI data}}

\section*{Abstract}

In this chapter, we apply the Bayesian model selection approach for fMRI group data analysis developed in Chapter~\ref{chap:bayesian} to a real fMRI dataset.

To validate our approach, we chose a paradigm based on extensively studied cognitive tasks (number and language processing), involving known brain regions. We show that our procedure successfully recovers in each case the complete functional network.

We also compare two different versions of our approach, both with and without modeling spatial uncertainty, along with the SPM-like approach, described in Chapter~\ref{chap:group}. In this way, we illustrate the shortcomings of standard voxel-based approaches which rely on the thresholding of a statistical map, and how these are overcome by the procedure we propose.

\section{Data analysis}

The data used here is extracted from the Localizer database \cite{Pinel07}. We used the same cohort of $38$~subjects as in Chapter~\ref{chap:unsupervised} and refer to Section~\ref{sec:fmri} for a detailed description.

\subsection{Individual data processing}

Individual data analyses were conducted following the standard pipeline described in Section~\ref{sec:single},
using SPM5 (\href{http://www.fil.ion.ucl.ac.uk/spm/}{\tt http://www.fil.ion.ucl.ac.uk/spm/}). Data were submitted successively to motion correction, slice timing and normalization to the~MNI template, and spatial smoothing using a $5\times 5\times 5$~mm$^3$ FWHM Gaussian filter. For each subject, BOLD contrast images were obtained from a fixed-effect analysis on all sessions.

\subsection{Methods compared}

\begin{figure}[ht!]
\begin{center}
\includegraphics[width=0.5\textwidth]{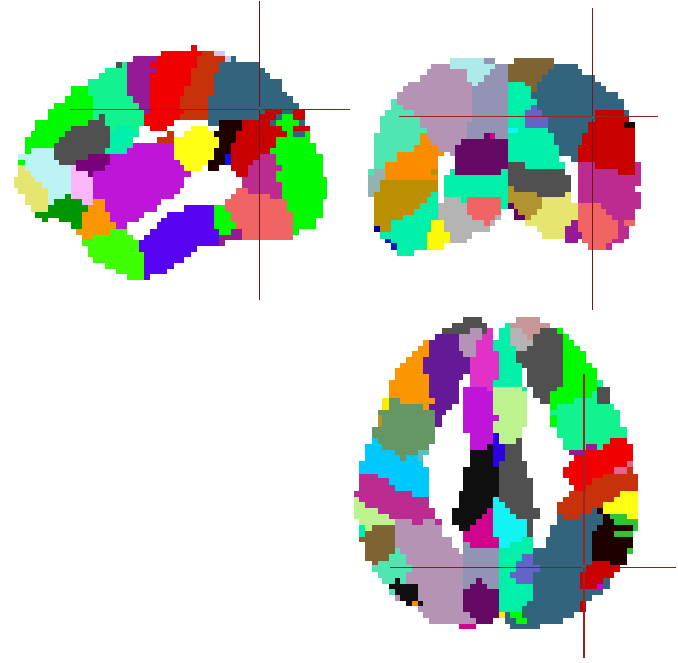}
\end{center}
\caption{\label{fig:CSA} The cortical sulci atlas}
\end{figure}

For each studied contrast, we used the method developed in Chapter~\ref{chap:bayesian} to select the functional network most probably involved in the cognitive task under investigation, based on a fixed brain parcellation. We used to this end the cortical sulci atlas (CSA) developed in \cite{Perrot08}, derived from the anatomical images of $63$ subjects and comprising $125$ regions which correspond to subdivisions of cortical sulci (see Figure~\ref{fig:CSA}).

More specifically, we used the posterior mode approximation described in Section~\ref{sec:independence} to compute the Bayes factor $B_j$ to test the presence of a nonzero mean activation, for each region~$j.$ The algorithm was tuned as in the simulation study in Section~\ref{sec:simu_3D_approx}. As explained in Section~\ref{sec:additional_penalty}, these Bayes factors where modified through the use of an additional penalty, according to~(\ref{eq:stabilization}). The factor calibrating this penalty was taken equal to~$c=0.05,$ which was found optimal from the simulation study in Section~\ref{sec:calibration}. We compared the results obtained in the model with and without spatial uncertainty, the latter corresponding to the special case where $\sigma_S^2 = 0,$ and the elementary displacements are frozen to zero.

We also included the results of the SPM-like approach, described in Section~\ref{sec:mass}. In this case, the group analyses were restricted to the intersection of all subjects' whole-brain masks, comprising $43\, 367$~voxels (no mask was used to restrict the analysis using our Bayesian model selection approach). First, a $t$-score map was computed from the $37$ individual estimated effect maps. Then, a permutation test was used to compute three different cluster-forming thresholds, tuned to control the per-comparison error rate (PCER) respectively at $10^{-2}$, $10^{-3}$ and $10^{-4}$ uncorrected. For each threshold, a second permutation test was used to determine the critical cluster size to guarantee a FWER control of $5\%$ (see Section~\ref{sec:cluster}). Each detected cluster was labeled according to the region containing its maximum $t$-score; this is one of the procedures suggested in \cite{Tzourio-Mazoyer02}.

\subsection{Summarizing the inference}\label{sec:display}

We chose to summarize the results of each procedure using the following statistics. The first one is the posterior estimate of the mean population effect, averaged across all possible networks, obtained as 
$$
\hat{\bs\mu} 
= \sum_{\bs\gamma} \mathbb E[ \bs\mu | \br y, \hat{\bs\theta}_{\bs\gamma}, \hat{\br w} ] \tilde\pi(\bs\gamma | \br y, \hat{\br w}),
$$
using the notations introduced in Section~\ref{sec:independence}. Here we note $\tilde\pi(\bs\gamma | \br y, \hat{\br w})$ using a tilde since we use the conservative approximation in Section~\ref{sec:additional_penalty}.

Secondly, we computed the posterior estimate of regional means, also averaged across all possible networks:
$$
\hat{\bs\eta} 
= \sum_{\bs\gamma} \hat{\bs\eta}_{\bs\gamma} \tilde\pi(\bs\gamma | \br y, \hat{\br w}),
$$
where $\hat{\bs\eta}_{\bs\gamma} = \arg\max_{\bs\eta} \pi(\bs\eta | \br y, \hat{\br w}, \bs\gamma).$ 

Finally, the functional network selected by our algorithm is summed up by the map of approximate posterior probabilities~$\tilde P_j$ that the regions~$j$ are involved (as defined by (\ref{eq:post_prob_approx2})). For clarity, we have only represented regions~$j$ detected as involved ($\tilde P_j > 0.5$), and with a positive regional mean estimate $\hat \eta_j,$ {\em i.e.}, regions detected as active for the task under study.

\section{Number processing task}

\begin{figure}[ht!]
\begin{center}
% \begin{tabular}{ccc}
% \multicolumn{3}{c}{Spatial Uncertainty}\\
% % \includegraphics[height=4cm]{Chapter6/palette_m.png}
% \includegraphics[width=0.33\textwidth]{Chapter6/con29_spatial_mean_m.png} &
% \includegraphics[width=0.33\textwidth]{Chapter6/con29_spatial_m_mean.png} &
% \includegraphics[width=0.33\textwidth]{Chapter6/con29_spatial_P.png}\vspace{-1cm}\\
% $\hat{\bs\mu}$ & $\hat{\bs\eta}$ & $\tilde P_j\br 1_{\tilde P_j > 0.5}\br 1_{\eta_j > 0}$\vspace{1cm}\\
% \hline
% \multicolumn{3}{c}{No Spatial Uncertainty}\\
% \includegraphics[width=0.33\textwidth]{Chapter6/con29_mean_m.png} &
% \includegraphics[width=0.33\textwidth]{Chapter6/con29_m_mean.png} &
% \includegraphics[width=0.33\textwidth]{Chapter6/con29_P.png}\vspace{-1cm}\\
% $\hat{\bs\mu}$ & $\hat{\bs\eta}$ & $\tilde P_j\br 1_{\tilde P_j > 0.5}\br 1_{\eta_j > 0}$\\
% \end{tabular}
Spatial uncertainty\\
\includegraphics[width=\textwidth]{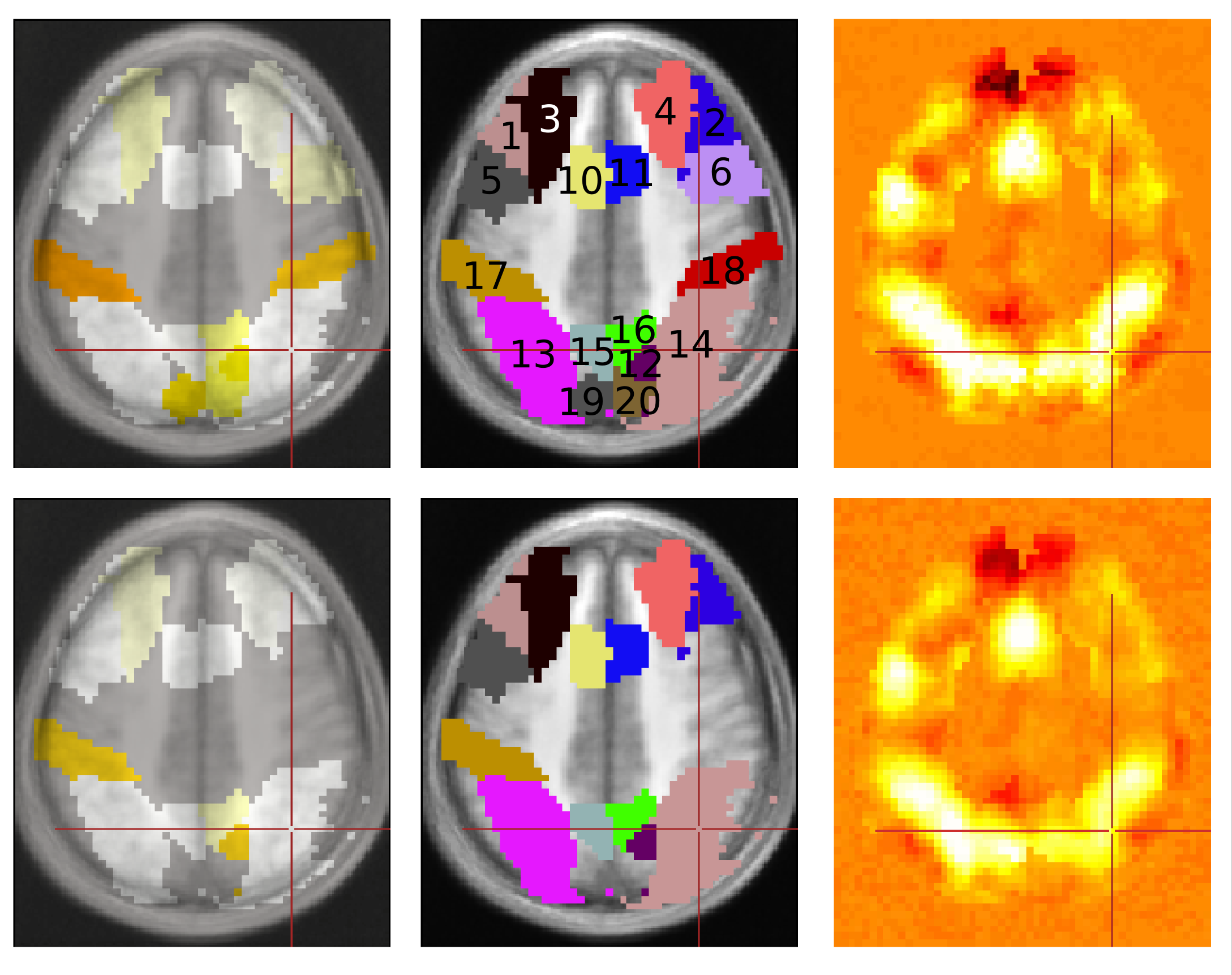}\\
No spatial uncertainty
\end{center}
\caption{\label{fig:results_con29} Number processing task, results of the model selection approach in axial slice $z=37{\rm mm}$ in Talairach space, (top) using the model with spatial uncertainty,
(bottom) with no spatial uncertainty. From left to right: approximate posterior probability map $\tilde P_j,$ restricted to regions detected as activated ($\tilde P_j>0.5$, $\hat \eta_j > 0$); labels of detected regions (numbers correspond to region index in Tables~\ref{tab:regions_con29_frontal} and \ref{tab:regions_con29_parietal}); posterior estimate of mean effect map $\hat{\bs\mu}$ (see Section~\ref{sec:display}). The first two maps are overlayed on the mean anatomical image of all subjects.}
\end{figure}

\begin{table}[!h]
\begin{center}
\begin{tabular}{lrr|rr}
Sulcus/Fissure & $\tilde P_j$ & $\hat{\eta}_j$ 
& $\tilde P_j$ & $\hat{\eta}_j$ \\
& \multicolumn{2}{c}{Spatial uncertainty}
& \multicolumn{2}{|c}{No spatial uncertainty}\\
\hline
\multicolumn{5}{c}{Frontal lobe}\\
\hline
1: Left middle frontal             &     1.00   &      2.85
& 1.00 &        2.89 \\
2: Right middle frontal*            &     1.00   &      3.47
& 1.00 &        3.20 \\
\hline
3: Left superior frontal           &     0.89   &     2.21
& 0.92 &       2.11 \\
4: Right superior frontal          &     0.96   &     1.87
& 0.98 &       1.85 \\
\hline
5: Left inferior frontal*           &     1.00   &      4.48
& 1.00 &        4.04 \\
\hline
6: Left middle precentral          &     0.99   &     4.53
& 1.00 &        4.50 \\
7: Right middle precentral         &     0.90   &      1.99
& \textit{ 0.38} &        \textit{ 1.85}  \\
\hline
8: Left inferior precentral        &     1.00   &      6.08
& 1.00 &        5.54 \\
9: Right inferior precentral       &     0.92   &     2.83
& 0.58 &       2.54 \\
\hline
10: Left anterior cingular          &     1.00   &      3.62
& 1.00 &        3.33 \\
11: Right anterior cingular         &     1.00   &      4.26
& 1.00 &        4.02 
\end{tabular}\\
\end{center}
\caption{\label{tab:regions_con29_frontal} Number processing task, regions detected in the frontal lobe using the Bayesian model selection approach. Reported regions have a posterior probability of being involved 
in the task greater than $0.5$, and constitute the most probable 
functional network given the data.  $\hat{\eta}_j$ is the posterior 
estimate of the regional mean effect.  Asterisks (*) mark regions 
that are found significant at $5\%$ by the SPM-like approach (corrected
cluster-level inference, cluster-forming threshold set to ${\rm
FPR}=10^{-3}$), and correspond to the middle activation map in
Figure~\ref{fig:clusters_con29}.}
\end{table}

\begin{table}[!ht]
\begin{center}
\begin{tabular}{lrr|rr}
Sulcus/Fissure & $\tilde P_j$ & $\hat{\eta}_j$ 
& $\tilde P_j$ & $\hat{\eta}_j$ \\
& \multicolumn{2}{c}{Spatial uncertainty}
& \multicolumn{2}{|c}{No spatial uncertainty}\\
\hline
\multicolumn{5}{c}{Parietal Lobe}\\
\hline
12: Right transverse parietal       &     0.72   &     6.67
& 0.64 &       5.28 \\
\hline
13: Left intra-parietal*             &     1.00   &      5.31
& 1.00 &        4.89 \\
14: Right intra-parietal            &     1.00   &      3.50
& 1.00 &        2.95 \\
\hline
15: Left precuneus                  &     0.99   &     6.09
& 1.00 &        5.81 \\
16: Right precuneus                 &     0.82   &     4.10
& 0.88 &       3.74 \\
\hline
17: Left postcentral intraparietal  &     0.53   &     2.35
& 0.64 &       2.14 \\
18: Right postcentral intraparietal &     0.61   &     2.07
& \textit{ 0.38} &       \textit{ 1.90} \\
\hline
19: Left parieto-occipital          &     0.68   &     1.75
& \textit{ 0.35} &       \textit{ 1.77} \\
20: Right parieto-occipital         &     0.78   &     1.53
& \textit{ 0.00} &       \textit{ 1.25} \\
\hline
\multicolumn{5}{c}{Other}\\
\hline
21: Right callosal                  &     0.97   &     2.10
& 0.59 &       1.99 
\end{tabular}
\end{center}
\caption{\label{tab:regions_con29_parietal} Number processing task, regions detected in the parietal lobe using the Bayesian model selection approach. Reported regions have a posterior probability of being involved 
in the task greater than $0.5$, and constitute the most probable 
functional network given the data.  $\hat{\eta}_j$ is the posterior 
estimate of the regional mean effect.  Asterisks (*) mark regions 
that are found significant at $5\%$ by the SPM-like approach (corrected
cluster-level inference, cluster-forming threshold set to ${\rm
FPR}=10^{-3}$), and correspond to the middle activation map in
Figure~\ref{fig:clusters_con29}.}
\end{table}

\begin{figure}[!ht]
  \begin{center}
  \begin{tabular}{ccc}
    \includegraphics[width=0.3\textwidth] {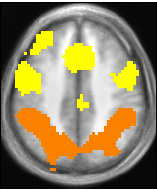} &
    \includegraphics[width=0.3\textwidth] {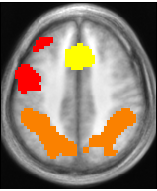} &
    \includegraphics[width=0.3\textwidth] {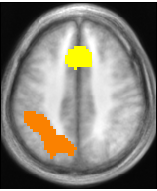} \\
    PCER = $10^{-2}$ & PCER = $10^{-3}$ & PCER = $10^{-4}$\\
  \end{tabular}
  \end{center}
    \caption{\label{fig:clusters_con29} Clusters detected at different cluster-forming thresholds, for the number processing task, in axial slices $z = 37{\rm mm}$ in Talairach, overlayed on the subjects' mean anatomical image in the background). The threshold is tuned to control the per-comparison error rate (PCER) respectively at $10^{-2}$, $10^{-3}$ and $10^{-4}$ uncorrected. Each cluster surviving the FWER controlling-threshold at $5\%$ is represented with a specific color.}
\end{figure}

As can be seen from Figure~\ref{fig:results_con29}, right, the estimated mean effect map $\hat{\bs\mu}$ is more contrasted using the model with spatial uncertainty than without, and slightly less noisy in the background. However, the regularizing effect observed previously (see Section~\ref{sec:fmri_data}), in a context where $\bs\mu$ was estimated by {\em marginalizing} out the elementary displacements $\br w,$ is absent here, since $\bs\mu$ is estimated {\em conditional} on a particular value~$\hat{\br w}.$

The posterior probability and region label maps, illustrated in Figure~\ref{fig:results_con29} suggest that the network detected using both models is very similar, though more regions are detected using the model with spatial uncertainty, indicating an increased sensitivity.

The complete list of regions detected as activated ($\tilde P_j > 0.5$ and $\hat\eta > 0$) is given in Tables~\ref{tab:regions_con29_frontal} and \ref{tab:regions_con29_parietal}. Our method successfully detected the bilateral intra-parietal and fronto-cingular networks known to be active during number processing~\cite{Chochon99,Dehaene03}. Interestingly, the bilateral precuneus sulci where also detected. Although not considered as part of the core numerical system, the precuneus has been linked to memory access and a wide range of high-level tasks~\cite{Cavanna06}.

The networks detected with and without spatial uncertainty are very similar. However, $4$~more regions were detected with spatial uncertainty; in particular, no activations were detected in the parieto-occipital region when neglecting spatial uncertainty. This is consistent with the fact that the estimated regional means~$\hat{\bs\mu}$ are systematically higher under spatial uncertainty, and the higher contrast observed on the estimated mean effect map.

In contrast, only three activated clusters were detected by the SPM-like approach at the chosen cluster-forming threshold. Each cluster contained over a thousand voxels, and extended over several atlas regions, hence merging several functionally distinct areas. Also, no activations were detected in the right frontal area. Using different thresholds could not solve these problems, as illustrated in Figure~\ref{fig:clusters_con29}.

\section{Language processing task}

\begin{figure}[!ht]
\begin{center}
% \begin{tabular}{ccc}
% \multicolumn{3}{c}{Spatial Uncertainty}\\
% % \includegraphics[height=4cm]{Chapter6/palette_m.png}
% \includegraphics[width=0.33\textwidth]{Chapter6/con31_spatial_mean_m.png} &
% \includegraphics[width=0.33\textwidth]{Chapter6/con31_spatial_m_mean.png} &
% \includegraphics[width=0.33\textwidth]{Chapter6/con31_spatial_P.png}\vspace{-1cm}\\
% $\hat{\bs\mu}$ & $\hat{\bs\eta}$ & $\tilde P_j\br 1_{\tilde P_j > 0.5}\br 1_{\eta_j > 0}$\vspace{1cm}\\
% \hline
% \multicolumn{3}{c}{No Spatial Uncertainty}\\
% \includegraphics[width=0.33\textwidth]{Chapter6/con31_mean_m.png} &
% \includegraphics[width=0.33\textwidth]{Chapter6/con31_m_mean.png} &
% \includegraphics[width=0.33\textwidth]{Chapter6/con31_P.png}\vspace{-1cm}\\
% $\hat{\bs\mu}$ & $\hat{\bs\eta}$ & $\tilde P_j\br 1_{\tilde P_j > 0.5}\br 1_{\eta_j > 0}$\\
% \end{tabular}
Spatial uncertainty\\
\includegraphics[width=\textwidth]{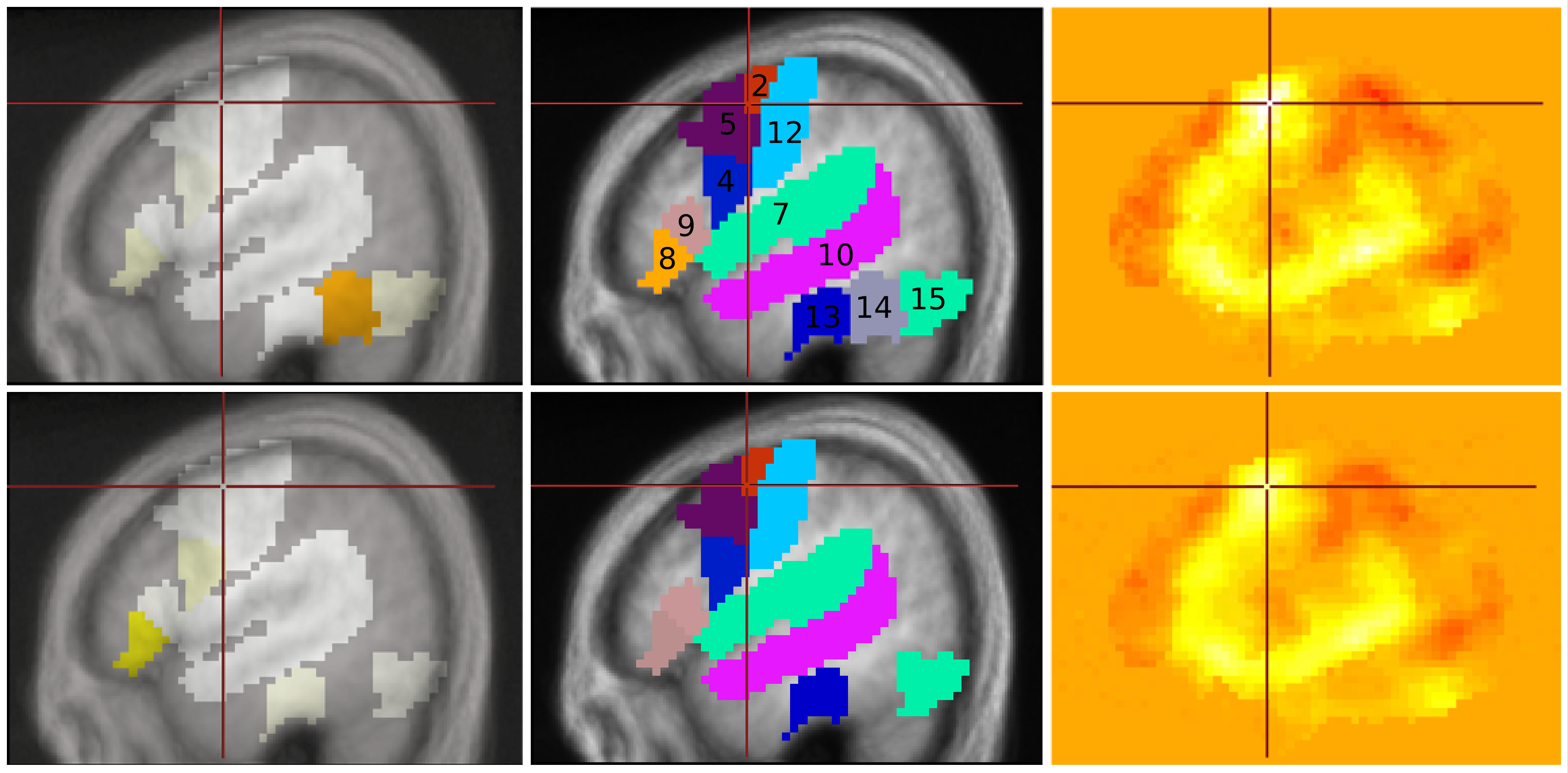}\\
No spatial uncertainty
\end{center}
\caption{\label{fig:results_con31} Language processing task, results of the model selection approach, in sagittal slice $x=-46{\rm mm}$ in Talairach space, (top) using the model with spatial uncertainty, (bottom) with no spatial uncertainty. From left to right: approximate posterior probabilities map $\tilde P_j,$ restricted to regions detected as activated ($\tilde P_j>0.5$, $\hat \eta_j > 0$); labels of detected regions (numbers correspond to region index in Table~\ref{tab:regions_con31}); posterior estimate of mean effect map $\hat{\bs\mu}$ (see Section~\ref{sec:display}). The first two maps are shown above the mean anatomical image of all subjects.}
\end{figure}

\begin{table}[!ht]
\begin{center}
\begin{tabular}{lrr|rr}
Sulcus/Fissure & $\tilde P_j$ & $\hat{\eta}_j$ 
& $\tilde P_j$ & $\hat{\eta}_j$ \\
& \multicolumn{2}{c}{Spatial uncertainty}
& \multicolumn{2}{|c}{No spatial uncertainty}\\
\hline
1: Left middle precentral    &        \textit{ 0.05}    &    \textit{ 2.19}
&         0.52  &      1.43 \\
2: Left superior precentral  &        0.99 &       6.62 
&         0.99  &      5.54 \\
3: Right superior precentral &        0.78 &       2.78 
&         \textit{ 0.00}  &      \textit{ 2.23}    \\
4: Left inferior precentral  &        0.97 &       3.64 
&         0.91  &     3.33 \\
5: Left middle precentral    &        0.99 &       5.11 
&         0.99  &     4.74 \\
\hline
6: Left paracentral          &        0.93 &       4.26 
&         0.86  &     3.38 \\
\hline
7: Left posterior sylvian    &        1.00 &       2.41 
&         1.00   &     2.19 \\
8: Left anterior sylvian     &        0.92 &       5.34 
&         0.72  &     4.11 \\
9: Left superior sylvian     &        1.00 &       5.64 
&         0.98  &     5.07 \\
\hline
10: Left superior temporal*    &        1.00 &       7.44 
&        1.00    &    6.99 \\
11: Right superior temporal*   &        1.00 &       4.5  
&        1.00    &    3.94 \\
\hline
12: Left central              &        1.00 &       2.0  
&        1.00    &    1.74 \\
\hline
13: Left anterior collateral  &        1.00 &       2.47 
&        0.96   &    2.39 \\
14: Left middle collateral    &        0.56 &       2.77 
&        \textit{ 0.17}   &    \textit{ 2.51} \\
15: Left posterior collateral *&        0.94 &       3.48
&        0.98   &    3.28
\end{tabular}
\end{center}
\caption{\label{tab:regions_con31} Language processing task, regions detected using the Bayesian model selection approach. Reported regions have a posterior probability of being involved 
in the task greater than $0.5$, and constitute the most probable 
functional network given the data.  $\hat{\eta}_j$ is the posterior 
estimate of the regional mean effect.  Asterisks (*) mark regions 
that are found significant at $5\%$ by the SPM-like approach (corrected
cluster-level inference, cluster-forming threshold set to ${\rm
FPR}=10^{-3}$), and correspond to the middle activation map in
Figure~\ref{fig:clusters_con31}.}
\end{table}

\begin{figure}[!h]
  \begin{center}
     \includegraphics[width=\textwidth] {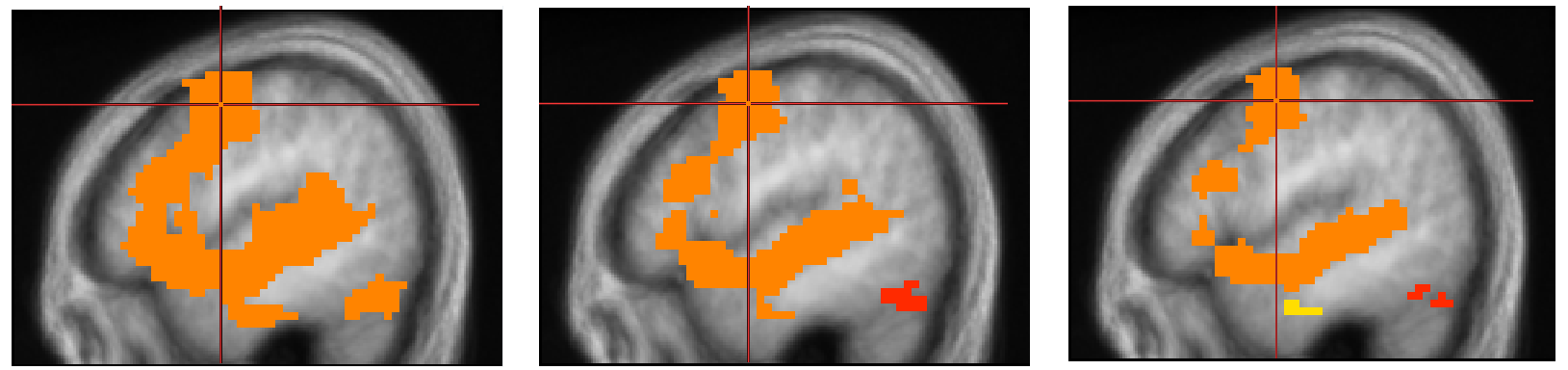} \\
    PCER = $10^{-2}$ \hspace{2.5 cm} PCER = $10^{-3}$ \hspace{2.5 cm} PCER = $10^{-4}$\\
  \end{center}
    \caption{\label{fig:clusters_con31} Clusters detected at different cluster-forming thresholds, for the language processing task, in the sagittal slice $x = -46{\rm mm}$ in Talairach,overlayed on the subjects' mean anatomical image in the background). The threshold is tuned to control the per-comparison error rate (PCER) respectively at $10^{-2}$, $10^{-3}$ and $10^{-4}$ uncorrected. Each cluster surviving the FWER controlling-threshold at $5\%$ is represented with a specific color.}
\end{figure}

As in the previous case, a gain in the contrast of the estimated mean effect map $\hat{\bs\mu}$ associated with the model under spatial uncertainty can be observed in Figure~\ref{fig:results_con31}, right. Apart from this, the regions detected in the displayed slice using both models were very similar, with a single additional region detected under spatial uncertainty.

The network detected by our method, shown in Table~\ref{tab:regions_con31} was consistent with the known organization of language processing in the cerebral cortex, as described in \cite{Pinel07} for instance. This involves the superior temporal sulcus, with a dominance of the left hemisphere (which can be linked to the higher estimate of the regional mean observed here), and the left frontal areas (such as the precentral sulcus detected here) and the supplementary motor area, part of the paracentral area detected here. Finally, activations were detected in the collateral fissure, which borders the lingual and the fusiform gyrus, both known to be involved in the visual recognition of words.

Again, there were some minor differences in the networks detected using both models, and the estimated regional means were generally higher in the model with spatial uncertainty. More importantly, the background, which was included in the list of regions, was detected as negatively active ($\tilde P_j = 1,$ $\hat{\bs\eta} = -0.01$) in the model without spatial uncertainty, suggesting that some activations were projected outside the brain volume due to registration errors. This error was corrected when modeling spatially uncertainties.

In contrast, the SPM-like approach detected three clusters at the selected threshold (see Figure~\ref{fig:clusters_con31}, middle). One of them (not represented) extended approximately over the right superior temporal sulcus, as defined in the CSA. Another one (in red), extended over several occipito-temporal parcels. Finally, a large cluster (in orange) containing over $1\,000$ voxels encompassed many different atlas regions, suggesting that it contained several functionally distinct areas. As in the case of number processing, this segmentation issue was not solved by varying the cluster-forming threshold.

% \section{IMAGEN Database}
% \subsection{Random Thresholding of Individual Activation Maps}
% \subsection{Assessing Center and Scanner Effects}
% \subsection{Comparison of Group Inference Methods}
% 
% \section{Random Thresholding for Feature Selection}
% \subsection{Mental State Classification Experiment}
% \subsection{Leave-One-Out Performance}
% \subsection{Comparison To Cross-Validated Threshold}

\section{Conclusion}

In this chapter, we have validated the Bayesian model selection approach for fMRI group data analysis developed in Chapter~\ref{chap:bayesian} using a real dataset, by correctly recovering the brain functional networks associated to basic number and language processing, according to the description found in previous works. Furthermore, we have illustrated certain shortcomings of the classical, SPM-like approach, which fails to properly segment distinct functional regions, or misses them altogether, depending on the choice of the cluster-forming threshold.

Similar results were obtained using the models with and without spatial uncertainty. A probable explanation is the use of an additional penalty on data overfit in both cases, which prevented the false detection of inactive regions. However, the posterior mode approximation in the model with spatial uncertainty did improve the results, in that it allowed to detect additional regions, and enhanced the contrast of the estimated mean effect maps. This, together with the higher estimated regional mean effects, suggest that the ``functional registration step'', which consists in displacing the individual images according to the most probable displacements {\em a posteriori}, provided a better alignment of the activated regions of the different subjects.

% Conclusion and Perspectives
\cleardoublepage
\chapter{Conclusion}\label{chap:conclusion}
\lhead{\emph{Conclusion}}

\section{Main Results}

Throughout this thesis, we have developed a new approach for the statistical analysis of multi-subject fMRI data, whose aim is to detect the cerebral regions involved in a certain cognitive task. Our objective was to address jointly certain limitations of the standard SPM-like approach, which are: the dependence on an arbitrary threshold to define clusters of potential activity; the lack of control over false negative risks; and the assumption that individual images are in perfect alignment.

In a first contribution, we revisited the adaptive thresholding technique developed in \cite{Lavielle07} by removing its dependence on a window parameter, making it more stable at low signal to noise ratios. This method provides an answer to the problem of choosing a detection threshold, while implicitly balancing false positive and false negative risks, by minimizing a model selection criterion rather than a multiple comparison type I error rate. It gave satisfying results when applied to individual and group activation maps, compared to Gamma-Gaussian mixture modeling \cite{Beckmann03b}. This technique may be best adapted to within-subject analysis however since it aims to detect individual voxels rather than regions, which are ultimately the objects of interest in group analysis.
% However, this approach is not entirely satisfactory for fMRI group data inference, since it aims at detecting individual voxels rather than regions, which ultimately constitute the objects of interest in terms of neuroscience. Furthermore, using it to define a cluster-forming threshold would be problematic, as the resulting procedure would not verify the subset pivotality condition which guarantees strong control on the family wise error rate. Finally, it requires a spatial independence assumption, which may be difficult to verify in a model which accounts for the imperfect match across subjects. For these reasons, this technique may be best adapted to threshold individual activation maps, or to define regions of interest.

The second contribution of this thesis consists in relaxing the assumption of perfect match between the estimated effect maps of the different subjects. To this end, we generalized the classical mass-univariate (voxelwise) model for fMRI group data analysis by incorporating a set of hidden variables, representing the unknown registration errors, modeled as random deformation fields. In contrast, all previous approaches dealing with spatial uncertainty in fMRI data were feature-based, meaning that they aimed to match high-level features extracted from the individual activation maps. 
% Our approach is thus conceptually simpler, and avoids the problematic choice of interest features. 
We proposed to estimate the map of group mean effects in a Bayesian setting by its posterior mean, showed the consistency of the joint posterior density of all model parameters, and designed a Metropolis-within Gibbs (MH-Gibbs) algorithm \cite{Tierney94} to draw samples from it, and compute Monte-Carlo estimates of the mean effect map.

Our simulation studies evidenced a stretching effect of the estimated activation pattern when the registration errors where unaccounted for, which caused neighboring activations to be merged. We also showed that this stretching effect could be substantially reduced when registration errors were modeled using our approach. When applied to real fMRI data, our method yielded estimates of the group effect map under spatial uncertainty that were both smoother and more contrasted than under no spatial uncertainty, an effect that could not be reproduced by linear isotropic smoothing. These encouraging results were obtained in spite of the slow mixing of our MH-Gibbs algorithm. In particular, the amplitude of the spatial displacements was under-estimated, suggesting some space for improvement.

Finally, we proposed a new paradigm for ROI-based fMRI group data analysis addressing jointly the three above-mentioned limitations of the SPM-like approach. Based on a Bayesian model selection framework, regions involved in the task under study are selected according to the posterior probabilities of a nonzero mean activation, given a pre-defined parcellation of the search volume into functionally homogeneous regions. Thus our approach is threshold-free, while allowing to incorporate prior information, provided that the parcellation is sensible. By controlling a Bayesian risk, our approach balances false positive and false negative risks, with weights than can be tuned depending on the application domain. Importantly, it is based on the same spatial uncertainty model as in Chapter~\ref{chap:modeling}, and thus accounts for the mis-alignment of individual images, due to inevitable registration errors. As a consequence, for each subject, the membership of a voxel to a given region is probabilistic rather than deterministic. This effectively allows to de-weight the contribution of activations to a region's mean level of activity, when they have been accidently projected into it by mis-registration.

This approach requires to evaluate the marginal likelihood of each model, defined as a partition of regions into involved and inactive. These marginal likelihoods are not available in closed form, but can be evaluated numerically. We chose to use Chib's method \cite{Chib95,Chib01}. This required the specification of a MCMC-SAEM algorithm \cite{Kuhn04}, adapted from the MH-Gibbs sampler above, to derive the MAP estimates of the model parameters, and a simulated annealing scheme to obtain the posterior mode of the displacement field density, conditional on these parameters.

Results on both simulated and real fMRI data show that the previously evidenced stretching effect may cause inactive regions to be contaminated by neighboring activations and be detected as active by mistake. This bias toward false positives is reduced when modeling spatial uncertainties. However, the marginal likelihood estimate in the model with spatial uncertainty turned out to be numerically instable, presumably because of the slow mixing observed when sampling the displacement fields. We proposed an approximate procedure, which consisted in fixing the displacement fields to their conditional MAP value, found by the above-mentioned simulated annealing algorithm. This approximation proved effective in stabilizing numerically the output of the algorithm, but did not entirely remove the bias toward false positives observed when neglecting spatial displacements. We compensated this residual bias by including an additional penalty on model fit, resulting in a lower bound on the probability of each region being involved in the task at hand. This final procedure was validated on both simulated and real fMRI datasets.

\section{Perspectives}

Many promising directions can be envisioned for future work, some of which have been mentioned earlier. 
% In the following, we start by discussing possible technical ameliorations concerning the modeling and sampling of spatial displacements. 
To start  with, the Metropolis-Hasting algorithm used to sample the displacement fields has a high rejection rate and mixes very slowly, as discussed in Appendix~\ref{app:MCMC}, and Sections~\ref{sec:conclusion2}, \ref{sec:simu}, and \ref{sec:discussion}. This results in the numerical instability of our model selection procedure, and needs to be dealt with, for instance by adopting alternative proposal densities to the isotropic and stationary Gaussian random distribution we use here. Indeed, the posterior density of the displacements is likely to be constrained along certain preferential directions, especially in highly contrasted regions of the mean effect map $\bs\mu,$ such as the interface between an active parcel and an inactive one. Also, we expect the posterior density to be more peaked in these regions than in less contrasted areas, such as the background. Generalizing the proposal's covariance structure, and allowing it to vary across control points, would therefore seem a reasonable direction toward which extending our work. Integrating out analytically $\br x$ and $\bs\mu$ instead of simulating them could also promote mixing of the Markov chain and constitute another promising line of work.
% These modifications could in fact be applied to the deformation field model itself, whose definition in Section~\ref{sec:deformation} may be over-simplistic, and explain in part the poor behavior of our sampling scheme.

Another issue requiring further investigation concerns the additional penalty fit we found necessary to include in our approximate procedure. This requires the tuning of a certain scale factor~$c,$ which we have done using simulated datasets. Though this was found to work well on a real-life application, it is necessary to study carefully the stability of the optimal value on a wider range of simulated datasets, in order to determine its possible dependence on certain critical variables, such as sample size or inter-subject variance.

In a broader perspective, the framework we propose here for fMRI group inference can be extended to address a wide variety of situations. For instance, we have focused on the characterization of the activation pattern of a single population of subjects. As previously discussed in Section~\ref{sec:between}, our model could be extended to compare different populations, such as a certain category of patients and healthy subjects, or to correlate fMRI activations with certain interest covariates. This means specifying the mean population effect, or, in terms of the regional response model~(\ref{eq:regional}), the mean regional effect, as a linear term in the regressor variables of interest. There is an increasing need for such models, particularly in studies combining neuroimaging and genetics, such as the IMAGEN project \href{http://www.imagen-europe.com/}{http://www.imagen-europe.com/}, which constitute a new and promising trend in neurosciences.

Another exciting prospect, mentioned in Section~\ref{sec:regionalized}, is to extend the model selection framework to infer the functional connectivity of the specified parcels. Following \cite{Bowman08}, this would imply modeling the covariance structure of the regional means. Identification of functional networks in this setting would involve both selecting the regions involved in the task at hand, and the pairs of regions functionally correlated. The ensuing model selection problem would be much more challenging than the one addressed in the present work, the number of possible models involved being increased by the number of potential interactions (in contrast, \cite{Bowman08} simply estimates the covariance matrix of the regional means, without performing any further inference). Furthermore, reducing this complexity would no longer be possible, even by adopting an approximation such as the one introduced here (see Section~\ref{sec:independence}).
% Some recent contributions on graphical model selection, such as \cite{Verzelen08}, would certainly be useful to attain this goal.

In another direction, our approach can be generalized to probabilistic instead of deterministic parcellations. This would especially make sense when using anatomical atlases, some of which are probabilistic to account for the inter-subject anatomical variability, such as the the CSA atlas \cite{Perrot08} (in Chapter~\ref{chap:case}, we used a deterministic version of this atlas, based on the most probable label for each voxel). Interestingly, this means that the labels defining the membership of each voxel would have a joint prior distribution, given by the probabilistic parcellation, and consequently also a posterior distribution. In other terms, a Bayesian estimate of the parcellation itself would be available, informed by the prior parcellation. Thus, our framework could be used to revisit group-level parcellation approaches, such as developed in \cite{Flandin04,Thirion06f}. It is also an alternative answer to the choice of a parcellation, which we discuss in \ref{sec:parcellations}.

Finally, the methodology we have developed is well-adapted to the analysis of surface-based data (see Section~\ref{sec:surface}), for the following reasons. First, because cortical structures are better matched across subjects using surface-based registration than classical volume-based affine registration \cite{Fischl99b}, we may expect better defined anatomical ROIs. The superiority of surface-based registration may not be as obvious when compared to nonlinear volume-based registration methods, such as those compared in \cite{Klein09}. However, to our knowledge, no comparison of surface versus volume based nonlinear registration has been conducted up to now.

On the other hand, we have seen that these approaches require projecting the fMRI data of each subject on its cortical surface, which cannot be done in a straightforward way, making the localization of activations on the cortical surface uncertain. Thus, the spatial uncertainty model we have developed to account for registration errors would still be necessary to account for these projection uncertainties, even though registration errors may be less important. Furthermore, the simulation study in Section~\ref{sec:simu_2D_approx} suggests a better behavior of our posterior mode approximation on 2D datasets than on 3D datasets, making this application to surface data a promising prospect.

%% ----------------------------------------------------------------
% Now begin the Appendices, including them as separate files

\addtocontents{toc}{\vspace{2em}} % Add a gap in the Contents, for aesthetics

\appendix % Cue to tell LaTeX that the following 'chapters' are Appendices

% \input{Notations}

% Permutation test Theory
% \input{Permutation}

\eject
\pagestyle{empty}
$ $ \eject\pagestyle{fancy}
\pagestyle{fancy}
\chapter{Elements of multiple testing theory}\label{app:multiple}
\lhead{\emph{Elements of multiple testing theory}}

\section{Generalities on multiple testing}\label{sec:generalities}
% Maybe better in  an appendix?

Consider the problem of testing simultaneously $m$ distinct null hypotheses. In the context of neuroimaging, there can be one hypothesis per voxel, as defined in Section~\ref{sec:test_group_effect}, or one hypothesis per cluster identified above a certain threshold. For all $k = 1, \ldots, m,$ the test is based on a certain decision statistic $T_k.$ $\mathcal H_k = 0$ indicates that the $k$-th null hypothesis is true, and $\mathcal H_k = 1$ that it is false.

The quantities of interest for multiple testing are then summarized in Table~\ref{tab:error}, following \cite{Benjamini95}. We note $\mathcal M_0 = \{k: \mathcal H_k = 0\}$ and $\mathcal M_1 = \{k: \mathcal H_k = 1\}$ the sets of true null and false null hypotheses, respectively, and their number: $m_0 = |\mathcal M_0|, m_1 = |\mathcal M_1|,$ which are unknown. $\mathcal M = \{1, \ldots, m\} = \mathcal M_0 \bigcup \mathcal M_1$ is the number of tested hypotheses. The only observed variables are the number of rejected null hypotheses $R,$ and the number of accepted null hypotheses, $R - m.$ The goal of any multiple testing procedure is to minimize both the number $V$ of type I errors, or false positives, and the number $T$ of type II errors, or false negatives.

\begin{table}[h]
\begin{center}
\begin{tabular}{l|cc|l}
\multicolumn{1}{c}{\ } & \multicolumn{1}{c}{$\#$ accepted} & \multicolumn{1}{c}{$\#$ rejected} & \multicolumn{1}{c}{\ } \\ \cline{2-3}
$\#$ True null hypotheses & $U$ & $V$ & $m_0$ \\
$\#$ False null hypotheses & $T$ & $S$ & $m_1$ \\ \cline{2-3}
\multicolumn{1}{c}{t\ } & \multicolumn{1}{c}{$W$} & \multicolumn{1}{c}{$R$} & \multicolumn{1}{c}{$m$}
\end{tabular}
\end{center}
\caption{Number of errors in a multiple testing problem. \label{tab:error}}
\end{table}

\subsubsection{Type I error rates}\label{sec:error}

A common strategy for limiting the number of errors is to maximize the statistical power of the tests while controlling a certain type I error rate at a given level $\alpha.$ The error rates most often used, as defined in \cite{Ge03}, are the following:

\begin{itemize}
\item \emph{False positive rate} ($\mathrm{FPR}$). It is the expectation of the proportion of type I errors among all the tests:
\begin{equation}
\mathrm{FPR} = E(V)/m
\end{equation}
\item \emph{Family-wise error rate}($\mathrm{FWER}$). It is the probability of at least one type I error:
\begin{equation}
\mathrm{FWER} = P[V > 0]
\end{equation}
\item \emph{False discovery rate}($\mathrm{FDR}$). It is the expectation of the proportion of type I errors among all rejected null hypotheses. When no null hypotheses have been rejected, this proportion is set to 0, yielding:
\begin{equation}
\mathrm{FDR} = E \left[ \frac{V}{R}1_{R>0} \right]
\end{equation}
\end{itemize}

These quantities can be compared through the following inequality \cite{Ge03}:
$$
\mathrm{FPR} \leq \mathrm{FDR} \leq \mathrm{FWER}.
$$
Thus, if the FWER is controlled at a given level $\alpha,$ {\em i.e.}, if $\mathrm{FWER} \leq \alpha,$ then the other rates are also automatically controlled at the same level. FWER is the most stringent criterion possible for multiple testing. As such, it is very much used in neuroimaging, where a high level of confidence is required to report a brain region as involved in a certain task. Conversely, the FPR involves no correction at all for multiple comparisons, since it is controlled at a given level $\alpha$ as soon as the individual hypoteses $\mathcal H_k = 0$ are tested at level $\alpha.$

\subsubsection{Exact, weak and strong control}\label{sec:control}

The error rates defined in the previous section are implicitely defined conditionally on the intersection of all null hypotheses: $\mathcal H_{\mathcal M_0} = \bigcap_{k \in \mathcal M_0} \{\mathcal H_k = 0\}$, as noted in \cite{Ge03}. Control on a given error rate conditional on the true intersection $\mathcal H_{\mathcal M_0}$ of null hypotheses is called {\em exact control}. For instance, a multiple comparison procedure has exact control on the FWER if it can control the quantity: $P[V > 0 | \mathcal H_{\mathcal M_0}]$ at any given level $\alpha.$

However the set $\mathcal M_0$ is unknown, so the error rates are generally computed, and controlled, under the global null hypothesis $\mathcal H_{\mathcal M} = \bigcap_{k \in \mathcal M} \{\mathcal H_k = 0\},$ referred to as a {\em weak control}. It is important to note that weak control alone does not in general imply exact control. Hence, a multiple testing procedure which has only weak control is in general not a valid procedure, since its results hold only under the assumption that all null hypotheses are true.

Finally, {\em strong control} means control for every possible choice of $\mathcal M_0.$ For instance, strong control of the FWER means to be able to control the quantity $\max_{\mathcal M' \subseteq \mathcal M} P[V > 0 | \mathcal H_{\mathcal M'}]$ at any given level $\alpha.$ Strong control implies both weak and exact control, and is therefore sufficient to define a valid mutiple comparison procedure.

\subsubsection{Subset pivotality and $p$-values}\label{sec:pivotality}

Subset pivotality is a central property in multiple testing. It is used to ensure that a multiple comparison procedure having weak control over a certain error rate also has strong control. It is usually defined as follows \cite{Westfall93}:

\begin{definition}[Subset Pivotality]\label{def:pivotality}
The joint distribution of the test statistics\\ $(T_1, \ldots, T_m)$ is said to have the {\em subset pivotality} condition if for all subset $\mathcal K \subset \mathcal M,$ the joint distribution of $(T_k)_{k \in \mathcal K}$ is the same under the global null hypothesis $\mathcal H_{\mathcal M} = \bigcap_{k \in \mathcal M} \{\mathcal H_k = 0\}$ as under the restriction $\mathcal H_{\mathcal K} = \bigcap_{k \in \mathcal K} \{\mathcal H_k = 0\}.$
\end{definition}

An immediate consequence of this definition is that, under subset pivotality, the joint distribution of $(T_k)_{k \in \mathcal K}$ is the same under any intersection of null hypotheses including $\mathcal H_{\mathcal K}.$ Informally, subset pivotality implies that the test of each given null hypothesis $\mathcal H_k = 0$ can be done independently of the status of all other hypotheses $\mathcal H_{k'},$ which is a quite natural requirement. For instance, if each null hypothesis $\mathcal H_k = 0$ concerns a data vector $D_k,$ if the $D_k$'s are independent, and $T_k$ is a function of $D_k$ only, then the subset pivotality property is trivially verified.

% Another case where subset pivotality arises is when the joint distribution of the $D_k'$s is multivariate Gaussian, and the null hypothesis $\mathcal H_k = 0$ is expressed in terms of the expectation $E[D_k].$ If the covariance matrix of the $D_k'$s is the same under all configurations of true and false null hypotheses, then any test statistic $T_k = T_k(D_k)$ verifies subset pivotality. Furthermore, in this case smoothing of the data map 

Otherwise, there is no general characterization of subset pivotality. We will see that most multiple testing procedures that have strong control rely on this property.

To illustrate the importance of this notion, consider the $p$-values, which are another useful tool in the multiple testing setting. They are traditonally defined as follows:

\begin{definition}[$p$-values]
Note $(t_1, \ldots, t_m)$ the observed values of the test statistics $(T_1, \ldots, T_m)$. For $k = 1, \ldots, m,$ The {\em $p$-value} $p_k$ associated with $t_k$ is given by:
\begin{equation}\label{eq:p-value}
p_k = P[T_k > t_k | \mathcal H_k = 0].
\end{equation}
Thus, $p_k$ is the lowest level $\alpha$ at which $\mathcal H_k = 0$ is rejected on the basis of $t_k.$
% \begin{equation}\label{eq:p-value_alt}
% p_k = \inf \{\alpha > 0; P[T_k > t_k | \mathcal H_k = 0] \leq \alpha\}
% \end{equation}
\end{definition}

Note that, under subset pivotality, the quantity defined in (\ref{eq:p-value}) is well-defined, since the probability in the right term takes the same value under any intersection of null hypotheses including $\mathcal H_k = 0,$ and in particular under the global null. On the other hand, if subset pivotality is not verified, then the probability of $T_k > t_k$ depends not only on $\mathcal H_k,$ but on the state of potentially all other hypotheses $\mathcal H_{k'},$ hence the quantity in (\ref{eq:p-value}) has more than one possible value, and is consequently ill-defined. This is an important point, though scarcely mentioned in the multiple testing llitterature. It shows that multiple comparison procedures based on $p$-values often rely implicitely on the subset pivotality property.

\begin{proposition}{\bf Strong control of the maxT procedure}

If the joint distribution of the test statistics $(T_k)_{1\leq k\leq d}$ verifies the subset pivotality property (Definition~\ref{def:pivotality}), The maxT procedure, defined in Section~\ref{sec:voxel}, has strong control over the family wise error rate, {\em i.e.}:
\begin{eqnarray}
FWER &\leq& P\left[\max_{k \in \mathcal M} T_k > u | \mathcal H_{\mathcal M}\right].
\end{eqnarray}
\end{proposition}

\paragraph{Proof.}

This follows directly from the definitions:

\begin{eqnarray}
FWER &=& P[V > 0 | H_{\mathcal M_0}] \nonumber\\
     &=& P[\exists k \in \mathcal M_0,\, T_k > u | \mathcal H_{\mathcal M_0}] \nonumber\\
     &=& P\left[\max_{k \in \mathcal M_0} T_k > u | \mathcal H_{\mathcal M_0}\right] \nonumber\\
     &=& P\left[\max_{k \in \mathcal M_0} T_k > u | \mathcal H_{\mathcal M}\right] \nonumber\\
     &\leq& P\left[\max_{k \in \mathcal M} T_k > u | \mathcal H_{\mathcal M}\right], \nonumber\\
\end{eqnarray}

where the the fourth equality holds under subset pivotality $\square$

\section{Strong control of the maxT and maximum cluster size tests}

Based on the definitions in Section~\ref{sec:voxel}, we now show that the test on the maximum suprathreshold cluster also has strong control:

\begin{proposition}{\bf Strong control of the maximum cluster size test}
If the joint distribution of the test statistics $(T_k)_{1\leq k\leq d}$ verifies the subset pivotality property, then the test on the maximum cluster size has strong control over the family-wise error rate:
\begin{eqnarray}
FWER &\leq& P[\max_{C_i \subseteq \mathcal M} \sharp C_i > N | \mathcal H_{\mathcal M}]. \nonumber
\end{eqnarray}
\end{proposition}

\paragraph{Proof.}

The cluster-level family-wise error rate (\ref{eq:cluster_FWER}) can be expressed as:
\begin{eqnarray}
FWER &=& P[\max_{C_i \subseteq \mathcal M_0} \sharp C_i > N | \mathcal H_{\mathcal M_0}]\nonumber\\
     &\leq& P[\max_{C_i \subseteq \mathcal M} \sharp C_i\cap\mathcal M_0 > N | \mathcal H_{\mathcal M_0}].\nonumber\\
\end{eqnarray}
This last event depends only $(T_k)_{k \in \mathcal M_0},$ whereas the first one depends on whether $C_i \subseteq \mathcal M_0$ for each suprathreshold clusters $C_i,$ an event depending on statistics $T_{k'}$ in all neighboring voxels $k'$ of $C_i,$ some of whom may be outside $\mathcal M_0$ (a voxel is said to be neighboring $C_i$ if it is neighbor to a voxel contained in $C_i$).

Hence, we may apply subset pivotality, yielding:
\begin{eqnarray}
FWER &\leq& P[\max_{C_i \subseteq \mathcal M} \sharp C_i\cap\mathcal M_0 > N | \mathcal H_{\mathcal M}].\nonumber\\
     &\leq& P[\max_{C_i \subseteq \mathcal M} \sharp C_i | \mathcal H_{\mathcal M}]\square\nonumber\\
\end{eqnarray}

\cleardoublepage
\chapter{Proof of the consistency of the random threshold procedure}\label{app:proof}
\lhead{\emph{Proof of the consistency of the random threshold procedure}}

Following \cite{Lavielle07}, we first recall some notations. Set $u_i = y_i$ for $i \in I_{k_n^{\star}}$ and $v_i = y_i$ for $i \notin I_{k_n^{\star}};$ notice that $(v_i)$ is a sample from the distribution $F_{\epsilon}.$ Let $(u_{(i)})_{1\leq i\leq k_n^{\star}}$ and $(v_{(i)})_{1\leq i\leq n - k_n^{\star}}$ be the sequences $(|u_i|)$ and $(|v_i|)$ in decreasing order. Let $\Omega_n$ be the subset of $\Omega$ where $v_{(1)} < \alpha_n / 2$ and $u_{(k_n^{\star})} > \alpha_n / 2.$

A first lemma in \cite{Lavielle07} shows that $P(\Omega_n) \to 1,$ \emph{i.e.}, the collections $(u_{(i)})$ and $(v_{(i)})$ are stochastically in order with high probability. The proof can then be restricted to $\Omega_n.$

Now, let $\mathbb{E}_k(T_{k,j})$ and $Q_{k,j}$ be defined as in Equation~(\ref{eq:procedure}). Using Proposition~\ref{prop:cond_mean}, we have:

\begin{eqnarray}
\mathbb{E}_k(T_{k,j}) & = & j ( 1 + \sum_{i = j + 1}^{n - k} 1 / i ); \nonumber\\
Q_{k,j} & = & \frac{\mathbb{E}_k(T_{k,j})}{\mathbb{E}_k(T_{k,n - k})} T_{k,n - k} \nonumber\\
 & = & B_{k,j,n} T_{k,n - k}.\nonumber
\end{eqnarray}

Also, let $a_i = \mathbb{E}_0(X_{(i)}) = \sum_{\ell = i}^n 1/\ell.$ Equation~(\ref{eq:consistency}) can be shown separately for $k > k_n^{\star}$ and $k < k_n^{\star}.$ Since the two cases are treated similarly, we will restrict ourselves here to the case $k > k_n^{\star}.$ On $\Omega_n:$

\begin{eqnarray}
T_{k,j} - Q_{k,j} 
& = & T_{k,j} - B_{k,j,n} T_{k,n - k} \nonumber\\
& = & \left( T_{k,j} - \mathbb{E}_{k_n^{\star}}(T_{k,j}) \right) - B_{k,j,n} \left( T_{k,n - k} - \mathbb{E}_{k_n^{\star}}(T_{k,n - k}) \right)\nonumber\\
&   & + \mathbb{E}_{k_n^{\star}}(T_{k,j}) - B_{k,j,n} \mathbb{E}_{k_n^{\star}}(T_{k,n - k}) \nonumber\\
 & = & R_{k,j} + S_{k,j}.\nonumber
\end{eqnarray}

Thus $T_{k,j} - Q_{k,j}$ is decomposed into a random part $R_{k,j}$ and a deterministic part $S_{k,j}.$ Over $\Omega_n,$ $R_{k,j}$ is a function of $v_{(k)}, \ldots, v_{(n - k_n^{\star})}.$ Before going further, we now recall the following result:

Let $X_{(1)} \geq \ldots \geq X_{(n)}$ be an ordered sequence of independent $Exp(1)$ random variables. For $1 \leq j \leq n,$ let $T_j = \sum_{i = 1}^j X_{(i)}.$ Introduce for $t \in [0,1]$ the random process $d_n(t) = T_{[nt]} - \mathbb{E}(T_{[nt]} | T_n).$ Then it is shown in \cite{Lavielle07} that $\frac{1}{\sqrt{n}}d_n(t),$ as a process indexed on $t \in [0,1],$ converges in distribution to a certain zero mean Gaussian process $\Delta.$

To use this result, let $k = [tn]$ and $j = [sn],$ for $0 < t < 1 - c$ and $0 < s < 1 - t^{\star} - t,$ for $c$ in [{\bf AF3}]. Then $\frac{1}{\sqrt{n-k}}(T_{k,j} - Q_{k,j})\mathbf{1}_{\Omega_n} = \frac{1}{\sqrt{n-k}}(T_{[tn],[sn]} - Q_{[tn],[sn]})\mathbf{1}_{\Omega_n},$  as a process indexed by $(t,s) \in (0,1)^2,$ converges in distribution to the zero-mean Gaussian process:

$$
\Gamma_{t,s} = \sqrt{\frac{1 - t^{\star}}{1 - t}}\left[ \Delta \left( \frac{t + s - t^{\star}}{1 - t^{\star}} \right) - \Delta \left( \frac{t - t^{\star}}{1 - t^{\star}} \right) \right].
$$

similarly, $\frac{1}{\sqrt{n - k}}B_{k,j,n} \mathbb{E}_{k_n^{\star}}(T_{k,n - k})\mathbf{1}_{\Omega_n}$ converges in distribution to another zero-mean Gaussian process, and so does their sum, $\frac{1}{\sqrt{n - k}}R_{k,j}\mathbf{1}_{\Omega_n}.$

On the other hand, 
% as in \cite{Lavielle07},

\begin{eqnarray}\label{eq:cond_mean}
S_{k,j} & = & \sum_{i = 1}^{k - k_n^{\star}} (a_{i+j} - a_i + B_{k,j,n}(a_{i+n-k} - a_i)),
\end{eqnarray}

so that there exists a constant $\gamma > 0,$ which depends on $c$ in [{\bf AF3}], such that for all $n \geq 1,\ k_n^{\star} < k \leq n - K_n,$ we have $\sup_{1 \leq j \leq n - k} |S_{k,j}| \geq \gamma (k - k_n^{\star}).$ Finally we use the following inequality:

\begin{eqnarray}
\mathbb{P}_{k_n^{\star}} (\hat{k}_n - k_n^{\star} > n u_n) & \leq & \mathbb{P}( \eta_{k_n^{\star}} > \inf_{k - k_n^{\star} > n u_n} \eta_k ). \nonumber
\end{eqnarray}

From Equation~(\ref{eq:cond_mean}), $S_{k_n^{\star},j} = 0,$ hence it follows that:
\begin{eqnarray}
\sqrt{n - k_n^{\star}}\, \eta_{k_n^{\star}} 
&=& 
\sup_{1 \leq j \leq n - k_n^{\star}} |R_{k_n^{\star},j} + S_{k_n^{\star},j}| \nonumber\\
&=& 
\sup_{1 \leq j \leq n - k_n^{\star}} |R_{k_n^{\star},j}|  \nonumber\\
&\leq& \sup_{k \geq k_n^{\star}} \sup_{1 \leq j \leq n - k} |R_{k,j}|.\nonumber
\end{eqnarray}
% where the last inequality holds $a.e.$ 
On the other hand,
\begin{eqnarray}
\sqrt{n - k}\inf_{k - k_n^{\star} > n u_n} \eta_k 
& = & \inf_{k - k_n^{\star} > n u_n} \sup_{1 \leq j \leq n - k} |R_{k,j} + S_{k,j}| \nonumber\\
&\geq& \inf_{k - k_n^{\star} > n u_n} \sup_{1 \leq j \leq n - k} |S_{k,j}| - \sup_{k \geq k_n^{\star}} \sup_{1 \leq j \leq n - k} |R_{k,j}|, \nonumber
\end{eqnarray}
so that we have:
\begin{eqnarray}
\mathbb{P}_{k_n^{\star}} (\hat{k}_n - k_n^{\star} > n u_n) & \leq & \mathbb{P}(C \sup_{k \geq k_n^{\star}} \sup_{1 \leq j \leq n - k} |R_{k,j}| \geq \inf_{k - k_n^{\star} > n u_n} \sup_{1 \leq j \leq n - k} |S_{k,j}|) + \mathbb{P}(\Omega_n^c) \nonumber\\
                                                         & \leq & \mathbb{P}(C \sup_{k \geq k_n^{\star}} \sup_{1 \leq j \leq n - k} |R_{k,j}| \geq \gamma n u_n) + \mathbb{P}(\Omega_n^c), \nonumber
\end{eqnarray}
where $C$ is a constant which depends on $c$ in [{\bf AF3}]. This last probability vanishes as $n$ goes to infinity, due to the weak convergence of $R_{k,j}\mathbf{1}_{\Omega_n} \square$

\eject
\pagestyle{empty}
$ $ \eject\pagestyle{fancy}
\pagestyle{fancy}
\chapter{Identifiability in the model with spatial uncertainty}\label{app:identifiability}
\lhead{\emph{Identifiability in the full hierarchical model}}

% Two models are considered throughout this work. The first one is the generalization of the mass univariate model introduced in Chapter~\ref{chap:modeling}, and defined by (\ref{eq:within}), (\ref{eq:between2}), (\ref{eq:deformation}), and (\ref{eq:elementary}). The second one is the regional model introduced in Chapter~\ref{chap:bayesian}, obtained by adding the additional level (\ref{eq:regional}) to the first model.
% 
% There is a possible confusion between both models, as (\ref{eq:regional}) is used to define the prior distribution in Chapter~\ref{chap:modeling}. In fact, they are completely equivalent from a Bayesian viewpoint, since they share the same set of latent variables (parameters included), the same density conditional on these latent variables, and the same prior distribution. From a frequentist viewpoint however, the two models differ by their parameter set and their likelihood function. In particular, the observations are independent in the first model, whereas they are correlated in the second.
% 
% However, as the proof of identifiability is very similar in both models, we detail it only for the second, more complex model, noting that it can easily be adapted to the first.

We show here the identifiability of the parameters in the hierarchical model introduced in Chapter~\ref{chap:bayesian}, $\bs\theta = (\bs\eta, \bs\nu^2, \bs\sigma^2, \sigma_S^2).$ Identificability in the model in Chapter~\ref{chap:modeling} requires a slightly different proof, since the parameter vector is defined by $\bs\theta = (\bs\mu, \sigma^2, \sigma_S^2).$ We have not included it here however, as it uses very similar arguments.

\begin{Theorem}{\bf (Identifiability in the Regionalized Model)}
The full hierarchical model specified by (\ref{eq:within}), (\ref{eq:between2}), (\ref{eq:deformation}), (\ref{eq:elementary}), (\ref{eq:regional}) is identifiable.
\end{Theorem}

\paragraph{Proof.}

This result may be re-phrased by saying that, if $\bs \theta \neq \bs \theta',$ then $f(\br y | \bs \theta) \neq f(\br y | \bs \theta')$ in at least one point $\br y.$  As noted previously in Appendix~\ref{app:likelihood}, the conditional density $f(\br y | \br w, \bs \theta)$ is multivariate Gaussian. It depends on $\br w$ only through the definition of the voxelwise blocks
$$
I_k = \left\{(i,l)_{\begin{subarray}{l} 1\leq i \leq n\\ 1\leq l \leq d \end{subarray}} \Big\vert \varphi_i(l) = k \right\},
$$
containing for each~$k$ the indices of observations $y_{i,l}$ displaced to voxel~$k.$ The collection of all voxelwise blocks $\br I = (I_1, \ldots, I_d)$ constitutes a partition of the cartesian product $\{1, \ldots, n\}\times\{1, \ldots, d\}$ into $d$ subsets. Noting $\mathcal I$ the set of all possible such partitions, we see that the PDF can be re-written as the mixture:
$$
f(\br y | \bs \theta)
=
\sum_{\br I \in \mathcal I} 
\pi(\br I | \sigma_S^2)
f(\br y | \br I, \bs \eta, \bs \nu^2, \bs \sigma^2),
$$
where $\pi(\br I | \sigma_S^2) = \int_{\br w \in \br I} \pi(\br w | \sigma_S^2) d\br w$ is the probability of block configuration $\br I,$ and $f(\br y | \br I, \bs \eta, \bs \nu^2, \bs \sigma^2)$ is multivariate Gaussian, with mean and covariance matrix given by functions of $(\bs \eta, \br I)$ and $(\bs \sigma^2, \br I),$ respectively. Note that these functions depend on $\br w$ (through $\br I$), and also that they may not be one-to-one. For instance, the observations may well be all displaced into a single region $j_0,$ in which case the regional parameters $(\eta_j, \nu_j^2, \sigma_j^2)$ of all other regions have no effect on the data.

Because any finite family of distinct Gaussian PDFs is linearly independent, and the mixing weights $\pi(\br I | \sigma_S^2)$ in the above display are always strictly positive, the only way we can have $f(\br y | \bs \theta) = f(\br y | \bs \theta')$ in all data points $\br y$ is if there exists a permutation $\delta$ of $\mathcal I,$ such that for all $\br I \in \mathcal I$ and all $\br y \in \mathbb R^{nd}:$
\begin{eqnarray}
\pi(\br I | \sigma_S^2)
&=& 
\pi(\delta(\br I) | \sigma_S'^2)\nonumber\\
f(\br y | \br I, \bs \eta, \bs \nu^2, \bs \sigma^2)
&=&
f(\br y | \delta(\br I), \bs \eta', \bs \nu'^2, \bs \sigma'^2)\label{eq:cond2}.
\end{eqnarray}
This the analog of the `label-switching' phenomenon in classical finite mixture models. We now show that such label-switching is impossible, because the partitions $\br I$ are not equally probable. Note $\br I_0$ the most probable block configuration,
$$
\br I_0 
= \arg\max_{\br I \in \mathcal I} \pi(\br I | \sigma_S^2).
$$
Because $\br w$ is a zero-mean Gaussian with spherical covariance (\ref{eq:elementary}), this most probable configuration is obtained for $\br w = \br 0$, so that $\br I_0 = (I_k^0)_{1\leq k\leq d},$ where for all k:
$
I_k^0 = \{(1,k), \ldots, (n,k)\}.
$
Likewise, $\delta(\br I_0) = \br I_0,$ hence from \ref{eq:cond2} it follows that for all $\br y:$
$$
f(\br y | \br I_0, \bs \eta, \bs \nu^2, \bs \sigma^2)
=
f(\br y | \br I_0, \bs \eta', \bs \nu'^2, \bs \sigma'^2),
$$
which implies that $(\bs \eta, \bs \nu^2, \bs \sigma^2) = (\bs \eta',\bs  \nu'^2, \bs \sigma'^2).$

Finally, $\br I_0$ is a star domain in that if $\br w \in \br I_0,$ then the segment $\{\lambda \br w, \lambda \in (0,1)\}$ lies inside $\br I_0.$ Thus $\pi(\br I_0 | \sigma_S^2)$ is a strictly decreasing function of $\sigma_S^2,$ and since $\pi(\br I_0 | \sigma_S^2) = \pi(\br I_0 | \sigma_S'^2),$ we have $\sigma_S^2 = \sigma_S'^2\ \square$

\chapter{Sampling the posterior density using a Metropolis within Gibbs algorithm}\label{app:MCMC}
\lhead{\emph{Sampling the posterior density using a Metropolis within Gibbs algorithm}}

As described at the beginning of Section~\ref{sec:marginal}, the approach in \cite{Chib95} assumes that a MCMC algorithm has been devised to sample the posterior density of the model parameters
% , and also that a maximizer of this density can be found
. As expressed in (\ref{eq:marginal}), this density is obtained as a marginal of the joint posterior density:
\begin{eqnarray}\label{eq:joint_posterior}
\pi(\br z, \bs \theta | \br y, \bs\gamma)
&=& 
\pi(\br x, \br w, \bs \mu, \bs \eta, \bs \nu, \bs \sigma^2, \sigma_S^2 | \br y, \bs\gamma) \\
&\propto&
f(\br y | \br x) 
\pi(\br x | \bs \mu, \bs \sigma^2, \br w)
\pi(\bs \mu | \bs \eta, \bs \nu^2)
\pi(\br w | \sigma_S^2)
\pi(\bs \eta | \bs \nu^2)
\pi(\bs \nu^2, \bs\sigma^2, \sigma_S^2)\nonumber.
\end{eqnarray}
We use the Gibbs sampler algorithm \cite{Geman84} to generate a sequence of samples from this joint density by sampling successively each of the six following blocks: $\br x,$ $ \br w,$ $ \bs \mu,$ $ (\bs \eta, \bs \nu),$ $ \bs \sigma^2,$ and $\sigma_S^2,$ conditionally on all others. The conditional density of each block is obtained from the complete density (\ref{eq:joint_posterior}), by treating all other blocks as constants. These conditional densities are available in closed form, except for the elementary displacements $\br w,$ and are given as follows.

\paragraph{Hidden effects.}
For each subject~$i$ and each voxel~$k,$ such that the displaced voxel~$\varphi_i(k)$ is in region~$j,$ it follows from (\ref{eq:within2}) and (\ref{eq:between2}) that:
\begin{eqnarray}
x_{i,k} | \ldots
&\sim& 
\mathcal N\left(
\frac{\sigma_j^2\, y_{i,k} + s_{i,k}^2\, \mu_{\varphi_i(k)}}{\sigma_j^2 + s_{i,k}^2},\, 
\frac{\sigma_j^2\, s_{i,k}^2}{\sigma_j^2 + s_{i,k}^2}\right).
\end{eqnarray}

\paragraph{Population mean effect.} According to (\ref{eq:between2} and \ref{eq:regional}), for all voxel~$k$ in region~$j,$
\begin{eqnarray}
\mu_k | \ldots
&\sim& 
\mathcal N\left(
\frac{\nu_j^2\, m_k + s_k^2\, \eta_j}{\nu_j^2 + s_k^2},\,
\frac{\nu_j^2\, S_k^2}{\nu_j^2 + S_k^2}\right),
\end{eqnarray}
where $m_k = n_k^{-1} \sum_{\phi_i(l)=k} x_{i,k}$ and $S_k^2 = n_k^{-1} \sum_{\phi_i(l)=k} (x_{i,k} - m_k)^2$ are the empirical mean and variance of the effects centered on $\mu_k,$ and $n_k = \sharp\{(i,k); \phi_i(l)=k\}$ the number of these.

\paragraph{Regional parameters.} For each region~$j,$ (\ref{eq:regional}), (\ref{eq:prior_mean}) and (\ref{eq:prior_mean}) yield:
\begin{eqnarray}
\eta_j | \nu_j^2, \gamma_j = 0, \ldots
&\sim& 
\delta(0) ;\\
\eta_j | \nu_j^2, \gamma_j = 1, \ldots
&\sim& 
\mathcal N\left(
\frac{\sum_{\ell_k = j} \mu_k}{\lambda + d_j},\,
\frac{\nu_j^2}{\lambda + d_j}
\right) ;\\
\nu_j^2 | \ldots
&\sim& 
\mathcal {IG} \left(
\alpha + d_j,\,
\beta 
+ \frac{1}{2} \sum_{\ell_k = j} \mu_k^2 - \gamma_j \frac {\left( \sum_{\ell_k = j} \mu_k\right)^2}{2(\lambda + d_j)}
\right),
\end{eqnarray}
where $d_j = \sharp\{k; \ell_k = j\}$ is the size of region~$j.$

\paragraph{Population variance.} For each region~$j,$ the posterior density of $\sigma_j^2$ is computed from (\ref{eq:between2}) and (\ref{eq:prior_var}), resulting in:
\begin{eqnarray}
\sigma_j^2 | \ldots
&\sim& 
\mathcal {IG} 
\left( 
\alpha + \frac{1}{2} \sum_{\ell_k = j} n_k,\,
\beta + \frac{1}{2} \sum_{\ell_k = j} \sum_{\varphi_i(l) = k} (x_{i,l} - \mu_k)^2
\right).
\end{eqnarray}

\paragraph{Spatial variance.} The posterior density of the spatial variance $\sigma_S^2$ is given by (\ref{eq:deformation}) and (\ref{eq:prior_var}):
\begin{eqnarray}
\sigma_S^2 | \ldots
&\sim& 
\mathcal {IG} 
\left( 
\alpha + \frac {3nB} {2},\, \beta + \frac {\sum_{i,b} \parallel \br w_{i,b} \parallel^2} {2}
\right).
\end{eqnarray}

\paragraph{Elementary displacements.} For each subject~$i$ and each control point~$b,$ according to (\ref{eq:between2}) and (\ref{eq:deformation}):
\begin{eqnarray} 
\br w_{i,b} | \ldots
&\propto&
\pi(\br w_{i,b} | \sigma_S^2)
\pi(\br x_i | \br \mu, \br \sigma^2, \br w_i),
\end{eqnarray}
where the prior conditional densities of $\br w_{i,b}$ and $\br x_i$ are defined in (\ref{eq:elementary}) and (\ref{eq:between2}). This density cannot be computed analytically, let alone directly sampled. Indeed, $\pi(\br x_i | \br \mu, \br \sigma^2, \br w_i),$ as a function of $\br w_i,$ is piecewise constant, due to the discrete interpolation of the mean population map (see Section~\ref{sec:observation}). Instead, we sample each elementary displacement $\br w_{i,b}$ using the Metropolis-Hastings algorithm \cite{Hastings70}, wherein a candidate value $\br w'_{i,b}$ is sampled from any proposal density $q(\br w'_{i,b})$ (which may depend on the other variables), and accepted with probability:
\begin{eqnarray}\label{eq:acceptance}
\alpha 
&=& 
\min \left\{ 
1, 
\frac {\pi(\br w'_{i,b} | \sigma_S^2) \pi(\br x_i | \br \mu, \br \sigma^2, \br w'_i)}
{\pi(\br w_{i,b} | \sigma_S^2) \pi(\br x_i | \br \mu, \br \sigma^2, \br w_i)} 
\frac {q(\br w_{i,b})} {q(\br w'_{i,b})} \right\}.
\end{eqnarray}
In the above expression, $\br w'_i$ stands for the vector $\br w_i,$ where $\br w_{i,b}$ has been replaced by the candidate value $\br w'_{i,b}.$ This means that our Gibbs sampler is actually a {\em Metropolis within Gibbs} algorithm (see for instance \cite{Tierney94}). We used a random walk proposal $q(\br w'_{i,b}) = \mathcal N(\br w'_{i,b}; \br w_{i,b}, \sigma_{RW}^2 \br I_3),$ with a proposal variance $\sigma_{RW}^2$ tuned during the burn-in period to obtain an acceptance rate close to $0.1.$ A higher acceptance rate of $0.25,$ as advocated in \cite{Roberts97}, proved a bad idea in our case, as it resulted in such a low proposal variance $\sigma_{RW}^2$ that the Markov Chain remained almost unmoving.

% This algorithm gave satisfying results in practice. See for instance the simulation study in Section~\ref{sec:simulations}, where we used it to obtain a posterior sample from the population mean's posterior density $\pi(\bs \mu | \br y),$ which succesfully accounted for the data being spatially jittered (for this precise example, a single functional region was defined, so that the regional parameters $\eta, \nu^2, \sigma^2,$ where uniform across the search volume).

\eject
\pagestyle{empty}
$ $ \eject\pagestyle{fancy}
\cleardoublepage
\pagestyle{fancy}
\chapter{Maximization of the posterior density using a MCMC-SAEM algorithm}\label{app:SAEM}
\lhead{\emph{Maximization of the posterior density by a MCMC-SAEM algorithm}}

We start by recalling basic facts concerning the MCMC-SAEM coupling developed in \cite{Kuhn04}, as described hereafter. This extension of the SAEM algorithm allows to find the maximum likelihood, or the maximum a posteriori, in mixture models where the conditional distribution of the hidden variables can only be sampled by an MCMC algorithm, as is our case. The main prerequisite in \cite{Kuhn04} is that the complete likelihood belongs to the curved exponential family. In our case, this means that:
\begin{eqnarray}\label{eq:curved_expon}
f(\br y, \br z | \bs \theta)
&=&
f(\br y| \br z) \pi(\br z| \bs \theta) \nonumber\\
&=& 
\exp - \left\{ \psi(\bs \theta) + \Big\langle  S(\br y, \br z) ; \phi(\bs\theta) \Big\rangle \right\},
\end{eqnarray}
where $\langle  \cdot ; \cdot \rangle$ denotes the scalar product. Since we consider MAP estimation here, the above likelihood must be mutiplied by the prior density. Due to its conditionally conjugate form, its expression is very close to that of the likelihood:
\begin{eqnarray}\label{eq:prior_expon}
\pi(\bs\theta)
&=&
\exp - \left\{\psi_{\pi}(\bs \theta) + \Big\langle  s_{\pi} ; \phi(\bs\theta) \Big\rangle \right\}.
\end{eqnarray}
Second, it is assumed that the hidden variables $\br z$ can be simulated using a transition kernel $\pi_{\bs\theta}(\cdot | \cdot)$ whose stationary distribution is the conditional density $\pi(\br z | \br y, \bs \theta),$ for any value of $\bs\theta.$ Then the $k$-th iteration of the MCMC-SAEM algorithm contains the following steps:

\begin{itemize}

\item {\em Simulation.} Draw a new value $\br z_k$ from the current one $\br z_{k-1},$ according to: $\br z_k \sim \pi_{\bs\theta_{k-1}}( \br z_k | \br z_{k-1});$

\item {\em Stochastic Averaging.} Update the expected value $s^{k-1}$ of the sufficient statistics by computing the weighted sum: $s^{k} = s^{k-1} + c_k \left(S(\br y, \br z_k) - s^{k-1} \right);$

\item {\em Maximization.} Update the parameter value $\bs\theta_{k-1}$ by computing: \\
$\bs\theta_k 
= 
\arg\min_{\bs\theta} 
\left\{
\psi(\bs \theta) + \psi_{\pi}(\bs \theta) + \langle  s^k + s_{\pi} ; \phi(\bs\theta) \rangle
\right\}.$

\end{itemize}
The sequence of positive coefficients $(c_k)_{k \geq 0}$ must verify: $\sum_k c_k = \infty$ but $\sum_k c_k^2 < \infty.$ Then, under certain regularity conditions concerning the functions $\phi(\cdot), S(\cdot, \cdot),$ and $\psi(\cdot),$ the sequence $(\bs\theta_k)_{k\geq 0}$ converges to a local maximum of the posterior density. The rationale underlying the SAEM algoritm is that the stochastic averaging step provides a Monte-Carlo estimate of the expected complete likelihood $\mathbb E[f(\br y, \br z, \bs \theta) | \br y, \bs\theta_{k-1}]$ when it cannot be computed analytically, as in the E-step of a standard EM algorithm.

In our case, the likelihood and prior can be factorized across regions, except for the part depending on the spatial parameter $\sigma_S^2,$ so the log-product of the right members in (\ref{eq:curved_expon}) and (\ref{eq:prior_expon}) can conveniently be written as
\begin{equation}
\begin{array}{r}
-\log f(\br y, \br z, \bs \theta)
=
C(\br y, \br z) +
\psi_S(\sigma_S^2) + S_S(\br w) \phi_S(\sigma_S^2) 
\\ + 
\sum_{j=1}^N
\Big[
\psi_j(\bs\theta_j) +
\Big\langle  
S_j(\br z) ; \phi_j(\bs\theta_j) 
\Big\rangle 
\Big],
\end{array}
\end{equation}
where the term $C(\br y, \br z)$ is independent from $\bs\theta$ and therefore irrelevant for the maximization,
\begin{equation}
\begin{array}{ccc}
\psi_S(\sigma_S^2) = \left( \alpha + 1 + \frac{3nB}{2} \right) \log \sigma_S^2;\ 
&
S_S(\br w) = \beta + \frac{\sum_{i,b} \parallel \br w_{i,b} \parallel^2}{2};\ 
&
\phi_S(\sigma_S^2) = \sigma_S^{-2};
\end{array}
\end{equation}
and for all $j = 1, \ldots, N:$
\begin{eqnarray}
\psi_j(\bs\theta_j)
&=&
% \left(\alpha + 1 + \frac{\sum_{\ell_k = j} n_k}{2}\right) \log \sigma_j^2
(\alpha + 1) \log \sigma_j^2
+ \left( \alpha + 1 + \frac{d_j + \gamma_j}{2} \right) \log \nu_j^2
+ \gamma_j \frac{(d_j + \lambda_j) \eta_j^2}{2 \nu_j^2};\\
S_j(\br z)
&=&
\frac{1}{2}
\bigg(
\sum_{\ell_k = j} n_k;
\sum_{\ell_k = j} \sum_{\varphi_i(l) = k} (x_{i,l} - \mu_k)^2;\ 
2\beta +  \sum_{\ell_k = j} \mu_k^2;\ 
2\gamma_j\sum_{\ell_k = j}\mu_k
\bigg)\\
\phi_j(\bs\theta_j)
&=&
\bigg(
\log \sigma_j^2;\ 
\sigma_j^{-2};\ 
\nu_j^2;\ 
-\gamma_j\frac{\eta_j}{\nu_j^2}
\bigg).
\end{eqnarray}
Thus, the MCMC-SAEM algorithm starts with initial values $\br z_0, \bs \theta_0, s_S^{0}, (s_j^{0})_{1\leq j\leq N}$ and iterates the following steps for $k=1,\ldots, K:$

\begin{itemize}

\item {\em Simulation.} Update the hidden variables $\br z_k = (\br x_k, \br w_k, \bs \mu_k)$ by simulating them conditionnally on the current values $(\br z_{k-1}, \bs \theta_{k-1}),$ as described in Appendix~\ref{app:MCMC}.

\item {\em Stochastic Averaging.} Update the expected values $s_S^{k-1}, (s_j^{k-1})_{1\leq j\leq N}$ of the sufficient statistics by computing $s_S^{k} = s_S^{k-1} + c_k \left(S_S(\br w_k) - s_S^{k-1} \right),$ and for all $j:$ $s_j^{k} = s_j^{k-1} + c_k \left(S_j(\br z_k) - s_j^{k-1} \right);$

\item {\em Maximization.} Update the parameter value $\bs\theta_{k-1}$ according to:
\begin{eqnarray}
\sigma_{S,k}^2
&=&
\frac {s_S^{k}} {\alpha + 1 + \frac{3nB}{2}}
\\
\sigma_{j,k}^2
&=&
\frac {s_{j,2}^{k}} {\alpha + 1 + s_{j,1}^{k}}
\\
\nu_{j,k}^2
&=&
\frac {s_{j,3}^{k} - \frac{1}{2}\big( s_{j,4}^{k} \big)^2/(d_j + \lambda_j)} {\alpha + 1 + (d_j + \gamma_j)/2}
\\
\eta_{j,k}
&=&
\frac{s_{j,4}^{k}}{d_j + \lambda_j}
.
\end{eqnarray}

\end{itemize}

To avoid getting trapped in a local maximum far from the global maximum, the stochastic averaging coefficients are set to $c_k = 1$ for iterations $k = 1, \ldots, K_0,$ corresponding to a `burn-in' period, during which the algorithm explores the parameter space. Subsequently, convergence to a local maximum is obtained by choosing $c_{K_0 + k} = \frac {1} {k}$ for $k = 1, \ldots, K-K_0.$

\eject
\pagestyle{empty}
$ $ \eject\pagestyle{fancy}
\pagestyle{fancy}
\chapter{Likelihood expression conditional on the displacements}\label{app:likelihood}
\lhead{\emph{Likelihood conditional on the displacements}}

Having estimated $\hat{\bs \theta}$ using the SAEM algorithm above, we now address the computation of the likelihood for region~$j,$ conditional on the elementary displacements $\br w$ and on the indicator variable. This will be used to compute the marginal likelihood approximation given by (\ref{eq:BMI3}). The choice of a particular value $\br w^{\star}$ is addressed in the forthcoming Appendix~\ref{app:SA}, as it uses the expression for the conditional likelihood derived here.

From (\ref{eq:cond_regional_obs}), it follows that the observations can be separated in conditionally independent blocks, depending on the voxel~$k$ they are displaced to. Thus, noting $\tilde{\br y}_k = (y_{il}, \phi_i(l) = k)$ the vector of observations displaced to voxel~$k,$ sorted {\em e.g.} in lexicographic order, we have:
\begin{eqnarray}
f( \br y_{\mathcal V_j} | \br w, \hat{\bs \theta}_j)
&=&
\prod_{\ell_k = j}
f(\tilde{\br y}_k | \br w, \hat{\bs \theta}_j),\nonumber
\end{eqnarray}
where 
\begin{eqnarray}
\tilde{\br y}_k | \br w, \hat{\bs \theta}_j
&\sim&
\mathcal N( \eta_j \br 1_{n_k}, \bs \Sigma_k),\nonumber
\end{eqnarray}
and the covariance matrix $\bs \Sigma_k$ is equal to:
\begin{eqnarray}
\bs \Sigma_k
&=&
\displaystyle 
\nu_j^2 \br 1_{nk}' \br 1_{nk} + \sigma_j^2 \br I_{n_k} + {\rm diag} (s_{il}^2)_{i,l;\, \phi_i(l) = k}.\nonumber
\end{eqnarray}

Consequently, the log-conditional likelihood boils down to the sum for all voxels in region~$j$ of the explicit quantities:
\begin{equation}\label{eq:log_conditional}
\begin{array}{l}
\displaystyle \log f(\tilde{\br y}_k | \br w, \hat{\bs \theta}_j) \\
\qquad \displaystyle =
-\frac{nk}{2} \log (2\pi) 
-\frac{1}{2} \log |\bs \Sigma_k|
-\frac{1}{2}
(\tilde{\br y}_k - \eta_j \br 1_{n_k})'
\bs \Sigma_k^{-1}
(\tilde{\br y}_k - \eta_j \br 1_{n_k}).
\end{array}
\end{equation}
The determinant and inverse of $\bs \Sigma_k$ are easily computed thanks to the matrix determinant lemma, and the Shermann-Morrison lemma (see \cite{Golub96} for instance):
\begin{eqnarray}
| \br A + \br u \br u' |\label{eq:mtx_det_lemma}
&=&
(1 + \br u' \br A^{-1} \br u)|\br A|;\\
(\br A + \br u \br u')^{-1}\label{eq:Shermann}
&=&
\br A^{-1} - \frac{\br A^{-1} \br u \br u' \br A^{-1}}{1 + \br u' \br A^{-1} \br u},
\end{eqnarray}
valid for any invertible $n \times n$ matrix $\br A$ and any $n \times 1$ vector $\br u,$ such that \\$\br u' \br A^{-1} \br u > -1.$ 

We apply these results to $\br A = \sigma_j^2 \br I_{n_k} + {\rm diag} (s_{i,l}^2)_{i,l;\, \phi_i(l) = k}$ and $\br u = \nu_j \br 1_{nk}.$ Using (\ref{eq:mtx_det_lemma}), we obtain:
\begin{eqnarray}\label{eq:determinant}
|\bs \Sigma_k|
&=&
\bigg(
1 + \nu_j^2 \sum_{i,l;\, \phi_i(l) = k} \frac{1}{\sigma_j^2 + s_{il}^2}
\bigg)
\prod_{i,l;\, \phi_i(l) = k} (\sigma_j^2 + s_{il}^2).
\end{eqnarray}

Next, defining $\br X = \tilde{\br y}_k - \eta_j \br 1_{n_k},$ we apply (\ref{eq:Shermann}), yielding:
\begin{eqnarray}
\br X' (\br A + \br u \br u')^{-1}\br X
&=&
\br X' 
\left( 
\br A^{-1} - \frac{\br A^{-1} \br u \br u' \br A^{-1}}{1 + \br u' \br A^{-1} \br u} 
\right) 
\br X \nonumber\\
 &=&
\br X' \br A^{-1} \br X - \frac{(\br X' \br A^{-1} \br u)^2}{1 + \br u' \br A^{-1} \br u},\nonumber
\end{eqnarray}
so that
\begin{equation}\label{eq:quadratic}
\begin{array}{c}
(\tilde{\br y}_k - \eta_j \br 1_{n_k})'
\bs \Sigma_k^{-1}
(\tilde{\br y}_k - \eta_j \br 1_{n_k})
=
\displaystyle\sum_{i,l;\, \phi_i(l) = k} 
\frac {(y_{il} - \eta_j)^2} {\sigma_j^2 + s_{il}^2}
- \frac
{\displaystyle\bigg(
\sum_{i,l;\, \phi_i(l) = k}
\frac {y_{il} - \eta_j} {\sigma_j^2 + s_{il}^2} 
\bigg)^2}
{\displaystyle\nu_j^{-2} + \sum_{i,l;\, \phi_i(l) = k} \frac{1}{\sigma_j^2 + s_{il}^2}}.
\end{array}
\end{equation}
Hence the exact value of $\log f(\tilde{\br y}_k | \br w, \hat{\bs \theta}_j)$ can be computed following (\ref{eq:log_conditional}), (\ref{eq:determinant}) and (\ref{eq:quadratic}).

\eject
\pagestyle{empty}
$ $ \eject\pagestyle{fancy}
\cleardoublepage
\pagestyle{fancy}
\chapter{Most probable displacement field {\em a posteriori} by simulated annealing}\label{app:SA}
\lhead{\emph{Most probable displacement field {\em a posteriori} by simulated annealing}}

As explained in Section~\ref{sec:marginal}, the marginal likelihood is approximated by conditioning on a certain value of the elementary displacements $\br w,$ defined as the most probable value given the data and the estimated parameter $\hat\theta:$
\begin{eqnarray}\label{eq:MAP_displacement}
\hat{\br w} &=& \arg\max_{\br w} \pi( \br w | \br y, \hat{\bs \theta}).
\end{eqnarray}
The conditional posterior density is proportional to:
\begin{eqnarray}\label{eq:objective}
\pi( \br w | \br y, \hat{\bs \theta})
&\propto&
f(\br y | \br w, \hat{\bs \theta}) \pi(\br w | \sigma_ S^2),
\end{eqnarray}
given the {\em a priori} independence of $\br w$ from all other variables conditional on $\sigma_ S^2.$ Though the right member of (\ref{eq:objective}) can be computed explicitely, using the formulas derived in Section~\ref{app:likelihood}, its maximization with respect to $\br w$ is difficult, because of the high number of dimensions involved, and of the complexity of the objective function. In particular, it is neither differentiable nor convex, thus prohibiting the use of standard gradient methods. Instead, we use the Simulated Annealing (SA) algorithm \cite{Kirkpatrick83}, as described below, which is well-adapted to this setting.

Given the objective function $\pi( \br w | \br y, \hat{\bs \theta})$ we wish to maximize, we define for all $\alpha > 0$ the modified density $\pi_{\alpha} (\br w) \propto \pi^{\alpha}( \br w | \br y, \hat{\bs \theta}).$ These densities share the same modes, but differ in the contrast of their landscape, which increases with $\alpha.$ At the limit $\alpha \to \infty,$ $\pi_{\alpha}$ converges weakly to a combination of Dirac mass in its modes.

The SA algorithm works by simulating successive values $(\br w^t)_{t \geq 1}$ from transition kernels $K_{\alpha_t}(\cdot, \cdot)$ with stationary distributions $\pi_{\alpha_t},$ where $\alpha_t$ increases progressively, so that the resulting Markov chain at first explores many possible states, and then gets attracted with more and more strength toward a mode of $\pi_{\alpha_t}.$ This heuristic can be straightened by showing that for any {\em cooling schedule} $(\alpha_t)_{t\geq 1}$ such that $\alpha_t \to \infty,$ then $\br w^t$ converges almost  surely to the global maximum $\pi( \br w | \br y, \hat{\bs \theta}),$ as $t \to \infty$ \cite{Granville94}.

As in Appendix~\ref{app:MCMC}, each elementary displacement $\br w_{i,b}$ is updated in turn under the target density $\pi_{\alpha},$ using a M-H step, by simulating a candidate $\br w_{i,b}'$ from the random-walk proposal
\begin{eqnarray}
q_{\alpha}(\br w_{i,b}' | \br w_{i,b})
&=&
\mathcal N \left( \br w_{i,b}'; \br w_{i,b}, \alpha^{-1}\sigma_{RW}^2 \br I_3 \right).
\end{eqnarray}
The proposal is then accepted with probability
\begin{eqnarray}
\tilde\alpha 
&=&
\min
\left\{
1, \frac
{\pi_{\alpha}(\br w')}
{\pi_{\alpha}(\br w)}
\right\} \nonumber\\
&=&
\left\{
1, \frac
{f^{\alpha}(\br y | \br w', \hat{\bs \theta})}
{f^{\alpha}(\br y | \br w, \hat{\bs \theta})}
\frac
{\mathcal N (\br w_{i,b}'; \bs 0, \sigma_ S^2/\alpha \br I_3)}
{\mathcal N (\br w_{i,b}; \bs 0, \sigma_ S^2/\alpha \br I_3)}
\right\} \nonumber,
\end{eqnarray}
where $\br w'$ is obtained from the current value $\br w,$ replacing $\br w_{i,b}$ by $\br w_{i,b}',$ and $f^{\alpha}(\br y | \br w', \hat{\bs \theta})$ is computed from (\ref{eq:log_conditional}), (\ref{eq:determinant}) and (\ref{eq:quadratic}) in Section~\ref{app:likelihood}.

We used the cooling schedule: $\alpha_t = \tau^{-t},$ where $\tau \in (0,1)$ is the cooling rate. In practice, we found that setting $\tau = 99\%$ and letting the algorithm run for $T=100$ iterations gave satisfying results.

\cleardoublepage
\chapter{Likelihood under spatial uncertainty, by Chib's method}\label{app:Chib}
\lhead{\emph{Likelihood under spatial uncertainty by Chib's method}}

Following the notations introduced in Section~\ref{sec:likelihood_spatial}, we start by showing how to compute the reduced posterior ordinates $\pi(\br w_{ib}^\ast | \br w_{-ib}^\ast, \bs \theta^\ast, \br y),$ for $i=1,\ldots, n$ and $b=1,\ldots,B,$ following \cite{Chib01}.
First, note that the full conditional density of $\br w_{ib}^\ast$ is given by:

$$
\pi(\br w_{ib}^\ast | \br w_{-ib}^\ast, \br w^{+ib}, \bs \theta^\ast, \br y)
\propto
f(\br y | \br w_{ib}^\ast, \br w_{-ib}^\ast, \br w^{+ib}, \bs \theta^\ast)
\pi(\br w_{ib}^\ast, \br w_{-ib}^\ast, \br w^{+ib} | \bs \theta^\ast),
$$
where $\br w^{+ib}$ denotes the collection of blocks outside $\br w_{-ib}$ and distinct from $\br w_{ib},$ that is, of blocks $\br w_{i'b'}$ for $i' \geq i,$ $b'  \geq b,$ and $(i',b')\neq(i,b).$ As described in Appendix~\ref{app:SA}, This conditional density can be sampled by the MH algorithm, with proposal
$$
q(\br w_{ib}, \br w'_{ib} | \br w_{-ib}^\ast, \br w^{+ib}, \bs \theta^\ast, \br y),
$$
and acceptance rate:

$$
\begin{array}{c}
\displaystyle
\alpha(\br w_{ib}, \br w'_{ib} | \br w_{-ib}^\ast, \br w^{+ib}, \bs \theta^\ast, \br y)\\
= 
\displaystyle
\min
\left\{
1,
\frac
{\pi(\br w'_{ib} | \br w_{-ib}^\ast, \br w^{+ib}, \bs \theta^\ast, \br y)}
{\pi(\br w_{ib} | \br w_{-ib}^\ast, \br w^{+ib}, \bs \theta^\ast, \br y)}
\times
\frac
{q(\br w'_{ib}, \br w_{ib} | \br w_{-ib}^\ast, \br w^{+ib}, \bs \theta^\ast, \br y)}
{q(\br w_{ib}, \br w'_{ib} | \br w_{-ib}^\ast, \br w^{+ib}, \bs \theta^\ast, \br y)}
\right\}
\end{array}
$$
(in fact, given that the proposal density $q$ is symmetric, the ratio to the right in the above display simplifies to $1$).

It can be verified by direct calculation that the transition kernel 

$$
\begin{array}{c}
\pi(\br w_{ib}^\ast, \br w_{ib} | \br w_{-ib}^\ast, \br w^{+ib}, \bs \theta^\ast, \br y) \\
= 
q(\br w_{ib}^\ast, \br w_{ib} | \br w_{-ib}^\ast, \br w^{+ib}, \bs \theta^\ast, \br y) 
\alpha(\br w_{ib}^\ast, \br w_{ib} | \br w_{-ib}^\ast, \br w^{+ib}, \bs \theta^\ast, \br y)
\end{array}
$$
verifies the {\em local reversibility condition}:

\begin{equation}\label{eq:local_reversibility}
\begin{array}{c}
\pi(\br w_{ib}^\ast, \br w_{ib} | \br w_{-ib}^\ast, \br w^{+ib}, \bs \theta^\ast, \br y)
\pi(\br w_{ib}^\ast | \br w_{-ib}^\ast, \br w^{+ib}, \bs \theta^\ast, \br y)\\
=
\pi(\br w_{ib}, \br w_{ib}^\ast | \br w_{-ib}^\ast, \br w^{+ib}, \bs \theta^\ast, \br y)
\pi(\br w_{ib} | \br w_{-ib}^\ast, \br w^{+ib}, \bs \theta^\ast, \br y).
\end{array}
\end{equation}

Multiply both sides of this equation by $\pi(\br w^{+ib} | \br w_{-ib}^\ast, \bs \theta^\ast, \br y),$ and integrate over both $\br w^{+ib}$ and $\br w_{ib},$  to obtain:

$$
\begin{array}{c}
\int \pi(\br w_{ib}^\ast, \br w_{ib} | \br w_{-ib}^\ast, \br w^{+ib}, \bs \theta^\ast, \br y)
\pi(\br w_{ib}^\ast, \br w^{+ib} | \br w_{-ib}^\ast, \bs \theta^\ast, \br y)
\br d\br w^{+ib} \br d\br w_{ib} \\
=
\int \pi(\br w_{ib}, \br w_{ib}^\ast | \br w_{-ib}^\ast, \br w^{+ib}, \bs \theta^\ast, \br y)
\pi(\br w_{ib}, \br w^{+ib} | \br w_{-ib}^\ast, \bs \theta^\ast, \br y)
\br d\br w^{+ib} \br d\br w_{ib}.
\end{array}
$$

Next, express $\pi(\br w_{ib}^\ast, \br w^{+ib} | \br w_{-ib}^\ast, \bs \theta^\ast, \br y)$ as:

$
\pi(\br w_{ib}^\ast | \br w_{-ib}^\ast, \bs \theta^\ast, \br y)
\pi(\br w^{+ib} | \br w_{-ib}^\ast, \br w_{ib}^\ast, \bs \theta^\ast, \br y),
$
and apply (\ref{eq:local_reversibility}), yielding:

\begin{equation}\label{eq:posterior_MCMC}
\pi(\br w_{ib}^\ast | \br w_{-ib}^\ast, \bs \theta^\ast, \br y) \\
=
\frac
{\mathbb E_1[
q(\br w_{ib}, \br w_{ib}^\ast | \br w_{-ib}^\ast, \br w^{+ib}, \bs \theta^\ast, \br y) 
\alpha(\br w_{ib}, \br w_{ib}^\ast | \br w_{-ib}^\ast, \br w^{+ib}, \bs \theta^\ast, \br y)]}
{\mathbb E_2[
\alpha(\br w_{ib}^\ast, \br w_{ib} | \br w_{-ib}^\ast, \br w^{+ib}, \bs \theta^\ast, \br y)]},
\end{equation}
where $\mathbb E_1$ is the expectation with respect to the conditional posterior 
$
\pi(\br w_{ib}, \br w^{+ib} | \br w_{-ib}^\ast, \bs \theta^\ast, \br y),
$
and $\mathbb E_2$ with respect to 
$
q(\br w_{ib}^\ast, \br w_{ib} | \br w_{-ib}^\ast, \br w^{+ib}, \bs \theta^\ast, \br y)
\pi(\br w^{+ib} | \br w_{-ib}^\ast, \br w_{ib}^\ast, \bs \theta^\ast, \br y).
$

Both integrals can be estimated from the output of reduced multiple-block MH runs, as follows:

\begin{itemize}

\item [{\bf 1.}] To estimate the denominator, sample the conditional posterior:
$
\pi(\br w_{ib}, \br w^{+ib} | \br w_{-ib}^\ast, \bs \theta^\ast, \br y).
$
Let $(\br w_{ib}^{(g)}, \br w^{+ib, (g)})_{1\leq g\leq G}$ stand for the generated sample.

\item [{\bf 2.}] Next, include $\br w_{-ib}^\ast$ in the conditioning set, and generate a sample $(\br w^{+ib, (j)})_{1\leq j\leq J}$ from the reduced conditional posterior
$
\pi(\br w^{+ib} | \br w_{-ib}^\ast, \br w_{ib}^\ast, \bs \theta^\ast, \br y).
$
For each $j,$ draw $\br w_{ib}^{(j)}$ from the proposal density $q(\br w_{ib}^\ast, \br w_{ib}^{(j)} | \br w_{-ib}^\ast, \br {w^{+ib}}^{(j)}, \bs \theta^\ast, \br y).$ Then, form the sample $(\br w_{ib}^{(j)}, \br w^{+ib, (j)})_{1\leq j\leq J}.$

\item [{\bf 3.}] Estimate the conditional posterior density in (\ref{eq:posterior_MCMC}) by:

$$
\begin{array}{c}
\displaystyle 
\hat \pi(\br w_{ib}^\ast | \br w_{-ib}^\ast, \bs \theta^\ast, \br y)\\
\displaystyle 
=\\
\displaystyle 
\frac
{G^{-1}\sum_{g=1}^G
q(\br w_{ib}^{(g)}, \br w_{ib}^\ast | \br w_{-ib}^\ast, \br w^{+ib, (g)}, \bs \theta^\ast, \br y) 
\alpha(\br w_{ib}^{(g)}, \br w_{ib}^\ast | \br w_{-ib}^\ast, \br w^{+ib, (g)}, \bs \theta^\ast, \br y)}
{J^{-1}\sum_{j=1}^J
\alpha(\br w_{ib}^\ast, \br w_{ib}^{(j)} | \br w_{-ib}^\ast, \br w^{+ib, (j)}, \bs \theta^\ast, \br y)}.
\end{array}
$$

\end{itemize}

Repeat steps {\bf 1 -- 3} for $i=1,\ldots,n$ and $b=1,\ldots,B.$ Finally, estimate the likelihood in the log scale as:
$$
\log \hat f(\br y | \bs \theta^\ast, \bs\gamma)
= \log f(\br y | \br w^\ast, \bs \theta^\ast, \bs\gamma)
+ \log \pi(\br w^\ast | \bs \theta^\ast)
- \sum_{i=1}^n \sum_{b=1}^B
\log \hat \pi(\br w_{ib}^\ast | \br w_{-ib}^\ast, \bs \theta^\ast, \br y).
$$

It can be noted that the values $(\br w^{+ib, (j)})_{1\leq j\leq J}$ generated in step~{\bf 2} are also produced in step~{\bf 1} of the next reduced run. This means that $n\times B$ reduced runs are necessary instead of $2n\times B.$ In practical application, we found that the successive reduced runs of the multiple block MH algorithm required fine tuning; indeed, the variability of the sampled transition kernel values~$\pi(\br w_{ib}^{(g)}, \br w_{ib}^\ast | \br w_{-ib}^\ast, \br w^{+ib, (g)}, \bs \theta^\ast, \br y)$ increased with the number of hidden variables included in the conditioning set~$\bs \theta^\ast.$ A possible explanation to this is that the modes of the reduced conditional distributions were more and more distant from the mode~$\br w^\ast$ of the full conditional, where the kernel was evaluated. We found empirically that using a number of iterations for each run inversely proportional to the number of sampled blocks worked best.

\addtocontents{toc}{\vspace{2em}}  % Add a gap in the Contents, for aesthetics
\backmatter

%% ----------------------------------------------------------------
\label{Bibliography}
\lhead{\emph{Bibliography}}  % Change the left side page header to "Bibliography"
\cleardoublepage

% \bibliography{revuedef,biben,baseTot,lnaopubli}

% \printindex

\end{document}